\documentclass[10pt,twoside]{book}
\usepackage{makeidx}
\usepackage{epic,eepic}
\usepackage{amssymb,amsmath,graphicx,bbm,pstricks}
\usepackage{rotating}\usepackage{epsfig} 
\usepackage{floatflt} \usepackage{supertabular} 
\usepackage{times}
\usepackage[bf,small]{caption}
\usepackage{fancyhdr} 
\input{epsf}
\newcommand{\dket}[1]{\vert #1 \rangle\rangle}
\newcommand{\dbra}[1]{\langle\langle #1 \vert}
\newcommand{\ket}[1]{\vert #1 \rangle}
\newcommand{\bra}[1]{\langle #1 \vert}

\newcommand{\braket}[2]{\langle #1 \vert #2 \rangle}

\newcommand{\mbfrac}{\mbox{$\frac12$}}

\newcommand{\mL}{\mathcal{L}}
\newcommand{\mD}{\mathcal{D}}

\newcommand{\real}[1]{\Re\mbox{e}[#1]}
\newcommand{\immag}[1]{\Im\mbox{m}[#1]}

\newcommand{\pexp}[1]{\exp\left\{ #1 \right\}}

\newcommand{\bmsigma}{\boldsymbol\sigma}

\newcommand{\tinyinf}{\mbox{\tiny $(\infty)$}}

\newcommand{\oalpha}{\overline{\alpha}}
\newcommand{\ox}{\overline{x}}
\newcommand{\oy}{\overline{y}}

\newcommand{\bmZero}{{\boldsymbol 0}}
\newcommand{\bmLambda}{\boldsymbol \Lambda}
\newcommand{\bmOmega}{\boldsymbol \Omega}
\newcommand{\bmalpha}{\boldsymbol \alpha}
\newcommand{\bmlambda}{\boldsymbol \lambda}
\newcommand{\bmgamma}{\boldsymbol \gamma}
\newcommand{\bmw}{\boldsymbol w}
\newcommand{\bmSigma}{\boldsymbol \Sigma}
\newcommand{\bmDelta}{\boldsymbol \Delta}

\newcommand{\bme}{\boldsymbol e}

\newcommand{\bmA}{{\boldsymbol A}}
\newcommand{\bmB}{{\boldsymbol B}}
\newcommand{\bmC}{{\boldsymbol C}}
\newcommand{\bmD}{{\boldsymbol D}}
\newcommand{\bmE}{{\boldsymbol E}}
\newcommand{\E}{{\boldsymbol E}}
\newcommand{\bmJ}{{\boldsymbol J}}
\newcommand{\bmM}{{\boldsymbol M}}
\newcommand{\bmO}{{\boldsymbol O}}
\newcommand{\bmR}{{\boldsymbol R}}
\newcommand{\bmS}{{\boldsymbol S}}
\newcommand{\bmT}{{\boldsymbol T}}
\newcommand{\bmU}{{\boldsymbol U}}
\newcommand{\bmV}{{\boldsymbol V}}
\newcommand{\bmX}{{\boldsymbol X}}
\newcommand{\bmY}{{\boldsymbol Y}}
\newcommand{\bmW}{{\boldsymbol W}}
\newcommand{\bmz}{{\boldsymbol z}}

\newcommand{\bbGamma}{{\rm I}\!\Gamma}

\newcommand{\calpha}{\alpha^{*}}
\newcommand{\cbeta}{\beta^{*}}
\newcommand{\cxi}{\xi^{*}}
\newcommand{\czeta}{\zeta^{*}}
\newcommand{\cv}{v^{*}}
\newcommand{\cw}{w^{*}}

\newcommand{\mcc}[1]{\parbox{2cm}{\begin{center} #1 \end{center}}}

\newcommand{\ntwb}{N_\lambda}
\newcommand{\thpm}{\theta_{\pm}}
\newcommand{\phpm}{\phi_{\pm}}
\newcommand{\tinyTWB}{\hbox{\tiny TWB}}

\newcommand{\rmSU}{{\rm SU}}
\newcommand{\rma}{{\rm a}}
\newcommand{\rmb}{{\rm b}}

\newcommand{\Gnoise}{{\cal G}_{\bmDelta}}
\newcommand{\be}{\begin{equation}}
\newcommand{\ee}{\end{equation}}
\newcommand{\bea}{\begin{eqnarray}}
\newcommand{\eea}{\end{eqnarray}}
\newtheorem{teo}{Theorem}

\newcommand{\Sp}{{\rm Sp}}

\newcommand{\cJ}{{\cal J}}
\newcommand{\cB}{{\cal B}}

\newcommand{\sT}{\scriptscriptstyle T}
\newcommand{\sH}{{\scriptscriptstyle H}}
\newcommand{\sPS}{{\scriptscriptstyle PS}}
\newcommand{\sDP}{{\scriptscriptstyle DP}}

\newcommand{\rr}{\mathbb{R}}

\newcommand{\cc}{\mathbb{C}}
\newcommand{\ii}{\mathbbm{1}}
\newcommand{\iid}{\mathbb{I}}
\newcommand{\jj}{\mathbb{J}}
\newcommand{\JJ}{\mathbb{J}}

\newcommand{\gr}[1]{\boldsymbol{#1}}

\newcommand{\refeq}[1]{Eq.~(\ref{#1})}
\newcommand{\eq}[1]{Eq.~(\ref{#1})}

\renewcommand{\det}{{\rm Det}\,}

\def\re#1{\Re\hbox{e}[#1]}
\def\im#1{\Im\hbox{m}[#1]}

\def\kket#1{|#1\rangle\rangle}
\def\bbra#1{\langle\langle #1|}

\makeindex
\begin{document}
\pagenumbering{roman}
\pagestyle{empty}
\chapter*{Preface}
\setcounter{page}{-1}
\addcontentsline{toc}{chapter}{{}Preface}{}

\rput(8,4.8){\psframebox[framearc=.2]{\begin{tabular}{c}
 \\ [-1ex]
{\huge Gaussian states in continuous variable}\\ [1ex]
{\huge quantum information}\\
 \\ 
{\Large Alessandro Ferraro, Stefano Olivares, Matteo G.~A.~Paris}\\[1ex]
\end{tabular}
}}
\rput(7.5,-15){\psshadowbox{
\begin{tabular}{c}
 \\
{\Large Published as\hfill}\\
 \\
{\LARGE \em Gaussian states in quantum information}\\
 \\
{\Large ISBN 88-7088-483-X (Bibliopolis, Napoli, 2005)}\\ [1ex]
{\Large \tt http://www.bibliopolis.it}\\
$ $
\end{tabular}
}}
These notes originated out of a set of lectures in Quantum Optics 
and Quantum Information given by one of us (MGAP) at the 
University of Napoli and the University of Milano. 
A quite broad set of issues are covered, ranging from elementary 
concepts to current research topics, and from fundamental concepts
to applications. A special emphasis has been given to the phase 
space analysis of quantum dynamics and to the role of Gaussian states 
in continuous variable quantum information.
\par
We thank Giuseppe Marmo for his invitation to write
these lecture notes and for his kind assistance in 
the various stages of this project. 
\par
MGAP would like to thank Mauro D'Ariano for the exciting 
introduction he gave me to this fields, and Rodolfo Bonifacio, 
who gave me the possibility of establishing a research 
group at the University of Milano. MGAP also thanks Maria Bondani 
and Alberto Porzio for the continuing discussions on quantum optics 
over these years.
\par
Many colleagues contributed in several ways to the materials in 
this volume. In particular we thank Alessio Serafini, Nicola 
Piovella, Mary Cola, Andrea Rossi, Fabrizio Illuminati, 
Konrad Banaszek, Salvatore Solimeno, Virginia D'Auria, Silvio 
De Siena, Alessandra Andreoni, Alessia Allevi, Emiliano Puddu, 
Antonino Chiummo, Paolo Perinotti, Lo\-ren\-zo Maccone, Paolo 
Lo Presti, Massimiliano Sacchi,  Jarda 
$\check{\mbox{R}}$eh$\acute{\mbox{a}} \check{\mbox{c}}$ek, 
Berge Englert, Paolo Tombesi, David Vitali, Stefano Mancini, 
Geza Giedke, Jaromir Fiur\'a$\check{\mbox{s}}$ek and Valen\-tina De Renzi. 
A special thank to Alessio Serafini for his careful reading and 
comments on various portions of the manuscript.
\par
One of us (SO) would like to remember here a friend, Mario Porta: my work
during these years is also due to your example in front of the difficulties
of life.
\vspace{4mm}
\begin{flushright}\noindent
Milano, December 2004 \\ $ $ \\ 
\hfill {\it Alessandro Ferraro}\\
\hfill {\it Stefano Olivares}\\
\hfill {\it Matteo G A Paris}\\
\end{flushright}

\pagestyle{myheadings}
\tableofcontents
\chapter*{List of symbols}
\pagestyle{myheadings}
\markboth{}{}
\addcontentsline{toc}{chapter}{{}List of symbols}
\begin{supertabular}{c l}

$i$ & imaginary unit \\ [1ex]

$x_\phi$ & quadrature operator \\ [1ex]

$\otimes$, $\oplus$ & tensor product, direct sum \\ [1ex]

$[ \cdot, \cdot ]$, $\{ \cdot, \cdot \}$ & commutator, anticommutator  \\ [1ex]

${\rm Tr}[\cdots]$ & trace \\ [1ex]

${\rm Tr}_{\cal H}[\cdots]$ & trace over the Hilbert space ${\cal H}$\\ [1ex]

${\rm Tr}_{n}[\cdots]$ & trace over the subsystem $n$\\ [1ex]

$\delta_{pq}$ & Kronecker's delta \\ [1ex]

$\delta^{(n)}(\cdots)$ & $n$-dimensional Dirac $\delta$-function \\ [1ex]

$\ii$, $\ii_n$ & identity matrix, $[\ii]_{pq} = \delta_{pq}$,
 $n\times n$ identity matrix\\ [1ex]

$\jj$ & parity matrix $[\jj]_{pq} = (-)^p \delta_{pq}$ \\ [1ex]

$\iid$ & identity operator \\ [1ex]

$\Pi$ & single-mode parity operator $\Pi=(-)^{a^\dag a}$ \\ [1ex]
$\gr{\Pi}$ & multimode parity operator 
$\gr{\Pi}=\otimes_k (-)^{a^\dag_k a_k}\equiv (-)^{\sum_k a^\dag_k a_k}$\\ [1ex]

$\bmOmega$, $\gr{J}$ & symplectic forms \\ [1ex]

${\rm Diag}(a_1,a_2,\ldots)$ & diagonal matrix with elements $a_k$,
$k=1,2,\ldots$ \\ [1ex]

${\rm Sp}(2n, {\mathbbm R})$ & real symplectic group with dimension $n(2n+1)$
\\ [1ex]

${\rm ISp}(2n, {\mathbbm R})$ & real inhomogeneus symplectic group
with dimension $n(2n+3)$ \\ [1ex]

${\rm SU}(n,m)$ & special unitary group with dimension $(n+m)^2-1$ \\ [1ex]

${\rm M}(n,\rr)$, ${\rm M}(n,\cc)$ & group of $n\times n$  matrices 
with real or complex elements \\ [1ex]

$(\cdots)^{\sT,*,\dag}$ & transposed, conjugated, adjoint\\ [1ex]

$(\cdots)^{\theta}$, $(\cdots)^{\scriptscriptstyle T_A}$ & partial
transposition (PT), PT with respect
to subsystem $A$\\ [1ex]

%
$\| O \|_{\rm op}$ & operator norm of $O$: the maximum eigenvalue of
$\sqrt{O^\dagger O}$\\ [1ex]

$\| O \|_{\rm tr}$ &  $\| O \|_{\rm tr} = {\rm Tr}\big[\sqrt{O^\dag
O}\big]$\\ [1ex]

${\rm Det}[\cdots]$ & determinant \\ [1ex]

$\dket{\ii}$, $\dket{\jj}$ & $\dket{\ii} = \sum_p \ket{p}\otimes\ket{p}$,
 $\dket{\jj} = \sum_p (-)^p \ket{p}\otimes\ket{p}$\\ [1ex]

$\mbox{{\bf :}}\cdots\mbox{{\rm\bf :}}$ & normal ordering of field
operators \\ [1ex]

$[\cdots]_{S}$ & symmetric ordering of field operators \\ [1ex]

$\bmE^{(\pm)}$ & positive, $\bmE^{(+)}$, and negative, $\bmE^{(-)}$, part
of the field $\bmE$ \\ [1ex]

$D(\cdots)$ & displacement operator \\ [1ex]

$U(\zeta)$ & two-mode mixing evolution operator \\ [1ex]

$U_{\phi}$ & beam splitter evolution operator, $U_{\phi}=U(\phi)$, $\phi
\in  \rr$ \\ [1ex]

$S(\xi)$, $S_2(\xi)$ & squeezing operator, two-mode squeezing operator \\ [1ex]

$\dket{\Lambda}$ & two-mode squeezed vacuum or twin-beam state (TWB) \\
[1ex]

$\chi[O](\cdots)$ & characteristic function of the operator $O$ \\ [1ex]

$Q(\cdots)$ & Husimi or $Q$-function \\ [1ex]

$W[O](\cdots)$ & Wigner function of the operator $O$\\ [1ex]

$W(\cdots)$ & Wigner function \\ [1ex]

$H_{k}(x)$ & Hermite polynomials \\ [1ex]

$L_n^d(\cdots)$, $L_n(\cdots)$ & Laguerre polynomials \\ [1ex]

${\cal L}[O]\varrho$ & $2 O \varrho O^\dag - O^\dag O \varrho - \varrho
O^\dag O$ \\ [1ex]

${\cal D}[O]\varrho$ & $2 O \varrho O - O O \varrho - \varrho
O O$ \\ [1ex]

${\cal R}[O](x,\phi)$ & kernel or pattern function for the operator $O$ \\
[1ex]

\end{supertabular}
\chapter*{Introduction}
\addcontentsline{toc}{chapter}{{}Introduction}
\chaptermark{Introduction}
In any protocol aimed at manipulate or transmit information,
symbols are encoded in states of some physical system such as
a polarized photon or an atom. If these systems are allowed to
evolve according to the laws of quantum mechanics, novel kinds of
information processing become possible. These include quantum
cryptography, teleportation, exponential speedup of certain
computations and high-precision measurements. In a way, quantum
mechanics allows for information processing that could not be
performed classically. 
\par 
Indeed, in the last decade, we have witnessed a dramatic
development of quantum information theory, mostly motivated by
the perspectives of quantum-enhanced communication, measurement
and computation systems. Most of the concepts of quantum
information were initially developed for discrete quantum
variables, in particular quantum bits, which have become the
symbol of the recently born quantum information technology.  More
recently, much attention has been devoted to investigating the
use of continuous variable (CV) systems in quantum information
processing.  Continuous-spectrum quantum variables may be easier
to manipulate than quantum bits in order to perform various
quantum information processes.  This is the case of Gaussian
state of light, {\em e.g.} squeezed-coherent beams and
twin-beams, by means of linear optical circuits and homodyne
detection. Using CV one may carry out quantum teleportation and
quantum error correction. The concepts of quantum cloning and
entanglement purification have also been extended to CV, and
secure quantum communication protocols have been proposed.  
Furthermore, tests of quantum nonlocality using CV quantum 
states and measurements have been extensively analyzed.  
\par 
The key ingredient of quantum information is entanglement, which
has been recognized as the essential resource for quantum
computing, teleportation, and cryptographic protocols. Recently,
CV entanglement has been proved as a valuable tool also for
improving optical resolution, spectroscopy, interferometry,
tomography, and discrimination of quantum operations.
\par
A particularly useful class of CV states are the Gaussian states.
These states can be characterized theoretically in a convenient way,
and they can also be generated and manipulated experimentally in
a variety of physical systems, ranging from light fields to
atomic ensembles. In a quantum information setting, entangled
Gaussian states form the basis of proposals for teleportation,
cryptography and cloning. 
\par
In implementations of quantum information protocols one needs to 
share or transfer entanglement among distant partners, and therefore 
to transmit entangled states along physical channels. As a matter 
of fact, the propagation of entangled states and the influence of 
the environment unavoidably lead to degradation of entanglement, 
due to decoherence induced by losses and noise and by the consequent 
decreasing of purity. For Gaussian states and operations, separability
thresholds can be analytically derived, and their influence on 
the quality of  the information processing analyzed in details.
\par
In these notes we discuss various aspects of the use of Gaussian 
states in CV quantum information processing. We analyze in some 
details separability, nonlocality, evolution in noisy channels and
measurements, as well as applications like teleportation,
telecloning and state engineering performed using Gaussian states
and Gaussian measurements. Bipartite and tripartite systems are
studied in more details and special emphasis is placed on the
phase-space analysis of Gaussian states and operations.
\par
In Chapter \ref{ch:basics} we introduce basic concepts and notation 
used throughout the volume. In particular, Cartesian 
decompositions of mode operators and phase-space variables
are analyzed, as well as basic properties of displacement and
squeezing operators. Characteristic and Wigner
functions are introduced, and the role of symplectic 
transformations in the description of Gaussian operations
in the phase-space is emphasized.
\par
In Chapter \ref{ch:gs} Gaussian states are introduced and 
their general properties are investigated. Normal forms for 
the covariance matrices are derived. In Chapter \ref{SepGS}
we address the separability problem for Gaussian states 
and discuss necessary and sufficient conditions.
\par
In Chapter \ref{c6:gauss:chan} we address the evolution of a $n$-mode 
Gaussian state in a noisy channel where both dissipation and noise, 
thermal as  well as phase--sensitive (``squeezed'') noise, are 
present. At first, we focus our attention on the evolution of a single 
mode of radiation. Then, we extend our analysis to the evolution of a 
$n$-mode state, which  will be treated as the evolution in a global 
channel made of $n$ non interacting different channels. 
Evolution of purity and nonclassicality for single-mode states, 
as well as separability threshold for multipartite states are 
evaluated.
\par
In Chapter \ref{ch:detection} we describe a set of relevant 
measurements that can be performed on continuous variable 
(CV) systems. These include both single-mode, as direct detection
or homodyne detection, and two-mode (entangled) measurements 
as multiport homodyne or heterodyne detection. The use of 
conditional measurements to generate non Gaussian CV states is 
also discussed.
\par
Chapter \ref{ch:nonloc} is devoted to the issue of nonlocality 
for CV systems. Nonlocality tests based on CV measurements are 
reviewed and two-mode and three-mode nonlocality of Gaussian 
and non Gaussian states is analyzed.
\par
In Chapter \ref{ch:tele} we deal with the transfer and the 
distribution of quantum information, {\em i.e.} of the 
information contained in a quantum state. At first, we address 
teleportation, {\em i.e.} the entanglement-assisted transmission 
of an unknown quantum state from a sender to a receiver that are 
spatially separated. Then, we address telecloning, {\em i.e.}
the distribution of (approximated) copies of a quantum state
exploiting multipartite entanglement which is shared among 
all the involved parties. 
Finally, in Chapter \ref{ch:StEng} we analyze the use of 
conditional measurements on entangled state of radiation 
to engineer quantum states, {\em i.e.} to produce, manipulate, 
and transmit nonclassical light. 
In particular, we focus 
our attention on realistic measurement schemes, feasible with 
current technology.
\par\noindent
Throughout this volume we use natural units and assume $\hbar=c=1$.
\par\vspace{20pt}
\par\noindent
Comments and suggestions are welcome. They should be addressed to \\ 
\centerline {\tt matteo.paris@fisica.unimi.it}
\par\vspace{10pt}
\par\noindent
Corrections, additions and updates to the text and the bibliography, as well 
as exercises and solutions will be published
at \\ \centerline{{\tt http://qinf.fisica.unimi.it/\~{}paris/QLect.html}}

\pagestyle{fancy}
\renewcommand{\chaptermark}[1]{\markboth{\MakeUppercase{\chaptername}\ 
\thechapter:\ #1}{}}
\renewcommand{\sectionmark}[1]{\markright{\thesection\ \hspace{.8mm} #1}}
\setcounter{page}{1}
\pagenumbering{arabic}
\chapter{Preliminary notions}\label{ch:basics}
In this Chapter we introduce basic concepts and notation 
used throughout the volume. In particular, Cartesian 
decompositions of mode operators and phase-space variables
are analyzed, as well as basic properties of displacement and
squeezing operators \cite{SamRMP}. Characteristic and Wigner
functions are introduced, and the role of symplectic 
transformations in the description of Gaussian operations
in the phase-space is emphasized \cite{JensRev}. 
\section[Systems made of $n$ bosons]{Systems made of $\boldsymbol{n}$ bosons}
\label{s:Nbos}
\sectionmark{Systems made of $\boldsymbol{n}$ bosons}
Let us consider a system made of $n$  bosons described by the mode
operators $a_k$, $k=1,\ldots,n$, with commutation relations
$[a_k,a^\dag_l]=\delta_{kl}$.  The Hilbert space of the system ${\cal
H}=\otimes_{k=1}^n \: {\cal F}_{k}$ is the tensor product of the infinite
dimensional Fock spaces ${\cal F}_{k}$ of the $n$ modes, each spanned by
the number basis $\{\ket{m}_k\}_{m \in {\mathbb N}}$, {\em i.e.} by the
eigenstates of the number operator $a_k^{\dag}a_k$. The free Hamiltonian of
the system (non interacting modes) is given by $H=\sum_{k=1}^n
(a_k^{\dag}a_k + \frac12)$.  Position- and momentum-like operators for each
mode are defined through the Cartesian decomposition of the mode operators
$a_k=\kappa_1 (q_k+ i p_k )$ with $\kappa_1 \in \rr$, {\em i.e.}
\begin{eqnarray}
q_k = \frac{1}{2\kappa_1}\,(a_k + a^\dag_k)\,, \qquad \qquad
p_k = \frac{1}{2i\kappa_1}\,(a_k - a^\dag_k)\,.
\label{defQP}\;
\end{eqnarray}
The corresponding commutation relations are given by
\begin{eqnarray}
[q_k,p_l]=\frac{i}{2\kappa_1^2}\, \delta_{kl}
\label{defCom}\;.
\end{eqnarray}
\index{canonical operators}
Canonical position and momentum operator are obtained for
$\kappa_1=2^{-1/2}$, while the quantum optical convention corresponds to
the choice $\kappa_1=1$. Introducing the vector of operators $\boldsymbol{
R}=(q_1,p_1,\ldots,q_n,p_n)^{\sT}$, \eq{defCom} rewrites as
\begin{eqnarray}
[R_k,R_l]=\frac{i}{2\kappa_1^2}\, \Omega_{kl}
\label{defCom1}\;,
\end{eqnarray}
where $\Omega_{kl}$ are the elements of the symplectic matrix 
\be
\gr{\Omega}=\bigoplus_{k=1}^{n}\gr{\omega}\:, \qquad
\gr{\omega}= \left(\begin{array}{cc}0&1\\ -1&0
\end{array}\right)\,. \label{defOM}
\ee
\index{symplectic!forms}
\index{commutation relations}
By a different grouping of the operators as
$\boldsymbol{S}=
(q_1,\ldots,q_n,p_1,\ldots,p_n)^{\sT}$, commutation relations rewrite
as
\begin{eqnarray}
[S_k,S_l]=-\frac{i}{2\kappa_1^2}\, J_{kl}
\label{defCom1S}\;,
\end{eqnarray}
where $J_{kl}$
are the elements of the $2n \times 2n$ symplectic antisymmetric matrix
\be
\gr{J}= \left(\begin{array}{cc} \boldsymbol{0} &-
\ii_n \\
\ii_n
&\boldsymbol{0} \end{array}\right)\:, \label{defJ}
\ee
$\ii_n$ being the $n\times n$ identity matrix. Both notations are
extensively used in the literature, and will be employed in the 
present volume.
\par
Analogously, for a quantum state of $n$ bosons the {\em covariance matrix} 
is defined in the following ways 
\index{covariance matrix}
\begin{subequations}
\label{defCOV}
\begin{align}
\sigma_{kl} \equiv [\bmsigma]_{kl} &=  \frac12 \langle \{R_k,R_l\} \rangle - 
\langle  R_l \rangle  
\langle  R_k \rangle\,,\\
V_{kl} \equiv [\bmV]_{kl} &= \frac12 \langle \{S_k,S_l\} \rangle - 
\langle  S_l \rangle  
\langle  S_k \rangle\;,
\end{align}
\end{subequations}
where $\{A,B\}=AB+BA$ denotes the anticommutator and 
$\langle O \rangle \equiv \overline{O} = \hbox{Tr}[\varrho\: O]$ is
the expectation value of the operator $O$, with $\varrho$ being 
the density matrix of the system. Uncertainty relations among canonical 
operators impose a constraint on the covariance matrix, corresponding to the
inequalities  \cite{simon_old}
\begin{eqnarray} \boldsymbol{\sigma} +
\frac{i}{4\kappa_1^2} \boldsymbol{\Omega} \geq 0\,, \qquad
\boldsymbol{V} - \frac{i}{4\kappa_1^2} \boldsymbol{J} \geq 0
\label{HeisSG}\;. \end{eqnarray}
Ineqs.~(\ref{HeisSG}) follow from the uncertainty relations for the mode
operators, and express, in a compact form, the positivity of the density 
matrix $\varrho$.
The vacuum state of $n$ bosons is characterized by the covariance
matrix $\boldsymbol{\sigma} = \boldsymbol{V} = (4\kappa^2_1)^{-1}
\ii_{2n}$, while for a state at thermal equilibrium, {\em i.e.}
$\nu = \bigotimes_{k=1}^n \: \nu_k$ with 
\begin{eqnarray}
\nu_k = \frac{e^{-\beta a^\dag_k a_k}}{\hbox{Tr}[e^{-\beta a^\dag_k a_k}]}
=\frac{1}{1+{N}_k} \sum_{m=0}^\infty \left(\frac{{N}_k}{1
+{N}_k}\right)^m\: | m \rangle_k {}_k\langle m |
\label{th}\;,
\end{eqnarray}
we have
\begin{subequations}
\begin{align}
\boldsymbol{\sigma}_{\nu}
&= \frac1{4\kappa_1^2}\: {\rm Diag}\left(2 {N}_1 +1,2
{N}_1 + 1, \dots, \dots,
2 {N}_n+1,2 {N}_n+1\right)\,, \\
\boldsymbol{V}_{\nu} &= \frac{1}{4\kappa_1^2}\:
{\rm Diag}\left(2 {N}_1 +1, \dots, 2 {N}_n+1,
2 {N}_1+1,\dots, 2 {N}_n+1\right) \label{sgth}\;,
\end{align}
\end{subequations}
where ${\rm Diag}(a_1,a_2,\ldots)$ denotes a diagonal
matrix with elements $a_k$, $k=1,2,\ldots$ and ${N}_k= 
(e^\beta-1)^{-1}$ is the
average number of thermal quanta at the equilibrium in the $k$-th mode . 
\par
The two vectors of operators $\gr{R}$ and $\gr{S}$ 
are related each other by a simple $2n \times 2n$ permutation matrix 
$\gr{S}=\gr{P} \:\gr{R}$, whose elements are given by
\be 
P_{kl} = \left\{ \begin{array}{cc}
\delta_{k,2l-1} & k \leq n \\ [1ex]
\delta_{n+k,2l} & l \leq n 
\end{array}\right.
\label{ElPerm}\:,
\ee
$\delta_{k,l}$ being the Kr\"onecker delta.
Correspondingly, the two forms of the covariance matrix, 
as well as the symplectic forms for the two orderings, 
are connected as $$
\gr{V} = \gr{P} \: \gr{\sigma} \gr{P}^{\sT}\,, \qquad
\gr{J} = -\gr{P} \: \gr{\Omega} \gr{P}^{\sT}
\:.$$
The average number of quanta in a system of $n$ bosons is 
given by $\sum_{k=1}^n \langle a^\dag_k a_k \rangle$. In terms of 
the Cartesian operators and covariance matrices we have 
\begin{eqnarray}
\sum_{k=1}^n \langle a^\dag_k a_k \rangle = 
\kappa_1^2 \sum_{l=1}^{2n}  \left(\sigma_{ll} + 
\overline{R}_l^2\right) - \frac{n}{2}
= \kappa_1^2 \sum_{l=1}^{2n}  \left(V_{ll} + 
\overline{S}_l^2\right) - \frac{n}{2}\:.
\label{AveCov}
\end{eqnarray}
\index{quadrature operators}
Eqs.~(\ref{defQP}) can be generalized to define the so-called 
{\em quadrature} operators of the field 
\begin{eqnarray}
x_{k\phi} =
\frac{1}{2\kappa_1}
\left(a_k e^{-i\phi}+ a^\dag_k e^{i\phi}\right)
\label{defquad}\;, \end{eqnarray} 
{\em i.e.} a generic linear combination of the mode operators 
weighted by phase factors.  Commutation relations read as follows
\begin{eqnarray}
[x_{k\phi},x_{l\psi}] = \frac{i}{2\kappa_1^2}\,\delta_{kl}\sin(\psi-\phi)
\label{CommQuad}\;.
\end{eqnarray}
Position- and momentum-like
operators are obtained for $\phi=0$ and $\phi=\pi/2$,
respectively. Eigenstates $|x\rangle_\phi$ of the field quadrature $x_\phi$ 
form a complete set $\forall \phi$, {\em i.e.} $\int_\rr dx |x\rangle_\phi 
{}_\phi\langle x| ={\mathbb I}$, and their expression in the number basis 
is given by
\begin{eqnarray}
|x\rangle_\phi = e^{-\kappa_1^2 x^2}\: 
\left(\frac{2\kappa_1^2}{\pi}\right)^{1/4}\:
\sum_{k=0}^\infty \frac{H_k(\sqrt{2}\kappa_1 x)}{2^{k/2} \sqrt{k!}}
\: e^{-i k \phi} |n\rangle
\label{eigenquad}\;,
\end{eqnarray}
$H_k(x)$ being the $k$-th Hermite polynomials.
\section{Matrix notations for bipartite systems}
\label{s:MatNot}
For pure states in a bipartite Hilbert space
${\cal H}_1 \otimes {\cal H}_2$ we will use the notation
\cite{lopresti}
\begin{equation}
\kket{\bmC}\doteq \sum_{kl} c_{kl}\: \ket{k}_1\otimes\ket{l}_2\,,
\qquad 
\bbra{\bmC}\doteq \sum_{kl} c_{kl}^*\: {}_1\bra{k}\otimes{}_2\bra{l}\,,
\label{defC}\;
\end{equation}
where $c_{kl}=[\bmC]_{kl}$ are the elements of the matrix $\bmC$ and
$\ket{k}_r$ is the standard basis of ${\cal H}_r$, $r=1,2$.
Notice that a given matrix $\bmA$ also individuates a linear operator
from ${\cal H}_1$ to ${\cal H}_2$, given by
$\bmA =  \sum_{kl} a_{kl} \ket{k}_1{}_2\bra{l}$.
In the following we will consider ${\cal H}_1$ and ${\cal H}_2$
both describing a bosonic mode. Thus we will refer only to
(infinite) square matrices and omit the indices for bras and kets.
We have the following identities
\begin{subequations}
\begin{align}
\bmA \otimes \bmB \kket{\bmC} = \kket{\bmA\bmC\bmB^{\sT}}\,,
\label{BR} &\qquad 
\bbra{\bmC} \bmA \otimes \bmB = \bbra{\bmA\bmC\bmB^{\sT}}\,, \\ 
\bbra{\bmA}\bmB\rangle\rangle &= \hbox{Tr}[\bmA^\dag \bmB]\,,
\label{BR2}
\end{align}
\end{subequations}
where $(\cdots)^{\sT}$ denotes transposition with respect to the
standard basis. Notice the ordering for the ``bra'' in (\ref{BR}).
Proof is straightforward by explicit calculations.
Notice that $\bmA\otimes \bmB = (\bmA\otimes \ii)(\ii\otimes \bmB)$, 
and therefore is enough to prove (\ref{BR})
for $\bmA\otimes \ii$ and $\ii\otimes \bmB$ respectively.
Normalization of state $\kket{\bmC}$ implies $\hbox{Tr}[\bmC^\dag \bmC]=1$.
Also useful are the following relations about partial traces
\begin{equation}
\hbox{Tr}_2 \left[\kket{\bmA}\bbra{\bmB}\right] = \bmA\bmB^\dag\,,
\qquad
\hbox{Tr}_1 \left[\kket{\bmA}\bbra{\bmB}\right] = \bmA^{\sT} \bmB^*
\label{PT}\,,
\end{equation}
where $(\cdots)^*$ denotes complex conjugation, and about partial
transposition
$$
\big(\kket{\bmC}\bbra{\bmC}\big)^{\theta} =
\big (\bmC \otimes \ii\big) \bmE \big (\bmC^\dag \otimes \ii\big)\:,
$$
where $\bmE=\sum_{kl} \ket{k}\bra{l}\otimes \ket{l}\bra{k}$ is the swap
operator. Notice that $\bmA\bmB^\dag$ and $\bmA^{\sT} \bmB^*$
in (\ref{PT}) should be meant as operators acting on
${\cal H}_1$ and ${\cal H}_2$ respectively.
Finally, we just remind that  $(\bmA^{\sT})^{\sT} =(\bmA^*)^* = 
(\bmA^\dag)^\dag = \bmA$,
and thus $\bmA^\dag = (\bmA^{\sT})^*=(\bmA^*)^{\sT}$,
$\bmA^{\sT} = (\bmA^\dag)^*=(\bmA^*)^\dag$, and
$\bmA^* = (\bmA^\dag)^{\sT}=(\bmA^{\sT})^\dag$.
\section{Symplectic transformations}
\index{symplectic!transformations}
\label{s:symp}
Let us first consider a classical system of $n$ particles described by
coordinates $(q_1,\ldots,q_n)$ and conjugated momenta $(p_1,\ldots,p_n)$.
If $H$ is the Hamiltonian of the system, the equation of motion are given by
\begin{eqnarray}
\dot q_k = \frac{\partial H}{\partial p_k}\,,
\quad \dot p_k = - \frac{\partial H}{\partial q_k}\,, \qquad (k=1,\ldots,n)
\label{HamEq}\;
\end{eqnarray}
where $\dot{x}$ denotes time derivative. The Hamilton equations can be
summarized as
\begin{eqnarray}
\dot R_k = \Omega_{kl}\, \frac{\partial H}{\partial R_l}\,, \qquad
\dot S_k = - J_{kl}\, \frac{\partial H}{\partial S_l}\,,
\label{HamEq1}\;
\end{eqnarray}
where $\boldsymbol{R}$ and $\boldsymbol{S}$ are vectors of coordinates
ordered as the vectors of canonical operators in Section \ref{s:Nbos}, whereas
$\boldsymbol{\Omega}$ and $\boldsymbol{J}$ are the symplectic matrices
defined in \eq{defOM} and \eq{defJ}, respectively.
The transformations of coordinates $\boldsymbol{R}'=\gr{F}\boldsymbol{R}$,
$\boldsymbol{S}'=\gr{Q}\boldsymbol{S}$ are described by matrices
\begin{eqnarray}
F_{kl}=\frac{\partial R'_{k}}{\partial R_l}\,, \qquad
Q_{kl}=\frac{\partial S'_{k}}{\partial S_l}
\label{MatPQ}\;,
\end{eqnarray}
and lead to
\begin{eqnarray}
\dot R'_k = F_{ks}\Omega_{st}F_{lt}\, \frac{\partial H}{\partial R_l}\,, 
\qquad
\dot S'_k = - Q_{ks}J_{st}Q_{lt}\, \frac{\partial H}{\partial R_l}
\label{HamEq2}\;.
\end{eqnarray}
Equations of motion thus remain invariant iff
\begin{eqnarray}
\boldsymbol{F}\, \boldsymbol{\Omega} \boldsymbol{F}^{\sT} =
\boldsymbol{\Omega}\,,
\qquad
\boldsymbol{Q}\, \boldsymbol{J} \boldsymbol{Q}^{\sT} = \boldsymbol{J}
\label{SympCond}\;,
\end{eqnarray}
which characterize symplectic transformations and, in turn, describe the
canonical transformations of coordinates.
Notice that the identity matrix and the symplectic forms themselves
satisfies Eq.~(\ref{SympCond}).
\par
Let us now focus our attention on a quantum system of $n$ bosons, 
described by the mode operators $\gr{R}$ or $\gr{S}$. A mode 
transformation $\gr{R}'= \gr{F} \gr{R}$ or $\gr{S}'= \gr{Q} \gr{S}$ 
leaves the kinematics invariant if it preserves canonical commutation 
relations (\ref{defCom1}) or (\ref{defCom1S}). In turn, this means  
that the $2n\times 2n$ matrices $\gr{F}$ and $\gr{Q}$ should satisfy
the symplectic condition $(\ref{SympCond})$. Since 
$\gr{\Omega}^{\sT}=\gr{\Omega}^{-1}=-\gr{\Omega}$ from
$(\ref{SympCond})$ one has that $\det[\gr{F}]^2=1$\footnote{\footnotesize 
Actually $\det[\gr{F}]=+1$ and never $-1$. This result may be obtained 
by showing that if $e$ is an eigenvalue of a symplectic matrix, than 
also $e^{-1}$ is an eigenvalue \cite{littlej}.} and therefore 
$\gr{F}^{-1}$ exists. Moreover, it is straightforward to show
that if $\gr{F}$, $\gr{F}_1$ and $\gr{F}_2$ are symplectic
then also $\gr{F}^{-1}$, $\gr{F}^{\sT}$ and $\gr{F}_1\gr{F}_2$ 
are symplectic, with $\gr{F}^{-1} = \gr{\Omega}\, \gr{F}^{\sT} 
\gr{\Omega}^{-1}$. Analogue formulas are valid for the 
$\gr{J}$-ordering. Therefore, the set of $2n \times 2n$ real 
matrices satisfying (\ref{SympCond}) form a group, the so-called 
{\em symplectic group} ${\rm Sp} (2n, {\mathbb R})$ with dimension 
$n(2n+1)$. Together with phase-space translation, it forms the 
{\em affine} (inhomogeneous) symplectic group ${\rm ISp} (2n, {\mathbb R})$.
If we write a $2n \times 2n$ symplectic matrix in the block form
\begin{eqnarray}
\gr{F} = \left(\begin{matrix}\gr{A} & \gr{B} \\ 
\gr{C} & \gr{D}\end{matrix}\right)
\label{Sblock}\;,
\end{eqnarray}
with $\gr{A}$, $\gr{B}$, $\gr{C}$, and $\gr{D}$ $n\times n$ matrices, 
then the symplectic conditions rewrites as the following
(equivalent) conditions
\begin{eqnarray}
\left\{ 
\begin{array}{l} 
\gr{A}\gr{D}^{\sT}-\gr{B}\gr{C}^{\sT} = \ii \\
\gr{A}\gr{B}^{\sT}=\gr{B}\gr{A}^{\sT} \\
\gr{C}\gr{D}^{\sT}=\gr{D}\gr{C}^{\sT}
\end{array}        
\right.\,,
\qquad
\left\{ 
\begin{array}{l} 
\gr{A}^{\sT}\gr{D}-\gr{C}^{\sT}\gr{B} = \ii \\
\gr{A}^{\sT}\gr{C}=\gr{C}^{\sT}\gr{A} \\
\gr{B}\gr{D}^{\sT}=\gr{D}^{\sT}\gr{B}
\end{array}        
\right.\label{SympBlock}\:.
\end{eqnarray}
The matrices $\gr{A}\gr{B}^{\sT}$, $\gr{C}\gr{D}^{\sT}$, $\gr{A}^{\sT}\gr{C}$, 
and $\gr{B}^{\sT}\gr{D}$ are symmetric and the inverse of the matrix
$\gr{F}$ writes as follows
\begin{eqnarray}
\gr{F}^{-1} = \left(\begin{matrix}\gr{D}^{\sT} & -\gr{B}^{\sT} \\ 
-\gr{C}^{\sT} & \gr{A}^{\sT}\end{matrix}\right)
\label{Invblock}\;.
\end{eqnarray}
For a generic real matrix the polar decomposition is given by $\gr{F}=
\gr{T}\gr{O}$ where $\gr{T}$ is symmetric and $\gr{O}$ orthogonal. If 
$\gr{F}\in {\rm Sp}(2n,\rr)$ then also  $\gr{T},\gr{O}\in {\rm Sp}(2n,\rr)$.
A matrix $\gr{O}$ which is symplectic and orthogonal writes as 
\begin{eqnarray}
\gr{O} = \left(\begin{matrix} \gr{X} & \gr{Y} \\ 
-\gr{Y} & \gr{X}
\end{matrix}\right)\,, \qquad 
\begin{array}{c}
\bmX\bmX^{\sT} + \bmY\bmY^{\sT} = \ii \\
\bmX\bmY^{\sT} - \bmY\bmX^{\sT} = \boldsymbol{0}
\end{array}
\label{OrtSymp}\;,
\end{eqnarray}
which implies that $\bmU = \bmX + i \bmY$ is a unitary $n\times n$
complex matrix. The converse is also true, {\em i.e.} any unitary 
$n\times n$ complex matrix generates a symplectic matrix in 
${\rm Sp} (2n,\rr)$ when written in real notation as in Eq. 
(\ref{OrtSymp}). \par
A useful decomposition of a generic symplectic transformation
$\gr{F}\in {\rm Sp}(2n, {\mathbb R})$ is the so-called 
{\em Euler decomposition}
\index{Euler decomposition}
\be
\gr{F}=\gr{O}
\begin{pmatrix}
\gr{D} & \boldsymbol{0} \\
\boldsymbol{0} & \gr{D}^{-1}
\end{pmatrix}
\gr{O'}\;,
\label{c1:EulerDecomp}
\ee
where $\gr{O}$ and $\gr{O'}$ are orthogonal and symplectic matrices,
while $ \gr{D}$ is a positive diagonal matrix. About the real symplectic 
group in quantum mechanics see Refs.~\cite{SimonSympl,littlej}, 
for details on the single mode case see Ref.~\cite{aw}.  
\section{Linear and bilinear interactions of modes}
\index{bilinear interactions}
\label{s:LinBil}
Interaction Hamiltonians that are linear and bilinear
in the field modes play a major role in quantum information
processing with continuous variables. On one hand, they can
be realized experimentally by parametric processes in
quantum optical \cite{mandel,klysko} and condensate 
\cite{meystre,PCB03,telebec,CPP04,khang} systems. On the other hand, 
they generate the whole set of symplectic transformations. 
According to the linearity of mode evolution, quantum
optical implementations of these transformations
is often referred to as quantum information processing
with {\em linear} optics. It should be noticed, however,
that their realization necessarily involves parametric
interactions in nonlinear media.
The most general Hamiltonian of this type can be written
as follows
\begin{eqnarray}
H=\sum_{k=1}^{n} g_{k}^{(1)}\, a_k^\dag +
\sum_{k>l=1}^{n} g_{kl}^{(2)}\, a_k^\dag a_l+
\sum_{k,l=1}^{n} g_{kl}^{(3)}\, a_k^\dag a^\dag_l + h.c.
\label{LinH}\;.
\end{eqnarray}
Transformations induced by Hamiltonians in \eq{LinH} correspond to 
unitary representation of the affine symplectic group ${\rm ISp}(2n,\rr)$, 
{\em i.e.} the so-called metaplectic representation.
Although the group theoretical structure is not particularly relevant 
for our purposes algebraic methods will be extensively used.
\par
Hamiltonians of the form (\ref{LinH}) contain three main
{\em building blocks}, which represents the generators of 
the corresponding unitary evolutions.
The first block, containing terms of the form
$H \propto g^{(1)}\:a^\dag + h. c. $,  is linear in the field modes. The
corresponding unitary transformations are the set of {\em displacement
operators}. Their properties will be analyzed in details in Section
\ref{ss:displa}. The second block, which contains terms of the form
$g^{(2)} a^\dag b  + h.c.$, describes linear mixing of
the modes, as the coupling realized for two modes of radiation in a
beam splitter. The dynamics of such a {\em passive} device (the
total number of quanta is conserved) will be described in
Section \ref{ss:bs}. This block also contains terms of the form 
$g^{(2)} a^\dag a$, which describes the free evolution of the modes.
In most cases these terms can be eliminated by choosing a suitable interaction
picture.  Finally, the third kind of interaction is
represented by Hamiltonians of the form $g^{(3)} a^{\dag 2} + h.c.$ and 
$g^{(3)} a^{\dag} b^\dag + h.c.$ 
which describe single-mode and two-mode squeezing respectively.
Their dynamics, which corresponds to that of degenerate
and nondegenerate parametric amplifier in quantum optics, will be
analyzed in Sections \ref{ss:squeeze} and \ref{ss:opa} respectively.
Finally, in Section \ref{ss:Hpq} we briefly analyze the multimode 
dynamics induced
by a relevant subset of Hamiltonians in Eq. (\ref{LinH}), corresponding
to the unitary representation of the group ${\rm SU}(p,q)$.
\par
Mode transformations imposed by Hamiltonians (\ref{LinH}) can be 
generally written as 
\begin{eqnarray}
\gr{R} \rightarrow \gr{F} \gr{R} + \gr{d}_{\scriptscriptstyle\bf R}\,, \qquad 
\gr{S} \rightarrow \gr{Q} \gr{S} + \gr{d}_{\scriptscriptstyle\bf S}
\label{modeLinH}\;,
\end{eqnarray}
where the $\gr{d}$'s are real vectors and $\gr{F}$, $\gr{Q}$ 
symplectic transformations.
\index{symplectic!transformations}
Changing the orderings we have 
$\gr{d}_{\scriptscriptstyle\bf S}=
\gr{P}\:\gr{d}_{\scriptscriptstyle\bf R}$ and $\gr{Q}=\gr{P}\gr{F}\gr{P}$,  
$\gr{P}$ being the permutation matrix (\ref{ElPerm}). 
Covariance matrices evolve accordingly 
\begin{eqnarray}
\gr{\sigma} \rightarrow \gr{F}\, \bmsigma\, \gr{F}^{\sT}\,, \qquad 
\gr{V} \rightarrow {\gr{Q}}\, \bmsigma\, {\gr{Q}}^{\sT}
\label{covLinH}\;.
\end{eqnarray}
Remarkably, the converse is also true, {\em i.e.} any symplectic
transformation of the form (\ref{modeLinH}) is generated by a 
unitary transformation induced by Hamiltonians of the form 
(\ref{LinH}) \cite{SimonSympl}.
In this context, the physical implication of the Euler 
decomposition (\ref{c1:EulerDecomp}) is that every symplectic 
transformation may be implemented by means of
two passive devices and by single mode squeezers \cite{Bra99}.
\par As we will also see in Chapter \ref{ch:gs} the set of 
transformations coming from Hamiltonians (\ref{LinH}) individuates 
the class of unitary Gaussian operations, {\em i.e.} unitaries that
transform Gaussian states into Gaussian states.
\subsection{Displacement operator}\label{ss:displa}
\index{displacement operator}
The displacement operator for $n$ bosons is defined as
\begin{eqnarray}
D(\boldsymbol{\lambda}) = \bigotimes_{k=1}^n D_k(\lambda_k)
\label{defDcmplx}\;
\end{eqnarray}
where $\gr{\lambda}$ is the column vector $\boldsymbol{\lambda} =
(\lambda_1,\ldots,\lambda_n)^{\sT}$, $\lambda_k \in \cc$,
$k=1,\ldots,n$ and $D_k(\lambda_k)=\exp \{\lambda_k a^\dag_k -
\lambda_k^* a_k \}$, are single-mode displacement operators; notice
the definition of the row vector $\gr{\lambda}^\dag =
(\lambda_1^*,\ldots, \lambda_n^*)$.
\par
Introducing Cartesian coordinates as $\lambda_k = \kappa_3 (\rma_k + i \rmb_k)$
we have $D(\boldsymbol{\lambda}) \equiv D(\boldsymbol{\Lambda}) \equiv
D(\boldsymbol{K})$ where
\begin{align}
D(\boldsymbol{\Lambda}) = \exp \left\{2i\kappa_1\kappa_3
\boldsymbol{R}^{\sT} \boldsymbol{\Omega} \boldsymbol{\Lambda} \right\}\,, 
\qquad
D(\boldsymbol{K}) = \exp \left\{- 2i\kappa_1\kappa_3
\boldsymbol{S}^{\sT} \boldsymbol{J} \boldsymbol{K} \right\}\,,
\label{defD_real}\;
\end{align}
and
\begin{align}
\boldsymbol{\Lambda} = (\rma_1,\rmb_1,\ldots,\rma_n,\rmb_n)^{\sT}\,, \qquad
\boldsymbol{K} = (\rma_1,\ldots,\rma_n,\rmb_1,\ldots,\rmb_n)^{\sT}\,,
\label{defVecs}\;
\end{align}
are vectors in $\rr^{2n}$ ($\kappa_1$ has been introduced in
Section \ref{s:Nbos}). Canonical coordinates corresponds to
$\kappa_1=\kappa_3=2^{-1/2}$ while a common
choice in quantum optics is $\kappa_1=1$, $\kappa_3=1/2$.
The two parameters are not independent on each other
and should satisfy the constraints $2\kappa_1\kappa_3=1$
(see also Section \ref{s:Wchi}). In the following, in order to simplify
notations and to encompass both cases, we will use complex notation
wherever is possible.
\par
Displacement operator takes its name after the action on the
mode operators
\begin{eqnarray}
D^\dag (\boldsymbol{\lambda})\, a_k\, D(\boldsymbol{\lambda})
= a_k + \lambda_k \quad (k=1,\ldots,n)
\label{displa_mode}\;.
\end{eqnarray}
The corresponding Cartesian expressions are given by
\begin{align}
D^\dag (\boldsymbol{\Lambda})\, \boldsymbol{R}\, D(\boldsymbol{\Lambda})
= \boldsymbol{R} + \boldsymbol{\Lambda}\,, \qquad
D^\dag (\boldsymbol{K})\, \boldsymbol{S}\, D(\boldsymbol{K})
= \boldsymbol{S} + \boldsymbol{K}
\label{displa_XY}\;.
\end{align}
The set of displacement operators $D(\boldsymbol{\lambda})$ with
$\boldsymbol{\lambda}\in \cc^n$ is {\em complete}, in the
sense that any operators $O$ on ${\cal H}$ can be written as
\index{characteristic function!Glauber formula}
\index{Glauber formula}
\begin{eqnarray}
O = \int_{\cc^n} \frac{d^{2n}\boldsymbol{\lambda}}{\pi^n}\,
\hbox{Tr}\left[O\,D(\boldsymbol{\lambda}) \right]\,
D^\dag (\boldsymbol{\lambda}) \label{GlauberF}\;,
\end{eqnarray}
where
\be
\chi[O](\boldsymbol{\lambda}) =
\hbox{Tr}\left[O\,D(\boldsymbol{\lambda}) \right]
\label{defChiO}
\ee
is the so-called characteristic function of the operator $O$, 
which will be analyzed in more details in Section \ref{s:Wchi}.
Eq.~(\ref{GlauberF}) is often referred to as Glauber formula
\cite{cahill}.
The corresponding Cartesian expressions are straightforward
\begin{subequations}
\label{GlauberF2}
\begin{align}
O &= \int_{\rr^{2n}} \frac{\kappa_3^{2n}d^{2n}\boldsymbol{\Lambda}}{\pi^n}\:
\chi[O](\gr{\Lambda})\: D^\dag (\boldsymbol{\Lambda})\:, \\ 
O &= \int_{\rr^{2n}} \frac{\kappa_3^{2n}d^{2n}\boldsymbol{K}}{\pi^n}\: \chi[O](\gr{K})\:
D^\dag (\boldsymbol{K})\:,
\end{align}
\end{subequations}
with
$d^{2n}\boldsymbol{\Lambda}=
d^{2n}\boldsymbol{K}= \prod_{k=1}^n
d\rma_k\, d\rmb_k$ and
\bea
\chi[O](\boldsymbol{\Lambda}) =
\hbox{Tr}\left[O\:D(\boldsymbol{\Lambda}) \right]\,,
\qquad
\chi[O](\boldsymbol{K}) =
\hbox{Tr}\left[O\:D(\boldsymbol{K}) \right]
\label{defChiO2}\:.
\eea
\par
For the single-mode displacement operator the following
properties are immediate consequences of the definition. Let $\lambda,
\lambda_1,\lambda_2\in \cc$, then
\begin{align}
D^{\dag} (\lambda) = D(-\lambda)\,, \qquad D^* (\lambda) &= D (\lambda^*)\,,
\qquad D^{\sT} (\lambda) = D (-\lambda^*)\:,\\
\hbox{Tr}\left[D(\lambda)\right] &= \pi\, \delta^{(2)}(\lambda) \:,
\label{TrD} \\
D(\lambda_1)D(\lambda_2) =
D(\lambda_1&+\lambda_2)\: \exp\left\{\mbox{$\frac12$}(\lambda_1 \lambda_2^*
- \lambda_1^*\lambda_2)\right\}
\label{compD}\;.
\end{align}
The 2-dimensional complex $\delta$-function in \eq{TrD} is defined
as
\begin{eqnarray}
\delta^{(2)}(z)=\int_{\cc} \frac{d^2\lambda}{\pi^2}\:
\exp\left\{\lambda^* z - z^* \lambda\right\}=\int_{\cc} \frac{d^2\lambda}{\pi^2}\:
\exp\left\{i(\lambda^* z + z^* \lambda\right)\}\:.
\label{defDeltaCmpl}
\end{eqnarray}
Setting $\lambda=\rma+ i\rmb$ and using \eq{compD} we have
\begin{subequations}
\begin{align}
D^*(\lambda ) D(z) D(\lambda )&= D(z+2\rma) \exp\{-2i\rmb(\rma+\re{z})\}\,,\\
D^\dag (\lambda ) D(z)  D(\lambda )&= D(z) \exp\{z\lambda^*- z^* \lambda\}\,,
\\
D(\lambda ) D(z) D(\lambda ) &= D(z+2\lambda )\,,
\\
D^{\sT}(\lambda ) D(z) D(\lambda ) &= D(z+2i\rmb) \exp\{2i\rma(\rmb+\im{z})\}\,,\:.
\label{compD1}
\end{align}
\end{subequations}
\index{displacement operator!matrix elements} 
Matrix elements in the Fock (number) basis are given by
\begin{subequations}
\begin{align}
\langle n+d | D(\alpha) | n\rangle &=
\sqrt{\frac{n!}{(n+d)!}}\: e^{-\frac12 |\alpha|^2}\:
\alpha^d\: L_n^d (|\alpha|^2)\,, \\
\langle n | D(\alpha) | n+d\rangle &= 
\sqrt{\frac{n!}{(n+d)!}}\: e^{-\frac12 |\alpha|^2}\:
(-\alpha^*)^{d}\: L_n^d (|\alpha|^2)\,, \\
\langle n | D(\alpha) | n\rangle &= e^{-\frac12 |\alpha|^2}\: L_n (|\alpha|^2)
\label{dnm}\;,
\end{align}
\end{subequations}
$L_n^d(x)$ being Laguerre polynomials.
\par
\index{coherent states}
The displacement operator is strictly connected with coherent
states. For a single mode coherent states are defined as the eigenstates
of the mode operator, {\em i.e.} $a|\alpha\rangle = \alpha |\alpha\rangle$,
where $\alpha \in \cc$ is a complex number. The expansion in terms of Fock
states reads as follows
\begin{eqnarray}
|\alpha\rangle = e^{-\frac12 |\alpha|^2} \sum_{k=0}^\infty
\frac{\alpha^k}{\sqrt{k!}}\: |k\rangle
\label{coh}\:.
\end{eqnarray}
Using Eq.~(\ref{displa_mode}) it can be shown that coherent states may
be defined also as $|\alpha\rangle = D(\alpha)|0\rangle$, {\em i.e.}
the unitary evolution of the vacuum through the displacement operator.
Properties of coherent states, {\em e.g.} overcompleteness and nonorthogonality,
thus follows from that of displacement operator.
The expansion (\ref{coh}) in the
number state basis is recovered from the definition
$|\alpha\rangle = D(\alpha)|0\rangle$ by the normal ordering of the
displacement
\be D(\alpha) = e^{\alpha a^\dag} e^{-\frac12 |\alpha|^2}
e^{-\alpha^* a}\label{Dbch}\,,
\ee
and by explicit calculations.
\index{coherent states!multimode}
Multimode coherent states are defined accordingly as
$|\gr{\alpha}\rangle = D(\gr{\alpha}) |
\gr{0}\rangle$ where
$|\gr{\alpha}\rangle$ denotes the product state $\otimes_k
|\alpha_k\rangle$. Coherent states are (equal) {\em minimum
uncertainty} states, {\em i.e.} fulfill (\ref{HeisSG}) with equality
sign and, in addition, with uncertainties that are equal for
position- and momentum-like operators. In other words, the covariance
matrix of a coherent states coincides with that of the vacuum
state $\sigma_{kk}=V_{kk} = (4\kappa_1^2)^{-1}$, $\forall k=1,\ldots,n$.
\par
The following formula connects displacement operator with functions
of the number operator, 
\begin{eqnarray}
\nu^{a^\dag a} = \int_{\mathbb C} \frac{d^2 z}{\pi(1-\nu)}\:
\exp\left\{-\frac12 \frac{1+\nu}{1-\nu}\, |z|^2\right\}\: D(z)
\label{other}\;,
\end{eqnarray}
with special cases 
\begin{align}
|0\rangle\langle 0|  = \int_{\mathbb C} \frac{d^2 z}{\pi}\:
\exp\left\{-\mbox{$\frac12$} |z|^2\right\}\: D(z)\,, \qquad
(-1)^{a^\dag a} = \int_{\mathbb C} \frac{d^2 z}{2\pi}\:D(z)
\label{spe}\;.
\end{align}
Proof is straightforward upon using the normal ordering (\ref{Dbch})
for the displacement and expanding the exponentials before integration.
\par
From \eq{GlauberF}, for any operator $O$, we have
\begin{subequations}
\begin{align}
&\hbox{Tr}\left[O^\dag \:
D(z)\right] = \hbox{Tr}\left[O\:D^\dag(z)\right]\,, \\
&\hbox{Tr}\left[O^* \: D(z)\right] = \hbox{Tr}\left[O\:D^* (z)\right]\,, \\
&\hbox{Tr}\left[O^{\sT} \: D(z)\right] = \hbox{Tr}\left[O\:D^{\sT}(z)\right]
\label{propChi}\:.
\end{align}
\end{subequations}
Using Eqs.~(\ref{propChi}), (\ref{compD1}) and (\ref{spe}),
other single mode relations can be proved
\begin{subequations}
\begin{align}
\int_{\mathbb C} \frac{d^2 z}{\pi}\:
D(z)\: O\:  D (z) &= \Pi \:
\hbox{Tr}[\Pi \:O]\,, \\
\int_{\mathbb C} \frac{d^2 z}{\pi}\:
D(z)\: O\:  D^\dag (z) &= \hbox{Tr}\left[O\right]\,{\mathbb I}\,,
\label{s2}\;
\\
\int_{\mathbb C} \frac{d^2 z}{\pi}\:
D(z)\: O\:  D^* (z) &= O^{\sT}\,,
\\
\int_{\mathbb C} \frac{d^2 z}{\pi}\:
D(z)\: O\:  D^{\sT} (z) &= O^*\,.
\end{align}
\end{subequations}
where $\Pi = (-)^{a^\dag a}$ {\em i.e.} $\Pi=\sum_p (-)^p |p\rangle\langle
p|$ is the parity operator.
Using the notation set out in Section \ref{s:MatNot},
we introduce the two-mode states $\kket{D(z)} = D(z)\otimes {\mathbb I}
\kket{\ii}$
with $\kket{\ii}=\sum_p \ket{p}\otimes\ket{p}$. Then we have the completeness
relation
\begin{eqnarray}
\int_{\mathbb C} \frac{d^2 z}{\pi}\:
\kket{D(z)}\bbra{D(z)} =  {\mathbb I} \otimes {\mathbb I}
\label{complDZ}\;.
\end{eqnarray}
Other two-mode relations
\begin{subequations}
\begin{align}
\int_{\mathbb C} \frac{d^2 z}{\pi}\: D(z)\otimes D^*(z) &=
\kket{\ii}\bbra{\ii} = \sum_{p,q}
\ket{p}\bra{q}\otimes \ket{p}\bra{q}\,, \\
\int_{\mathbb C} \frac{d^2 z}{\pi}\: D(z)\otimes D^{\sT}(z) &=
\kket{\jj}\bbra{\jj} = \sum_{p,q} (-)^{p+q}\:
\ket{p}\bra{q}\otimes \ket{p}\bra{q}\,, \\
\int_{\mathbb C} \frac{d^2 z}{\pi}\: D(z)\otimes D (z) &= F =
\left(\kket{\jj}\bbra{\jj}\right)^\theta\,, \\
\int_{\mathbb C} \frac{d^2 z}{\pi}\: D(z)\otimes D^\dag(z) &= E =
\left(\kket{\ii}\bbra{\ii}\right)^\theta
\label{s3}\;,
\end{align}
\end{subequations}
where $\kket{\JJ}=\sum_p (-)^p\:
|p\rangle\otimes |p\rangle$, $(\cdots)^\theta$ denotes partial transposition, and
$E$ and $F$ are the swap operator and the parity-swap operator
respectively, the latter being defined as
\begin{eqnarray}
F = \sum_{p,q} (-)^{p+q}\:\ket{p}\bra{q} \otimes  \ket{q}\bra{p}
\label{defParSwap}\;.
\end{eqnarray}
The action of $E$ and $F$ on a generic two-mode state is given by
\begin{align}
E \big(\:\ket{\psi}_1 \otimes \ket{\varphi}_2\:\big) &= \ket{\varphi}_1
\otimes \ket{\psi}_2 \\
F \big(\:\ket{\psi}_1 \otimes \ket{\varphi}_2\:\big) &= (-)^{a^\dag_1 a_1}
\ket{\varphi}_1 \otimes(-)^{a^\dag_2 a_2} \ket{\psi}_2\:.
\end{align}
Notice that the operator asso\-ciated to bipartite state $|\jj\rangle\rangle$ 
is the parity operator defined above. Finally, notice that using 
properties of Hermite polynomials, it is easy to show that 
\begin{eqnarray}
\int_\rr dx\:  |x\rangle_\phi |x\rangle_\phi = \sum_{n=0}^{\infty} e^{-2in\phi}
|n\rangle |n\rangle \equiv |{\mathbbm F}_\phi\rangle\rangle
\label{xtwb}\;, 
\end{eqnarray}
{\em e.g.}
$|\ii\rangle\rangle = |{\mathbbm F}_0\rangle\rangle = \int_\rr dx\:  
|x\rangle_0 |x\rangle_0$ 
\cite{msa}
and 
$|\jj\rangle\rangle = |{\mathbbm F}_{\frac{\pi}{2}}\rangle\rangle = 
\int_\rr dx\:  |x\rangle_{\frac{\pi}{2}} |x\rangle_{\frac{\pi}{2}}$.
\subsection{Two-mode mixing}\label{ss:bs}
\index{beam splitter}
\index{two-mode mixing}
\index{SU(2) interaction}
The simplest example of two-mode interaction is the linear mixing
described by Hamiltonian terms of the form $H\propto a^\dag b + b^\dag a$.
For two modes of the radiation field it corresponds to a beam splitter,
{\em i.e.} to the interaction taking place in a linear optical medium such
as a dielectric plate.
The evolution operator can be recast in the form
\begin{eqnarray}
U(\zeta) = \exp\left\{\zeta a^\dag b - \zeta^* a b^\dag\right\}\:,
\label{ubs}\;
\end{eqnarray}
where the coupling $\zeta = \phi\, e^{i\theta} \in {\mathbb C}$ is
proportional to the interaction length (time) and to the linear
susceptibility of the medium.
Using the Schwinger two-mode boson representation of $\rmSU(2)$ algebra
\cite{schwinger},
{\em i.e.} $J_+= a^\dag b$, $J_-=(J_+)^\dag=ab^\dag$ and $J_3=
\frac12[J_+,J_-]=\frac12 (a^\dag a - b^\dag b)$,
it is possible to {\em disentangle} the evolution operator 
\cite{magnus,wilcox67,witschel}, thus achieving 
the normal ordering either in the mode $a$ or in the mode $b$
\begin{align}
U(\zeta) &= \exp\left\{ \zeta J_+ - \zeta^* J_- \right\}\nonumber\\
&=\exp\left\{\mbox{$\frac{\zeta}{|\zeta|}$} \tan |\zeta|\, J_+ \right\}
\exp\left\{\log (1+ \tan|\zeta|^2)\, J_3\right\}
\exp\left\{- \mbox{$\frac{\zeta^*}{|\zeta|}$} \tan |\zeta|\, J_- \right\}
\nonumber \\
&= \pexp{e^{i\theta}\tan\phi\, a^\dag b}\:
\left(\cos^2\phi\right)^{b^\dag b - a^\dag a}\:
\pexp{-e^{-i\theta}\tan\phi\, a b^\dag } \nonumber\\
&= \pexp{-e^{-i\theta}\tan\phi\, a b^\dag }\:
\left(\cos^2\phi\right)^{a^\dag a - b^\dag b}\:
\pexp{e^{i\theta}\tan\phi\, a^\dag b}\:
\label{BCHJubs}\;.
\end{align}
Eq.~(\ref{BCHJubs}) are often written introducing the 
quantity $\tau=\cos^2\phi$, which is referred to as the  
{\em transmissivity} of the beam splitter.
Mode evolution under a unitary action can be obtained using
the Hausdorff recursion formula
\begin{align}
e^{\alpha A} B\, e^{-\alpha A} &= B + \alpha\:[A,B]
+\frac{\alpha^2}{2!}\:[A,[A,B]] + 
\frac{\alpha^3}{3!} [A,[A,[A,B]]] + \ldots\\
&= \sum_k \frac{\alpha^k}{k!}\: \{\!\{ A^k,B \}\!\}  \equiv B_\alpha \:, 
\label{hsd}
\end{align}
where $\{\!\{A,B\}\!\}=[A,B]$ and $\{\!\{A^k,B\}\!\}=[A,\{\!\{A^{k-1},B\}\!\}]$.
Eq.~(\ref{hsd}) generalizes to $e^{\alpha A} B^n e^{-\alpha A} = B_\alpha^n$
and $e^{\alpha A} e^{B} e^{-\alpha A} = e^{B_\alpha}$ \cite{froh}.
The Heisenberg evolution of modes $a$ and $b$ under the action of $U(\zeta)$
is thus given
\begin{eqnarray}
U^\dag(\zeta) \left(\begin{array}{c}a \\
 b\end{array}\right) U(\zeta)=
\gr{B}_\zeta \:
\left(\begin{array}{c}a \\ b\end{array}\right)
\label{EvolMode}\:,
\end{eqnarray}
where the unitary matrix $\gr{B}_\zeta$ is given by
\begin{eqnarray}
\gr{B}_\zeta = \left(\begin{array}{cc} \cos\phi& e^{i\theta}\sin\phi\\ [1ex]
- e^{-i\theta}\sin\phi& \cos\phi \end{array}\right)
\label{Matrixbs}\;.
\end{eqnarray}
Correspondingly, we have 
$U^\dag (\zeta)\, \gr{S}\, U( \zeta) = \gr{N}_\zeta \gr{S}$  and 
$U^\dag (\zeta)\, \gr{R}\, U( \zeta) = \gr{N}'_\zeta \gr{R}$, 
where $\gr{N}'_\zeta = \gr{P}_{23}\, \gr{N}_\zeta\, \gr{P}_{23}\, \gr{S}$.
The $4\times 4$ orthogonal symplectic matrix, obtained from (\ref{Matrixbs})
as described in Eq.~(\ref{OrtSymp}), is given by
\begin{eqnarray}
\gr{N}_\zeta=\left(\begin{array}{cc}
\re{\gr{B}_\zeta} &  - \im{\gr{B}_\zeta} \\ [1ex]
 \im{\gr{B}_\zeta} & \re{\gr{B}_\zeta}
\end{array}\right)
\label{RealMatrixbs}\;,
\end{eqnarray}
and describes the symplectic transformation of two-mode mixing, whereas 
$\gr{P}_{23}$ is the permutation matrix
\begin{eqnarray}
\gr{P}_{23}=\left(\begin{array}{cccc}
1 &0 &0 &0\\
0 &0 &1 &0\\
0 &1 &0 &0\\
0 &0 &0 &1
\end{array}\right)\,.
\end{eqnarray}
The two-mode covariance matrices evolve accordingly, {\em i.e.} as  
$\gr{\sigma} \rightarrow \gr{N}'_\zeta\, \bmsigma\, \gr{N}_\zeta^{\prime\sT}$
and 
$\gr{V} \rightarrow \gr{N}_\zeta\,\bmV\, {\gr{N}_\zeta}^{\sT} $.
If $\varrho$ is the two-mode density matrix before the mixer and 
$\varrho'=U(\zeta)\,\varrho\, U^\dag(\zeta)$ that of the evolved state
it is straightforward to show, using Eq.~(\ref{EvolMode}), that 
\begin{align}
\langle a^\dag a \rangle_{\varrho'}&= 
\langle a^\dag a \rangle_{\varrho}\cos^2\phi +
\langle b^\dag b \rangle_{\varrho}\sin^2\phi + 
\langle \mbfrac (a^\dag b\:e^{i\theta}+ b^\dag a\: e^{-i\theta})
\rangle_{\varrho}\sin (2\phi)\,,
\\
\langle b^\dag b \rangle_{\varrho'}&= 
\langle a^\dag a \rangle_{\varrho}\sin^2\phi +
\langle b^\dag b \rangle_{\varrho}\cos^2\phi -
\langle \mbfrac (a^\dag b\:e^{i\theta}+ b^\dag a\:e^{-i\theta})
\rangle_{\varrho}\sin (2\phi)\,,
\end{align}
and therefore
\begin{align}
\langle a^\dag a + b^\dag b \rangle_{\varrho'}&= 
\langle a^\dag a + b^\dag b \rangle_{\varrho}\,,  \label{mixN}\\
\langle a^\dag a - b^\dag b \rangle_{\varrho'}&= 
\langle a^\dag b\: e^{i\theta}+ b^\dag a\: e^{-i\theta}
\rangle_{\varrho}\sin (2\phi)
\label{mixD}\:.
\end{align}
Eq.~(\ref{mixN}) says that the total number of quanta in the two modes
is a constant of motion: this is usually summarized by saying that a
two-mode mixer is a {\em passive} device. It also implies that the
vacuum is invariant under the action of $U(\zeta)$, {\em i.e.}
$U(\zeta) |\gr{0}\rangle = |\gr{0}\rangle$, where
$|\gr{0}\rangle=|0\rangle\otimes |0\rangle$.
The two-mode displacement operator evolve as follows
\begin{eqnarray}
U(\zeta)^\dag\, D(\gr{\lambda})\, U(\zeta) = D(\gr{B}_\zeta^\dag \gr{\lambda})
\label{EvolD}\;,
\end{eqnarray}
and thus the evolution of coherent states is given by
$U(\zeta) |\gr{\alpha}\rangle\rangle = |\gr{B}_\zeta^\dag 
\gr{\alpha}\rangle\rangle$.
Analogously, $U^\dag (\zeta)\, D(\gr{\Lambda})\, U(\zeta) =
D(\gr{N}_\zeta^{\prime -1} \gr{\Lambda})$ and $U^\dag (\zeta)\, D(\gr{K})\,
U(\zeta) = D({\gr{N}_\zeta}^{-1}  \gr{\Lambda})$. 
\subsection{Single-mode squeezing}\label{ss:squeeze}
\index{squeezing operator}
\index{SU(1,1) interactions}
We observe the phenomenon of {\em squeezing} when an observable
or a set of observables shows a second moment which is below the 
corresponding vacuum level. Historically, squeezing has been firstly 
introduced for quadrature operators \cite{yuen76}, which led to consider the 
squeezing operator analyzed in this section. Squeezing transformations 
correspond to Hamiltonians of the
form $H\propto (a^{\dag})^{2} + h.c. $. The evolution operator is
usually written as
\begin{eqnarray}
S (\xi) = \exp\left\{\mbfrac \xi (a^{\dag})^{2} - \mbfrac \xi^* a^2\right\}\:,
\label{usq}
\end{eqnarray}
corresponding to mode evolution given by
\begin{eqnarray}
S^\dag(\xi) \: a \: S(\xi) = \mu a + \nu a^\dag\,, \qquad
S^\dag(\xi) \: a^\dag \: S(\xi) = \mu a^\dag + \nu^* a 
\label{modesq}\;,
\end{eqnarray}
where $\mu \in {\rr}$, $\nu \in \cc$, $\mu=\cosh r$, $\nu = e^{i\psi} 
\sinh r$, $\xi=r e^{i\psi}$. 
Using the two-boson representation of the ${\rm SU} (1,1)$ algebra 
$K_+ = \frac12\, a^{\dag 2}$, $K_-=(K_+)^\dag$, $K_3=-\frac12[K_+,K_-]=
\frac12 (a^\dag a + \frac12)$, it is possible to {\em disentangle} $S(\xi)$, 
achieving normal orderings of mode operators
\begin{align}
S(\xi) &= \pexp{\xi K_+ - \xi^* K_-} \nonumber\\
&= \pexp{\mbox{$\frac{\xi}{|\xi|}$}\,K_+}\:
\pexp{\log(1-\tanh^2 |\xi|)K_3} \pexp{-\mbox{$\frac{\xi^*}{|\xi|}$}\,K_-}
\nonumber \\
&= \pexp{\mbox{$\frac{\nu}{2\mu}$}\, (a^{\dag})^{2}}
\mu^{-(a^\dag a +\frac12)}
\pexp{-\mbox{$\frac{\nu^*}{2\mu}$}\,a^{2}}\:,
\end{align}
from which one also obtain the action of the squeezing operator
on the vacuum state $|\xi\rangle=S (\xi) |0\rangle$.
The state $|\xi\rangle$ is the known as {\em squeezed vacuum} state.
Expansion over the number basis contains only even components {\em i.e.}
\begin{eqnarray}
|\xi\rangle= \frac{1}{\sqrt{\mu}} \sum_{k=0}^{\infty} 
\left(\frac{\nu}{2\mu}\right)^k \: \frac{\sqrt{(2k)!}}{k!} \: |2k\rangle
\label{sqvacdef}\;.
\end{eqnarray}
Despite its name, the squeezed vacuum is not empty and the mean photon
number is given by $\langle \xi | a^\dag a | \xi\rangle = |\nu|^2$, whereas
the expecta\-tion value of qua\-dra\-ture opera\-tor vanishes $\langle \xi
| x_\theta| \xi\rangle = 0$, $\forall \theta$. Quadrature variance $\Delta
x_\theta^2$ is thus given by 
\begin{eqnarray}
\Delta x_\theta^2 = \langle \xi | x_\theta^2 | \xi \rangle = 
\frac{1}{4\kappa_1^2}\left[e^{2r} \cos^2(\theta-\mbfrac\psi) + 
e^{-2r} \sin^2(\theta-\psi/2) \right]\:.
\end{eqnarray}
Squeezed vacuum is thus a minimum uncertainty state for 
the pair of observables $x_{\psi/2}$ and $x_{\psi/2+\pi/2}$, 
for which we have  
$\Delta x_{\psi/2}^2 = (4\kappa_1^2)^{-1} e^{2r}$ and 
$\Delta x_{\psi/2+\pi/2}^2 = (4\kappa_1^2)^{-1} e^{-2r}$, respectively.
Applying the displacement operator to the squeezed vacuum
one obtain the class of {\em squeezed states} $|\alpha,\xi\rangle
=D(\alpha)S(\xi)|0\rangle$. Squeezed states are still 
minimum uncertainty states for the pair of observables 
$x_{\psi/2}$ and $x_{\psi/2+\pi/2}$. However, the photon
distribution is no longer characterized by the odd-number 
suppression of the squeezed vacuum.
Notice that the evolution of the displacement operator is given 
by $ S^\dag (\xi) D(\lambda) S(\xi) = D(\mu\lambda - \nu \lambda^*)$,
and that $S(\xi)D(\alpha)=D(\mu\alpha+\nu\alpha^*)S(\xi)$. Therefore, 
application of the squeezing operator to coherent states leads to
a squeezed state of the form
$ S(\xi)|\alpha\rangle = |\mu\alpha+\nu\alpha^*,\xi\rangle$.
\par
Properties of quantum states obtained by squeezing number \cite{mskim89} 
and thermal state \cite{marians} have been extensively studied. 
In general, if $\varrho'=S(\xi)\varrho S(\xi)$ is the state after
the squeezer, the total number of photon is given by 
\begin{eqnarray}
\langle a^\dag a \rangle_{\varrho'} = 
\sinh^2 r + (2 \sinh^2 r +1)\langle a^\dag a \rangle_{\varrho} + 
\sinh (2r) \langle a^2\: e^{-i\psi} + a^{\dag 2}\: e^{i\psi} 
\rangle_{\varrho} \label{sqN}\:.
\end{eqnarray}
Mode evolutions in Cartesian representation are given by $\gr{R} \rightarrow 
\gr{\Sigma}_\xi \gr{R}$ and $\gr{\sigma} \rightarrow 
\gr{\Sigma}_\xi\, \gr{\sigma}\, \gr{\Sigma}_\xi^{\sT}$ ($\gr{S}\equiv
\gr{R}$ and  $\gr{\sigma}\equiv\gr{V}$ since we have 
only one mode) where the symplectic squeezing matrix is given by
\begin{eqnarray}
\gr{\Sigma}_\xi =  \mu \ii_2 + \gr{R}_\xi \qquad 
\gr{R}_\xi =
\left(\begin{array}{cc} \re{\nu} & \im{\nu} \\ [1ex]
\im{\nu} & - \re{\nu}\end{array}\right)
\label{Smatrix}\;.
\end{eqnarray}
\subsection{Two-mode squeezing}\label{ss:opa}
\index{two-mode squeezing}
\index{SU(1,1) interactions}
\index{twin-beam}
Two-mode squeezing transformations correspond to Hamiltonians of the
form $H\propto a^\dag b^\dag + h.c. $. The evolution operator 
is written as
\begin{eqnarray}
S_2 (\xi) = \exp\left\{\xi a^{\dag} b^{\dag} - \xi^* a b\right\}\:,
\label{usq2}
\end{eqnarray}
where the complex coupling $\xi$ is again written as $\xi=re^{i\psi}$.
The corresponding two mode evolution is given by
\begin{eqnarray}
S_2^\dag(\xi) \: 
\left(\begin{array}{c}a \\ b^\dag \end{array}\right) 
\: S_2(\xi) = 
\gr{S}_{2\xi}
\left(\begin{array}{c}a \\ b^\dag \end{array}\right) 
\:,\label{modesq2}
\end{eqnarray}
where $\gr{S}_{2\xi}$ denotes the matrix 
\begin{eqnarray}
\gr{S}_{2\xi} = \left(\begin{array}{cc}\mu & \nu \\ 
\nu^* & \mu\end{array}\right)\;. \label{matrixsq2}
\end{eqnarray}
As for single mode squeezing we have $\mu=\cosh r$ and 
$\nu = e^{i\psi} \sinh r$.
A different two-boson realization of the ${\rm SU} (1,1)$ algebra, namely
$K_+ = a^\dag b^\dag$, $K_-=(K_+)^\dag$, $K_3 = -\frac12 [K_+,K_-]=\frac12
(a^\dag a + b^\dag b + 1)$, allows to put $S_2(\xi)$ in the normal ordering
for both the modes 
\begin{eqnarray}
S_2(\xi) = \pexp{\mbox{$\frac{\nu}{\mu}$}\, a^\dag b^\dag} \: 
\mu^{\frac12 (a^\dag a + b^\dag b + 1)} \:
\pexp{-\mbox{$\frac{\nu^*}{\mu}$}\, a b}
\label{bchusq2}\;.
\end{eqnarray}
A two-mode squeezer is an {\em active} devices, {\em i.e.} it adds
energy to the incoming state. According to Eqs.~(\ref{modesq2}) and 
(\ref{matrixsq2}), with $\varrho'=S_2 (\xi)\, \varrho\, S^\dag_2(\xi)$ 
we have 
\begin{align}
&\langle a^\dag a \rangle_{\varrho'} =
\cosh^2 r \langle a ^\dag a \rangle_{\varrho} + 
\sinh^2 r (1+ \langle b^\dag b \rangle_{\varrho}) \nonumber\\
&\hspace{4cm}
+\mbfrac  \sinh (2r)  \langle a\:b\:e^{-i\psi} + a^\dag b^\dag \:
e^{i\psi}\rangle_{\varrho}\,,
\\
&\langle b^\dag b \rangle_{\varrho'} =
\sinh^2 r (1+ \langle a ^\dag a \rangle_{\varrho}) + 
\cosh^2 r \langle b^\dag b \rangle_{\varrho} \nonumber\\
&\hspace{4cm}
+\mbfrac  \sinh (2r) 
\langle a\:b\:e^{-i\psi} + a^\dag b^\dag \: e^{i\psi}\rangle_{\varrho}
\:, 
\end{align}
and therefore
\begin{align}
\langle a^\dag a + b^\dag b \rangle_{\varrho'} &=
2 \sinh^2 r\: (1+ \langle a^\dag a + b^\dag b \rangle_{\varrho})
\nonumber\\
&\hspace{2cm}
+ \sinh (2r)\: \langle a\:b\:e^{-i\psi} + a^\dag b^\dag \: 
e^{i\psi}\rangle_{\varrho}\,, \label{sq2N}
\\
\langle a^\dag a - b^\dag b \rangle_{\varrho'} &=
\langle a^\dag a - b^\dag b \rangle_{\varrho} \label{sq2D} \:. 
\end{align}
The difference in the mean photon number is thus a constant of motion.
The action of $S_2(\xi)$ on the vacuum can be evaluated starting 
from Eq.~(\ref{bchusq2}). The resulting state is given by
\begin{eqnarray}
S_2(\xi)|\gr{0}\rangle = \frac{1}{\sqrt{\mu}} \sum_{k=0}^\infty
\left(\frac{\nu}{\mu}\right)^k \: |k\rangle \otimes |k\rangle
\label{twbdef}\;
\end{eqnarray}
and is known as {\em two-mode squeezed vacuum} or {\em twin-beam state}
(TWB).  The second denomination refers to the fact that TWB shows perfect
correlation in the photon number, {\em i.e} is an eigenstate of the photon
number difference $a^\dag a - b^\dag b$, which is a constant of motion. 
TWB will be also denoted as $|\Lambda
\rangle\rangle$ where, adopting the notation introduced in Section
\ref{s:MatNot}, $\Lambda$ is the infinite matrix $\Lambda
=\sqrt{1-|\lambda|^2}\, \lambda^{a^\dag a}$,  with $\lambda=\nu/\mu =
e^{i\psi}\tanh r$.  Often, by a proper choice of the reference phase, it
will be enough to consider  $\lambda$ as real.  On the other hand, the first 
name is connected to a duality under the action of two-mode mixing. Consider a
balanced mixer with evolution operator $U(\zeta=\frac{\pi}{4}\,
e^{i\theta})$, then we have 
\begin{eqnarray}
U^\dag(\mbox{$\frac\pi4$}\, e^{i\theta}) \:S_2(\xi)\:
U(\mbox{$\frac\pi4$}\,e^{i\theta})
= S(\xi e^{i\theta}) \otimes 
S(-\xi e^{-i\theta})
\label{duality}\;,
\end{eqnarray}
where $S(\xi)$ are single-mode squeezing operators acting on the
evolved mode out of the mixer. In other words, a TWB entering a 
balanced beam-splitter is transformed into a factorized states
composed of two squeezed vacuum with opposite squeezing phases \cite{joint}.
{\em Viceversa}, a TWB may be generated using single-mode squeezers and
a linear mixer \cite{furu}. 
Using Eq.~(\ref{xtwb}) we may also write 
$$ |\Lambda\rangle\rangle = \sqrt{1-|\lambda|^2}|\lambda|^{a^\dag a} \: 
|{\mathbbm F}_\psi \rangle\rangle = 
\sqrt{1-|\lambda|^2}|\lambda|^{b^\dag b} \: 
|{\mathbbm F}_\psi \rangle\rangle \:.  $$
Finally, the symplectic transformation 
associated to the two-mode squeezer is represented 
by the block matrix $\gr{\Sigma}_{2\xi}$. We have 
$S_{2\xi}\gr{R}S^\dag_{2\xi} = \gr{\Sigma}_{2\xi}\gr{R}$
and
$S_{2\xi}\gr{S}S^\dag_{2\xi} = \gr{P}_{23}\gr{\Sigma}_{2\xi}\gr{P}_{23}
\gr{S}$ with
\begin{eqnarray}
\gr{\Sigma}_{2\xi} =\left(
\begin{array}{cc} 
 \mu \ii_2 & \gr{R}_\xi\\ [1ex]
 \gr{R}_\xi & \mu \ii_2 
\end{array}
\right)\,, \qquad 
\gr{\Sigma}_{2\xi}^{-1} =\left(
\begin{array}{cc} 
 \mu \ii_2 & -\gr{R}_\xi^{\sT}\\ [1ex]
 -\gr{R}_\xi^{\sT} & \mu \ii_2 
\end{array}
\right) 
\label{Qxi}\;,
\end{eqnarray}
where $\gr{R}_\xi$ is defined as in (\ref{Smatrix}), and the inverse is
evaluated using Eq.~(\ref{Invblock}).
\subsection[Multimode interactions: ${\rm SU}(p,q)$ Hamiltonians]{Multimode
interactions: ${\rm\bf SU} \boldsymbol{(p,q)}$ Hamiltonians}\label{ss:Hpq}
\index{SU(p,q) interactions}
\index{multimode interactions}
Let us consider the set of Hamiltonians expressed by 
\begin{equation}
H_{pq}=\sum_{l<k=1}^{p} \gamma_{kl}^{(1)}\, a_k a^{\dag}_l
+ \sum_{l<k=1}^{q} \gamma_{kl}^{(2)} \, b_k b^{\dag}_l
+ \sum_{k=1}^{p}\sum_{l=1}^{q} \gamma_{kl}^{(3)}\, a_k b_l + h. c.
\label{LinHpq}\;,
\end{equation}
where we have partitioned the modes in two groups $a_k$, $k=1,\ldots,p$
and $b_l$, $l=1,\ldots,q$, where $p+q=n$, with the properties that the
interactions among modes of the two groups takes places only
through terms of the form $a_k b_l + h. c.$. 
Hamiltonians (\ref{LinHpq}) form a subset of Hamiltonians of 
the form (\ref{LinH}). The conserved quantity is the difference $D$ 
between the total mean photon number of the $a$ modes and the 
$b$ modes, in formula 
\begin{eqnarray}
D= \sum_{k=1}^p a^\dag_k a_k - \sum_{l=1}^q b^\dag_l b_l
\label{pqD}\;
\end{eqnarray}
The transformations induced by Hamiltonians (\ref{LinH}) correspond 
to the unitary representation of the ${\rm SU}(p,q)$ algebra \cite{Puri94}. 
Therefore, the set of states obtained from the vacuum coincides with 
the set of ${\rm SU}(p,q)$ coherent states {\em i.e.}
\index{coherent states!SU(p,q)}
\begin{eqnarray}
|\gr{C}_{pq}\rangle\equiv
\pexp{-i H_{pq} t} |\gr{0}\rangle = \pexp{\sum_{k=1}^p\sum_{l=1}^q 
\beta_{kl} a_k^\dag b^\dag_l - h.c.} |\gr{0}\rangle
\label{pqcoh}\;,
\end{eqnarray}
where $\beta_{kl}$ are complex numbers parametrizing the state.
Upon defining 
$$\alpha_{kl} = \beta_{kl}\frac{\tanh\left(\sum_{r=1}^p 
|\beta_{rl}|^2\right)}{\sum_{r=1}^p |\beta_{rl}|^2}\:,$$
$|\gr{C}_{pq}\rangle$ in Eq.~(\ref{pqcoh}) can be explicitly 
written as 
\begin{align}
|\gr{C}_{pq}\rangle = \sum_{\{\gr{m}\}} \sum_{\{\gr{t}\}}
\prod_{k=1}^p \prod_{l=1}^q
\: \alpha_{kl}^{t_{kl}} 
\frac{\sqrt{m_k ! \left(\sum_{r=1}^{p} t_{rl}\right) \mbox{\large !}}}{t_{kl}!}
\: \left|\left.\{\gr{m}\}; \sum_{r=1}^p t_{r1}, 
\sum_{r=1}^p t_{r2}, ..., \sum_{r=1}^p t_{rq} \right\rangle\right.
\label{cumb}\;
\end{align}
where $t_{kq}=m_k - \sum_{h=1}^{q-1} p_{kh}$, 
$\{\gr{m}\}=\{m_1,m_2,...,m_p\}$ and the sums over $\gr{m}$ and 
$\gr{t}$  are extended over natural numbers. 
In the special case $q=1$, Eq.~(\ref{cumb}) reduces to a simpler form, 
we have that $|\gr{C}_{p1}\rangle\equiv |\gr{C}_{p}\rangle$ is given by
\begin{align}
|\gr{C}_p\rangle =& \sqrt{N_p}\sum_{\{\gr{m}\}} 
\frac{\alpha_1^{m_1} \alpha_2^{m_2}... \alpha_p^{m_p}\: \sqrt{(m_1+m_2+...
+m_p)!}}{\sqrt{m_1! m_2! ... m_p!}}\: 
|\{\gr{m}\}; \sum_{k=1}^p n_p \rangle
\label{Cp1}\;
\end{align}
where $N_p= 1-\sum_{k=1}^p |\alpha_k|^2$ is a normalization factor.
\section{Characteristic function and Wigner function}
\index{characteristic function}
\index{Wigner function}
\label{s:Wchi}
The characteristic function of a generic operator $O$ has been
introduced  in Eq.~(\ref{defChiO}). For a quantum state $\varrho$
we have $\chi [\varrho](\boldsymbol{\lambda}) = \hbox{Tr}\left[\varrho\:
D(\boldsymbol{\lambda})\right]$. 
In the following, for the sake of simplicity, we will sometime 
omit the explicit dependence on $\varrho$. The characteristic function
$\chi(\gr{\lambda})$ is also known as the moment-generating function
of the signal $\varrho$, since its derivatives in the origin of the complex
plane generates symmetrically ordered moments of mode operators. 
\index{operator ordering}
In formula
\begin{eqnarray}
\left.
(-)^q \frac{\partial^{p+q}}{\partial \lambda_k^p\partial \lambda_l^{*q}}
\chi (\gr{\lambda})\right\vert_{\gr{\lambda}=\gr{0}}=
\hbox{Tr} \left[\varrho\: \left[(a_k^{\dag})^{p} a^q _l\right]_S \right]
\label{Mgf1}\:.
\end{eqnarray}
For the first non trivial moments we have $[a^\dag a]_S
=\frac12 (a^\dag a + a a^\dag)$, $
[a a^{\dag 2}]_S=
\frac13 (a^{\dag 2} a + a a^{\dag 2} + a^\dag a a^\dag)$, 
$[a^\dag a^{2}]_S= \frac13 (a^{2} a^\dag  + a^\dag a^{2} + a^\dag a)$
\cite{Carmichael}.
In order to evaluate the symmetrically ordered form of generic
moments, one should expand the exponential in the displacement
operator 
\begin{align}
D(\lambda) &= \sum_{k=0}^\infty \frac{1}{k!} (\lambda a^\dag - \lambda^* a)^k = 
\sum_{k=0}^\infty \frac{1}{k!} \sum_{l=0}^k \binom{k}{l} 
\lambda^k \lambda^*_l [a^{\dag k} a^l]_S \nonumber \\ &= \sum_{k=0}^\infty
\sum_{l=0}^\infty \frac{\lambda^k \lambda^*_l}{k!\, l!}
[a^{\dag k} a^l]_S
\label{expandD}\;.
\end{align}
Using  Eqs.~(\ref{GlauberF}), (\ref{defChiO}) and (\ref{TrD}) it can 
be shown (see Section \ref{ss:trace} for details) that for any pair of 
generic operators acting on the Hilbert space of $n$ modes we have
\be
\hbox{Tr}\left[O_1\: O_2 \right] = \frac{1}{\pi^{n}}
\int_{\cc^n} d^{2n}\!\boldsymbol{\lambda}\: \chi[O_1] (\gr{\lambda})
\: \chi[O_2] (-\gr{\lambda})\:, \label{Xtrace}
\ee
which allows to evaluate a quantum trace as a phase-space
integral in terms of the characteristic function.
Other properties of the characteristic function follow from the 
definition, for example we have $\chi[O](0) = \hbox{Tr}[O]$ and 
\begin{align}
\int_{\cc^n} \frac{d^{2n}\gr{\lambda}}{(2\pi)^n}\: \chi[O](\gr{\lambda}) = 
\hbox{Tr}\left[O\:\gr{\Pi}\right] \qquad
\int_{\cc^n} \frac{d^{2n}\gr{\lambda}}{\pi^n}\:
\big|\chi[O](\gr{\lambda})\big|^2 = \hbox{Tr}[O^2]
\label{OpropChi}\;,
\end{align}
where $\gr{\Pi}= \otimes_{k=1}^n(-)^{a^{\dag}_k a_k}= 
(-)^{\sum_{k=1}^n a^{\dag}_k a_k}$ is the tensor product of
the parity operator for each mode.
\par
The so-called Wigner function of the operator $O$, and in particular the
Wigner function associated to the quantum state $\varrho$, is defined as the
Fourier transform of the characteristic function as follows
\bea
W[O](\gr{\alpha}) = \int_{\cc^n} \frac{d^{2n}\!\gr{\lambda}}{\pi^{2n}}\:
\exp\left\{\gr{\lambda}^\dag \gr{\alpha}+ \gr{\alpha}^\dag \gr{\lambda}
\right\} \: \chi[O](\gr{\lambda})\:.
\label{defW}
\eea
The Wigner function of density matrix $\varrho$,
namely $W[\varrho](\bmalpha)$, is a {\em quasiprobability} 
for the quantum state. Using the formula on the right of Eqs.~(\ref{OpropChi})
we have that $\chi[\varrho](\bmlambda)$ is a square integrable function
for any quantum state $\varrho$. Therefore, the Wigner function is a well 
behaved function for any quantum state. In other words, although it may assume 
negative values, it is bounded and regular and can be used to evaluate 
expectation values of symmetrically ordered moments. Starting from
Eq.~(\ref{Mgf1}) and using properties of the Fourier transform it is
straightforward to prove that 
\begin{eqnarray}
\int_{\cc^n} d^{2n}\gr{\alpha} \: W[\varrho](\gr{\alpha})\:
\alpha ^k (\alpha^{*})^l = 
\hbox{Tr}\left[\varrho\: \left[(a^{\dag})^{l} a^k\right]_S \right] \:.
\end{eqnarray}
More generally (see Section \ref{ss:trace}) we have that 
\be\label{Wtrace}
\hbox{Tr}\left[O_1\: O_2 \right] = \pi^{n}
\int_{\cc^n} d^{2n}\!\boldsymbol{\alpha}\: W[O_1] (\gr{\alpha})
\: W[O_2] (\gr{\alpha})\:.
\ee
Notice that the identity operator for $n$ modes has 
a Wigner function given by $W[{\mathbbm I}] (\gr{\alpha})=
\pi^{-n}$. Indeed we have 
$\hbox{Tr}[O]=\int_{\cc^n} d^{2n}\gr{\alpha} \: W[O](\gr{\alpha})$.
The analogue of Eq.~(\ref{GlauberF}) reads as follows 
\begin{align}
O= 2^n \int_{\cc^n} d^{2n} \gr{\alpha}\: W[O](\gr{\alpha}) 
\: D(\gr{\alpha}) \: \gr{\Pi}\:D^\dag(\gr{\alpha}) 
\label{WigF} \:,
\end{align}
from which follows a trace form for the Wigner function 
\index{Wigner function!trace rule}
\begin{eqnarray}
W[O](\gr{\alpha}) = \left(\frac{2}{\pi}\right)^n
\hbox{Tr}\left[O\:D(\gr{\alpha})\:\gr{\Pi}\:D^\dag(\gr{\alpha})\right]\:
\label{WdefTr}\;.
\end{eqnarray}
Other forms of the Eqs.~(\ref{WigF}) and (\ref{WdefTr}) can be obtained
by means of the identity $ D(\gr{\alpha})\gr{\Pi} D^\dag(\gr{\alpha}) = 
D(2\gr{\alpha})\gr{\Pi}=\gr{\Pi}D^\dag(2\gr{\alpha})$.
\par
The Wigner function in Cartesian coordinates is also obtained from the
corresponding characteristic function by Fourier transform. 
Let us define the vectors
\begin{align}
\boldsymbol{X} = (x_1,y_1,\ldots,x_n,y_n)^{\sT}\,,
 \qquad
\boldsymbol{Y} = (x_1,\ldots,x_n,y_1,\ldots, y_n)^{\sT}
\label{defVecsXY}\;,
\end{align}
where $\alpha_k=\kappa_2(x_k+iy_k)$. Notice that the scaling
coefficients $\kappa_2$ and $\kappa_3$ are not independent one each other, but should
satisfy $2\kappa_2\kappa_3=1$. To show this, consider the $n$-mode
extension of \refeq{defDeltaCmpl}
\be 
\delta^{(2n)}(\bmalpha)=\int_{\cc^{\,n}}
\frac{d^{2n}\bmlambda}{\pi^{2n}}\: \exp \left\{i(\bmlambda^* \bmalpha + \bmalpha^*
  \bmlambda)\right\} 
\label{defDeltaCmplNDim}
\ee
from which follows that
\begin{align}
\delta^{(2n)}(\gr{X}) &= \int_{\rr^{2n}} \frac{d^{2n}
\boldsymbol{\Lambda}}{(2\pi)^{2n}}\, (2\kappa_2\kappa_3)^{2n} \: 
\exp\left\{ 2 i \kappa_2 \kappa_3
\boldsymbol{\Lambda}^{\sT} \boldsymbol{X}\right\}\,
\label{defDeltaCartAux}\;.
\end{align}
The identity
\be
\delta(x)=\int_{\rr}\frac{da}{2\pi}e^{iax}
\label{DeltaMono}
\ee
implies then that $2\kappa_2\kappa_3=1$, as we claimed above. The
corresponding definition of the Fourier transform allows to obtain the
Wigner function in Cartesian coordinates as
\begin{subequations}
\begin{align}
W[O](\gr{X}) &= \int_{\rr^{2n}}
\frac{d^{2n} \boldsymbol{\Lambda}}{(2\pi)^{2n}} \:
\exp\left\{ i \boldsymbol{\Lambda}^{\sT}
\boldsymbol{X}\right\} \chi[O] (\gr{\Lambda})
 \label{defWCarta}\,, \\
W[O](\gr{Y}) &= \int_{\rr^{2n}}
\frac{d^{2n} \boldsymbol{K}}{(2\pi)^{2n}} \:
\exp\left\{ i \boldsymbol{K}^{\sT}
\boldsymbol{Y} \right\} \chi[O](\gr{K})
\label{defWCartb}\;.
\end{align}
\label{defWCart}
\end{subequations}
\par
Notice that in the literature different definitions equivalent to
\refeq{defDeltaCmplNDim} of the $n$-mode complex $\delta$-function
are widely used, which correspond to a change of coordinates in
\refeq{defWCart}. As an example, if one consider \cite{cahill}
\be 
\delta^{(2n)}(\bmalpha)=\int_{\cc^{\,n}}
\frac{d^{2n}\bmlambda}{\pi^{2n}}\: \exp\left\{\bmlambda^* \bmalpha - \bmalpha^*
  \bmlambda\right\} 
\label{defDeltaCmplNDim2}
\ee
it follows that
\begin{align}
W[O](\gr{X}) &= \int_{\rr^{2n}}
\frac{d^{2n} \boldsymbol{\Lambda}}{(2\pi)^{2n}} \:
\exp\left\{ i \boldsymbol{\Lambda}^{\sT} \bmOmega
\boldsymbol{X}\right\} \chi[O] (\bmOmega^{\sT}\gr{\Lambda})
 \label{defWCarta2}\,,
\end{align}
the same observation being valid for $W[O](\gr{Y})$.
\par
\index{characteristic function!evolution}
\index{Wigner function!evolution}
Let us now analyze the evolution of the characteristic
and the Wigner functions under the action of unitary operations
coming from linear Hamiltonians of the form (\ref{LinH}). 
If $\varrho$ is the state of the modes {\em before} a device 
described by the unitary $U$, the characteristic and the Wigner 
function of the state {\em after} the device $\varrho^\prime = 
U \varrho\, U^\dag$ can be computed using the Heisenberg evolution of 
the displacement operator. The action of the displacement operator 
itself corresponds to a simple translation in the phase space. 
Using Eq.~(\ref{compD1}) we have 
\begin{subequations}
\begin{align}
\chi[D(\gr{z})\,\varrho\, D^\dag(\gr{z})](\gr{\lambda}) &=
\chi [\varrho] (\gr{\lambda})
\: \pexp{\gr{z}^\dag \gr{\lambda}  - \gr{\lambda}^\dag \gr{z}} \label{Xdispla}\:,
\\
W[D(\bmz)\,\varrho\, D^\dag(\bmz)](\gr{\lambda}) &=
W[\varrho] (\gr{\alpha}-\gr{z})
\:. \label{Wdispla}
\end{align}
\end{subequations}
In the notation of Eq.~(\ref{modeLinH}), $\gr{Q}=\iid$ and $\gr{d}=\gr{z}$, 
thus we have no change in the covariance matrices.
In general, for the interactions described by Hamiltonians of the form
(\ref{LinH}) and, excluding displacements, we have
\begin{subequations} \label{scalar}
\begin{align}
&\begin{array}{l}
\chi[U \varrho\,U^{\dag} ](\gr{\Lambda}) =
\chi [\varrho] (\gr{P}_{23} \gr{F}^{-1} \gr{P}_{23}\gr{\Lambda})\,,\\[1ex]
\chi[U \varrho\,U^{\dag} ](\gr{K}) =
\chi [\varrho] (\gr{F}^{-1}  \gr{K})\,,
\end{array}\\
&\begin{array}{l}
W[ U \varrho\,U^{\dag} ](\gr{X}) =
W [\varrho] (\gr{P}_{23} \gr{F}^{-1}  \gr{P}_{23} \gr{X})\,,\\[1ex]
W[U \varrho\,U^{\dag}  ](\gr{Y}) =
W [\varrho] (\gr{F}^{-1} \gr{Y}) \,,
\end{array}
\end{align} 
\end{subequations}
where $\gr{F}$ is the symplectic transformation associated
the unitary $U$. Eqs.~(\ref{scalar}) say that the characteristic
and the Wigner functions transform as a {\em scalars} under the
action of $U$. For two-mode mixing, single-mode squeezing and two-mode 
squeezing the symplectic matrices are  given by in
Eqs.~(\ref{RealMatrixbs}), (\ref{Smatrix}) and (\ref{Qxi}) respectively.
\par  
In summary, the introduction of the Wigner function allows to describe
quantum dynamics of physical systems in terms of phase-space
quasi-distribution, without referring to the wave-function or the density
matrix of the system.  Quantum dynamics may be viewed as the evolution of a
phase-space distribution, the main difference being the fact that the Wigner
function is only a {\em quasi-distribution}, {\em i.e.} it is bounded and
normalized but it may assume negative values.  Unitary evolutions induced
by bilinear Hamiltonians correspond to symplectic transformations of mode
operators and, in turn, of the phase-space coordinates.  Evolution of the
characteristic and the Wigner functions then corresponds to transformation
(\ref{scalar}), whereas non-unitary evolution induced by interaction with
the environment will be analyzed in details in Chapter \ref{c6:gauss:chan}.
\subsection{Trace rule in the phase space}\label{ss:trace}
\index{Wigner function!trace rule}
\index{characteristic function!trace rule}
The introduction of the characteristic and the Wigner functions 
allows to evaluate operators' traces as integrals in the phase space. 
This is useful in order to evaluate correlation functions and the 
statistics of a measurement 
since we are mostly dealing with Gaussian states and, as we will see in 
Chapter \ref{ch:detection}, also many detectors are described by 
Gaussian operators. In this Section we
explicitly derive Eqs.~(\ref{Wtrace}) and (\ref{Xtrace}),
for the trace of two generic operators
in terms of their characteristics or Wigner function.
The starting points are the Glauber expansions of an operator in terms
of the characteristic or the Wigner functions, {\em i.e.} formulas
(\ref{GlauberF}) and (\ref{WigF}).
For the characteristic function we have 
\begin{align}
\hbox{Tr}[O_1\:O_2] &= 
\int_{\cc^n} \frac{d^{2n}\gr{\lambda}_1}{\pi^{n}}\: \chi[O_1](\gr{\lambda}_1)\:
\int_{\cc^n} \frac{d^{2n}\gr{\lambda}_2}{\pi^{n}}\: \chi[O_2](\gr{\lambda}_1)\:
\hbox{Tr}[D(\gr{\lambda}_1) \: D(\gr{\lambda}_2)]\,, \nonumber \\
&= 
\int_{\cc^{2n}}  
\frac{d^{2n}\gr{\lambda}_1}{\pi^{n}} 
\frac{d^{2n}\gr{\lambda}_2}{\pi^{n}}\: 
\chi[O_1](\gr{\lambda}_1)\:
\chi[O_2](\gr{\lambda}_1)\nonumber\\
&\hspace{4cm}\times\hbox{Tr}[D(\gr{\lambda}_1+\gr{\lambda}_2)]
\pexp{\bmlambda_1^\dag\bmlambda_2-\bmlambda_2^\dag\bmlambda_1}\,,\nonumber \\
&=  
\int_{\cc^n} \frac{d^{2n}\gr{\lambda}}{\pi^{n}}\: \chi[O_1](\gr{\lambda})\:
\chi[O_2](-\gr{\lambda})\:,
\end{align}
where we have used the trace rule for the displacement 
$\hbox{Tr}[D(\gr{\gamma})]=\pi^n \delta^{(2n)} (\gr{\gamma})$.
For the Wigner function we have
\begin{align}
\hbox{Tr}[O_1\:O_2] &=
2^{2n} \int_{\cc^n} d^{2n} \gr{\alpha}_1\: W[O_1](\gr{\alpha}_1)
\int_{\cc^n} d^{2n} \gr{\alpha}_2\: W[O_2](\gr{\alpha}_2) \nonumber \\
&\hspace{5.25cm}\times \hbox{Tr}\left[D(2\gr{\alpha}_1)
\gr{\Pi}\gr{\Pi}
D(-2\gr{\alpha}_2)\right]\,, \nonumber \\
&=
2^{2n} \int_{\cc^{2n}} d^{2n} \gr{\alpha}_1 d^{2n} \gr{\alpha}_2
W[O_1](\gr{\alpha}_1)W[O_2](\gr{\alpha}_2)\:
\nonumber \\
&\hspace{3.25cm}\times
\hbox{Tr}[D(2\gr{\alpha}_1-2\gr{\alpha}_2)]
\pexp{2\bmalpha_1^\dag\bmalpha_2-2\bmalpha_2^\dag\bmalpha_1}\,, \nonumber \\
&= \pi^n\int_{\cc^n} d^{2n} \gr{\alpha}\: 
W[O_1](\gr{\alpha}) W[O_2](\gr{\alpha}) \:,
\end{align}
where we have used the relations $\gr{\Pi}^2={\mathbbm I}$ and 
$\delta^{(2n)}(a\gr{\gamma}) = |a|^{-2n}\delta^{(2n)}(\gr{\gamma})$ 
with $a\in\rr$.
\subsection[A remark about parameters $\kappa$]{A remark
about parameters $\boldsymbol{\kappa}$}\label{kappa:remark}
In order to encompass the different notations used in the literature
to pass from complex to Cartesian notation, we have
introduced the three parameters $\kappa_h$, $h=1,2,3$, in the decomposition of
the mode operator, the phase-space coordinates and the reciprocal
phase-space coordinates respectively. We
report here again their meaning
\begin{align}
a_k = \kappa_1 (q_k+i p_k)\,, \qquad
\alpha_k = \kappa_2 (x_k+i y_k)\,, \qquad
\lambda_k = \kappa_3 (\rma_k+i \rmb_k)
\label{ReDefKs}\:.
\end{align}
The three parameters are not independent on each other and should
satisfy the relations $2\kappa_1\kappa_3 = 2 \kappa_2 \kappa_3 = 1$,
{\em i.e.} $\kappa_1=\kappa_2 = (2\kappa_3)^{-1}$.  The so-called
canonical representation corresponds to the choice
$\kappa_1=\kappa_2=\kappa_3=2^{-1/2}$, while the quantum optical
convention corresponds to $\kappa_1=\kappa_2=1$, $\kappa_3=1/2$.  We
have already seen that $2 \kappa_2 \kappa_3 = 1$; in order
to prove that $2 \kappa_1 \kappa_3 = 1$, 
it is enough to consider the vacuum state of a
single mode and evaluate the second moment of the ``position''
operator $\langle q^2 \rangle =\hbox{Tr}\left[\varrho\: q^2\right]$,
which coincides with the variance $\langle \Delta q^2 \rangle$, since
the first moment $\langle q\rangle=0$ vanishes. Starting from the
commutation relation $[q,p]=( 2\kappa_1^2)^{-1}$ it is straightforward
to show that the vacuum is a minimum uncertainty state with \be
\langle \Delta q^2 \rangle = (4\kappa_1^2)^{-1}\:.\label{dq1} \ee On
the other hand, the Wigner and the characteristic functions of a
single-mode vacuum state are given by
\begin{align}
W_0(x,y) = \frac{2}{\pi} \exp \left\{ - 2 \kappa_2^2 (x^2+y^2) \right\}\,,
\quad
\chi_0(\rma,\rmb) =
\exp \left\{ - \mbfrac  \kappa_3^2 (\rma^2+\rmb^2) \right\}\:.
\end{align}
Therefore, using the properties of $W$ as quasiprobability, and of
$\chi$ as moment generating function, respectively, we have
\begin{subequations}
\begin{align}
\langle \Delta q^2 \rangle &=
\int_{\rr^2} dx\:dy\: \kappa_2^2\: x^2 \: W_0(x,y) =
(4\kappa_2^2)^{-1} \label{dq2}\,,  \\
\langle \Delta q^2 \rangle &= -
\left.\frac{\partial^2}{\partial_\rma^2}\,
\chi_0(\rma,\rmb)\right|_{\rma=\rmb=0} =
\kappa_3^2\:, \label{dq3}
\end{align}
\end{subequations}
from which the thesis follows, upon using
Eqs.~(\ref{dq1}), (\ref{dq2}) and (\ref{dq3}) and assuming
positivity of the parameters. Now, thanks to these results and denoting 
by $\gr{\sigma}_0=\gr{V}_0= (4\kappa_1^2)^{-1}\ii_{2n} =
(4\kappa_2^2)^{-1}\ii_{2n}= \kappa_3^2\,\ii_{2n}$ 
the covariance matrix of the $n$-mode vacuum, we have that the 
characteristic and the Wigner functions can be rewritten as 
\begin{align}
\chi_0(\gr{\Lambda}) =
\pexp{- \mbfrac \gr{\Lambda}^{\sT} \gr{\sigma}_0 \gr{\Lambda}}\,,\qquad
\chi_0(\gr{K}) = \pexp{- \mbfrac \gr{K}^{\sT} \gr{V}_0 \gr{K}}\:,
\end{align}
and
\begin{subequations}
\begin{align}
W_0(\gr{X}) &= \frac{\pexp{- \mbfrac \gr{X}^{\sT} \gr{\sigma}_0^{-1} 
\gr{X}}}{(2 \pi)^n \kappa_2^{2n} \sqrt{\det [\gr{\sigma}_0]}}=
\left(\frac{2}{\pi}\right)^n 
\pexp{- \frac12 \gr{X}^{\sT} \gr{\sigma}_0^{-1} \gr{X}}\,,  \\ 
W_0(\gr{Y}) &=
\frac{\pexp{- \mbfrac \gr{Y}^{\sT} \gr{V}_0^{-1} 
\gr{Y}}}{(2 \pi)^n \kappa_2^{2n} \sqrt{\det [\gr{V}_0]}}
=\left(\frac{2}{\pi}\right)^n 
\pexp{- \mbfrac \gr{Y}^{\sT} \gr{V}_0^{-1} \gr{Y}}\,,
\end{align}
\end{subequations}
respectively, independently on the choice of parameters $\kappa_h$.
This form of the characteristic and Wigner function individuates the
so-called class of {\em Gaussian states}. The simplest example of 
Gaussian state is indeed the vacuum state. Thermal, coherent as well 
as squeezed states are other examples. The whole class of Gaussian states 
will be analyzed in detail in Chapter \ref{ch:gs}. 

\chapter{Gaussian states}\label{ch:gs}
\index{Gaussian states}
Gaussian states are at the heart of quantum information processing
with continuous variables. The basic reason is that the vacuum state 
of quantum electrodynamics is itself a Gaussian state. This
observation, in combination with the fact that the quantum evolutions
achievable with current technology are described by Hamiltonian
operators at most bilinear in the quantum fields, accounts for the
fact that the states commonly produced in laboratories are Gaussian.
In fact, as we have already pointed out, bilinear evolutions preserve
the Gaussian character of the vacuum state. Furthermore, recall that
the operation of tracing out a mode from a multipartite Gaussian state
preserves the Gaussian character too, and the same observation, as we will
see in the Chapter \ref{c6:gauss:chan}, is valid when the evolution of a state in a standard 
noisy channel is considered.  
\section{Definition and general properties}\label{c3:def}
A state $\varrho$ of a continuous variable system with $n$ degrees of
freedom is called Gaussian if its Wigner function, or equivalently its
characteristic function, is Gaussian, {\em i.e.} in the notation 
introduced in Chapter \ref{ch:basics}:
\begin{equation}\label{c3:WAlpha}
W[\varrho](\bmalpha)=\frac{\exp\{-\frac12 (\bmalpha -
\overline{\bmalpha})^{\sT}\,
\bmsigma_{\alpha}^{-1}\,(\bmalpha - \overline{\bmalpha})\}}
{(2\pi)^n\sqrt{{\rm Det}[\bmsigma_{\alpha}]}}
\end{equation}
with $\bmalpha = \kappa_2\,\bmX$, $\overline{\bmalpha} = \kappa_2\,
\overline{\bmX}$, where $\overline{\bmX}$ is the vector of the
quadratures' average values. The matrix $\bmsigma_{\alpha}^{-1}$ is
related to the covariance matrix $\bmsigma$ defined in
Eqs.~(\ref{defCOV}) by $\bmsigma_{\alpha}=\kappa_2^2\,\bmsigma$. In
Cartesian coordinates we have:
\begin{equation}\label{c3:WX}
W[\varrho](\bmX) =
\frac{\exp\left\{ -\frac12 (\bmX-\overline{\bmX})^{\sT} \bmsigma^{-1}
(\bmX-\overline{\bmX}) \right\}}
{(2\pi)^n\,\kappa_2^{2n}\,\sqrt{\mbox{Det}[\bmsigma]}}\,,
\end{equation}
or equivalently:
\begin{equation}\label{c3:WY}
W[\varrho](\bmY) =
\frac{\exp\left\{ -\frac12 (\bmY-\overline{\bmY})^{\sT} \bmV^{-1}
(\bmY-\overline{\bmY}) \right\}}
{(2\pi)^n\,\kappa_2^{2n}\,\sqrt{\mbox{Det}[\bmV]}}\,.
\end{equation}
\index{Gaussian states!Wigner function}
Correspondingly, the characteristic function is given 
by~\footnote{Recall that for every symmetric positive-definite 
matrix $\boldsymbol{Q}\in{\rm M}(n,\rr)$ the following identity holds \par
  \vspace{0.2cm} $\;\;\;\;\;\;\;\;\;\;\;\;
  \displaystyle{\int}_{\!\rr^n}d^n\bmX\exp \left\{ -\mbfrac
    \bmX^{\sT}\boldsymbol{Q}^{-1}\bmX +i\bmLambda^{\sT}\bmX
  \right\}=\sqrt{(2\pi)^n\mbox{Det}[\boldsymbol{Q}]}\exp \left\{
    -\mbfrac \bmLambda^{\sT}\boldsymbol{Q}\bmLambda \right\}$.}
\be
\chi_0(\gr{\Lambda}) = 
\pexp{- \mbfrac \gr{\Lambda}^{\sT} \gr{\sigma} \gr{\Lambda}
+i\gr{\Lambda}^{\sT}\overline{\bmX}}\,,\;
\chi_0(\gr{K}) = \pexp{- \mbfrac \gr{K}^{\sT} \gr{V}\gr{K}
+i\gr{K}^{\sT}\overline{\bmY}}\:.
\ee
In the following, since we are mostly interested in the entanglement
properties of the state, the vector $\overline{\bmX}$ (or
$\overline{\bmY}$) will be put to zero. Indeed, entanglement is not changed
by local operations and the vectors $\overline{\bmX}$ (or
$\overline{\bmY}$) can be changed arbitrarily by phase-space translations,
which are in turn local operations. Gaussian states are then entirely
characterized by the covariance matrix $\bmsigma$ (or $\bmV$). This is a
relevant property of Gaussian states since it means that typical issues of
continuous variables quantum information theory, which are generally
difficult to handle in an infinite Hilbert space, can be faced up with the
help of finite matrix theory.
\par
Pure Gaussian states are easily characterized. Indeed, recalling that
for any operator $O_k$, which admits a well defined Wigner function
$W_k(\bmalpha)$, we can write $\hbox{Tr}[O_1\,O_2]$ in terms
of the overlap between Wigner function [see Eq.~(\ref{Wtrace})]
it follows that the {\em purity} $\mu=\hbox{Tr}[\varrho^2]$ of a
Gaussian state is given by:
\index{Gaussian states!purity}
\index{purity}
\begin{equation}\label{c3:purity:gauss:N}
\mu(\bmsigma) = \pi^n\,\kappa_2^{2n} \int_{{\mathbb R}^{2n}}
d^{2n}{\bmX}\, W^{2}(\bmX) =
\frac{1}
{(2\kappa_2)^{2n}\,\sqrt{\mbox{Det}[\bmsigma]}}\,.
\end{equation}
Hence a Gaussian state is pure if and only if
\be
\label{c3:PurityCond}
\mbox{Det}[\bmsigma]=(2\kappa_2)^{-4n}. 
\ee
\par
Another remarkable feature of pure Gaussian states is that they are
the only pure states endowed with a {\em positive} Wigner function
\cite{Hud74,LB95}.  In order to prove the statement we consider system 
with only one degree of freedom. The extension to $n$ degrees of 
freedom is straightforward.
Let us consider the Husimi function
$Q(\alpha)=\pi^{-1}|\braket{\psi}{\alpha}|^2$ where $|\alpha\rangle$ is 
a coherent state, which is related to the Wigner function
as 
\be \label{c3:QandW}
Q(\alpha)=\frac2\pi \int_{\cc}\,
d^2\beta\, W(\beta) \pexp{ -2|\alpha-\beta|^2 } \;.
\ee
Eq. (\ref{c3:QandW}) implies that if 
$Q(\alpha_0)=0$ for at least one $\alpha_0$ then $W(\alpha)$ must have
negative regions, because the convolution involves a Gaussian strictly
positive integrand. But the only pure states characterized by a strictly
positive Husimi function turns out to be Gaussian ones. Indeed,
consider a generic pure state expanded in Fock basis as
$\ket{\psi}=\sum c_n \ket{n}$ and define the function
$f(\alpha)=e^{\frac12 |\alpha|^2}\braket{\psi}{\alpha}=\sum c_n^*
\frac{\alpha^n}{\sqrt{n!}}$. Clearly, $f(\alpha)$ is an analytic
function of growth order less than or equal to 2 which will have zeros
if and only if $Q(\alpha)$ has zeros. Hence it is possible to apply
Hadamard's theorem \cite{SZ71}, which states that any function that is
analytic on the complex plane, has no zeros, and is restricted in
growth to be of order 2 or less must be a Gaussian function. It
follows that the $Q(\alpha)$ and $W(\alpha)$ functions are Gaussian.
\par
Gaussian states are particularly important from an applicative point
of view because they can be generated using only the linear and 
bilinear interactions introduced in Section
\ref{s:LinBil}. Indeed, the following theorem, due to Williamson,
ensures that every covariance matrix (every real symmetric matrix
positive definite) can be diagonalized through a symplectic
transformation \cite{Wil36},
\index{Williamson theorem}
\begin{teo}\label{williamson}{\em\bf (Williamson)}
Given $\bmV\in {\rm M}(2n,{\mathbb R})$, $\bmV^{\sT}=\bmV$,
$\bmV>0$ there exist $\bmS\in {\rm Sp}(2n,{\mathbb R})$ and
$\bmD\in {\rm M}(n,{\mathbb R})$ diagonal and positive
defined such that:
\be
\label{c3:SympDiagV} \bmV=\bmS^{\sT}
\left(
\begin{array}{cc}
\bmD & \bmZero \\
\bmZero & \bmD 
\end{array}
\right) \bmS \,.
\ee
Matrices $\bmS$ and $\bmD$ are unique, up to a permutation of the elements of
$\bmD$. 
\end{teo}
\noindent{\em Proof.}
\par
By inspection it is straightforward to see that Eq.~(\ref{c3:SympDiagV})
implies that $$\bmS=(\bmD\oplus \bmD)^{-1/2}\,\bmO\,\bmV^{-1/2}\,,$$
with $\bmO$ orthogonal. Requiring symplecticity to matrix $\bmS$ means that
\be \label{c3:proofSD} 
\bmO\bmV^{-1/2}\bmJ\bmV^{-1/2}\bmO^{\sT}=\left(
\begin{array}{cc}
\bmZero & \bmD^{-1} \\
-\bmD^{-1} & \bmZero  
\end{array}
\right) \,, \ee 
$\bmJ$ being defined in Eq.~(\ref{defJ}).
Since $\bmV$ and $\bmJ$ are symmetric and antisymmetric, respectively, it
follows that $\bmV^{-1/2}\bmJ\bmV^{-1/2}$ is antisymmetric, hence there
exist a unique $\bmO$ such that Eq.~(\ref{c3:proofSD}) holds. $\square$
\vspace{.5cm}
\par
\index{symplectic!eigenvalues}
The elements $d_k$ of $\bmD=\hbox{Diag}(d_1,\ldots,d_n)$ are called
symplectic eigenvalues and can be calculated from the spectrum of
$i\bmJ\bmV$, while matrix $\bmS$ is said to perform a {\em symplectic
diagonalization}. Changing to $\gr{\Omega}$-ordering, {\em i.e.} in terms 
of the covariance matrix $\bmsigma$ defined in Eq.~(\ref{defCOV}), the 
decomposition (\ref{c3:SympDiagV}) reads as follows
 \be
\label{c3:SympDiagSG}
\bmsigma=\bmS^{\sT} \bmW\,\bmS
\ee
where $\bmW=\bigoplus_{k=1}^n d_k\,\ii_2$, $\ii_2$ being the $2\times2$
identity matrix. 
\par
The physical statement implied by decompositions 
(\ref{c3:SympDiagV}) and (\ref{c3:SympDiagSG}) is that every Gaussian 
state $\varrho$ can be obtained from a thermal state $\nu$, described 
by a diagonal covariance matrix, by performing the unitary 
transformation $U_{\bmS}$ associated to the symplectic matrix 
$\bmS$, which in turn can be generated by linear and bilinear 
interactions. In formula,
\be
\label{c3:GenericGS}
\varrho=U_{\bmS}\,\nu\, U_{\bmS}^\dag \;,
\ee
where $\nu=\nu_1\otimes\dots\otimes\nu_n$ is a product of thermal
states $\nu_k$ of the form (\ref{th}) for each mode, 
with parameters $\beta_k$ given by
\be \beta_k =\ln\left[\frac{d_k+1+
    (2\kappa_2)^{-2}}{d_k - (2\kappa_2)^{-2}}\right]\;, 
    \ee 
in terms of the symplectic eigenvalues $d_k$. Correspondingly, the
mean number of photons is given by $N_k = d_k - (2\kappa_2)^{-2}$.
The decomposition (\ref{c3:SympDiagV}) allows to recast the
uncertainty principle (\ref{HeisSG}), which is invariant under
${\rm Sp}(2n,{\mathbb R})$, into the following form
\be \label{c3:SympEigUncert}
d_k\ge (2\kappa_2)^{-2} \;.
\ee
Pure Gaussian states are obtained only
if $\nu$ is pure, which means that $\nu_k=\ket0\bra0$, $\forall k$,
{\em i.e.} $d_k=(2\kappa_2)^{-2}$. Hence a condition equivalent to 
Eq.~(\ref{c3:PurityCond}) for the purity of a Gaussian
state is that its covariance matrix may be written as 
\be
\label{c3:PurityCond2}
\bmV = (2\kappa_2)^{-4n}\: \bmS\bmS^{\sT} \; .
\ee
Furthermore it is clear from Eq.~(\ref{c3:SympEigUncert}) that pure
Gaussian states, for which one has that $d_k=(2\kappa_2)^{-2} \;
\forall k$ \footnote{This is an immediate consequence of
  \refeq{c3:GenericGS} together with purity condition
  $\hbox{Tr}[\varrho^2]=1$.}, are minimum uncertainty states with
respect to suitable quadratures.
\section{Single-mode Gaussian states}\label{c3:1m}
\index{Gaussian states!single-mode}
The simplest class of Gaussian states involves a single mode. In this case
decomposition (\ref{c3:GenericGS}) reads as follows \cite{Ada95}:
\begin{eqnarray}
\varrho= D(\oalpha)S(\xi)\,\nu\,S^{\dag}(\xi)
D^{\dag}(\oalpha) \, ,
\label{c3:rho:1m}\;
\end{eqnarray}
where $\oalpha=\kappa_2 (\ox+i\oy)$ (for the rest of the section we
put $\kappa_2=2^{-1/2}$ ), $\nu$ is a thermal state with
average photon number $N$, $D(\oalpha)$ denotes the
displacement operator and $S(\xi)$ with $\xi=r\,e^{i\varphi}$
the squeezing operator.  A convenient parametrization of Gaussian
states can be achieved expressing the covariance matrix $\bmsigma$ as
a function of ${N}$, $r$, $\varphi$, which have a direct
phenomenological interpretation.  In fact, 
following Chapter \ref{ch:basics}, {\em i.e.} applying the phase-space
representation of squeezing \cite{BR97,WM94}, we have that for the state 
(\ref{c3:rho:1m}) the covariance matrix is given by $\bmsigma=\gr{\Sigma}_\xi^{\sT}
\bmsigma_{\nu} \gr{\Sigma}_\xi$ where $\gr{\sigma}_{\nu}$
is the covariance matrix (\ref{sgth}) of a thermal state and $\gr{\Sigma}_\xi$
the symplectic squeezing matrix. The explicit expression of the covariance
matrix elements is given by
\begin{subequations}
\label{c3:vars}
\begin{align}
\sigma_{11}&= \frac{2N
+1}{2}\: \Big[\!\cosh(2r)+\sinh(2r)\cos(\varphi)\Big] \:, \\
\sigma_{22}&= \frac{2N
+1}{2}\:\Big[\!\cosh(2r)-\sinh(2r)\cos(\varphi)\Big] \:, \\
\sigma_{12}&=\sigma_{21}=-\frac{2N +1}{2}\:
\sinh(2r)\sin(\varphi) \:,
\end{align}
\end{subequations}
and, from Eq.~(\ref{c3:purity:gauss:N}), it follows that \cite{marians,Dod02}
$ \mu=(2 N +1)^{-1}$, which means that the purity of a generic Gaussian state
depends only on the average number of thermal photons, as one should
expect since displacement and squeezing are unitary operations hence
they do not affect the trace involved in the definition of purity. The
same observation is valid when one considers the {\em von Neumann entropy}
$S_V$ of a generic single mode Gaussian state, defined in general as
\be
\label{c3:S_V}
S_V(\varrho)=-\hbox{Tr}[\varrho \ln \varrho] \;.
\ee
Indeed, one has
\be
\label{c3:S_V1m}
S_V(\varrho)=N
\ln \left(\frac{N +1}{N}\right) +
\ln\left(N+1\right) =
\frac{1-\mu}{2\mu}\ln\left(\frac{1+\mu}{1-\mu}\right)-
\ln\left(\frac{2\mu}{1+\mu}\right)\;.
\ee 
Eq.~(\ref{c3:S_V1m}), firstly achieved in Ref.~\cite{Aga71},
shows that the von Neumann entropy is a monotonically increasing
function of the {\em linear entropy} (defined as $S_L=1-\mu$), so that both
of them lead to the same characterization of mixedness, a fact
peculiar of Gaussian states involving only one single mode.
\par
Examples of the most important families of single mode Gaussian states
are immediately derived considering Eq.~(\ref{c3:rho:1m}). Thermal
states $\nu$ are re-gained for $\oalpha=r=\varphi=0$, coherent states
for $r=\varphi=N=0$, while squeezed vacuum states are recovered
for $\oalpha=N=0$. For $N=0$ we have the vacuum and 
coherent states covariance matrix. The covariance matrix associated with 
the real squeezed vacuum state is recovered for $\varphi=0$ and is given 
by $\bmsigma=\frac12 \hbox{Diag}(e^{-2r},e^{2r})$.
\section{Bipartite systems} \label{c3:2m}
\index{Gaussian states!two-mode}
Bipartite systems are the simplest scenario where to investigate the
fundamental issue of the entanglement in quantum information. In order
to study the entanglement properties of bipartite Gaussian systems it
is very useful to introduce normal forms to represent them. This
section is for the most part devoted to this purpose. The main concept
to be introduced in order to derive useful normal forms is that of
{\em local equivalence}. Two states $\varrho_1$ and $\varrho_2$ of a
bipartite system ${\cal H}_A\otimes{\cal H}_B$ are locally equivalent
if there exist two unitary transformations $U_A$ and $U_B$ acting on
${\cal H}_A$ and ${\cal H}_B$ respectively, such that
$\varrho_2=U_A\otimes U_B\:\varrho_1U_A^\dagger\otimes U_B^\dagger $.
The extension to multipartite systems is straightforward.
\par
Let us start introducing the following
\begin{teo}\label{c3:SingValues}{\em\bf (Singular values decomposition)}
Given $\bmC \in {\rm M}(n,{\mathbb C})$ then there exist two unitary
matrices $\bmU$ and $\bmV$, such that
$$
\bmC = \bmU\, \bmSigma\, \bmV\,,
$$
$\bmSigma \equiv {\rm Diag}(\sqrt{p_1},\ldots,\sqrt{p_n})$, where
$p_k$ ($k=1,\dots,n$) are the eigenvalues of the positive operator
$\bmC^\dag \bmC$.
\end{teo}
\noindent{\em Proof.}
\par
Let $\bmV$ be the unitary matrix that diagonalizes $\bmC^\dag \bmC$; we have
\begin{align}
\bmV\,\bmC^\dag \bmC\,\bmV^\dag &= \bmSigma^2 \nonumber\\
\bmC^\dag \bmC &= \bmV^\dag\, \bmSigma^2\, \bmV\nonumber\\
\bmC^{\dag} \bmC &=
\bmV^\dag\, \bmSigma\,
\bmU^\dag \bmU\,
\bmSigma\, \bmV\,,\nonumber
\end{align}
and, from the last equality, one has $\bmC^\dag = \bmV^\dag\,
\bmSigma\, \bmU^\dag$ and $\bmC = \bmU\, \bmSigma\, \bmV$, provided that $\bmU = \bmC (\bmSigma\, \bmV)^{-1}$ (for a detailed proof see Ref.~\cite{HJ85}). $\square$
\par
\vspace{.5cm}
Let us now consider a generic bipartite state
\begin{equation}
\dket{\bmC} = \sum_{h,k} c_{hk}\ket{\Phi_h}\ket{\Psi_k}\,.
\end{equation}
\index{singular values decomposition}
Thanks to the singular values decomposition Theorem \ref{c3:SingValues},
the coefficients' matrix $\bmC$ can be rewritten as
$\bmC = \bmU\, \bmSigma\, \bmV$, so that
\begin{equation}
c_{hk} = \sum_{r,s} u_{hr}\,\sigma_{rs}\, v_{sk}\,,
\end{equation}
where $\sigma_{rs} = \sqrt{p_r}\,\delta_{r,s}$. In this way the bipartite
state $\dket{\bmC}$ reads
\begin{equation}\label{schmidt}
\dket{\bmC} = \sum_{k} \sqrt{p_k}\, \ket{\phi_k}\ket{\psi_k}\,,
\end{equation}
with
\begin{equation}
\ket{\phi_k} \doteq \sum_s v_{sk} \ket{\Phi_s}\,,
\qquad \ket{\psi_k} \doteq \sum_s u_{ks} \ket{\Psi_s}\,.
\end{equation}
Note that $\braket{\psi_h}{\psi_k}=\delta_{h,k}$ and
$\braket{\phi_h}{\phi_k}=\delta_{h,k}$. Eq.~(\ref{schmidt}) is known
as ``Schmidt decomposition'', while the coefficients $\sqrt{p_k}$ are
called ``Schmidt coefficients''. By construction the latter are
unique.
\par
Let us consider now Gaussian {\em pure states} for $m+n$ canonical systems
partitioned into two sets $A=\{A_1,\dots,A_m\}$ and
$B=\{B_1,\dots,B_n\}$ in their Schmidt form
\begin{equation}\label{c3:schmidt}
    \ket{\psi}_{AB} = \sum_k \sqrt{p_k}
    \ket{\phi_k}_A\ket{\varphi_k}_B\,.
\end{equation}
\index{Schmidt decomposition} 
In general the Schmidt decomposition has an ``irreducible'' structure:
generally speaking, Eq.~(\ref{c3:schmidt}) cannot be brought into a
simpler form just by means of local transformations on set $A$ and $B$. In
the case of Gaussian bipartite systems however a remarkably simpler form
can be found \cite{BR03,HW01,GEC+03}. As a matter of fact, a Gaussian pure
state $\ \ket{\psi}_{AB}$ may always be written as
\begin{equation}\label{c3:NF2pPure}
\ket{\psi}_{AB} =
\ket{\widetilde{\psi}_1}_{{\widetilde A}_1{\widetilde B}_1}\ldots
\ket{\widetilde{\psi}_s}_{{\widetilde A}_s{\widetilde B}_s}
\ket{0}_{{\widetilde A}_v}
\ket{0}_{{\widetilde B}_v}\,,
\end{equation}
where $\widetilde{A}=\{\widetilde{A}_1\ldots,\widetilde{A}_m\}$ and
$\widetilde{B}=\{ \widetilde{B}_1,\ldots,\widetilde{B}_n\}$ are new sets of
modes obtained from $A$ and $B$ respectively through local linear canonical
transformations, the states $\ket{\widetilde{\psi}_k}$ are two-mode
squeezed states for modes $k=1,\dots,s$, for some $s\le\hbox{min}[m,n]$ and
$\ket{0}_{{\widetilde A}_v}$ and $\ket{0}_{{\widetilde B}_v}$ are products
of vacuum states of the remaining modes. In order to prove
Eq.~(\ref{c3:NF2pPure}) we consider the partial density matrices obtained
from the Schmidt decomposition (\ref{c3:schmidt})
\begin{equation}
    \varrho_A =\sum_k p_k
    \ket{\phi_k} \bra{\phi_k} \,, \ \ \ \varrho_B =\sum_k p_k
    \ket{\varphi_k} \bra{\varphi_k}\, .
\end{equation}
Since $\varrho_A$ and $\varrho_B$ are Gaussian, they can be brought to
Williamson normal form (\ref{c3:GenericGS}) through local linear canonical
transformations. Suppose that there are $s$ modes in $A$ and $t$ modes in
$B$ with symplectic eigenvalue $d \neq (2\kappa_2)^{-2}$. Since the
remaining modes factor out from the respective density matrices as
projection operators onto the vacuum state, we may factor $\ket{\psi}_{AB}$
as $\ket{\widetilde{\psi}}_{AB}\ket{0}_{{\widetilde A}_v}
\ket{0}_{{\widetilde B}_v}$ where $\ket{\widetilde{\psi}}_{AB}$ is a
generic entangled state for modes $\widetilde{A}_1,\ldots,\widetilde{A}_s$
and $\widetilde{B}_1,\ldots,\widetilde{B}_t$. The partial density matrices of
the state $\ket{\widetilde{\psi}}_{AB}$ may be written as         
\be
\widetilde{\varrho}_A =
\sum_{\vec{n}_A}\frac{e^{-\vec{\beta}_A \cdot
\vec{n}_A}}{\hbox{Tr}\left[e^{-\vec{\beta}_A \cdot {\vec{N}_A}}\right]}
\ket{\vec{n}_A}\bra{\vec{n}_A} \,, \ \ \ \ \widetilde{\varrho}_B =
\sum_{\vec{n}_B}\frac{e^{-\vec{\beta}_B \cdot
\vec{n}_B}}{\hbox{Tr}\left[e^{-\vec{\beta}_B \cdot {\vec{N}_B}}\right]}
\ket{\vec{n}_B}\bra{\vec{n}_B}\, ,
\ee 
where we have used the notation
$$
\vec{c}_{A}
=\left(c_{\widetilde{A}_1},\ldots,
c_{\widetilde{A}_s}\right)^{\sT}\,,\qquad
\vec{c}_{B}
=\left(c_{\widetilde{B}_1},\ldots,
c_{\widetilde{B}_t}\right)^{\sT}\,,
$$
hence $\vec{n}_A$ and
$\vec{n}_B$ represent occupation number distributions on each side,
$\vec{N}_A$ and $\vec{N}_B$ are the number operators, and
$\vec{\beta}_A$ and $\vec{\beta}_B$ represent the distribution of
thermal parameters. In order to have the same rank and the same
eigenvalues for the two partial density matrix, as imposed by Schmidt
decomposition, there must exist a one-to-one pairing between the
occupation number distributions $\vec{n}_A$ and $\vec{n}_B$, such that
$\vec{\beta}_A\cdot\vec{n}_A = \vec{\beta}_B\cdot\vec{n}_B$. It turns
out that this is possible only if $s=t$ and $\vec{n}_A=\vec{n}_B$,
provided that $\vec{\beta}_A=\vec{\beta}_B$ (for a detailed proof see
Ref.~\cite{BR03}). Hence, reconstructing the Schmidt decomposition of
$\ket{\widetilde \psi}_{AB}$ from ${\widetilde \varrho}_A$ and ${\widetilde
\varrho}_B$ we see that the form (\ref{c3:NF2pPure}) is recovered for
$\ket\psi_{AB}$.
\par
Let us consider now the case of a generic bipartite {\em mixed state}. Due to
the fact that the tensor product structure of the Hilbert space
translates into a direct sum on the phase space, the generic
covariance matrix of a bipartite $m+n$ modes system is a $2m+2n$
square matrix which can be written as follows:
\be
\label{c3:CM2pGeneric}
\bmsigma = 
\left(
\begin{array}{cc}
\bmA   & \bmC \\
\bmC^{\sT} & \bmB
\end{array}
\right) \;,
\ee 
Here $\bmA$ and $\bmB$ are $2m$ and $2n$ covariance matrices
associated to the reduced state of system $A$ and $B$, respectively,
while the $2m\times2n$ matrix $\bmC$ describes the correlations between
the two subsystems.  Applying again the concept of local equivalence we can
straightforwardly find a normal form for matrix (\ref{c3:CM2pGeneric}). A
generic local transformation $\bmS_\bmA \oplus \bmS_\bmB$, with
$\bmS_{\bmA}\in{\rm Sp}(2m,{\mathbb R})$ and
$\bmS_{\bmB}\in{\rm Sp}(2n,{\mathbb R})$, acts on $\bmsigma$ as
\be
\bmA\rightarrow \bmS_\bmA\, \bmA\, \bmS_\bmA^{\sT}\;,
\qquad \bmB\rightarrow \bmS_\bmB\, \bmB\, \bmS_\bmB^{\sT}\;, 
 \qquad \bmC\rightarrow \bmS_\bmA\, \bmC\, \bmS_\bmB^{\sT}\;.
\ee
\index{symplectic!invariants}
Notice that four local invariants [{\em i.e.} invariant with respect to
transformation belonging to the subgroup ${\rm Sp}(2m,{\mathbb R})\otimes
{\rm Sp}(2n,{\mathbb R}) \subset {\rm Sp}(2m+2n,{\mathbb R})$]
can immediately be identified: $I_1=\det[\bmA]$, $I_2=\det[\bmB]$,
$I_3=\det[\bmC]$, $I_4=\det[\bmsigma]$.
Now, the Theorem \ref{williamson} allows to choose $\bmS_\bmA$ and $\bmS_\bmB$
such to perform a symplectic diagonalization of matrices $\bmA$ and $\bmB$
[see Eq.~(\ref{c3:SympDiagSG})], namely
\be
\bmS_\bmA\, \bmA\, \bmS_\bmA^{\sT} =
\bmW_{\!\!\bmA} = \bigoplus_{k=1}^{m} d_{\bmA,k} \, \ii_2\,,
\qquad
\bmS_\bmB\, \bmB\, \bmS_\bmB^{\sT} = \bmW_{\!\!\bmB} = 
\bigoplus_{k=1}^{n} d_{\bmB,k} \, \ii_2 \,,
\ee
where $\bmW_{\!\!\bmA(\bmB)}$ is a diagonal matrix. Thus any covariance matrix
$\bmsigma$ of a bipartite $m\times n$ system can be brought into the form
\be
\label{c3:NF2pMixAux}
\bmsigma = 
\left(
\begin{array}{cc}
 \bmW_{\!\!\bmA} & \bmE \\ [1ex]
 \bmE^{\sT} & \bmW_{\!\!\bmB}
\end{array}
\right) \;,  \ee
where $\bmE=\bmS_\bmA\, \bmC\, \bmS_\bmB^{\sT}$. A further simplification
concerns the case $m=n$ if we focus our attention on the $2 \times 2$
diagonal blocks of $\bmE$, which we call $\bmE^h$, with $h=1,\ldots,n$.
Matrices $\bmE^h$, being real $2 \times 2$ matrices, admit a singular
value decomposition by suitable orthogonal (and symplectic) matrices
$\bmO_\bmA^h$ and $\bmO_\bmB^h$:
$\widetilde{\bmE}\,\!^h = \bmO_\bmA^h \bmE^h \,
(\bmO_\bmB^h)^{\sT}$. Such $\bmO^h_{\bmA(\bmB)}$'s
transformations do not affect matrices $\bmW_{\!\!\bmA}$ and
$\bmW_{\!\!\bmB}$, being
their diagonal blocks proportional to the identity matrix. Collecting this
observations we can write the following normal form for a generic $n \times
n$ covariance matrix
\index{Gaussian states!two-mode!normal form}
\index{covariance matrix!canonical form!two-mode}
\be
\label{c3:NF2pMix}
\bmsigma = 
\left(
\begin{array}{cc}
 \bmW_{\!\!\bmA} &  {\widetilde \bmE}\\[1ex]
 {\widetilde \bmE}^{\sT} & \bmW_{\!\!\bmB}
\end{array}
\right) \;,  \ee
where 
\be
 {\widetilde \bmE} = 
\left(
\begin{array}{ccccc}
 e_{1,1} &    0    & \ldots &  e_{1,2n-1} &  e_{1,2n} \\
 0      &  e_{2,2} & \ldots &  e_{2,2n-1} &  e_{2,2n} \\
 \vdots &&&& \vdots \\
 e_{2n-1,1} &  e_{2n-1,2} & \ldots  &  e_{2n-1,2n-1} &  0     \\
 e_{2n,1}   &  e_{2n,2}   & \ldots  &     0        &  e_{2n,2n} 
\end{array}
\right) \;.  \ee 
\par
Due to the relevance in what will follow, we write
explicitly the normal form (\ref{c3:NF2pMix}) for the case of a
$1\times 1$ system. It reads as follows:
\be
\label{c3:NF2mMix}
\bmsigma = 
\left(
\begin{array}{cccc}
 a   &  0   &  c_1 &  0   \\
 0   &  a   &  0   &  c_2 \\
 c_1 &  0   &  b   &  0   \\
 0   &  c_2 &  0   &  b 
\end{array}
\right) \;, \ee 
where the correlations $a$, $b$, $c_1$, and $c_2$ are
determined by the four local symplectic invariants $I_1=a^2$,
$I_2=b^2$, $I_3=c_1c_2$ and $I_4=(ab-c_1^2)(ab-c_2^2)$. When $a=b$ the
state is called symmetric. The normal form (\ref{c3:NF2mMix}) allows
to recast the uncertainty principle (\ref{HeisSG}) in a manifestly
${\rm Sp}(2,{\mathbb R})\otimes {\rm Sp}(2,{\mathbb R})$ invariant form
\cite{Sim00}: \be
\label{c3:HeisInvForm}
I_1+I_2+2I_3 \le 8\kappa_1^2I_4+1/(8\kappa_1^2)\,.
\ee 
In order to prove this result it is sufficient to note that it is true for
the normal form (\ref{c3:NF2mMix}), then the invariance of
Ineq.~(\ref{c3:HeisInvForm}) ensures its validity for every covariance
matrix.
\par
Finally we observe that Ineq.~(\ref{c3:HeisInvForm})
can be recast in an even simpler form, in the $1\times 1$ modes case.
Indeed the symplectic eigenvalues of a
generic covariance matrix can be computed in terms of the symplectic
invariants (we put $\kappa_1=2^{-1/2}$) \cite{SIS03}:
 \be
\label{c3:SympEig2m}
\sqrt{2}d_{\pm}=\left[I_1+I_2+2I_3\pm\sqrt{(I_1+I_2+2I_3)^2-4I_4}\right]^{1/2} \;,
\ee
[defining $d_{1}=d_{+}$ and $d_{2}=d_{-}$ in the relation
(\ref{c3:SympDiagV})].
The uncertainty relation (\ref{c3:SympEigUncert}) then reads
\be
\label{c3:HeisSympEig} 
d_-\ge 2^{-1} \;.  \ee
\index{purity}
Therefore, we may see that purity of two mode states corresponds to
$I_4=1/16$ and $I_1+I_2+2I_3=1/2$, where the first relation follows
from \refeq{c3:PurityCond} and the second one from the fact that a
pure Gaussian state has minimum uncertainty. Notice also that
bipartite pure states necessarily have a symmetric normal form ({\em
  i.e.}, $a=b$), as can be seen by equating the partial entropies $S_V$
calculated from \refeq{c3:NF2mMix}.
\par
Eq.~(\ref{c3:SympEig2m}) allows also to express the von Neumann
entropy $S_V$ of Eq.~(\ref{c3:S_V}) in a very simple form. Indeed the
entropy of a generic two-mode state is equal to the entropy of the
two-mode thermal state obtained from it by a symplectic
diagonalization, which in turn corresponds to a unitary operation on the level
of the density operator $\varrho$ hence not affecting the trace appearing
in the definition (\ref{c3:S_V}). Exploiting Eq.~(\ref{c3:S_V1m}) and
the additivity of the von Neumann entropy for tensor product states,
one immediately obtains:
 \be
\label{c3:S_V2m}
S_V(\bmsigma)=f(d_+)+f(d_-) \;,
\ee
where $f(d)=(d+\frac12)\ln(d+\frac12)-(d-\frac12)\ln(d-\frac12)$. 
\par
A relevant subclass of Gaussian states is constituted by
the two-mode squeezed thermal states\footnote{For a general
  parametrization of an arbitrary bipartite Gaussian state,
  by means of a proper symplectic diagonalization, see Ref.~\cite{SIS03}.}
\be
\label{c3:thermalsq}
\varrho = S_{2}(\zeta)\,\varrho_\nu\,S_{2}^{\dag}(\zeta)\,, \quad
\varrho_\nu =\nu_{\hbox{\tiny A}}\otimes\nu_{\hbox{\tiny B}}\,,
\ee
where $\nu_k$, $k=A,B$, are thermal states with mean photon number
$N_1$ and $N_2$ respectively, whose covariance
matrix $\bmsigma_\nu$ is given in \refeq{sgth}.  Following Chapter
\ref{ch:basics}, {\em i.e.}, applying the phase-space representation
of squeezing, we have that for the state (\ref{c3:thermalsq}) the
covariance matrix is given by $\bmsigma=\gr{\Sigma}_{2\xi}^{\sT}
\bmsigma_\nu \gr{\Sigma}_{2\xi}$, where $\gr{\Sigma}_{2\xi}$ is the
symplectic two-mode squeezing matrix given in \refeq{Qxi}. In formula,
\begin{equation}
\bmsigma = \frac{1}{4\kappa_1^2}\begin{pmatrix}
A\, \mathbbm{1}_2&  C\, \boldsymbol{R}_\xi \\
C\, \boldsymbol{R}_\xi & B\, \mathbbm{1}_2 
\end{pmatrix}\label{c:matrix}
\end{equation}
$\boldsymbol{R}_\xi$ being defined in \refeq{Qxi} and
\begin{subequations}\label{c3:thermalsqCov}
\begin{align}
A &\equiv A(r, N_1, N_2) =
\cosh (2r) + 2\,{N_1}\,{\cosh^2 r} +
2\,{N_2}\,{\sinh^2 r}\,,\\
B &\equiv B(r, N_1, N_2) =
\cosh (2r) + 2\,{N_1}\,{\sinh^2 r} +
2\,{N_2}\,{\cosh^2 r}\,,\\
C &\equiv C(r, N_1, N_2) =
\left( 1 + {N_1} + {N_2} \right)\,\sinh (2r) \,.
\end{align}
\end{subequations}
The TWB state $|\Lambda\rangle\rangle$ described in Section \ref{ss:opa}
is recovered when $\varrho_\nu$ is the vacuum state
(namely, $N_1 = N_2 = 0$) and $\varphi = 0$, leading
to
\begin{equation}
\label{c3:TWBsigma}
A = B = \cosh(2r)\,, \qquad
C = \sinh(2r) \,.
\end{equation}
\section{Tripartite systems}\label{s:3mGS}
\index{Gaussian states!three-mode}
In the last Section of this Chapter we deal with the case of
three-mode tripartite systems, {\em i.e.} $1 \times 1 \times 1$
systems. The generic covariance matrix of a three-mode system can be
written as follows
\be
\label{c3:GenSG3m}
\bmsigma=
\left(
\begin{array}{ccc}
\bmsigma_{11} & \bmsigma_{12} &  \bmsigma_{13} \\
\bmsigma_{21} & \bmsigma_{22} &  \bmsigma_{23} \\
\bmsigma_{31} & \bmsigma_{32} &  \bmsigma_{33} 
\end{array}
\right)\,,
\ee 
where each $\bmsigma_{hk}$ is a real $2 \times 2$ matrix.
Exploiting the local invariance introduced above and following the
strategy that led to the normal form (\ref{c3:NF2pMix}) for the bipartite
case, it is possible to find a local invariant form also for matrix
(\ref{c3:GenSG3m}) \cite{WLY04}.
\par
In the following we will consider a local symplectic
transformation belonging to the subgroup $\Sp(2,{\mathbb R})\otimes
\Sp(2,{\mathbb R})\otimes \Sp(2,{\mathbb R}) \subset \Sp(6,{\mathbb R})$,
referred to as $\bmS=\bmS_1 \oplus \bmS_2 \oplus \bmS_3$.
The action of $\bmS$ on the covariance matrix $\bmsigma$ is given by
\be
(\bmS \bmsigma \bmS^{\sT})_{hk}=
\bigg\{
\begin{array}{c}
\bmS_h \bmsigma_{hh} \bmS^{\sT}_h  \qquad \hbox{for} \; h=k  \\
\bmS_h \bmsigma_{hk} \bmS^{\sT}_k  \qquad \hbox{for} \; h\ne k
\end{array}\,.
\ee 
Performing, now, a symplectic diagonalization, we can
reduce the diagonal blocks as
 \be \bmS_h \bmsigma_{hh} \bmS^{\sT}_h = a_h\,\ii_2 \;.
\ee 
Concerning the remaining three blocks, one may follow the procedure
that led to Eq.~(\ref{c3:NF2pMix}). In fact it is still possible to
find three orthogonal symplectic transformation $\bmO_1$, $\bmO_2$ and
$\bmO_3$ able to put two of the three blocks in a diagonal and in a
triangular form, respectively, leaving unchanged the diagonal
blocks. The covariance matrix (\ref{c3:GenSG3m}) can then be recast
into the following normal form
 \be
\label{c3:NF3m}
\bmsigma=
\left(
\begin{array}{cccccc}
a_1 & 0 & b_1 & 0 & b_6 & b_7 \\
0 & a_1 & 0 & b_2 & b_8 & b_9 \\
b_1 & 0 & a_2 & 0 & b_3 & b_4 \\
0 & b_2 & 0 & a_2 & 0 & b_5 \\
b_6 & b_8 & b_3 & 0 & a_3 & 0 \\
b_7 & b_9 & b_4 & b_5 & 0 & a_3 \\
\end{array}
\right) \;, 
\ee
where we can identify 12 independent parameters.
\index{Gaussian states!three-mode!normal form}
\index{covariance matrix!canonical form!three-mode}
\par 
Three-mode tripartite systems have been studied in different contests,
from quantum optics \cite{LB00,FPB+04}, to condensate physics
\cite{PCB03}. A study was also performed in which the mode of a
vibrational degree of freedom of a macroscopic object such as a mirror
has been considered \cite{PMV+03}. As examples we consider here the
two classes of states generated by means of the all optical systems
proposed in \cite{LB00} and \cite{FPB+04}. The first generation scheme
is a very natural and scalable way to produce multimode entanglement
using only passive optical elements and single squeezers, while the
second one is the simplest way to produce three mode entanglement
using a single nonlinear optical device. They both can be achieved
experimentally \cite{ATY+03,BAP+04}. As concern the first class of
states, it is generated with the aid of three single mode squeezed
states combined in a ``tritter'' (a three mode generalization of a
beam-splitter). The evolution is then ruled by a sequence of single
and two mode quadratic Hamiltonians. As a consequence, being generated
from vacuum, the three-mode entangled state is Gaussian, and its
covariance matrix is given by (for the rest of this section we set
$\kappa_2=2^{-1/2}$):
\begin{equation}
{\bmV}_{3}= \frac12\left(
\begin{array}{cccccc}
 {\cal R}_+  &  {\cal S} &  {\cal S}  &  0        &  0        &  0  \\
 {\cal S}  &  {\cal R}_+ &  {\cal S}  &  0        &  0        &  0  \\
 {\cal S}  &  {\cal S} &  {\cal R}_+  &  0        &  0        &  0         \\
 0         &  0        &  0         &  {\cal R}_- & -{\cal S} & -{\cal S}  \\
 0         &  0        &  0         & -{\cal S} &  {\cal R}_- & -{\cal S}  \\
 0         &  0        &  0         & -{\cal S} & -{\cal S} &  {\cal R}_-  \\
\end{array}
\right)
\label{c3:matC}\;,
\end{equation}
where
\begin{align}
{\cal R}_\pm=\cosh (2r)\pm\frac13\sinh (2r)\,, \quad
{\cal S}=-\frac23 \sinh (2r)\,,
\end{align}
and $r$ is the squeezing parameter (with equal squeezing in all initial
modes).  
\par 
The second class of tripartite entangled states is generated in a
single non linear crystal through a special case of Hamiltonian $H_{pq}$
in \refeq{LinHpq}, namely
\begin{equation}
H_{\rm int} =\gamma_1 a_1^\dag a^{\dag}_3 + \gamma_2 a_2^{\dag} a_3 + h.c.
\label{intH}\;,
\end{equation}
which describes two interlinked bilinear interactions taking place
among three modes of the radiation field coupled with the support of
two parametric pumps. It can be realized in $\chi^{(2)}$ media by a
suitable configuration exposed in Ref.~\cite{BAP+04}. The effective
coupling constants $\gamma_k$, $k=1,2$, of the two parametric
processes are proportional to the nonlinear susceptibilities and the
pump intensities. As already seen in Section \ref{ss:Hpq}, if we take
the vacuum $|{\bf 0}\rangle\equiv |0\rangle_1 \otimes |0\rangle_2
\otimes |0\rangle_3$ as the initial state, the evolved state
$|T\rangle=e^{-iH_{\rm int}t}|{\bf 0}\rangle $ belongs to the class of
the coherent states of ${\rm SU}(2,1)$ and it reads [see \refeq{Cp1}]
\begin{equation}
|T\rangle = \frac{1}{\sqrt{1+N_1}} \sum_{p,q=0}^{\infty}
\left(\frac{N_2}{1+N_1}\right)^{p/2} 
\left(\frac{N_3}{1+N_1}\right)^{q/2}
e^{-i(p\phi_2+q\phi_3)} 
\sqrt{\frac{(p+q)!}{p! q!}}\: |p+q,p,q\rangle
\label{T}\;,
\end{equation}
where $N_k(t)=\langle a^\dag_k(t)\, a_k(t)\rangle$ represent the average
number of photons in the $k$-th mode and $\phi_k$ are phase factors.
Notice that the latter may be eliminated by proper local unitary
transformations $U_2$ and $U_3$ on modes $a_2$ and $a_3$, namely
$U_{k}=\exp{\{i\phi_{k}a^\dagger_{k} a_{k}\}}$, $k=2,3$. The
symmetry of the Hamiltonian (\ref{intH}) implies that $N_1 =
N_2+N_3$, where
\begin{equation}
N_2 = \frac{|\gamma_1|^2 |\gamma_2|^2}{\Omega ^4}
\left[\cos(\Omega t)-1 \right]^2\,, \qquad
N_3 = \frac{|\gamma_1|^2}{\Omega ^2} \sin^2(\Omega t) \;,
\label{c3:Ndit}
\end{equation}
with $\Omega = \sqrt{|\gamma_2|^2 -|\gamma_1|^2}$. Also for this
second class, being the initial state Gaussian and the Hamiltonian
quadratic, the evolved states will be Gaussian. The explicit
expression of its covariance matrix reads as follows
\begin{equation}
{\bmV}_{T}= \left(
\begin{array}{cccccc}
 {\cal F}_1  &  {\cal A}_2 &  {\cal A}_3  &  0  & -{\cal B}_2 & -{\cal B}_3 \\
 {\cal A}_2  &  {\cal F}_2 &  {\cal C}    & -{\cal B}_2 &  0  &  {\cal D} \\
 {\cal A}_3  &  {\cal C}   &  {\cal F}_3  & -{\cal B}_3 & -{\cal D} &  0    \\
 0  & -{\cal B}_2 & -{\cal B}_3  &  {\cal F}_1 &  -{\cal A}_2 &  -{\cal A}_3  \\
-{\cal B}_2  &   0  & -{\cal D}  &  -{\cal A}_2 &  {\cal F}_2 &  {\cal C}  \\
-{\cal B}_3  &  {\cal D} &  0    &  -{\cal A}_3 &  {\cal C} &  {\cal F}_3  \\
\end{array}
\right)
\label{c3:matV}\;,
\end{equation}
where ${\cal F}_k = N_k +\frac12$ and 
\begin{align}
&{\cal A}_k =  \sqrt{N_k(1+N_1)} \cos\phi_k\,, 
&{\cal B}_k =  \sqrt{N_k(1+N_1)} \sin\phi_k\,, \nonumber\\
&{\cal C} =  \sqrt{N_2 N_3}\cos(\phi_2-\phi_3)\,, 
&{\cal D} =  \sqrt{N_2 N_3}\sin(\phi_2-\phi_3)\,. \nonumber
\end{align}
As already noticed, the covariance matrix (\ref{c3:matV}) may be
simplified by local transformations setting $\phi_2=\phi_3=0$.
Finally, if the Hamiltonian (\ref{intH}) acts on the thermal state
$\varrho_\nu =\nu_{\hbox{\tiny A}}\otimes\nu_{\hbox{\tiny
B}}\otimes\nu_{\hbox{\tiny C}}$, with equal mean thermal photon
number $N$ on each mode, we obtain the following
covariance matrix
\be
\label{c3:matVThermal}
\bmV_{T,{\rm th}}=(2 N+1)\bmV_{T} \;.
\ee

\chapter{Separability of Gaussian states}\label{SepGS}
\index{Gaussian states!separability}
Entanglement is perhaps the most genuine ``quantum'' property that a
physical system may possess. It occurs in composite systems as a
consequence of the superposition principle and of the fact that
the Hilbert space that describes a composite quantum system is the
tensor product of the Hilbert spaces associated to each
subsystems. In particular, if the entangled subsystems are
spatially separated nonlocality properties may arise, showing a very
deep departure from classical physics. 
\par
A non-entangled state is
called {\em separable}. Considering a bipartite quantum system ${\cal
H}={\cal H}_A\otimes{\cal H}_B$ (the generalization to
multipartite systems is immediate), a separable state $\varrho$
is defined as a convex combination of product states, namely 
\cite{wer89}: 
\be
\label{c4:DefSep} \varrho=\sum_k p_k \varrho_k^{(A)} \otimes
\varrho_k^{(B)} 
\ee 
\index{separability!definition}
where $p_k\ge0$, $\sum_k p_k=1$, and
$\varrho_k^{(A)}$, $\varrho_k^{(B)}$ belong to ${\cal H}_A$,
${\cal H}_B$ respectively.  The physical meaning of such a
definition is that a separable state can be prepared by means of
operations acting on the two subsystems separately ({\em i.e. 
local operations}), possibly coordinated by classical communication 
between the two subsystems\footnote{\footnotesize 
Quantum operations obtained by local actions plus
classical communication is usually referred to as LOCC 
operations.}  The correlations present, if any, in a separable 
state should be attributed to this communication and hence are 
of purely classical origin.  As a consequence no Bell inequality can be 
violated and no enhancement of computational power can be expected.
\par
The separability problem, that is recognizing whether a given state is
separable or not, is a challenging question still open in quantum
information theory. In this chapter a review of the separability
criteria developed to date will be presented, in particular for
what concern Gaussian states. We will profusely use the results
obtained in Chapter \ref{ch:gs} regarding the normal forms in which
Gaussian states can be transformed.
\section{Bipartite pure states}\label{c4:2pPure}
Let us start by considering the simplest class of states, for which 
the separability problem can be straightforwardly solved, that is 
pure bipartite states belonging to a Hilbert space of arbitrary 
dimension. First of all, recall that such states can be transformed 
by local operations into the normal form given by the Schmidt 
decomposition (\ref{c3:schmidt}), namely
\begin{equation}\label{c4:schmidt}
    \ket{\psi}_{AB} = \sum_{k=1}^d \sqrt{p_k}\,
    \ket{\phi_k}_A\ket{\varphi_k}_B \;.
\end{equation}
\index{separability!bipartite pure states}
Therefore, since the Schmidt coefficients are unique, it follows that 
the {\em Schmidt rank} ({\em i.e.}, the number of Schmidt coefficients 
different from zero) is sufficient to discriminate between separable 
and entangled states. Indeed, a pure bipartite state is separable if 
and only if its Schmidt
rank is equal to $1$. On the opposite, a pure state is said to be {\em
  maximally entangled} if its Schmidt coefficients are all equal to
$d^{-1/2}$ (up to a phase factor). In order to understand 
this definition, consider the partial traces
$\varrho_A=\hbox{Tr}_B[\varrho]$ and $\varrho_B=\hbox{Tr}_A[\varrho]$
of the state in Eq.~(\ref{c4:schmidt}), where $\varrho=\ket{\psi}
\bra{\psi}$. From \refeq{c4:schmidt} it follows that 
\be
\label{c4:PartialTrSch}
\varrho_A=\sum_k p_k\, \ket{\phi_k}_A {}_{A}\bra{\phi_k}\,, \qquad
\varrho_B=\sum_k
p_k\, \ket{\varphi_k}_B {}_{B}\bra{\varphi_k} \;, 
\ee 
hence it is clear that the
partial traces of a maximally entangled state are the maximally chaotic
states in their respective Hilbert space. From \refeq{c4:PartialTrSch}
it also follows that the von Neumann entropies (\ref{c3:S_V}) of the 
partial traces  are equal one each other, in formula:
\be
\label{c4:partialS_V}
S_V^A=S_V^B=-\sum_k p_k\log_d p_k \;.
\ee
\index{separability!partial entropies}
It is possible to demonstrate that (\ref{c4:partialS_V}) is the 
unique measure of entanglement for pure bipartite states
\cite{po97}. It ranges from $0$, for separable
states, to $1$, for maximally entangled states.
\par
\index{entanglement}
Let us now address the case we are more interested in, that is
infinite dimensional systems. Consider for the moment a two-mode
bipartite system. Following the definition of maximally entangled
states given above, it is clear that the twin-beam state (TWB)
\begin{equation}\label{def:twb}
|\Lambda\rangle\rangle = \sqrt{1-\lambda^2}\,
\sum_{n=0}^{\infty}\,\lambda^n |n,n\rangle\,,
\end{equation}
where $\lambda = \tanh r$, $r$ being the TWB squeezing parameter, is
a maximally entangled state. In fact, its partial traces are thermal
states, {\em i.e.}, the maximally chaotic state of a single-mode
continuous variable system, with mean photon number equal to the mean
photon number in each mode of the TWB, namely 
$\langle a^\dagger_k a_k \rangle=|\lambda|^2/(1-
|\lambda|^2)=\sinh^2r$, $k=A,B$,
in the notation of Section \ref{ss:opa}. The unique measure of
entanglement is then given by \refeq{c3:S_V1m}, that is the von
Neumann entropy of a generic Gaussian single mode state. Remarkably,
these observations are sufficient to fully characterize 
the entanglement of any bipartite $m\times n$ pure Gaussian state.
Indeed in Section \ref{c3:2m} we have
demonstrated that such a system can be reduced to the product of TWB
states and single mode local state at each party. As a consequence the
bipartite entanglement of a Gaussian pure state is essentially a $1
\times 1$ entanglement. 
\par
We mention here that besides the separability criterion given by the 
Schmidt rank, for pure bipartite system another necessary and 
sufficient condition for the entanglement is provided by the 
violation of local realism, for some suitably chosen Bell 
inequality \cite{GP92}.
\section{Bipartite mixed states}\label{c4:2pMix}
\index{separability!bipartite mixed states}
The problem of separability shows its complexity as soon as we deal
with mixed states. For example, there exist states that do not violate
any inequality imposed by local realism, but yet cannot be constructed
by means of LOCC. The first example of such a state was given by
Werner \cite{wer89}. Despite the efforts, a general solution to the
problem of separability in the case of an arbitrary mixed state has
not been found yet. Most of the criteria proposed so far are
generally only necessary for separability, even if for some particular
classes of states they provide also necessary conditions for
entanglement. Fortunately, these particular cases are of great
relevance in view of the application to quantum information, in fact
they include $2 \times 2$ and $2 \times 3$ finite dimensional systems
and $m \times n$ and $1 \times 1 \times 1$ infinite dimensional systems
in case of Gaussian states.
\par
Most of the separability criteria relies on the key
observation that separability can be revealed with the aid of
positive but not completely positive maps. Let us explain these
point in more details. Every linear map $\varrho \mapsto {\cal
L}[\varrho]$, in order to be an admissible physical
transformation, has to be trace preserving and positive in the
sense that it maps positive semidefinite operators (statistical
operators) again onto positive semidefinite operators. However, a
physical transformation has not only to be positive: if we
apply the transformation only to one part of a composite system, and
leave the other parts unchanged, then the overall state after the
operation has to be described by a positive semi-definite operator
as well. In other words, all the possible extensions of the map
should be positive. Such a map is called {\em completely positive}
(CP). A map which is positive but not CP doesn't correspond to any
physical operation, nevertheless these maps have become an
important tool in the theory of entanglement. The reason for this
will be clear considering the most prominent example of such a
map, transposition ($T$). Transposition applied only to a part of
a composite system is called {\em partial transposition} (in the
following we will use the symbol $T$ with a subscript that
indicates the subsystem with respect to the transposition is
performed). Positivity under partial transposition has been introduced in
entanglement theory by Peres \cite{Per96} as a necessary
condition for separability. In fact, consider a separable state as
defined in \refeq{c4:DefSep} and apply a transposition only to
elements of the first subsystem $A$.  Then we have:
\be
\label{c4:ParTrans} \varrho^{\scriptscriptstyle T_A}=\sum_k p_k
\big(\varrho_k^{\scriptstyle (A)}\big)^{\sT} \otimes 
\varrho_k^{\scriptstyle (B)} \;.
\ee 
Since the transposed matrix
$\big(\varrho_k^{(A)}\big)^{\sT}=\big(\varrho_k^{(A)}\big)^*$ is 
non-negative and with unit trace it is a legitimate density 
matrix itself. It follows that none of the eigenvalues of
$\varrho^{\scriptscriptstyle T_A}$
is negative if $\varrho$ is separable. This criterion is often
referred to as {\tt ppt} criterion ({\em positivity} under 
{\em partial transposition}).
\index{separability!ppt condition}
\index{partial transposition}
\index{ppt condition}
Of course, partial transposition with respect to the second subsystem
$B$ yields the same result.
\par
If we consider systems of arbitrary dimensions {\tt ppt} criterion is not
sufficient for separability, but it turns out to be necessary and
sufficient for systems consisting of two qubits \cite{HHH96}, that
is a system described in the Hilbert space ${\cal H}={\mathbb C}^2
\otimes {\mathbb C}^2$. This is due to the fact that there exist
a general necessary and sufficient criterion for separability
which saying that a state $\varrho$ is separable if and only if
for all positive maps ${\cal L}$, defined on subsystem ${\cal
H}_A$,   $({\cal L}\otimes {\mathbb I})[\varrho]$ is a
semi-positive defined operator \cite{HHH96}. Due to the limited knowledge about
positive maps in arbitrary dimension this criterion turn out to be
inapplicable in general. Nevertheless, in the case of ${\mathbb
C}^2$ it is known that all the positive maps can be decomposed as
${\cal P}_1+{\cal P}_2 T$, where ${\cal P}_1$, ${\cal P}_2$
are CP maps \cite{Str63}. Hence the sufficiency
of partial transposition criterion for two qubits follows.
\par
The {\tt ppt} criterion turns out to hold also for ${\cal H}={\mathbb C}^2
\otimes {\mathbb C}^3$ systems, but for higher dimensions no criterion
valid for every density operator $\varrho$ is known. Indeed, in general there
exist entangled states with positive partial transpose, the so called
{\em bound entangled states}. The first example of such a state was
given in Ref.~\cite{Hor97}.
\par
At first sight it is not clear how the {\tt ppt} criterion, developed
for discrete variable systems, can be translated to continuous
variables. Furthermore, considering that {\tt ppt} criterion ceases to
be sufficient for separability as the dimensions of the system
increases, one might expect that it will provide only a necessary
condition for separability in case of continuous variables. In
fact, Simon \cite{Sim00} showed that for arbitrary continuous
variable case this conjecture is true. However, Simon also
demonstrated that for $1 \times 1$ Gaussian states the {\tt ppt}
criterion represents also a sufficient condition for separability.
\par
\index{partial transposition!continuous variable systems}
Simon's approach relies on the observation that transposition
translates to mirror reflection in a continuous variables scenario. In
fact, since density operators are Hermitian, transposition
corresponds to complex conjugation. Then, by taking into account 
that complex conjugation corresponds to time reversal of the 
Schroedinger equation, it is clear that, in terms of continuous 
variables, transposition corresponds to a sign change of the momentum
variables, {\em i.e.} mirror reflection. In formula,
\be
\label{c4:TransPhSpace}
\bmR \rightarrow \bmDelta \bmR\,, \qquad \bmS\rightarrow \bmLambda\, \bmS\,,
\ee
where
 \be \bmDelta = {\rm Diag}(1,-1,\dots,1,-1)\,, \qquad \bmLambda
= {\mathbbm 1}_n\oplus (-{\mathbbm 1}_n) \;.  
\ee 
The action of transposition on the covariance matrices
$\bmV$ and $\bmsigma$ of a generic state reads as follows: 
$\bmV \rightarrow \bmLambda\bmV\bmLambda$ and $\bmsigma 
\rightarrow \bmDelta \bmsigma\bmDelta$, respectively. For a
bipartite system ${\cal H}={\cal H}_A\otimes{\cal H}_B$ partial
transposition with respect to system $A$ will be performed on the
phase space through the action of the matrices $\bmLambda_A=\bmLambda
\oplus {\mathbbm 1}$ and $\bmDelta_A=\bmDelta \oplus {\mathbbm 1}$, 
where the first factor of the tensor product refers to subsystem 
$A$ and the second one to $B$.
Following now the strategy pursued above in case of discrete
variables, a necessary condition for separability is that the partial
transposed operator is semi-positive definite, which in terms of
covariance matrix is now reflected to the following uncertainty
relation
\begin{eqnarray}
\bmDelta_A \bmsigma\bmDelta_A +
\frac{i}{4\kappa_1^2}\, \boldsymbol{\Omega} \geq 0\,, \qquad
\bmLambda_A\bmV\bmLambda_A - \frac{i}{4\kappa_1^2}\, \boldsymbol{J} \geq 0
\label{c4:HeisSG}\;.
\end{eqnarray}
We may write these conditions also in the equivalent form
\begin{equation}
 \bmsigma \ge
-\frac{i}{4\kappa_1^2}\, {\widetilde {\boldsymbol \Omega}}_A\,, \qquad
\bmV \ge \frac{i}{4\kappa_1^2}\, {\widetilde {\boldsymbol J}}_A
\label{c4:HeisSGReversed}\;,
\end{equation}
where ${\widetilde {\boldsymbol
\Omega}}_A=\bmDelta_A\boldsymbol{\Omega}\bmDelta_A$ and
${\widetilde {\boldsymbol
J}}_A=\bmLambda_A\boldsymbol{J}\bmLambda_A$.
\par
\index{separability!continuous variable systems}
\index{ppt condition!continuous variable systems}
Let us consider, in particular, the case of $1 \times 1$ Gaussian states. 
We have already seen that, by virtue of the normal form (\ref{c3:NF2mMix}),
relation \refeq{c4:HeisSG} has the simple local symplectic invariant
form given by \refeq{c3:HeisInvForm}. Recalling the definition of the
four invariants given in Section \ref{c3:2m} we have
\be {\tilde I}_1=I_1\,, \qquad {\tilde I}_2=I_2\,,
\qquad {\tilde I}_3=-I_3\,, \qquad {\tilde I}_4=I_4\,,
\ee
where ${\tilde I}_k$ are referred to matrix $\bmDelta_A \bmsigma\bmDelta_A$,
while $I_k$ to $\bmsigma$. Notice that of course these relations would not have
changed if we had chosen to transpose $\bmsigma$ with respect to the
second subsystem $B$. Hence, a separable Gaussian state must obey not
only to Ineq.~(\ref{c3:HeisInvForm}) but also to 
the same inequality with a minus sign in front of $I_3$. This leads to a more
restrictive uncertainty relation. Together with (\ref{c3:HeisInvForm}) 
they summarize as follows
\be
\label{c4:PTInvForm}
I_1+I_2+2|I_3| \le 8\kappa_1^2I_4+\frac{1}{8\kappa_1^2} \;.
\ee
Moreover, notice that for states with $I_3\ge0$, this relation is subsumed by
the physical constrain given by the uncertainty relation
(\ref{c3:HeisInvForm}). Relation \refeq{c4:PTInvForm}, being invariant
under local symplectic transformations, does not depend on the normal
form (\ref{c3:NF2mMix}), nevertheless it is worthwhile to rewrite it
in case of a correlation matrix given in the normal form \refeq{c3:NF2mMix}. In
fact, \refeq{c4:PTInvForm} then simplifies to:
\be
\label{c4:PTSymplified}
8\kappa_1^2(ab-c_1^2)(ab-c_2^2)\ge a^2+b^2+2|c_1c_2|-\frac{1}{8\kappa_1^2} \;.
\ee
\par
As pointed out in Chapter \ref{ch:gs}, the uncertainty relation 
for a covariance matrix can be summarized by a condition imposed on 
its minimum symplectic eigenvalue.
Hence, in terms of the symplectic eigenvalues  ${\tilde d}_{\pm}$
of the partially transposed covariance matrix the {\tt ppt} criterion
becomes
\be
\label{c4:nptSympEig} {\tilde d}_{-}\ge
(2\kappa_1)^{-2} \;.
\ee
\par
Viewed somewhat differently, the {\tt ppt} criterion can be translated
also in term of expectation values of variances of properly chosen
operators. In fact, it is equivalent to the statement that for
every four-vectors $\bme$ and $\bme'$ the following Inequality is
true: \be \label{c4:SimNPTVariances} \langle[\Delta
w(\bme)]^2\rangle + \langle[\Delta w(\bme')]^2\rangle \ge
\frac{1}{2\kappa_1^2}(|\mathfrak J(\bme_A,\bme'_A)| + |\mathfrak
J(\bme_B,\bme'_B)|) \;, \ee where $w(\bme)=\bme^{\sT}\bmR$, and we defined
$\bme=(e_1,e_2,e_3,e_4)$, $\bme_A=(e_1,e_2)$, $\bme_B=(e_3,e_4)$,
$\mathfrak J(\bme_{k},\bme'_{k})=\bme_{k}^{\sT}\,\boldsymbol
J\,\bme'_{k}$, with $k=A,B$.
\par
Here we have shown that the {\tt ppt} criterion is necessary for
separability, as concern its sufficiency we remand to the original
paper by Simon \cite{Sim00}.
\par
Another necessary and sufficient criterion for the case of
two-mode bipartite Gaussian states has been developed in Ref.
\cite{DGC+00}, following a strategy independent of partial
transposition. It relies upon a normal form slightly different
from [\refeq{c3:NF2mMix}]
 \be
\label{c4:NF2mMixDuan}
\bmsigma =
\left(
\begin{array}{cccc}
 a_1   &  0   &  d_1 &  0   \\
 0   &  a_2   &  0   &  d_2 \\
 d_1 &  0   &  b_1   &  0   \\
 0   &  d_2 &  0   &  b_2
\end{array}
\right) \;, \ee
where
\begin{subequations} 
\begin{align}
\frac{a_1-1/4}{b_1-1/4}&=\frac{a_2-1/4}{b_2-1/4}\,, \\
|d_1|-|d_2|&=\sqrt{(a_1-1/4)(b_1-1/4)}-
\sqrt{(a_2-1/4)(b_2-1/4)}
 \;. 
\end{align}
\end{subequations}
Every two mode covariance matrix can be put in this normal form by
combining first a transformation into the normal form
(\ref{c3:NF2mMix}), then two appropriate local squeezing operations. In
terms of the elements of (\ref{c4:NF2mMixDuan}) the criterion reads as follows:
\be
\label{c4:DuanCritNecSuff}
\langle(\Delta u_0)^2 \rangle +\langle(\Delta v_0)^2 \rangle \ge
\frac{1}{2\kappa_1^2}\Big(a_0^2+\frac{1}{a_0^2}\Big) \;, \ee
where $\Delta u_0$ indicate the
variance of the operator $u_0$, and
\begin{align}
u_0=a_0q_1-\frac{d_1}{|d_1|a_0}q_2\,,  \qquad
v_0=a_0p_1-\frac{d_2}{|d_2|a_0}p_2 \,, \quad
a_0^2=\sqrt{\frac{a_1-1/4}{b_1-1/4}} \,.
\end{align}
\par
Without the assumption of Gaussian states, an approach based only on
the Heisenberg uncertainty relation of position and momentum and on
the Cauchy-Schwarz inequality, leads to an inequality similar to
\refeq{c4:DuanCritNecSuff}. It only represents a necessary condition
for the separability of arbitrary states, and it states that for any two
pairs of operators $A_{k}$ and $B_{k}$, with $k=x,p$, such that
$[A_x,A_p]=[B_x,B_p]=i/(2\kappa_1^2)$, if
$\varrho$ is separable then
 \be
\label{c4:DuanCritNec}
\langle(\Delta u)^2 \rangle +\langle(\Delta v)^2 \rangle \ge
\frac{1}{2\kappa_1^2}\Big(a^2+\frac{1}{a^2}\Big) \;,
\qquad (\forall a>0)
\ee
where
\be
u=a A_x\mp \frac1a B_x\,, \qquad
v=a A_p\pm \frac1a B_p
 \;. \ee
\par
In order to demonstrate the equivalence between the necessary
condition given by Simon's and Duan {\em et al.} criteria let us
compare Ineqs.~(\ref{c4:SimNPTVariances}) and (\ref{c4:DuanCritNec}).
We follow the argument given in Ref.  \cite{Gie01}. It is immediate to
see that when Ineq.~(\ref{c4:SimNPTVariances}) is respected then so is
Ineq.~(\ref{c4:DuanCritNec}). In fact, for any given $a$ it is
sufficient to consider Ineq.~(\ref{c4:SimNPTVariances}) for
$\bme=(a,0,\pm a^{-1},0)$ and $\bme'=(0,a,0,\mp a^{-1})$, and to
identify $q_{A}\equiv A_x$, $p_{A}\equiv A_p$ and $q_{B}\equiv B_x$,
$p_{B}\equiv B_p$. The
reverse, can be seen as follows: denoting with $\bme$ and $\bme'$ the
two vectors for which Ineq.~(\ref{c4:SimNPTVariances}) is
violated, then there exist $a$, $\lambda$ and a pair of symplectic
transformations $S_{A},S_{B}\in {\rm Sp}(2,\mathbb R)$ such that
$S_A(a,0)=\lambda \bme_A$, $S_A(0,a)=\lambda \bme'_A$
$S_B(a^{-1},0)=\lambda \bme_B$ $S_B(0,a^{-1})=\lambda \bme'_B$, namely
\begin{subequations}
\begin{align}
S_A& =\frac{\lambda}{a}
\begin{pmatrix}
e_1 & e_1' \\
e_2 & e_2'
\end{pmatrix}\,,
&
S_B& =\lambda a
\begin{pmatrix}
e_3 & e_3' \\
e_4 & e_4'
\end{pmatrix}\,,
 \\[1ex]
\lambda& =\frac{a}{e_1e_2'-e_2e_1'}\,,
&
a& =\sqrt{\frac{e_1e_2'-e_2e_1'}{e_3e_4'-e_4e_3'}} \;.
\end{align}
\end{subequations}
The existence of $\lambda$ and $a$ is ensured by the fact that a
violation of Ineq.~(\ref{c4:SimNPTVariances}) implies that $\mathfrak
J(\bme_{A},\bme'_{A})\mathfrak J(\bme_{B},\bme'_{B})\le 0$ [otherwise
Ineq.~(\ref{c4:SimNPTVariances}) would correspond to the
uncertainty principle, consequently it should be respected by any
state].  Notice now that if Ineq.~(\ref{c4:SimNPTVariances}) is
violated for $\bme$ and $\bme'$, so is for $\lambda\bme$ and
$\lambda\bme'$, implying that
 \be
\label{last:formulax} 
\langle[\Delta
w(\lambda\bme)]^2\rangle + \langle[\Delta w(\lambda\bme')]^2\rangle
\ge \frac{1}{2\kappa_1^2}\big(|\mathfrak J(\lambda\bme_A,\lambda\bme'_A)| +
|\mathfrak J(\lambda\bme_B,\lambda\bme'_B)|\big) \;.
\ee
By inspection of the left hand side of the Ineq.~(\ref{last:formulax}) one can
identify
$A_x=w(S_A(1,0)^{\sT})$, $A_p=w(S_A(0,1)^{\sT})$,
$B_x=\pm w(S_B(1,0)^{\sT})$, $B_p=\mp w(S_B(0,1)^{\sT})$,
where
for simplicity we indicated $w(S_A)(1,0)^{\sT}\equiv w((S_A\oplus \mathbb
I)(1,0,0,0)^{\sT})$ and so on.  Being $S_{A}$ and $S_{B}$ symplectic,
the operators
introduced satisfy the commutation relation
$[A_x,A_p]=[B_x,B_p]=i/(2\kappa_1^2)$ and $\mathfrak
J(\lambda\bme_A,\lambda\bme'_A)=a^2$, $\mathfrak
J(\lambda\bme_B,\lambda\bme'_B)=a^{-2}$. Consequently 
Ineq.~(\ref{c4:DuanCritNec}) is violated.
\par
Other criteria based on variances of suitable operators can be found
in Refs.~\cite{Tan99,GMV+03,LF03}. These criteria, though only
necessary for separability, are worthwhile in view of an experimental
implementation. In fact, in order to apply criteria (\ref{c4:HeisSG})
or (\ref{c4:DuanCritNecSuff}) it is necessary to measure all the
entries of the covariance matrix. Although this is achievable, 
{\em e.g.} by quantum tomography, it may experimentally demanding. 
The criteria in Refs.~\cite{Tan99,GMV+03,LF03,TW03} allow instead to witness
entanglement measuring only the variances of appropriate linear
combinations of all the modes involved. An experimental implementation
of such a criterion can be found in Ref.~\cite{ATY+03}.
\par
As an example consider the TWB, whose covariance matrix is given in
\refeq{c3:TWBsigma}. It is immediate to see that the criterion given
by \refeq{c4:PTSymplified} implies that $\sinh^2(2r)<0$, which is
violated for every squeezing parameter $r$. The application of
criterion \refeq{c4:DuanCritNecSuff} is also straightforward.
\par
Concerning more than one mode for each party, it is possible to
demonstrate that the criterion given by Ineq.~(\ref{c4:HeisSG}) gives a
necessary and sufficient condition for separability only for the case
of $1 \times n$ modes \cite{WW01}. The simplest example where the
criterion ceases to be sufficient for separability involves a $2
\times 2$ system, where bound entangled states can be found.  For a general
$n \times m$ Gaussian state there is also a necessary and sufficient
condition, which states that a covariance matrix $\bmsigma$ correspond
to a separable state if and only if there exist a pair of correlation matrices
$\bmsigma_A$ and $\bmsigma_B$, relative to subsystems $A$ and $B$
respectively, such that the following inequality holds \cite{WW01}:
\be 
\label{c4:WWdecomposition}\bmsigma\ge\bmsigma_A\oplus\bmsigma_B \;.  
\ee
Unfortunately this criterion is difficult to handle in practice,
due to the problem of finding such a pair of correlation matrices. A
more practical solution has been given in Ref.~\cite{GKL+01}. It gives
an operational criterion based on a nonlinear map, rather on the usual
linear partial transposition map, hence independent of {\tt ppt} criterion.
Consider a generic covariance matrix $\bmsigma_0$, decomposed as usual
in the following blocks: 
\be
\label{c4:SigmaBlocks}
\bmsigma_0=
\begin{pmatrix}
\bmA_0 & \bmC_0 \\
\bmC_0^{\sT} & \bmB_0
\end{pmatrix} \;.
\ee
Define now a sequence of matrices $\{\bmsigma_k\}$,
$k=0,\dots,\infty$, of the form (\ref{c4:SigmaBlocks}), according to
the following rule: if $\bmsigma_k$ is not a covariance matrix [{\em
i.e.}, if $\bmsigma_k\not\ge-i(4\kappa_1^2)^{-1} {\boldsymbol
\Omega}$] then $\bmsigma_{k+1}=0$, otherwise
\begin{subequations}
\label{c4:mapGiedkeMxN}
\begin{align}
\bmA_{k+1}&=\bmB_{k+1}=\bmA_k-\real{\bmD_k}\\
\bmC_{k+1}&=-\immag{\bmD_k}
\end{align}
\end{subequations}
where   $\bmD_k\equiv \bmC_k[\bmB_k+i(4\kappa_1^2)^{-1} {\boldsymbol
\Omega}]^{-1}\bmC_k$ (the inverse should be meant as pseudo-inverse).
The importance of this sequence is that $\bmsigma_0$ is separable if
and only if $\bmsigma_k$ is a valid separable covariance matrix. Then
the necessary and sufficient separability criterion states that if,
for some $k>1$
\begin{enumerate} 
\item $\bmA_k\not\ge-i(4\kappa_1^2)^{-1} {\boldsymbol \Omega}$,
then $\bmsigma_0$ is not separable;
\item $\bmA_k- \|\bmC_k\|_{\rm op}\,{\mathbbm 1}\ge-i(4\kappa_1^2)^{-1}
{\boldsymbol \Omega}$, then $\bmsigma_0$ is separable.;
\end{enumerate}
$\|O\|_{\rm op}$ stands for the operator norm of $O$, {\em i.e.} the
maximum eigenvalue of $\sqrt{O^\dagger O}$.  Thus, one just has to iterate
the map (\ref{c4:mapGiedkeMxN}) until he finds that either $\bmA_k$ is no
longer a covariance matrix or $\bmA_k-\|\bmC_k\|_{\rm op}\,{\mathbbm 1}$ is a
covariance matrix. Moreover it is possible to demonstrate that these
conditions occur after a finite number of steps, and that in case of a
separable $\bmsigma_0$ decomposition (\ref{c4:WWdecomposition}) can be
explicitly constructed. We finally mention that recently it has been shown
\cite{SAI04} that {\tt ppt} criterion is necessary and sufficient for a
subclass of $m\times n$ Gaussian states, namely the bisymmetric ones. The
latter are defined as $ m\times n$ Gaussian states invariant under local
mode permutations on subsystems $A$ and $B$. This result is based on the
observation that bisymmetric states are locally equivalent to the tensor
product of a two-mode entangled state and of $m+n-2$ uncorrelated
single-mode states.
\par
\index{entanglement!measures}
\index{entanglement!negativity}
As for the quantification of entanglement, no fully satisfactory
measure is known at present for arbitrary mixed two-mode Gaussian
states. There are various measures available such as the entanglement
of distillation and of formation \cite{BVS+96}. They quantify the
entanglement of a state in terms of the pure state entanglement that
can be distilled out of it and the one that is needed to prepare it,
respectively. Another computable measure of entanglement is the
``logarithmic negativity'' based on the negativity of the partial
transpose \cite{VW02}. Physically it is related to the robustness of
the entanglement when the state under consideration evolves in a noisy
environment. The negativity of a quantum state $\varrho$ is defined as
\be
{\cal N}(\varrho)=\frac{\|\varrho^{\scriptscriptstyle T_A} \|_{\rm tr}-1}{2}\: ,
\ee
where $\| O\|_{\rm tr}\equiv\,{\rm Tr}\big[\sqrt{ O^{\dag} O}\big]$ stands
for the trace norm of an operator $O$. The quantity ${\cal N} (\varrho)$ is
equal to $|\sum_{k}\lambda_{k}|$, the modulus of the sum of the negative
eigenvalues of $\varrho^{\scriptscriptstyle T_A}$, and it quantifies the
extent to which $\varrho^{\scriptscriptstyle T_A}$ fails to be positive.
Strictly related to $\cal N$ is the logarithmic negativity $E_{\cal N}$,
defined as $E_{\cal N}\equiv \ln\|\varrho^{\scriptscriptstyle T_A}\|_{\rm
tr}$. The negativity has been proved to be convex and monotone under LOCC
\cite{VW02}. For two-mode Gaussian states it can be easily shown that the
negativity is a simple function of $\tilde{d}_{-}$, which is thus itself an
(increasing) entanglement monotone; one has in fact \cite{VW02}
\be
E_{\cal N}(\bmsigma)=\max\left\{0,
-\ln{\big[(2\kappa_2)^2\tilde{d}_{-}\big]}\right\}\: .
\ee
This is a decreasing function of the smallest partially
transposed symplectic eigenvalue $\tilde{d}_{-}$. Thus, recalling
\refeq{c4:nptSympEig}, the eigenvalue $\tilde{d}_{-}$ completely
qualifies and quantifies the entanglement of a two-mode Gaussian state
$\bmsigma$.
\section{Tripartite states}\label{c4:3party}
\index{separability!tripartite states}
When systems composed by $n>2$ parties are considered, the
separability issue becomes more involved. An immediate observation
concerns the fact that situations can occur in which some parties
of the total system may be entangled one each other but separable
from the rest of the system. Thus, a classification of all the
possible situations must be firstly considered. We adopt the
classification introduced in Ref.~\cite{DCT99} which exploits all
the possible ways to group the $n$ parties into $m\leq n$ subsets,
which are then themselves considered each as a single party. Now,
it has to be determined whether the resulting $m$-party state can
be written as a mixture of $m$-party product states.  The complete
record of the $m$-separability of all these states then
characterizes the entanglement of the $n$-party state.  Let us
investigate in particular the case we are more interested in, that
is tripartite systems. For these systems, we need to consider four
cases, namely the three bipartite cases in which $AB$, $AC$, or
$BC$ are grouped together, and the tripartite case
in which all $A$, $B$, and $C$ are separate. In total, we have the
following five different entanglement classes:
\begin{description}
\item[Class 1] (\emph{Fully inseparable states} or \emph{genuinely
    entangled states}) States which are not separable for any grouping
  of the parties.
\item[Class 2] (\emph{1-party biseparable states}) States which are
  separable if two of the parties are grouped together, but
  inseparable with respect to the other groupings. In general, such a
  state can be written as $\sum_h p_h\, \varrho_h^{(r)} \otimes
  \varrho^{(s\,t)}_h$ for one party $r$.
\item[Class 3] (\emph{2-party biseparable states}) States which are
  separable with respect to two of the three bipartite splits but
  inseparable with respect to the third, {\em i.e.} they can be
  written as $\sum_h p_h \, \varrho_h^{(r)} \otimes
  \varrho^{(s\,t)}_h$ for
  two parties $r$.
\item[Class 4] (\emph{3-party biseparable states}) States which are
  separable with respect to all three bipartite splits but cannot be
  written as a mixture of tripartite product states.
\item[Class 5] (\emph{fully separable}) States that can be written as a
  mixture of tripartite product states, $\sum_h p_h \,
\varrho^{(A)}_h \otimes \varrho^{(B)}_h\otimes \varrho^{(C)}_h$.
\end{description}
Needless to say, the most interesting class is the first one. In fact
fully inseparable states are necessary to implement genuinely
multipartite quantum information protocols able to increase the
performances with respect to classical ones \cite{LB00}.
\par
In general, it is hard to identify the class to which a given state
belong. The problem arises even in case of pure states, because a
Schmidt decomposition doesn't exist in general. The state vector then
cannot be written as a single sum over orthonormal basis state.
Concerning discrete variable systems, it is known that are only two
inequivalent classes of pure fully inseparable three-qubits states,
namely the GHZ \cite{GHS+90} and the W states \cite{DVC00}
\begin{align} 
\ket{\hbox{GHZ}}= (\ket{000}+\ket{111})/\sqrt2 
\qquad
\ket{\hbox{W}}= (\ket{100}+\ket{010}+\ket{001})/\sqrt 3
\;. 
\end{align}
In other words, any pure fully inseparable three-qubit
state can be transformed via stochastic LOCC (where stochastic
means that the transformation occurs with non-zero probability) to
either the GHZ or the W state. Hence a satisfactory knowledge for
this case has been achieved.
\par
When arbitrary mixed states are considered there is no general
necessary and sufficient criterion to ensure genuine entanglement.
The difficult of the subject is well exemplified if one exploits
the issue of nonlocality considering multi-party Bell
inequalities. Indeed, violations of such inequalities ensures only
that the state under investigation is partially entangled.
Reversely, fully inseparable states do not necessarily violate
multi-party Bell inequalities. As an example, the pure genuinely
$n$-party entangled state
\be
\ket{\psi}=\cos\alpha|0\dots0\rangle+\sin\alpha|1\dots1\rangle
\ee
for $\sin\alpha \le 2^{-(n-1)/2}$ does not violate any
$n$-party Bell inequality \cite{Bel64}, if $n$ is odd, and does not
violate Mermin-Klyshko inequalities \cite{Mer90,Kly93,GBP98} for
any $n$.
\par
Nevertheless, regarding the case of tripartite three-mode Gaussian
states, the separability has been completely solved.  Extending
Simon's {\tt ppt} approach Giedke {\em et al.} \cite{GKL+01b} gave a
simple criterion that allows to determine which class a given state
belong to. Hence genuine entanglement, if present, can be unambiguously
identified. Observing that for these systems the only partially
separable forms are those with a bipartite splitting of $1 \times 2$
modes, it follows that the {\tt ppt} criterion is necessary and sufficient.
We have the following equivalences:
\begin{subequations}
\begin{align}
\bmsigma \not\ge -\frac{i}{4\kappa_1^2} {\widetilde {\boldsymbol \Omega}}_A
\,,\;
\bmsigma \not\ge -\frac{i}{4\kappa_1^2} {\widetilde {\boldsymbol
\Omega}}_B\,,\;
\bmsigma \not\ge -\frac{i}{4\kappa_1^2} {\widetilde
{\boldsymbol \Omega}}_C
&\Leftrightarrow \mbox{Class 1}\\
\bmsigma \not\ge -\frac{i}{4\kappa_1^2} {\widetilde {\boldsymbol \Omega}}_A
\,,\; \bmsigma \not\ge -\frac{i}{4\kappa_1^2} {\widetilde {\boldsymbol
\Omega}}_B\,,\; \bmsigma \ge -\frac{i}{4\kappa_1^2} {\widetilde {\boldsymbol
\Omega}}_C
&\Leftrightarrow \mbox{Class 2}\\
\bmsigma \not\ge -\frac{i}{4\kappa_1^2} {\widetilde {\boldsymbol \Omega}}_A
\,,\; \bmsigma \ge -\frac{i}{4\kappa_1^2} {\widetilde {\boldsymbol
\Omega}}_B\,,\; \bmsigma \ge -\frac{i}{4\kappa_1^2} {\widetilde {\boldsymbol
\Omega}}_C
&\Leftrightarrow \mbox{Class 3}\\
\bmsigma \ge -\frac{i}{4\kappa_1^2} {\widetilde {\boldsymbol \Omega}}_A
\,,\; \bmsigma \ge -\frac{i}{4\kappa_1^2} {\widetilde {\boldsymbol
\Omega}}_B\,,\; \bmsigma \ge -\frac{i}{4\kappa_1^2} {\widetilde {\boldsymbol
\Omega}}_C &\Leftrightarrow \mbox{Class 4 or 5}
\end{align}
\label{c4:npt3m}
\end{subequations}
\index{separability! tripartite Gaussian states}
Analogue formulas may be written for the covariance matrix $\bmV$.
Notice that in classes 2 and 3 all the permutations of the indices
$A$, $B$, and $C$ must be considered. Classes 4 and 5 cannot be
distinguished via the {\tt ppt} criterion. An additional criterion has
been given in Ref.~\cite{GKL+01b} to distinguish between these two
classes. It is based on the consideration that necessary and
sufficient for full separability is the existence of three single
mode covariance matrices $\bmsigma^1_A$, $\bmsigma^1_B$, $\bmsigma^1_C$ such
that \be \label{c4:3mSepWW} 
\bmsigma\ge \bmsigma^1_A\oplus\bmsigma^1_B\oplus\bmsigma^1_C \;. \ee
Obviously, for the identification of fully inseparable states,
only class 1 has to be distinguished from the rest, thus the {\tt ppt}
criterion alone suffices.
\par
As examples, consider the states given in Section \ref{s:3mGS}.  The
separability issue of state (\ref{c3:matC}) has been addressed in
Refs.~\cite{LB00,GKL+01b}. In particular, in Ref.~\cite{GKL+01b} the
authors analyzed a generalization of state (\ref{c3:matC}), in which
some noise has been added. The state considered is described by the
covariance matrix $\bmsigma_{3,\mu}=\bmsigma_{3}+\frac{\mu}{2} \ii$.
Depending on the value of the squeezing parameter $r$ and of the noise
coefficient $\mu$, $\bmsigma_{3,\mu}$ belongs either to class 1, 4 or 5.
If $\mu=0$ the state is fully inseparable for any value of $r$.  In
fact, applying the {\tt ppt} criterion we find that matrix
$\bmsigma_3+i(4\kappa_1^2)^{-1} {\widetilde {\boldsymbol \Omega}}_A$
always has a negative minimum eigenvalue $\lambda_{\rm min}$ given by
($\kappa_1=2^{-1/2}$)
\be
\lambda_{\rm min}=\cosh (2\,r) - \frac{1}{\sqrt{6}}{\sqrt{3 + 3\,\cosh (4\,r) + 
        8\,{\sqrt{2}}\,\sinh (2\,r)}} \;.
\ee
From the symmetry of the state full
inseparability follows. On the contrary, if $\mu\ge1$ then
Ineq.~(\ref{c4:3mSepWW}) is satisfied with
$\bmsigma^1_A=\bmsigma^1_B=\bmsigma^1_C= \frac12 \ii$, hence the
state is separable. A detailed inspection considering a fixed
squeezing parameter $r$ shows that two threshold value of the noise
$\mu_0$, $\mu_1$ can be identified, such that $\bmsigma_{3,\mu}$ is fully
inseparable for $\mu<\mu_0$ and separable for $\mu>\mu_1$. When
$\mu_0\le\mu\le\mu_1$ it belongs to class 4, hence it is an example of
a bound entangled state, having every partial transpose positive,
nevertheless being inseparable.
\par
Let us focus now on state (\ref{c3:matVThermal}). The symmetry of this
state under the exchange of modes $a_2$ and $a_3$ allows to study the
separability problem only for modes $a_1$ and $a_2$. Furthermore, as
already pointed out, we can set $\phi_2=\phi_3=0$ without affecting the
entanglement properties of the state under investigation.  Concerning the
first mode, from an explicit calculation of the minimum eigenvalue of
matrix $\bmV_{T,{\rm th}}-\frac{i}{2} {\widetilde {\boldsymbol J}}_A$ (we
consider again $\kappa_1=2^{-1/2}$) it follows that
\begin{equation}
\lambda^{\rm min}_1= N+(1+2 N)
\left[N_1-\sqrt{N_1(N_1+1)}\right]\ \;.
\label{c4:TThAvalMin}
\end{equation}
As a consequence mode $a_1$ is separable from the others when
\begin{equation}
N > N_1+\sqrt{N_1(N_1+1)} \;.
\label{c4:TThTrsh1}
\end{equation}
Calculating the characteristic polynomial of matrix $\bmV_{T,{\rm th}}
-\frac{i}{2} {\widetilde {\boldsymbol J}}_B$ one deals with the following
pair of cubic polynomials
\begin{multline}
q_1(\lambda,N_1,N_2,{N}) = 
\lambda^3 - 2\left[ 2(1+ N_1)+ N(3+4N_1)
\right]\lambda^2\\
+4\left[1+ N_2+2N_3+
 N(4+4N_2+6N_3+ N (3+4N_1))\right]\lambda\\
-8 N\left[1+N_2+ N(2+ N
+2N_2)\right] 
\;,
\end{multline}
\vspace{-1cm}
\begin{multline}
q_2(\lambda,N_1,N_2,{N}) =
\lambda^3
-2\left[1+2N_1+ N(3+4N_1) \right]\lambda^2
\\
+4\left[ N_2+2 N(1+N_1)+ N^2(3+4N_1)
\right]\lambda\\
-8(1+ N)( N^2-2N_2-2{N} N_2)
\;.
\end{multline}
While the first polynomials admits only positive roots, the second one
shows a negative root under a certain threshold. It is possible to
summarize the three separability thresholds of the three modes
involved in the following inequalities
\begin{equation}
N > N_k+\sqrt{N_k(N_k+1)} \;.
\label{c4:TThGenericThrs}
\end{equation}
If the inequality (\ref{c4:TThGenericThrs}) is satisfied for a given $k$, 
then  mode $a_k$ is separable. Clearly, it follows that the state 
$\ket{T}$ evolved from vacuum ({\em i.e.}, $ N=0$) is fully 
inseparable.
\par
When one deals with more than three parties and modes the separability
issue becomes more involved, even remaining in the framework of
Gaussian states. As an example, consider the case of four parties and
modes, labeled by $A$, $B$, $C$ and $D$. The one-mode bipartite
splittings can still be tested via the {\tt ppt} criterion, involving $1
\times 3$ modes forms. In the Gaussian language it is necessary and
sufficient to consider whether $\bmsigma \not\ge
-i(4\kappa_1^2)^{-1} {\widetilde {\boldsymbol \Omega}}_S$ (for
$S=\{A,B,C,D\}$). However, also bipartite splittings of the $2 \times
2$ mode type must be taken into account. We have already mentioned 
above that for this case the {\tt ppt} criterion ceases to be sufficient for
separability. Hence to rule out the possibility of bound entanglement
one have to rely on the operational criteria given in
Ref.~\cite{GKL+01}. In general, in order to confirm genuine $n$-party
entanglement, one has to rule out any possible partially separable
form. In principle, this can be accomplished by considering all
possible bipartite splittings and applying either the {\tt ppt} criterion or
the criterion from \cite{GKL+01}. Although a full theoretical
characterization including criteria for entanglement classification
has not been considered yet for more than three parties and modes, the
presence of genuine multipartite entanglement can be confirmed, once
the complete correlation matrix of the state is given.

\chapter{Gaussian states in noisy channels}\label{c6:gauss:chan}
\index{Gaussian states!propagation in noisy channels}
In this Chapter we address the evolution of a $n$-mode Gaussian state in
noisy channel where both dissipation and noise, thermal noise as 
well as phase--sensitive (``squeezed'') noise, are present. At first we 
focus our attention on the evolution of a single mode of radiation. Then
we extend our analysis to the evolution of a $n$-mode state, which  will 
be treated as the evolution in a global channel made of $n$ non interacting 
different channels. For the single mode case a thorough analysis may be found 
in \cite{DecoRev}.
\section{Master equation and Fokker-Planck equation}\label{c6:s:MastEq}
The propagation of a mode of radiation (the {\em system}) in a noisy
channel may be described as the interaction of the mode of interest with a
{\em reservoir} (bath) made of large number of external modes, which may be
the modes of the free field or the phonon modes of a solid. We denote by
$b_j$ such mode operators and assume a weak coupling $g_j$ between the
system and the bath modes. Interaction Hamiltonian is written as $H_I = a
B^\dag (t) e^{-i\omega_a t} + a^\dag B(t) e^{i\omega_a t}$, $B(t)=\sum_k g_k
b_k e^{-i \omega_k t}$ being the collective mode of the bath and $\omega_a$
the frequency of the system.  The global density matrix $R_t\equiv R$,
describing both the system and the bath at time $t$, evolves, in the
interaction picture, according to the equation  $\dot R = i[R,H_I]$, while
the reduced density operator $\varrho$ for the system only is obtained by
partial trace over the bath degrees of freedom. Upon a perturbative
expansion to second order and assuming a {\em Markovian} bath, {\em i.e.}
$\langle b(\omega_h)\,b(\omega_k)\rangle_R= M \delta(2\omega_a -
\omega_h-\omega_k)$ and $\langle b^\dag (\omega_h)\,b(\omega_k)\rangle_R= N
\delta(\omega_h-\omega_k)$ the dynamics of the reduced density matrix is
described by the following {\em Master} equation
\begin{equation}\label{c6:me:lind}
\dot{\varrho} = \frac{\Gamma}{2}\Big\{
(N+1) \mL [a] + N \mL [a^\dag]
- M^{*} \mD [a] - M \mD [a^\dag]
\Big\}\,\varrho\,,
\end{equation}
where $\Gamma$ is the overall damping rate, while $N\in\rr$ and $M\in\cc$ 
represent the effective photons number and the squeezing parameter of the bath 
respectively.  $\mL [O]\varrho=2 O\varrho O^{\dag}-O^{\dag}O\varrho 
- \varrho O^{\dag} O$ and $\mD [O]\varrho = 2 O\varrho O - O O \varrho -
  \varrho O O$ are {\em Lindblad superoperators}.
The terms proportional to $\mL [a]$ and to $\mL [a^\dag]$ describe
losses and linear, phase-insensitive, amplification processes,
respectively, while the terms proportional to $\mD [a]$ and $\mD [a^\dag]$
describe phase dependent fluctuations.  The positivity of the density
matrix imposes the constraint $|M|^2 \le N(N+1)$. At thermal
equilibrium, {\em i.e.} for $M=0$, $N$ coincides with the average number of
thermal photons in the bath at frequency $\omega_a$. 
\subsection{Single-mode Gaussian states in noisy
channels}\label{s:single:noise}
\index{noisy channels}
Let us now focus on Gaussian states and start with single mode states. 
The first step is to transform the Master equation (\ref{c6:me:lind}) 
into a Fokker-Planck equation for the Wigner function. Thanks to
Eq.~(\ref{WdefTr}) it is straightforward to verify the correspondence 
\begin{subequations}
\label{corresp1}
\begin{align}
a \varrho \rightarrow (\alpha + \mbfrac
\partial_{\alpha^*}) W[\varrho](\alpha)\,, & \qquad
a^\dag \varrho \rightarrow (\alpha^* - \mbfrac
\partial_{\alpha}) W[\varrho](\alpha)\,, \\
\varrho a \rightarrow (\alpha - \mbfrac
\partial_{\alpha^*}) W[\varrho](\alpha)\,,  &\qquad 
\varrho a^\dag \rightarrow (\alpha^* + \mbfrac
\partial_{\alpha}) W[\varrho](\alpha) \;.
\end{align}
\end{subequations}
Eqs.~(\ref{corresp1}), together with the composition rules 
$L[O_1 O_2]=L[O_1]\,L[O_2]$ and $R[O_1 O_2]=R[O_2]\,R[O_1]$, 
where $L$ and $R$ denote action on the density matrix 
from the left and from the right respectively, allows
to evaluate the differential representation of superoperators 
in Eq.~(\ref{c6:me:lind}). We have
\begin{subequations}
\begin{align}
& \mL [a] \varrho \rightarrow \left[ \partial_{\alpha}\alpha +
\partial_{\alpha^*}\alpha^* + \partial_{\alpha\alpha^*}^{2}\right]
W[\varrho](\alpha)\,, \\
&\mL [a^\dag] \varrho \rightarrow -\left[ \partial_{\alpha}\alpha +
\partial_{\alpha^*}\alpha^* - \partial_{\alpha\alpha^*}^{2}\right]
W[\varrho](\alpha)\,, \\
& \mD [a] \varrho \rightarrow - \partial_{\alpha^*\alpha^*}^{2}
W[\varrho](\alpha)\,, \quad
\mD [a^\dag] \varrho \rightarrow -\partial_{\alpha\alpha}^{2}
W[\varrho](\alpha)\label{c6:corr:alpha:D:adag}\:.
\end{align}
\end{subequations}
From now on we put $W(\alpha)\equiv W[\varrho](\alpha)$.
In this way, the Master equation (\ref{c6:me:lind}) transforms into
the following Fokker-Planck equation for the Wigner function
\begin{equation}
\partial_t W(\alpha) = \frac{\Gamma}{2}
\Big\{
\partial_{\alpha}\alpha + \partial_{\alpha^*}\alpha^* +
(2N+1)\partial_{\alpha\alpha^*}^2
+ M^* \partial_{\alpha^*\alpha^*}^2 + M \partial_{\alpha\alpha}^2
\Big\} W(\alpha)\,.\label{c6:me:alpha}
\end{equation}
Passing to Cartesian coordinates $\alpha = \kappa_2\,(x + i y)$, 
$ \partial_{\alpha} = (2\kappa_2)^{-1} (\partial_x - i \partial_y)$, 
we have
\begin{align*}
&\partial_{\alpha}\alpha + \partial_{\alpha^*}\alpha^* =
\partial_{x} x + \partial_{y} y\,,
&\partial_{\alpha\alpha^*}^2& = \frac{1}{4\kappa_2^2} (\partial_{xx}^2 +
\partial_{yy}^2)\,, \\
&\partial_{\alpha\alpha}^2 = \frac{1}{4\kappa_2^2} (\partial_{xx}^2
-2i\partial_{xy}^2-\partial_{yy}^2)\,,
&\partial_{\alpha^*\alpha^*}^2& = \frac{1}{4\kappa_2^2} (\partial_{xx}^2
+2i\partial_{xy}^2-\partial_{yy}^2)\:,
\end{align*}
and Eq.~(\ref{c6:me:alpha}) rewrites as 
\begin{multline}
\partial_t W(x,y) = \frac{\Gamma}{2}\bigg\{
\partial_{x} x + \partial_{y} y +
\frac{1}{4\kappa_2^2} ( 2N + 1)(\partial_{xx}^2 + \partial_{yy}^2)\\
+\frac{1}{2\kappa_2^2} \Big(\real{M}(\partial_{xx}^2 - \partial_{yy}^2)
+2\,\immag{M} \partial_{xy}^2 \Big)\bigg\} W(x,y)\,,\label{c6:me:xy}
\end{multline}
or, in a more compact form, as 
\begin{equation}
\partial_t W(\bmX) = \frac{\Gamma}{2} \bigg(\partial_{\bmX}^{\sT} X +
\partial_{\bmX}^{\sT} \bmsigma_{\infty} \partial_{\bmX} \bigg)
W(\bmX)\label{c6:me:single:cmpct}\,,
\end{equation}
where $\bmX \equiv
(x, y)^{\sT}$, $\partial_{\bmX} \equiv (\partial_x, \partial_y)^{\sT}$, and
we introduced the {\em diffusion} matrix  $\bmsigma_\infty$
\begin{equation}
\bmsigma_{\infty} = \frac{1}{2\kappa_2^2}\left(\begin{array}{cc}
\left(\frac12 +N\right)+\real{M} & \immag{M} \\[1ex]
\immag{M} & \left(\frac12 +N\right)-\real{M}
\end{array}\right) \,. \label{c6:diff:infinity}
\end{equation}
\index{Master equation}
\index{Fokker-Planck equation}
\index{Wigner function!Fokker-Planck equation for the}
The diffusion matrix is determined only by the bath parameters and, as we
will see, represents the asymptotic covariance matrix when the initial
state is Gaussian.
\par
The Wigner function at time $t$, $W_t(\gr{X})$, {\em i.e.} the general
solution of Eq.~(\ref{c6:me:single:cmpct}) can be expressed as the
following convolution
\begin{eqnarray}
W_t(\gr{X}) = \int_{\rr^2} d^2 \gr{Z}\: G_t(\gr{X}|\gr{Z})\:W_0(\gr{Z})
\label{solFP:single}\;
\end{eqnarray}
where  $W_0(\gr{X})$ is the initial Wigner
function and the propagator $G_t(\gr{X}|\gr{Z})$ is given by 
\begin{eqnarray}
G_t(\gr{X}|\gr{Z}) = \frac{
\pexp{-\mbfrac
(\gr{X}-e^{-\frac12 \Gamma t} \gr{Z})^{\sT} \,
\gr{\Sigma}_t^{-1}
(\gr{X}-e^{-\frac12 \Gamma t} \gr{Z})}
}
{2\pi\sqrt{\det [\gr{\Sigma}_t]}} 
\label{solFP:green}\;,
\end{eqnarray}
with $\gr{\Sigma}_t=(1-e^{-\Gamma t})\, \bmsigma_\infty$.
The solution (\ref{solFP:single}) holds for any initial $W_0(\gr{X})$.
For an initial Gaussian state, since the propagator is Gaussian, 
Eq.~(\ref{solFP:single}) says that an initial
Gaussian state mantains its character at any time.
This fact is usually summarized saying that the Master equation
(\ref{c6:me:lind}) induces a Gaussian map on the density matrix
of a single-mode.
\par
From now on, we put $\kappa_2=1$ and consider an initial 
Gaussian state. Using Eq.~(\ref{c6:me:single:cmpct}), the 
evolution of $\overline{\bmX}$ is given by
\begin{align}
\dot{\overline{\bmX}} &= \int_{\mathbb{R}^2} d^2\bmX\, \bmX\,\partial_t
W(\bmX) \nonumber \\
&=
\frac{\Gamma}{2}\int_{\mathbb{R}^2} d^2\bmX\, \bmX\,
\partial_{\bmX}^{\sT} \bmX\,W(\bmX) +
\frac{\Gamma}{2}\int_{\mathbb{R}^2} d^2\bmX\, \bmX\,
\partial_{\bmX}^{\sT} \bmsigma_{\infty} \partial_{\bmX}\, W(\bmX)
\,.\label{c6:s:fm}
\end{align}
The first integral is easily evaluated by parts, leading to 
$-\frac12 \Gamma$, while the second gives no contribution. 
Eq.~(\ref{c6:s:fm}) thus becomes
\begin{equation}
\dot{\overline{\bmX}} = - \frac{\Gamma}{2}\, \overline{\bmX}\,,
\label{dampX}
\end{equation}
{\em i.e.} $\overline{\bmX}$ is damped to zero.
\par
Now we address the evolution of the covariance matrix
$\bmsigma$. Since
\begin{align}
\dot{\sigma}_{xx}= \dot{\overline{x^2}} - 2\,
\overline{x}\,\dot{\overline{x}}\,, \qquad 
\dot{\sigma}_{yy} = \dot{\overline{y^2}} - 2\,
\overline{y}\,\dot{\overline{y}}\,, \qquad
\dot{\sigma}_{xy} = \dot{\overline{xy}} -
\overline{x}\,\dot{\overline{y}} - \dot{\overline{x}}\,\overline{y}\,,
\end{align}
we should only evaluate $\dot{\overline{x^2}}$, $\dot{\overline{y^2}}$
and $\dot{\overline{xy}}$. These evolve as follows
\begin{equation}
\left(
\begin{array}{c}
\dot{\overline{x^2}}\\ \dot{\overline{y^2}} \\ \dot{\overline{xy}}
\end{array}\right) = \int_{\mathbb{R}^2}d^2\bmX\,
\left(
\begin{array}{c}
x^2\\ y^2 \\ xy
\end{array}\right)
\partial_t W(\bmX) = 
- 2\left(
\begin{array}{c}
\overline{x^2}\\ \overline{y^2} \\ \overline{xy}
\end{array}\right)
+
2
\left(
\begin{array}{c}
\sigma_{xx}^{\tinyinf}\\ \sigma_{yy}^{\tinyinf} \\ \sigma_{xy}^{\tinyinf}
\end{array}\right)\,,
\label{evolve2mom}
\end{equation}
\index{covariance matrix!evolution}
where, in solving Eq.~(\ref{evolve2mom}), we have substituted 
(\ref{c6:me:single:cmpct}) and integrated 
by parts.
Therefore, the evolution equation for $\bmsigma$ simply reads
\begin{equation}
\dot{\bmsigma} = -\Gamma\, (\bmsigma - \bmsigma_{\infty})\,,
\end{equation}
which yields
\begin{equation}\label{c6:cov:evol}
\bmsigma(t) =  e^{-\Gamma t}\,\bmsigma(0) +
(1-e^{-\Gamma t})\,\bmsigma_{\infty}\,,
\end{equation}
in agreement with Eqs.~(\ref{solFP:single}) and (\ref{solFP:green}).
Eq.~(\ref{c6:cov:evol}) says that the evolution imposed 
by the Master equation is a Gaussian map with $\bmsigma_{\infty}$ as 
asymptotic covariance matrix.  $\bmsigma(t)$
satisfies the uncertainty relation (\ref{HeisSG})
iff these are satisfied by both $\bmsigma_{\infty}$
and $\bmsigma(0)$.
\index{Gaussian map}
\subsection{$\boldsymbol n$-mode Gaussian states in noisy channels}
\label{s:n:mode:noise}
In this Section we extend the above results to the evolution of an arbitrary 
$n$-mode Gaussian state in noisy channels. We assume no correlations among 
noise in the different channels. Therefore, the dynamics is governed by the 
Master equation
\begin{equation}
\dot \varrho = \sum_{h=1}^n\frac{\Gamma_h}{2}\Big\{
(N_{h}+1) \mL[a_{h}] + N_{h} \mL [a_{h}^{\dag}]
- M_{h}^{*} \mD[a_{h}] - M_{h} \mD[a_{h}^{\dag}] \Big\}\,\varrho\,,
\label{c6:rhoev}
\end{equation}
where $N_{h}$ and $M_{h}$ have the same meaning as in
Eq.~(\ref{c6:me:lind}) and each channel has a damping rate $\Gamma_h$. The
positivity of the density matrix imposes the constraint $|M_{h}|^{2} \le
N_{h}(N_{h}+1)$ $\forall h$. At thermal equilibrium, {\it i.e.}~for
$M_{h}=0$, the parameter $N_{h}$ coincides with the mean number of thermal
photons in the channel $h$.
\par
As for the single mode case, we can convert the Master equation
(\ref{c6:rhoev}) into a Fokker-Planck equation for the Wigner function. 
In compact notation we have
\begin{equation}
\partial_t W(\bmX) = \frac12 \Big(\partial_{\bmX}^{\sT} \bbGamma \bmX +
\partial_{\bmX}^{\sT} \bbGamma\, \bmsigma_{\infty} \partial_{\bmX} \Big)
W(\bmX)\label{c6:me:N:cmpct}\,,
\end{equation}
with $\bbGamma = \bigoplus_{h=1}^{n}\Gamma_h \ii_2$.
Eq.~(\ref{c6:me:N:cmpct}) is formally
identical to Eq.~(\ref{c6:me:single:cmpct}), but now $\bmX \equiv(x_1, y_1,
\ldots, x_n, y_n)^{\sT}$, $\partial_{\bmX}
\equiv (\partial_{x_1}, \partial_{y_1}, \ldots, \partial_{x_n},
\partial_{y_n})^{\sT}$ and the diffusion matrix is given by the direct 
sum
$\bmsigma_\infty = \bigoplus_{h=1}^{n} \bmsigma_{h,\infty}$
where 
\begin{equation}
\bmsigma_{h,\infty} = \frac{1}{2\kappa_2^2}\left(\begin{array}{cc}
\left(\frac12 +N_{h}\right)+\real{M_{h}} & \immag{M_{h}} \\ [1ex]
\immag{M_{h}} & \left(\frac12 +N_{h}\right)-\real{M_{h}}
\end{array}\right) \, \label{c6:canalino}
\end{equation}
is the asymptotic covariance matrix of the $h$-th channel.
The general solution of (\ref{c6:me:N:cmpct}) is an immediate
generalization of (\ref{solFP:single}) and therefore, also 
for the $n$-mode case, we have that Gaussian states remains Gaussian
at any time. For an initial $n$-mode Gaussian state of the form (\ref{c3:WX})
the Fokker--Planck equation (\ref{c6:me:N:cmpct}) corresponds to a set of
decoupled equations for the second moments that can be solved as for the
single mode case.
Notice that
the drift term always damps to $0$ the first statistical moments, {\em
i.e.}
\begin{equation}
\bmX(t) = {\mathbbm G}_t^{1/2} \, \boldsymbol{X}(0)
\quad \mbox{with} \quad
\mathbbm G_t =\, \bigoplus_{h=1}^{n}\, e^{-\Gamma_h t}\,\mathbbm 1_2\,.
\end{equation} 
The evolution imposed by the Master equation preserves the Gaussian 
character of the states. The covariance matrix at time $t$ is given by 
\begin{equation}\label{c6:cov:evol:N}
\bmsigma(t)
= {\mathbbm G}_t^{1/2}\,\bmsigma(0)\,{\mathbbm G}_t^{1/2}
+ ({\mathbbm 1} - {\mathbbm G}_t)\,
\bmsigma_\infty\,.
\end{equation}
Eq.~(\ref{c6:cov:evol:N}) describes the evolution of an initial Gaussian 
state $\bmsigma(0)$ into the Gaussian environment $\bmsigma_{\infty}$. 
Since Eq.~(\ref{c6:cov:evol:N}) is formally similar to Eq.~(\ref{c6:cov:evol}), 
the considerations we made about the evolved covariance matrix for the single 
mode also hold for the $n$-mode state.
\par
Sometimes it is useful to describe a system by means of its characteristic
function  $\chi(\bmLambda)$. Since the Wigner function is defined as
the FT of $\chi(\bmLambda)$ we have that $\chi(\bmLambda)$ obeys 
\begin{equation}
\partial_t \chi(\bmLambda) = -\sum_{h=1}^{n} \frac12 \left(
\bmLambda^{\sT} \bbGamma\partial_{\bmLambda} +
\bmLambda^{\sT} \bbGamma\, \widetilde{\bmsigma}_{\infty}
\bmLambda\right)\,\chi(\bmLambda)
\label{c6:me:N:cmpct:chi}\,,
\end{equation}
with $\partial_{\bmLambda} \equiv (\partial_{a_1},\partial_{b_1},\ldots,
\partial_{a_n},\partial_{b_n})^{\sT}$ and $\widetilde{\bmsigma}_{\infty} =
(\kappa_2/\kappa_3)^2\,\bmsigma_{\infty}$
Finally, Eq.~(\ref{c6:me:N:cmpct:chi}) can be integrated leading to the
solution
\begin{equation}
\chi(\bmLambda) = \bmLambda^{\sT} {\mathbbm G}_t^{1/2}\,
\exp\left\{-\mbfrac\, \bmLambda^{\sT}\, ({\mathbbm 1}-{\mathbbm G}_t)\,
\widetilde{\bmsigma}_{\infty}\,\bmLambda \right\}\, \chi_0(\bmLambda) \,,
\label{last}
\end{equation}
$\chi_0(\bmLambda)$ being the initial characteristic function.
Eq.~(\ref{last}) confirms that the evolution imposed by the Master equation
maintains the Gaussian character of states.
\index{covariance matrix!evolution}
\section{Gaussian noise}\label{gauss:noise}
\index{Gaussian noise}
In this Section we address the noise described by the map
\begin{equation}
{\Gnoise}(\varrho) = \int_{\cc^n} d^{2n}\bmgamma\,
\frac{\,\pexp{-\bmgamma^{\sT}\bmDelta^{-1}\,\bmgamma}}
{\pi^n\,\sqrt{{\rm Det}[\bmDelta]}}\,
D(\bmgamma)\,\varrho\, D^{\dag}(\bmgamma)\,,
\label{Gnoise:dmatrix}
\end{equation}
$\bmDelta$ being the covariance matrix characterizing the noise and
$D(\bmgamma)$ is the displacement operator (\ref{defDcmplx}). 
The map $\Gnoise$ is usually referred to as {\em Gaussian noise}.
Using Eq.~(\ref{Xdispla}), the characteristic function of the state 
${\Gnoise}(\varrho)$ is given by
\begin{align}
\chi\left[\Gnoise(\varrho)\right](\bmlambda) &= \int_{\cc^n} d^{2n}\bmgamma\,
\frac{\,\exp\{-\bmgamma^{\sT}\bmDelta^{-1}\,\bmgamma\}}
{\pi^n\,\sqrt{{\rm Det}[\bmDelta]}}\,e^{\bmgamma^{\dag} \bmlambda -
\bmlambda^{\dag} \bmgamma}\,
\chi[\varrho](\bmlambda)\nonumber\\
&=\chi[\varrho](\bmlambda)\,\pexp{-\bmlambda^{\sT} \bmDelta\, \bmlambda}\,,
\label{Gnoise:Xfn}
\end{align}
whereas, thanks to  Eq.~(\ref{Wdispla}), its Wigner function reads
\begin{equation}
W\left[\Gnoise(\varrho)\right](\bmalpha) = \int_{\cc^n} d^{2n}\bmgamma\,
\frac{\,\pexp{-(\bmgamma-\bmalpha)^{\sT}\bmDelta^{-1}\,(\bmgamma-\bmalpha)}}
{\pi^n\,\sqrt{{\rm Det}[\bmDelta]}}\,
W[\varrho](\bmgamma)\,,
\label{Gnoise:Wig}
\end{equation}
{\em i.e.} a Gaussian convolution of the original Wigner function. 
The average number of photons of a state passing through a 
Gaussian noise channel is obtained using Eq.~(\ref{AveCov})
\begin{equation}
\sum_{k=1}^{\infty} \langle a_{k}^{\dag}a_{k} \rangle_{\Gnoise(\varrho)}
= \sum_{k=1}^{\infty} \langle a_{k}^{\dag}a_{k} \rangle_\varrho
+ \sqrt{{\rm Det}[\bmDelta]}\:,
\end{equation}
$\sum_k \langle a_{k}^{\dag}a_{k} \rangle_\varrho$ being the
average number of photons in the absence of noise.
\par
When $W[\varrho](\bmalpha)$ itself describes a Gaussian state, {\em i.e.}
has the form Eq.~(\ref{c3:WAlpha}), then
$W\left[\Gnoise(\varrho)\right](\bmalpha)$ is Gaussian too, with covariance
matrix
\begin{equation}
\label{GN:equality:1}
\bmsigma_{\hbox{\tiny GN}} = \bmsigma_{\bmalpha} + \mbfrac\bmDelta\,.
\end{equation}
The Gaussian noise map can be seen as the solution of the Master 
equation (\ref{c6:rhoev}) in the limit of large thermal noise and
short interaction time. In order to derive this result, 
let us consider $\Gamma_h=\Gamma$, $N_h=N$ and $M=0$ $\forall h$ in the 
Eq.~(\ref{c6:me:N:cmpct}). Then, in the limit $\Gamma t\ll 1$ 
Eq.~(\ref{c6:cov:evol:N}) reads $
\bmsigma(t) = \bmsigma_{\bmalpha} + \Gamma t\,(\bmsigma_{\infty}-
\bmsigma_{\bmalpha})$,
which, assuming $N\gg 1$, namely considering $\bmsigma_{\bmalpha}$
negligible with respect to $\bmsigma_{\infty}$, reduces to
\begin{equation}
\label{GN:equality:2}
\bmsigma(t) = \bmsigma_{\bmalpha} + \Gamma t\,\bmsigma_{\infty}\,.
\end{equation}
By comparing (\ref{GN:equality:1}) and
(\ref{GN:equality:2}), one has that, for $N\gg1$ and $\Gamma t\ll 1$, the
evolution imposed by the Master equation (\ref{c6:rhoev}) is equivalent 
to an overall Gaussian noise with covariance matrix given by
\begin{equation}
\bmDelta = 2 \Gamma t\, \bmsigma_{\infty}\,.
\end{equation}
\section{Single-mode Gaussian states}
In this Section we address the evolution of single-mode Gaussian 
states in a noisy channel described by the Master equation 
(\ref{c6:me:lind}). In particular, in the following two Sections, 
we analyze the evolution of purity and nonclassicality as a function
of time and noise parameters.
\subsection{Evolution of purity}
\index{purity!evolution}
As we have seen in Chapter \ref{ch:gs}, the purity $\mu$ of a quantum
state $\varrho$ is defined as $\mu \equiv {\rm Tr}[\varrho^2]$. 
For continuous variable systems one has $0 < \mu \le 1$. Since $\mu$ is a
nonlinear function of the density matrix it cannot be the expectation 
value of an observable quantity. On the other hand, if collective 
measurements on two copies of the state are possible, then the purity 
may be directly measured \cite{EkeEtAl:PRL:01}. For instance, collective 
measurement of overlap and fidelity have been experimentally realized 
for qubits encoded into polarization states of photons 
\cite{Fil:PRA:02,HenEtAl:02}.
\par
Purity $\mu$ can be easily computed for Gaussian states. In fact, using
Eqs.~(\ref{Wtrace}) and Eq.~(\ref{c3:purity:gauss:N}), for an $n$-mode
Gaussian state we have $\mu(\bmsigma) =
[(2\kappa_2)^{2n}\sqrt{\det[\bmsigma]}]^{-1}$.
\par
Here we focus our attention only on the evolution of the purity in the case
of a single mode Gaussian state of \cite{ParEtAl:PRA:03};
the purity of a two-mode Gaussian state is studied in
Ref.~\cite{SerEtAl:PRA:04}. We assume an initial state 
with zero first moments $\overline{\gr{X}}=0$, {\em i.e.} 
a state of the form $S(r_0,\varphi_0)\, \nu \,
S^\dag (r_0,\varphi_0)$.  Using (\ref{dampX}) 
we conclude that $\overline{\gr{X}}_t=0$ $\forall t$ and that three
parameters are enough to describe the state at any time.
These may be either the three independent elements of the 
covariance matrix, the three parameters $r(t)$, $\varphi(t)$
and $ N(t)$, or as it will be the following the three
parameters $r(t)$, $\varphi(t)$ and $\mu(t)$.
\par
Let us first consider the case $M=0$, for which the initial state is damped
toward a thermal state with mean photon number $N$
\cite{BR97,marians}.
In this case $\varphi$ is constant in time and
does not enter in the expression of $\mu$.  
The quantities $\mu(t)$ and $r(t)$ in Eqs.~(\ref{purezza}) 
solve the following system of coupled equations
\begin{align}
\dot{\mu} = \Gamma \left(\mu-\frac{\mu^{2}\cosh(2r)}{\mu_{\infty}}
\right)\,, \qquad
\dot r = -\frac{\Gamma}{2}\frac{\mu}{\mu_{\infty}}\sinh(2r)
{\rm \; ,} 
\label{system}
\end{align}
which, in turn, can be directly found working out the basic evolution
equation $\dot \mu=2{\rm Tr}[\dot \varrho\: \varrho]$ as a phase--space
integral; $\mu_\infty$ is defined as 
$\mu_{\infty}\equiv (2N+1)^{-1}$. 
It is easy to see that, as $t\rightarrow \infty$,
$\mu(t)\rightarrow\mu_{\infty}$ and $r(t)\rightarrow 0$, as
one expects, since the channel damps (pumps) the initial state to a thermal
state with mean photon number $N$.  Therefore, the only constant solution
of Eq.~(\ref{system}) is $\mu=\mu_{\infty}$, $r=0$, {\em i.e.}~only initial
non--squeezed states are left unchanged by the evolution in the noisy
channel. The general solution of (\ref{system}) is given by
\begin{equation}
\mu(t) = \mu_{0} \Bigg[\frac{\mu_{0}^{2}}{\mu_{\infty}^{2}}
\left(1-{e}^{-\Gamma t}\right)^{2}+{e}^{-2\Gamma t}
+\frac{2 \mu_{0}}{\mu_{\infty}}\,{e}^{-\Gamma t}
\left(1-{e}^{-\Gamma t}\right)\cosh(2r_{0})\Bigg]^{-1/2} \: ,
\label{purezza}
\end{equation}
with
\begin{equation}
\cosh[2r(t)] =
\mu(t)\left(\frac{1-{\rm e}^{-\Gamma t}}{\mu_{\infty}}+
{\rm e}^{-\Gamma t}\frac{\cosh(2r_{0})}{\mu_{0}}\right){\rm
\, .} \label{squiz}
\end{equation}
Eq.~(\ref{purezza}) shows that $\mu(t)$ is a decreasing function of
$r_{0}$: in a non--squeezed channel ($M=0$), a squeezed state decoheres
more rapidly than a non-squeezed one. The optimal evolution for the purity,
obtained letting $r=0$ in Eq.~(\ref{purezza}), reads
\begin{equation}
\mu(t)=\frac{\mu_{0}\:
\mu_{\infty}}{\mu_{0}+{\rm e}^{-\Gamma t}(\mu_{\infty}-\mu_{0})}
\: . \label{optpur}
\end{equation}
Obviously, $\mu(t)$ is not necessarily a decreasing function of time: if
$\mu_{0} < \mu_{\infty}$ then the initial state will undergo a certain
amount of purification, asymptotically reaching the value $\mu_{\infty}$
which characterizes the channel.  In
addition, $\mu(t)$ is not a monotonic function for any choice of the
initial conditions.  Letting $\dot{\mu} = 0$ in Eq.~(\ref{system}), and
exploiting Eqs.~(\ref{purezza}) and (\ref{squiz}), one finds the following
condition for the appearance of a zero of $\dot \mu$ at finite positive
times: $\cosh(2r_{0}) > {\rm max} [\mu_{0}/\mu_{\infty},\,
\mu_{\infty}/\mu_{0}]$.  If this condition is satisfied, then $\mu(t)$
shows a local minimum.
\par
Let us now consider the case $M \neq 0$, corresponding to 
a squeezed thermal bath. The general solution for purity can be 
written as 
\begin{multline}
\mu(t)=\mu_{0}\bigg\{\frac{\mu_{0}^{2}}{\mu_{\infty}^{2}}\left(1-
{e}^{-\Gamma t}\right)^{2} \, + \,
{e}^{-2\Gamma t} + 2\frac{\mu_{0}}{\mu_{\infty}}
\left(1-{e}^{-\Gamma t}\right){e}^{-\Gamma t}\\
\times\Big[
\sinh(2r_{\infty})\sinh(2r_{0})\cos(2\varphi_{\infty}-2\varphi_{0})
+\cosh(2r_{\infty})\cosh(2r_{0})
\Big]\bigg\}^{-1/2}\:,  
\label{mubathsqueez}
\end{multline}
where we have already inserted the asymptotic values of the parameters 
$\mu$, $r$ and $\varphi$, {\em i.e.}
\begin{subequations}
\begin{align}
&\mu_{\infty}=\left[(2N+1)^{2}-4|M|^2 \right]^{-1/2} \\ 
&\cosh(2r_{\infty})= \sqrt{1+4\mu_{\infty}^{2}|M|^{2}}\,,
\\
&\tan(2\varphi_{\infty})=-\frac{\immag{M}}{\real{M}}\,.
\end{align}
\end{subequations}
These values characterize the squeezed channel. 
We see from Eq.~(\ref{mubathsqueez}) that $\mu(t)$ is a monotonically
decreasing function of the factor $\cos(2\varphi_{\infty}-2\varphi_{0})$,
which gives the only dependence on the initial phase $\varphi_{0}$ of the
squeezing.  Thus, for any given $\varphi_{\infty}$ characterizing the
squeezing of the bath, $\varphi_{0}=\varphi_{\infty}+\pi/ 2$ is the most
favorable value of the initial angle of squeezing, i.e.~the one which
allows the maximum purity at a given time.  For such a choice, $\mu(t)$
reduces to
\begin{multline}
\mu(t) = \mu_{0}\bigg\{\frac{\mu_{0}^{2}}{\mu_{\infty}^{2}}\left(1-
{e}^{-\Gamma t}\right)^{2}+
{e}^{-2\Gamma t}
+2\frac{\mu_{0}}{\mu_{\infty}}\cosh(2r_{\infty}-2r_{0})
\left(1-{e}^{-\Gamma t}\right){e}^{-\Gamma t}
\bigg\}^{-1/2}\:.
\end{multline}
This is a decreasing function of the factor $\cosh(2r_{\infty}-2r_{0})$, so
that the maximum value of the purity at a given time is achieved for the
choice $r_{0}=r_{\infty}$, and the evolution of the purity of a squeezed
state in a squeezed channel is identical to the evolution of the purity of
a non--squeezed state in a non--squeezed channel. 
\par
In conclusion, for a general channel characterized by
arbitrary $\mu_{\infty}$, $r_{\infty}$, $\varphi_{\infty}$ and $\Gamma$,
the initial Gaussian state for which purity is best preserved in time must
have a squeezing parameter $r_{0}=r_{\infty}$ and a squeezing angle
$\varphi_{0}=\varphi_{\infty}+\pi/ 2$, {\em i.e.} it must be anti-squeezed
(orthogonally squeezed) with respect to the bath.  The net effect for the
evolution of the purity is that the two orthogonal squeezings of the
initial state and of the bath cancel each other exactly, thus reproducing
the optimal purity evolution of an initial non--squeezed coherent state in
a non--squeezed thermal bath.
\subsection{Evolution of nonclassicality}
\index{nonclassicality}
As a measure of {\em nonclassicality} of the quantum state $\varrho$, the
quantity $\tau$, referred to as {\em nonclassical depth}, has been
proposed in Ref.~\cite{Lee:PRA:91}
\begin{equation}
\tau=\frac{1-\overline{s}}{2} \; ,
\end{equation}
where $\overline{s}$ is the maximum $s$ for
which the generalized quasiprobability function
\begin{equation}
W_{s}(\bmX) = \int_{\mathbb{R}^{2n}}\frac{d^{2n}\bmLambda}{\pi^{2n}}\,
\chi(\bmLambda)\,\pexp{i \bmLambda^{\sT} \bmX +
s \kappa_3 |\bmLambda|^2}\,,
\end{equation}
is a probability distribution, {\em i.e.} positive semidefinite and non
singular.  As one should expect, $\tau=1$ for number states and $\tau=0$
for coherent states. The nonclassical depth can be interpreted as the
minimum number of thermal photons which has to be added to a quantum state
in order to erase all the `quantum features' of the state.\footnote{This
statement can be made more rigorous by assuming that a given state owns
`quantum features' if and only if its $P$-representation is more singular
than a delta function (which is the case for coherent states)
\cite{Lee:PRA:91}.} While quite effective in detecting nonclassicality of
states, the nonclassical depth is not easily evaluated for relevant quantum
states, with the major exception of Gaussian states. In fact, for a
Gaussian state characterized by a covariance matrix $\bmsigma$, the
explicit expression for the nonclassical depth reads
\begin{equation}
\tau=\max\left[\frac{1-2u}{2},0\right]\,, \label{c6:ncgau}
\end{equation}
$u$ being the minimum of the eigenvalues of $\bmsigma$. In the case of a
single mode Gaussian state, this smallest eigenvalue turns out to be 
simply $u=e^{-2r}/\mu$ \cite{DecoRev}. In this way,
thanks to Eq.~(\ref{c6:ncgau}), we obtain the following expression for the
nonclassical depth:
\begin{equation}\label{c6:1ncgau}
\tau = \max\left[ \frac12 \left(1 - \frac{e^{-2r}}{\mu}\right), 0
\right]\, .
\end{equation}
Therefore, we define the quantity $\kappa(t)$ as
\begin{equation}
\kappa(t)=\frac{\cosh(2r_{0})}{\mu_{0}}\,{e}^{-\Gamma t}+
\frac{\cosh(2r_\infty)}{\mu_{\infty}}\left(1-
{e}^{-\Gamma t}\right)\:,
\end{equation} 
the time evolution of the nonclassical depth is given by
\begin{equation}
\tau(t)=\frac{1-\kappa(t)
+\sqrt{\kappa(t)^{2}-\mu(t)^{-2}}}{2} \,,
\end{equation}
which increases with both $\mu(t)$ and $\kappa(t)$.  The phase maximizing
$\tau(t)$ at any time is again $\varphi_0=\varphi_{\infty}+\pi/2$, as for
the purity.  The maximization of $\tau(t)$ in terms of the other parameters
of the initial state is the result of the competition of two different
effects: on the one hand a squeezing parameter $r_0$ matching the squeezing
$r_{\infty}$ maximizes the purity thus delaying the decrease of $\tau(t)$;
on the other hand, a bigger value of $r_0$ obviously yields a greater
initial $\tau(0)$.  Numerical analysis unambiguously shows \cite{DecoRev}
that, in non-squeezed baths, the nonclassical depth increases with
increasing squeezing $r_0$ and purity $\mu_0$, as one should expect.
\section{Two-mode Gaussian states}\label{prop:two:gauss}
\index{noisy channels!two-mode}
In this Section we address the separability of two-mode Gaussian 
states propagating in a noisy channel. In particular, we consider 
the effect of noise on the twin-beam state of two modes 
of radiation $\dket{\Lambda} = \sqrt{1-\lambda^2} \sum_h \lambda^h 
\ket{h}\ket{h}$, $|\lambda|<1<$, $\lambda\in\rr$, whose Gaussian Wigner 
function has the form (\ref{c3:WX}) with $n=2$, $\overline{\bmX}=0$, and the 
covariance matrix given by
\begin{equation}
\bmsigma_{\hbox{\tiny TWB}} = \frac{1}{4\kappa_2^2}
\left(
\begin{array}{cc}
{\cosh(2 r)\,\mathbbm 1}_2 & \sinh(2 r)\,\bmsigma_3 \\ [1ex]
\sinh(2 r)\,\bmsigma_3  & \cosh(2 r)\,{\mathbbm 1}_2
\end{array}
\right)\,,
\end{equation}
where $
\bmsigma_3 =\hbox{Diag}(1,-1)$ is 
a Pauli matrix, and $r = \tanh^{-1} \lambda$ the squeezing parameter
of the TWB.
\index{twin-beam!separability in a noisy channel}
\subsection{Separability thresholds}
\index{separability!threshold in a noisy channel}
\index{entanglement!degradation}
The Wigner function of a TWB is Gaussian and the evolution in a noisy
channel preserves such character, as we have seen in Section
\ref{c6:s:MastEq}. Therefore, we are able to characterize the entanglement
at any time and find conditions to preserve it after a given propagation
time or length. As we have seen in Chapter \ref{SepGS} a Gaussian state is
separable iff its covariance matrix satisfies the relation
$\mathbf{S}\equiv\bmsigma+ i(4\kappa_1^2)^{-1} {\widetilde
{\mathbf{\Omega}}}_A\geq0$. Let us now focus on the separability of the TWB
evolving in generalized Gaussian noisy channels described by the Master
equation (\ref{c6:rhoev}). The evolved covariance matrix is simply given by
Eq.~(\ref{c6:cov:evol:N}) with $\bmsigma(0)=\bmsigma_{\hbox{\tiny TWB}}$,
and, assuming $M$ as real, its explicit expression is
\begin{eqnarray}\label{c6:evol:twb:CM}
\mathbf{\bmsigma}(t)= \frac{1}{2\kappa_2^2} \left(
\begin{array}{cccc}
\Sigma_{1}^{2}+\Sigma_{3}^{2} & 0 & \Sigma_{1}^{2}-\Sigma_{3}^{2} & 0 \\
0 & \Sigma_{2}^{2}+\Sigma_{4}^{2} & 0 & \Sigma_{2}^{2}-\Sigma_{4}^{2} \\
\Sigma_{1}^{2}-\Sigma_{3}^{2} & 0 & \Sigma_{1}^{2}+\Sigma_{3}^{2} & 0 \\
0 & \Sigma_{2}^{2}-\Sigma_{4}^{2} & 0 &
\Sigma_{2}^{2}+\Sigma_{4}^{2}
\end{array} \right)\,,
\end{eqnarray}
where
\begin{equation}
\begin{array}{lll}
\Sigma_{1}^{2}=\sigma_{+}^{2} e^{-\Gamma t}+D_{+}^{2}(t)\,,
&\quad&
\Sigma_{2}^{2}=\sigma_{-}^{2} e^{-\Gamma t}+D_{-}^{2}(t)\,, \\
\mbox{}\\
\Sigma_{3}^{2}=\sigma_{-}^{2} e^{-\Gamma t}+D_{+}^{2}(t)\,,
&\quad&
\Sigma_{4}^{2}=\sigma_{+}^{2} e^{-\Gamma t}+D_{-}^{2}(t)\,,
\end{array}\label{sigma:1234}
\end{equation}
$\sigma_{\pm}^2 = \frac14\,e^{\pm2 r}$, and
\begin{eqnarray}\label{sq:D:pm}
D_{\pm}^{2}(t) = \frac{1 + 2 N \pm 2 M}{4} \left(1 -e^{-\Gamma
t}\right)\,.
\end{eqnarray}
In deriving Eq.~(\ref{c6:evol:twb:CM}) we have put $M_1=M_2=M$ and $N_1=N_2=N$.
The conditions (\ref{c4:HeisSGReversed}) are then satisfied when
\begin{eqnarray} \label{20}
\Sigma_{1}^{2}\:\Sigma_{4}^{2} \geq \frac{1}{16}\,,\quad\quad
\Sigma_{2}^{2}\:\Sigma_{3}^{2} \geq \frac{1}{16}\,,
\end{eqnarray}
which do not depend on the sign of $M$.
\par
From now on we put $\kappa_2 =1$.  If we assume the environment as composed
by a set of harmonic oscillators excited in a squeezed-thermal state of the
form $\varrho= S (\xi)\, \nu\, S^\dag (\xi)$,  we can rewrite the
parameters $N$ and  $M$ in terms of the squeezing and thermal number of
photons $N_{\rm s}=\sinh^2 \xi$ and $N_{\rm th}$, respectively. In this
way we get \cite{Gre:PRA:03}
\begin{equation}
\label{phys:par:MN}
M =  \left(1 + 2\,N_{\rm th}\right)\sqrt{N_{\rm s}(1 + N_{\rm s})}\,,
\qquad
N = N_{\rm th} + N_{\rm s}(1 + 2\,N_{\rm th})\:.
\end{equation}
Now, by solving inequalities (\ref{20}) with respect to time
$t$, we find that the two-mode state becomes separable for $t > t_{\rm s}$,
where the threshold time $t_{\rm s} = t_{\rm s}(r,\Gamma,N_{\rm th}, N_{\rm
s})$ is given by \cite{OliPar:PLA:03}
\begin{eqnarray}\label{tau:sep:squeezed}
t_{\rm s} = \frac{1}{\Gamma}\ln\left[ f + \frac{1}{1+2 N_{\rm th}}
\sqrt{f^2 + \frac{N_{\rm s}(1+N_{\rm s})} {N_{\rm th}(1+N_{\rm
th})}} \right]\,,
\end{eqnarray}
and we defined
\begin{eqnarray}
f \equiv f(r,N_{\rm th}, N_{\rm s}) =\frac{(1+2\,N_{\rm th})\:
\left[ 1+2\,N_{\rm th}-e^{-2\,r}(1+2\,N_{\rm  s}) \right]}
{4\,N_{\rm th}(1+N_{\rm th})}\,.
\end{eqnarray}
As one may expect, $t_{\rm s}$ decreases as $N_{\rm th}$ and
$N_{\rm s}$ increase.  Moreover, in the limit $N_{\rm\rm  s}
\rightarrow 0$, the threshold time (\ref{tau:sep:squeezed})
reduces to the case of a non squeezed bath, in formula
\cite{Pra:QuantPh:02,RosOli:JModOpt:04}
\begin{equation} \label{25}
t_{0} =  t_{\rm s}(r,\Gamma,N_{\rm th}, 0)
= \frac{1}{\Gamma} \ln \left[ 1+\frac{1 - e^{-2 \,
r}}{2\,N_{\rm th}} \right]\,,
\end{equation}
which is always longer than $t_{\rm s}$.
We conclude that coupling a TWB with
a squeezed-thermal bath destroys the correlations between the two
channels faster than the coupling with a non squeezed environment.
\par
One can also evaluate the threshold time for separability in the
case of an {\em out-of-phase} squeezed bath, {\em i.e.} 
for complex $M=|M|\,e^{i \theta}$.
The analytical expression is quite cumbersome and will not 
be reported here. However, in order to investigate the positivity 
of $\mathbf{S}$ as a function of $\theta$, it suffices to consider   
the characteristic polynomial $q_{\mathbf{S}}(x)$ associated to
$\mathbf{S}$, and study the sign of its roots. This polynomial has 
four {\em real} roots and three of them are always {\em positive}. 
The fourth becomes positive adding noise, and the threshold decreases
with varying $\theta$. In other words, the survival time becomes 
shorter \cite{RosOli:JModOpt:04}.  
\section{Three-mode Gaussian states}\label{c4:3modes}
\index{noisy channels!three-mode}
As an example of propagation in a three-mode noisy channel, we
investigate now the evolution of state $\ket{T}$ introduced in
Section \ref{s:3mGS}. We will refer to the model described in Section
\ref{s:n:mode:noise} and we will consider three noisy channels with the
same damping constant $\Gamma$, same thermal noise ${N}_{\rm th}$
and no phase-dependent fluctuation [$M_k=0$ in \refeq{c6:canalino}].
From now on we denote with $N_h$, $h=1,2,3$, the mean photon numbers
that characterize the state $\ket{T}$ via the Eqs.~(\ref{c3:Ndit}).
Accordingly to \refeq{c6:cov:evol:N}, rearranging the order of the entries,
the evolution of the covariance matrix $\bmV_T$ is given by the following
convex combination of $\bmV_T$ itself and of the stationary covariance
matrix $\bmV_\infty=({N}_{\rm th}+\mbfrac)\ii_6$ (for the rest of
the Section we put $\kappa_2=2^{-1/2}$):
\begin{equation}
\bmV(t)=e^{-\Gamma t}\,\bmV_T+ (1-e^{-\Gamma t})\,{\bf V}_\infty
\;.
\label{c4:3mCMEvol}
\end{equation}
Consider for the moment a pure dissipative environment, namely
${N}_{\rm th}=0$.  Applying the separability criterion
(\ref{c4:npt3m}), one can show that the covariance matrix $\bmV(t)$
describes a fully inseparable state for every time $t$. In fact, defining
$\bmV\!_K(t)=\bmV(t)-\frac{i}{2}{\widetilde \bmJ}_K$, with $K=A,B,C$
corresponding to channel (mode) $1$, $2$ or $3$, respectively, we have that
the minimum eigenvalue of $\bmV\!_A(t)$ is given by
\begin{equation}
\lambda^{\rm min}_A= 2e^{-\Gamma t}\left[N_1-\sqrt{N_1(N_1+1)}\right]\,.
\label{c4:LChAvalMin1}
\end{equation}
Clearly, $\lambda_A^{\rm min}$ is negative at every time $t$, implying
that mode $A$ is always inseparable from the others. Concerning mode
$B$, the characteristic polynomial of $\bmV\!_B(t)$ factorizes into two
cubic polynomials:
\begin{subequations}
\label{c4:LossyChCubic}
\begin{align}
&q_1(\lambda,\Gamma,N_1,N_2,N_3) = 
-\lambda^3+4\left[ 1+ e^{-\Gamma t}N_1 \right]\lambda^2 \nonumber\\
&\hspace{1cm}+4\left[ -1-e^{-\Gamma t}( 2N_2 + 3N_3 
- e^{-\Gamma t} N_1) \right]\lambda + 8e^{-\Gamma t}N_3(1-e^{-\Gamma t})\;,
\\
&q_2(\lambda,\Gamma,N_1,N_2,N_3) =
-\lambda^3 + 2\left[1+2e^{-\Gamma t}N_1 \right]\lambda^2 \nonumber\\
&\hspace{3cm}
+ 4\left[ -e^{-\Gamma t}(2N_2+N_3)+ e^{-2\Gamma t}N_1 \right]\lambda
- 8e^{-2\Gamma t}N_2 \;.
\end{align}
\end{subequations}
While the first polynomial has only positive roots, the second one
admits a negative root at every time. Due to the symmetry of state
$\ket{T}$ the same observation apply to mode $C$, hence full
inseparability follows. Notice that this result resembles the case of
the TWB state in a two-mode channel studied in the previous Section
[see \refeq{25} for ${N}_{\rm th}\rightarrow 0$].
\par 
When thermal noise is considered (${N}_{\rm th}\neq0$) separability
thresholds arise, again resembling the two-mode channel case. Concerning
mode in channel $A$, the minimum eigenvalue of matrix $\bmV\!_A(t)$ is
negative when 
\be
\label{c4:ThresholdTh1}
t < \frac{1}{\Gamma}
\ln\left(1+\frac{\sqrt{N_1(N_1+1)}-N_1}{{N}_{\rm th}}\right) \;.
\ee
Remarkably, this threshold is the same as the two-mode one given in
\refeq{25}, if one consider both of them as a function of the total
mean photon number of the TWB and of state $\ket{T}$ respectively.
This consideration confirms the robustness of the entanglement of the
tripartite state $\ket{T}$. Concerning mode $B$, the characteristic
polynomial of $\bmV\!_{B}(t)$ factorizes again into two cubic
polynomials. As above, one of the two have always positive roots,
while the other one admits a negative root for time $t$ below a
certain threshold, in formula:
\begin{multline}
-8e^{-2\Gamma t}N_2+8(e^{-\Gamma t}-1)e^{-\Gamma t}(e^{-\Gamma t} N_1
- 2N_2 - N_3){N}_{\rm th} \\
+8(e^{-\Gamma t}-1)^2(1+2e^{-\Gamma t}N_1){N}_{\rm th}^2
- 8(e^{-\Gamma t}-1)^3 {N}_{\rm th}^3<0 \;.
\label{c4:ThresholdTh2}
\end{multline}
Mode $C$ is thus subjected to an identical separability threshold, upon 
the replacement $N_2\leftrightarrow N_3$.

\chapter[Quantum measurements on continuous variable systems]{Quantum
measurements on continuous variable systems}\label{ch:detection}
\chaptermark{Quantum measurements on CV systems}
In this Chapter we describe some relevant measurements that
can be performed on continuous variable (CV) systems. These include
both single-mode and two-mode (entangled) measurements. As
single-mode measurements we will consider direct detection of
quanta through counters or {\em on/off} detectors and
homodyne detection for the measurement of the field quadratures.
As concerns two-mode entangled measurements, we will analyze
the joint measurement of the real and the imaginary part of the
normal operators $Z_\pm =a\pm b^\dag$, $a$ and $b$ being two modes of
the field, through double homodyne, six-port homodyne or
heterodyne-like detectors.  Throughout the Chapter we will mostly
refer to implementation obtained for the radiation field. This is
in order since it is in this context that they have been firstly
developed, and are available with current technology. However, it
should be mentioned that the schemes analyzed in this
Chapter can also be realized, or approximated, also for other
fields, as for example in atomic or condensate systems.  The
measurement schemes will be described in some details in order to
evaluate their positive operator-valued measure
(POVM), as well as the corresponding characteristic and Wigner 
functions, both in ideal conditions and in the presence of noise, 
{\em i.e.} of non-unit quantum efficiency of the detectors. 
In Section \ref{s:pom} we review the concept of POVM and its
relations with customary measurement of observables, whereas in 
Section \ref{s:mgf} we briefly review the concept 
of moment generating function. 
Direct detection of the field, either by counting
or by on/off detectors, is the subject of Section \ref{s:dir}
while homodyne detection of the field quadratures is analyzed
in Section \ref{s:hom}. Finally, the joint measurement
of $\re{Z}$ and $\im{Z}$ is analyzed in Section \ref{s:two}.
\section{Observables and POVM}\label{s:pom}
\index{POVM}
In order to gain information about a quantum state one has to
measure some observable. The measurement process unavoidably
involves some kind of interaction, which couples the mode under
examination (the signal) to one or more other modes of the field
(the probe). Therefore, one has to admit that, in general, the
measured observable is not defined on the sole Hilbert space of
the signal mode. Rather, it reflects properties of the global
state which results from the interaction among the signal mode
and the set of the probe modes. In some cases, it is possible to
get rid of the probe modes, such that the statistics of the
outcomes can be described in terms of an observable defined only
on the Hilbert space of the signal mode. As we will see, this is
the case of homodyne detection of a field quadrature. More generally,
eliminating the probe modes by partial trace, we are left with a
more general object, that is a spectral measure of an observable to
describe the statistics in terms of the signal's density matrix. 
Let us denote by ${\cal H}$ the Hilbert space of the signal, by
${\cal K}$ the Hilbert space of the probe modes, and by $X$ the
measured observables on ${\cal H}\otimes {\cal K}$. The spectral
measure of $X$ is given by  $x\rightarrow dE (x) = |x\rangle
\langle x| dx $ with $x \in {\cal X}\subset \rr$ (the spectrum of $X$) 
and $\langle x |x'\rangle = \delta(x-x')$\footnote{
\footnotesize For observables with a discrete spectrum the
spectral measure reads $k \rightarrow \Pi_k = |k\rangle\langle k|$ with
$\langle k | l \rangle = \delta_{kl}$}. The probability density 
of the outcomes is thus given by 
\begin{eqnarray}
p (x) = \hbox{Tr}[\varrho \otimes \sigma \: E (x)]
\label{pom1}\;,
\end{eqnarray}
where $\varrho$ and $\sigma$ are the initial preparations of 
the signal and the probe respectively and the trace is taken 
over all the Hilbert spaces. Eq.~(\ref{pom1}) can be written
as 
\begin{eqnarray}
p (x) = \hbox{Tr}_{\cal H}\Big[\varrho\:
\hbox{Tr}_{\cal K}\big[\sigma\: E (x)\big]\Big] =
\hbox{Tr}_{\cal H}[\varrho\:\Pi (x)]
\label{pom2}\;,
\end{eqnarray}
where $\Pi (x) \doteq \hbox{Tr}_{\cal K}[\sigma\: E (x)]$ is usually 
referred to as the positive operator-valued measure (POVM) of 
the measurement scheme\footnote{POVM are also sometimes referred 
to as probability operator measure (POM)}. From the definition we 
have that a POVM is a set $\left\{ \Pi(x)\right\}_{x\in{\cal X}}$ 
of positive, $\Pi (x)\geq 0$ (hence selfadjoint), and normalized 
$\int_{\cal X} dx\, \Pi (x) = \iid$ operators while, in general, 
they do not form a set of orthogonal projectors. \\ 
In summary, a detection process generally corresponds to the 
measurement of an obser\-va\-ble defined in the global Hilbert space 
of the signal and the probe. If we restrict our attention to the 
system only, the statistics of the outcomes is described by a POVM. 
The converse is also true, {\em i.e.} whenever a set of operators 
satisfying the axioms for a POVM is found, then  the following 
theorem assures that it can be seen as the measurement of an 
observable on a larger Hilbert space \cite{Naimark,Mlak}.
\par\vspace{.25cm}\noindent
{\bf Theorem (Naimark)} {\em If $\left\{ \Pi(x)\right\}_{x\in{\cal X}}$ 
is a POVM on ${\cal H}$
then there exist a Hilbert space ${\cal K}$, a spectral measure
$E(x)$ on ${\cal H} \otimes {\cal K}$ and a density operator $\sigma$ 
on ${\cal K}$ such that $\Pi(x) = \hbox{Tr}_{\cal K}[\sigma\: E(x)]$.}
\vspace{.25cm} \par\noindent
In addition, the number of these 
{\em Naimark extensions} is infinite, corresponding to the fact 
that a given POVM may results from different physical
implementations. We will see an example of this property 
in Section \ref{s:two}. 
\section{Moment generating function}\label{s:mgf}
\index{moment generating function}
For a detection scheme measuring the observable $X$, with 
eigenvalues $x\in {\cal X} \subset \rr$, the so-called moment 
generating function (MGF) is defined as 
\begin{eqnarray}
M_X (y) = \hbox{Tr}\big[R\: e^{iyX}\big]\,,
\label{defMgf}\;
\end{eqnarray}
where $R$ is the overall quantum state (signal plus probe) at 
the detector. MGF generates the moments of the measured quantity 
$X$ according to the formula 
\begin{eqnarray}
\langle X^n \rangle = (-i)^n \frac{\partial^n}{\partial y^n} \left. 
M_X (y)\right|_{y=0}
\label{momMGF}\;.
\end{eqnarray}
The MGF $M_X(y)$ also provide the distribution of the 
outcomes $p(x)$ through its Fourier 
transform. In fact, 
\begin{align}
\int_\rr \frac{d\mu}{2\pi}\: e^{-i \mu x} M_X(\mu) &= 
\hbox{Tr}\left[R \int_\rr \frac{d\mu}{2\pi}\, e^{i \mu (X-x)}\right]
\nonumber \\ &=
\hbox{Tr}\left[R\: \delta(X-x) \right] =
\hbox{Tr}\left[R\: |x\rangle\langle x|\right] = p(x)
\label{pxMGF}\;.
\end{align}
The density matrix $R$ in Eq.~(\ref{pxMGF}) 
should be meant as the global quantum state, system plus probe, 
at the input of the detector. In turn, the trace should be performed 
over the global Hilbert space ${\cal H}\otimes{\cal K}$ describing all 
the degrees of freedom of the detector. Comparing Eq.~(\ref{pxMGF}) 
with Eq.~(\ref{pom2}) we have that for $R=\varrho\otimes\sigma$ the 
POVM of the detector, obtained by tracing out the probes, can be 
expressed as follows 
\begin{eqnarray}
\Pi (x) = \hbox{Tr}_{\cal K}\left[\sigma \int_\rr \frac{d\mu}{2\pi} \,
e^{i \mu (X-x)}\right]
\label{pomMGF}\;. 
\end{eqnarray}
Eq.~(\ref{pxMGF}) can be generalized to the multidimensional case.
In particular,  any detector measuring a couple of commuting operators
$[X,Y]=0$ can be seen as measuring the complex normal operator
$Z=X+iY$. The MGF is defined as 
\begin{eqnarray}
M_Z(\lambda) = \hbox{Tr}\left[R\: e^{\lambda Z^\dag - \lambda^* Z}\right]
\label{defMGF2}\;, 
\end{eqnarray}
with $\lambda \in \cc$, whereas the probability distribution of the 
outcomes $\alpha\in\cc$ is obtained as the
complex Fourier transform
\begin{eqnarray}
p (\alpha) = \int_\cc \frac{d^2\lambda}{\pi^2}\, e^{\lambda^* \alpha - \alpha^*
\lambda}\: M_Z (\lambda)
\:.
\end{eqnarray}
Again, in Eq.~(\ref{defMGF2}), $R$ denotes the overall quantum state
at the detector; the POVM can be evaluated as 
\begin{eqnarray}
\Pi (\alpha) = \hbox{Tr}_{\cal K}\left[\sigma \int_\cc
\frac{d^2\lambda}{\pi^2} 
e^{\alpha (Z^\dag - \lambda^*)  - \alpha^*
(Z-\lambda)}\right]
\label{pomMGF2}\;. 
\end{eqnarray}
\section{Direct detection}\label{s:dir}
\index{direct detection}
By direct detection we mean the measurement of the quanta of the 
field, either by effective counting ({\em i.e.} discriminating among 
the number of incoming quanta) or just by revealing their presence or 
absence (on/off detection). In the following we analyze
in some details the detection of photons. Analogue schemes have been
developed for atomic systems.
\subsection{Photocounting}
\index{photodetection!photon counting}
Light is revealed by exploiting its interaction with
atoms/molecules or electrons in a solid: each photon
ionizes a single atom or promotes an electron to a conduction band,
and the resulting charge is then amplified to produce a measurable
pulse. In practice, however, available photodetectors are not
ideally counting all photons, and their performances are limited by a
non-unit quantum efficiency $\zeta$, namely only a fraction $\zeta$ of
the incoming photons lead to an electric signal, and ultimately to a
{\em count}: some photons are either reflected from the surface of the
detector, or are absorbed without being transformed into electric pulses.  
\par
Let us consider a light beam entering a photodetector of quantum
efficiency $\zeta$, {\em i.e.} a detector that transforms just a
fraction $\zeta$ of the incoming light pulse into electric signal. If
the detector is small with respect to the coherence length of
radiation and its window is open for a time interval $T$, then the
probability $p(m;T)$ of observing $m$ counts is a 
Poissonian distribution of the form \cite{kelley}
\begin{eqnarray}
p_m(T) = \hbox{Tr}\left[\varrho\,
\mbox{{\bf :}}\,{\frac {[\zeta I (T)T]^m }{m!}}
\pexp{-\zeta I (T)T}\mbox{{\bf :}} \right]        
\label{pc-gen}\;, 
\end{eqnarray}
where $\varrho $ is the quantum state of light, $\mbox{{\bf :}}\cdots
\mbox{{\bf :}}$ denotes the normal ordering of field operators, and $I(T)$
is the beam intensity
\begin{eqnarray}
 I (T)={\frac {2\epsilon_0 c}{T}}
 \int_0^Tdt\,
 \bmE^{(-)} (t)\cdot
\bmE^{(+)} (t) \label{intensity}\;,
\end{eqnarray}
given in terms of the positive, $\bmE^{(+)}$, and
negative, $\bmE^{(-)}$, frequency part of the electric
field operator.  The quantity $p(T)=\zeta\,
\hbox{Tr}\left[\varrho\, I (T) \right ]$ equals the probability
of a single count during the time interval $(T,T+dt)$.  Let us now focus
our attention to the case of the radiation field excited in a stationary
state of a single mode at frequency $\omega $. 
Eq.~(\ref{pc-gen}) can be rewritten as 
\begin{eqnarray}
p_m (\eta) = \hbox{Tr}\left[\varrho \,
\mbox{{\bf :}}\,{\frac {(\eta a^{\dag} a)^m}{m!}}
\pexp{-\eta a^{\dag} a}\mbox{{\bf :}} \right] 
\label{pc-singlemode}\;,
\end{eqnarray}
where the parameter $\eta$ denotes the 
overall {\em quantum efficiency} of the photodetector. By means of
the identities $
\mbox{{\bf :}}\,({a^\dag } a)^m \mbox{{\bf :}}
=({a^\dag } )^m a^m ={a^\dag } a({a^\dag } a-1)
\ldots({a^\dag } a-m+1)$ and 
$ \mbox{{\bf :}}e^{-x{a^\dag } a}\mbox{{\bf :}}=(1-x)^{{a^\dag } a}$ 
\cite{louisell},
one obtains 
\begin{eqnarray}
p_m(\eta) =\sum_{k=m}^{\infty} \varrho_{kk}
\binom{k}{m} \eta^m (1-\eta )^{k-m}
\label{conv_n}\;,
\end{eqnarray}
where $\varrho_{kk}\equiv \langle k |\varrho | k \rangle 
=p_k(\eta\equiv1)$.
Hence, for unit quantum efficiency, a photodetector measures the photon
number distribution of the state, whereas for non-unit quantum
efficiency the output distribution of counts is given by a Bernoulli
convolution of the ideal distribution.  
Eq.~(\ref{conv_n}) can be written as $p_m(\eta) =
\hbox{Tr}[\varrho\:\Pi_m(\eta)]$ where the POVM of the photocounter
is given by 
\begin{eqnarray}
\Pi_m (\eta) = \eta^m \sum_{k=m}^\infty (1-\eta)^{k-m} 
\binom{k}{m} |k \rangle\langle k|
\label{povm_pn}\;.
\end{eqnarray}
Notice that $\Pi_m (\eta)\geq 0$ and $\sum_m \Pi_m (\eta)=\iid$, but 
$[\Pi_m (\eta),\Pi_k (\eta)]\neq 0$, {\em i.e.} they do not form a set of
orthogonal projectors.
The corresponding characteristic and Wigner functions can be easily 
obtained from that of a number state $| k \rangle\langle k |$, namely
\begin{subequations}
\begin{align}
\label{chin}
\chi[| k \rangle\langle k |](\lambda)
&= \langle k | D(\lambda )|k\rangle = 
e^{-\frac12 |\lambda|^2} \: L_k(|\lambda |^2)\,, \\
W[| k \rangle\langle k |](\alpha)
&= \frac{2}{\pi} 
\langle k| (-)^{a^\dag a} D(2\alpha) | k \rangle = 
\frac{2}{\pi} (-)^k\: e^{-2|\alpha |^2}\: L_k (4|\alpha |^2)\:,
\label{wn}
\end{align}
\end{subequations}
where $L_k(x)$ is a Laguerre polynomials. We have
\begin{subequations}
\begin{align}
\chi[\Pi_m(\eta)](\lambda) &=
\frac1\eta \: L_m \left(\frac{|\lambda|^2}{\eta}\right) 
\exp\left\{-\frac{2-\eta}{2\eta}\,|\lambda|^2\right\}\,, \\
W[\Pi_m(\eta)](\alpha) &= \frac{2}{\pi} \frac{(-)^m \eta^m}{(2-\eta)^{1+m}}\:
L_m\left(\frac{4|\alpha |^2}{2-\eta}\right)\: 
\exp\left\{- \frac{2\eta}{2-\eta}\,|\alpha|^2\right\}\:.
\end{align}
\end{subequations}
\par
The effects of non-unit quantum efficiency on the statistics of a 
photodetector, namely Eqs.~(\ref{conv_n}) and (\ref{povm_pn}),
can be also described by means of a simple model in which the realistic 
(not fully efficient) photodetector is replaced with an ideal photodetector 
preceded by a beam splitter of transmissivity $\cos^2\phi$, 
with the second mode left in the vacuum state.
The reflected mode is absorbed, whereas the transmitted mode is photodetected 
with unit quantum efficiency.  The probability of measuring 
$m$ clicks in such a configuration is given by 
\begin{eqnarray}
p_m (\phi)  = \hbox{Tr}_{ab} \left[ U_\phi \,
\varrho\otimes|0\rangle \langle  0|  
U^{\dag}_\phi\,   |m\rangle \langle m| \otimes \iid \right] 
\label{conv_n1}\;,
\end{eqnarray}
where we denoted by $a$ and $b$ the two involved modes, $U_\phi$ is the
unitary evolution of the beam-splitter (see Section \ref{ss:bs}) and
$\varrho$ the initial preparation of the signal. Using the cyclic
properties of the (full) trace and, then, performing the partial trace over
the vacuum mode, we have 
\begin{align}
p_m (\phi)  &= \hbox{Tr}_{ab} \left[  
\varrho\otimes|0\rangle \langle  0|  \:
U^{\dag}_\phi   |m\rangle \langle m| \otimes \iid \: U_\phi \right] 
\nonumber \\
&= \hbox{Tr}_a \left[\varrho \: \langle 0| 
\: U^{\dag}_\phi   \: |m\rangle \langle m| \otimes \iid \: U_\phi\: 
|0\rangle\right] = \hbox{Tr}_a\left[\varrho\:\Pi_m (\cos^2\phi) \right]\,.
\end{align}
Eq.~(\ref{conv_n1}) reproduces the probability distribution of
Eq.~(\ref{conv_n}) with $\eta=\cos^2\phi$. We conclude that a photodetector
of quantum efficiency $\eta$ is equivalent to an ideal photodetector
preceded by a beam splitter of transmissivity $\eta$ 
which accounts for the overall losses of the detection process.
\par
If we have more than one mode impinging on a photocounter, we should 
take into account that each click may be due to a photon
coming from each of the mode. The resulting POVM assumes the 
form 
\begin{eqnarray}
\gr{\Pi}_m (\gr{\eta}) = \sum_{k_1=0}^\infty ... \sum_{k_n=0}^\infty
\Pi_{k_1}(\eta_1) \otimes ... \otimes \Pi_{k_n} (\eta_n)\:
\delta\left(\sum_{s=1}^n k_s - m\right)\,,
\end{eqnarray}
where we have also supposed that each mode may be detected
with a different quantum efficiency.
\subsection{On/off photodetectors}\label{ss:onoff}
\index{photodetection!on/off}
\begin{floatingfigure}[r]{0.45\textwidth}
\vspace{-.3cm}
\centerline{
\includegraphics[width=0.35\textwidth]{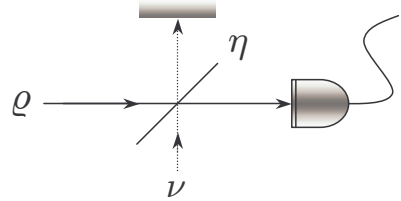}}
\vspace{-.3cm}
\caption{
Model of a realistic on/off photodetector with non-unit quantum efficiency 
$\eta$, and non-zero dark counts.}\label{f:on:off}
\end{floatingfigure}
\noindent
As mentioned above, in a photodetector each photon ionizes a 
single atom and, at least in principle, the resulting charge is 
amplified to produce a measurable pulse. 
Taking into account the quantum efficiency, we conclude that the 
resulting current is proportional to the incoming photon flux 
and thus we have a linear detector. On the other hand,
detectors operating at very low intensities resort to avalanche
process in order to transform a single ionization event into a
recordable pulse. This implies that one cannot discriminate between 
a single photon or many photons as the outcomes from such detectors 
are either a {\em click}, corresponding to any number of photons, or 
{\em nothing} which means that no photons have been revealed. 
These {\em Geiger}-like detectors are often referred to as on/off 
detectors. For unit quantum efficiency, the action of an on/off 
detector is described by the two-value POVM
$\{\Pi_0\doteq |0\rangle\langle 0|, \Pi_1 \doteq {\mathbb I} -  
\Pi_0\}$, which represents a partition of the Hilbert space of the signal.
In the realistic case, when an incoming photon is not detected 
with unit probability, the POVM is given by  
\begin{eqnarray}
\Pi_0 (\eta) = \sum_{k=0}^\infty(1-\eta)^k \: |k\rangle\langle
k|\,, \qquad  \Pi_1 (\eta) =  {\mathbb I} -  
\Pi_0(\eta)\label{onoffPOM}\;, 
\end{eqnarray}
with $\eta$ denoting quantum efficiency. 
The corresponding characteristic and the Wigner functions 
can be easily obtained from that of a number state [see Eqs.~(\ref{chin}) and 
(\ref{wn})]. We have 
\begin{subequations}
\begin{align}
\chi[\Pi_0(\eta)] (\lambda) &=
\frac{1}{\eta}\: \exp\left\{- \frac{2-\eta}{2\eta}\,
|\lambda|^2\right\}\,, \quad \chi[\Pi_1(\eta)] (\lambda)
= \pi \delta^{(2)} (\lambda) 
\label{ChiPi0} -\chi[\Pi_0(\eta)] (\lambda) \:,\\
W[\Pi_0(\eta)](\alpha) &=
\frac{1}{\pi} \frac{2}{2-\eta} 
\exp\left\{-\frac{2\eta}{2-\eta}\,|\alpha |^2\right\}\,,
\quad
W[\Pi_1(\eta)] (\alpha) = \frac1{\pi} -
W[\Pi_0(\eta)] (\alpha)
\label{WPi0}\;.
\end{align}
\end{subequations} 
\par 
Besides quantum efficiency, {\em i.e.} lost photons, the performance of a
realistic photodetector are also degraded by the presence of dark-count,
namely by ``clicks'' that do not correspond to any incoming photon. In
order to take into account both these effects we use the simple scheme
introduced in the previous Section and depicted in Fig.~\ref{f:on:off}. 
A real photodetector is modeled as an ideal photodetector (unit quantum
efficiency, no dark-count) preceded by a beam splitter of transmissivity
equal to the quantum efficiency $\eta$, whose second port is in an
auxiliary excited state $\nu$, which can be a thermal state, or a
phase-averaged coherent state, depending on the kind of background noise
(thermal or Poissonian) we would like to  describe.
When the second port of the beam splitter is the vacuum 
$\nu = |0\rangle\langle 0|$, we have no dark-counts and the 
POVM of the photodetector reduces to that of Eq.~(\ref{onoffPOM}). 
On the other hand, when the second port of the BS excited in a 
generic mixture $$\nu =
\sum_s \nu_{ss} |s\rangle\langle s|\:,$$ then the overall 
POVM describing the  {on/off} photodetection is expressed as 
the following generalized convolution
\begin{align}
\Pi_0^\nu(\eta) = 
\hbox{Tr}_b \left[ U_\phi \varrho \otimes \nu\: U^\dag_\phi\: \iid \otimes
|0\rangle\langle 0|\right]
=\sum_{n=0}^\infty (1-\eta)^n \sum_{s=0}^\infty \nu_{ss} \: 
\eta^s \: \binom{n+s}{s} \: 
|n\rangle\langle n| \label{gendark}\;,
\end{align}
whereas the characteristic and the Wigner functions read as follows
\begin{subequations}
\begin{align}
\chi[\Pi_0^\nu(\eta)] (\lambda) &=
\frac{1}{\eta} 
\exp\left\{- \frac{2-\eta}{\eta}\,|\lambda|^2\right\}
\sum_{s=0}^\infty \nu_{ss}\, L_s\left(
\frac{(1-\eta)|\lambda|^2}{2\eta}\right)\,,
\\
W[\Pi_0^\nu(\eta)] (\alpha) &=
\frac{2}{\pi}
\exp\left\{- \frac{2\eta}{2-\eta}\,|\alpha|^2\right\}
\sum_{s=0}^\infty \nu_{ss} \: \frac{\eta^s}{(2-\eta)^{1+s}}\: 
L_s\left( \frac{4 (\eta-1) |\alpha|^2}{2-\eta}\right) \:.
\end{align}
\end{subequations}
The density matrices of a thermal state and a phase-averaged 
coherent state (with $n_b$ mean photons) are given by 
\begin{align}
\nu_{t} = \frac{1}{n_b+1}
\sum_{s=0}^\infty \left(\frac{n_b}{n_b+1}\right)^s 
\: |s\rangle\langle s|\,, \qquad
\nu_{p} = e^{-n_b} \sum_{s=0}^\infty 
\frac{(n_b)^s}{s!}\,|s\rangle\langle s|  \;.
\label{darkstate}
\end{align}
\index{photodetection!dark counts}
In order to reproduce a background noise with mean photon number $N$ 
we consider the state $\nu$ with average photon number $n_b=N/(1-\eta)$. 
In this case we have 
\begin{subequations}
\label{darkpom}
\begin{align}
\Pi_0^{t}(\eta,N) &= \frac{1}{1+N} \sum_{n=0}^{\infty} \left( 1- 
\frac{\eta}{1+N}\right)^n
\:|n\rangle\langle n|\,, \\  
\Pi_0^{p}(\eta,N) &= e^{-N} \sum_{n=0}^{\infty} (1-\eta)^n 
\: L_n \left(-  \frac{\eta N}{1-\eta}\right) \:|n\rangle\langle n|\;,
\end{align}
\end{subequations}
where $t$ and $p$ denotes thermal and Poissonian respectively. 
The corresponding characteristic and Wigner function are given by 
\begin{subequations}
\begin{align}
\chi[\Pi_0^t(\eta,N)] (\lambda) &=
\frac{1}{\eta}\exp\left\{
- \frac{2(1+N)-\eta}{2\eta}\,|\lambda|^2 
\right\}\,,\\
\chi[\Pi_0^p(\eta,N)] (\lambda) &= 
\frac{1}{\eta} \exp\left\{ -\frac{2-\eta}{2\eta}\,|\lambda|^2 \right\}
I_0\left(2\sqrt{-\frac{N}{\eta}|\lambda|^2}\right)
\end{align}
\end{subequations}
and
\begin{subequations}
\label{WPidark}
\begin{align}
   W[\Pi_0^t(\eta,N)](\alpha)
   &= \frac{1}{\pi}\frac{2}{2(1+N)-\eta} \exp\left\{
 - \frac{2 \eta }{2(1+N)-\eta}\,|\alpha|^2   
   \right\}\,,\\
   W[\Pi_0^p(\eta,N)](\alpha)
   &= \frac{1}{\pi}\:\frac{2}{2-\eta} \exp\left\{
 - \frac{2 \eta}{2-\eta}(N+|\alpha |^2)  
   \right\}\: I_0 \left(\frac{4|\alpha |\sqrt{\eta N}}{2-\eta}\right)\;,
\end{align}
\end{subequations}
respectively, where $I_0(x)$ is the $0$-th modified Bessel function of
the first kind.

$ $
\section{Application: de-Gaussification by vacuum removal} \label{s:NGS}
\index{degaussification}
As we have already pointed out, Gaussian states are very important for
continuous variable quantum information. However, there are situations, as
for example in testing nonlocality with feasible measurements (see Chapter
\ref{ch:nonloc}), where one needs to go beyond Gaussian states. Indeed,
when the Gaussian character is lost, then immediately the Wigner function
of the state becomes negative, for pure states, hence stronger nonclassical
properties should emerge.  An effective method to ``de-Gaussify'' a state
is through a conditional measurement, and, in particular, by elimination of
its vacuum component leading to a state which is necessarily described by a
negative Wigner function.  In the next two Sections this strategy will be
applied both to the TWB and the tripartite state $|T\rangle$ given in
\refeq{T} through the on/off detection scheme introduced in the previous
Section.
\subsection{De-Gaussification of TWB: the IPS map}\label{s:degauss}
\index{degaussification!two-mode}
In this Section we address a de-Gaussification process onto
the twin-beam state (TWB) of two modes of radiation
$|\Lambda \rangle\rangle_{ab} = \sqrt{1-\lambda^2}\,
\sum_{n=0}^{\infty}\,\lambda^n |n,n\rangle_{ab}$, where
we assume the TWB parameter $\lambda = \tanh r$ as real, $r$ 
being referred to as the squeezing parameter. The corresponding 
Wigner function is given by 
\begin{equation}
W_{r}(\alpha,\beta) =
\frac{4}{\pi^2}\exp\{
-2 A (|\alpha|^2+|\beta|^2)
+ 2 B (\alpha\beta + \calpha\cbeta)\}\,,
\label{twb:wig}
\end{equation}
with $A \equiv A(r) = \cosh(2 r)$ and $B \equiv B(r) = \sinh (2 r)$.
\par
The de-Gaussification of a TWB can be achieved by subtracting
photons from both modes through on/off detection \cite{ips:tele,opatr,coch}. 
Since the scheme does not discriminate the number of subtracted photons,
we will refer to this process as to inconclusive photon subtraction (IPS).
\index{photon subtraction}
\begin{floatingfigure}[r]{.47\textwidth}
\begin{center}
\vspace{-.5cm}
\includegraphics[width=.4\textwidth]{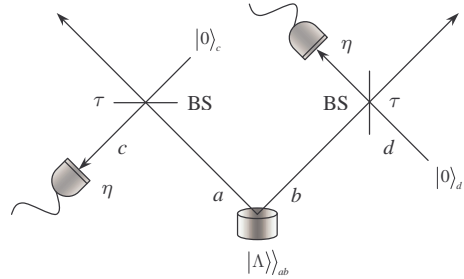}
\vspace{-.4cm}
\caption{Scheme of the IPS process.\label{f:IPS:scheme}}
\vspace{-.3cm}
\end{center}
\end{floatingfigure}
The IPS scheme is sketched in Fig.~\ref{f:IPS:scheme}. The
modes $a$ and $b$ of the TWB are mixed with vacuum modes at 
two unbalanced beam splitters (BS) with equal transmissivity 
$\tau = \cos^2\phi$; the  reflected modes $c$ and
$d$ are then revealed by avalanche photodetectors (APD) with equal
efficiency $\eta$. APD's can only discriminate the presence of
radiation from the vacuum. The positive operator-valued measure
(POVM) $\{\Pi_0(\eta),\Pi_1(\eta)\}$ of each detector is given 
in Eq.~(\ref{onoffPOM}). Overall, the conditional measurement
on modes $c$ and $d$, is described by the POVM (assuming 
equal quantum efficiency for the photodetectors)
\begin{subequations}
\label{povm11}
\begin{align}
{\Pi}_{00} (\eta) &= {\Pi}_{0,c} (\eta) \otimes {\Pi}_{0,d}
(\eta)\,,\qquad\qquad
{\Pi}_{01} (\eta) = {\Pi}_{0,c} (\eta) \otimes
{\Pi}_{1,d} (\eta)\:, \\
{\Pi}_{10} (\eta) &= {\Pi}_{1,c} (\eta) \otimes {\Pi}_{0,d}
(\eta)\,,\qquad\qquad
{\Pi}_{11} (\eta) = {\Pi}_{1,c} (\eta) \otimes
{\Pi}_{1,d} (\eta)\;.
\end{align}
\end{subequations}
When the two photodetectors jointly click, the conditioned output state
of modes $a$ and $b$ is given by \cite{ips:tele,ips:nonloc}
\begin{equation}
\mathcal{E}(R)
= \frac{\hbox{Tr}_{cd}\big[
U_{ac}(\phi)\otimes U_{bd}(\phi) \: R
\otimes |0\rangle_c{}_{c}\langle 0|
\otimes |0\rangle_d{}_{d}\langle 0|
\: U_{ac}^{\dag}(\phi)\otimes U_{bd}^{\dag}(\phi) \:
{\mathbb I}_a \otimes
{\mathbb I}_b \otimes
{\Pi}_{11} (\eta)
\big]}{p_{11}(r,\phi,\eta)}\:, \label{ptr}
\end{equation}
where $U_{ac}(\phi)=\exp\{-\phi(a^{\dag} c-a c^{\dag}) \}$ and
$U_{bd}(\phi)$ are the evolution operators of the beam splitters
and $R$ the density operator of the two-mode state entering
the beam splitters (in our case $R = \varrho_{\hbox{\tiny TWB}} =
|\Lambda\rangle\rangle_{ab}{}_{ba} \langle\langle \Lambda |$). The partial
trace on modes $c$ and $d$ can be explicitly evaluated, thus
arriving at the following decomposition of the IPS map 
\footnote{\footnotesize Eq.~(\ref{KE}) is indeed an operator-sum
representation of the IPS map: $\{p,q\}\equiv \theta$ should be
intended as a polyindex so that (\ref{KE}) reads
$\mathcal{E}(R)=\sum_\theta A_\theta R A^\dag_\theta$ with $A_\theta= 
[p_{11}(r,\phi,\eta)]^{-1/2} m_p(\phi,\eta)\:M_{pq}(\phi)$.}.
We have
\begin{eqnarray}
\mathcal{E}(R)
= \frac{1}{p_{11}(r,\phi,\eta)}\:
\sum_{p,q=1}^{\infty}\:m_p(\phi,\eta)\:M_{pq}(\phi)\: R \:
M_{pq}^{\dag}(\phi)\: m_q(\phi,\eta)\:\label{KE}
\end{eqnarray}
where
\begin{equation}
m_p(\phi,\eta) =
{\displaystyle \frac{\tan^{2p}\phi\,\,[1-(1-\eta)^p]}{p!}}\,,
\qquad
M_{pq}(\phi) = \frac{\mbox{}}{\mbox{}}
a^p b^q \, (\cos\phi)^{a^\dag a + b^\dag b}\,.
\end{equation}
Now we explicitly calculate the Wigner function of the state
$\varrho_{\hbox{\tiny{IPS}}} = \mathcal{E}(\varrho_{\hbox{\tiny TWB}})$,
which, as one may expect, is no longer Gaussian and positive-definite.
The state entering the two beam splitters is described by the Wigner
function
\begin{equation}
W_{r}^{\hbox{\tiny (in)}}(\alpha,\beta,\zeta,\xi) =
W_{r}(\alpha,\beta)\,
\frac{4}{\pi^2} \exp\left\{ -2|\zeta|^2 - 2|\xi|^2 \right\}\,,
\end{equation}
where the second factor at the right hand side represents the two vacuum
states of modes $c$ and $d$.
The action of the beam splitters on $W^{\hbox{\tiny (in)}}_{r}$ can be
summarized by the following change of variables (see Section \ref{ss:bs})
\begin{subequations}
\begin{align}
\alpha &\to \alpha\cos\phi + \zeta\sin\phi\,,\quad
\zeta  \to \zeta\cos\phi  - \alpha\sin\phi\,,\\
\beta  &\to \beta\cos\phi  + \xi\sin\phi\,,\quad
\xi    \to \xi\cos\phi    - \beta\sin\phi\,,
\end{align}
\end{subequations}
and the output state, after the beam splitters, is then given by
\begin{multline}
W_{r,\phi}^{\hbox{\tiny (out)}}(\alpha,\beta,\zeta,\xi) =
\frac{4}{\pi^2}\, W_{r,\phi}(\alpha,\beta)\,
\exp\left\{ -a |\xi|^2 + w \xi + \cw \cxi \right\} \\
\times\exp\big\{ -a |\zeta|^2 + (v + 2 B \xi \sin^2\phi)\zeta 
+ (\cv + 2 B \cxi \sin^2\phi)\czeta \big\}\,,
\end{multline}
where
\begin{equation}
W_{r,\phi}(\alpha,\beta) =
\frac{4}{\pi^2}\,
\exp\left\{ -b (|\alpha|^2 + |\beta|^2)
+ 2 B \cos^2\phi\, (\alpha\beta + \calpha\cbeta) \right\}
\end{equation}
and
\begin{subequations}
\begin{align}
a &\equiv a(r,\phi) = 2 (A \sin^2\phi + \cos^2\phi),\\
b &\equiv b(r,\phi) = 2 (A \cos^2\phi + \sin^2\phi)\,,\\
v &\equiv v(r,\phi) = 2 \cos\phi\, \sin\phi\, [(1-A)\calpha + B \beta],\\
w &\equiv w(r,\phi) = 2 \cos\phi\, \sin\phi\, [(1-A)\cbeta + B \alpha]\,.
\end{align}
\end{subequations}
\par
At this stage on/off detection is performed on modes
$c$ and $d$ (see Fig.~\ref{f:IPS:scheme}). We are interested in
the situation when both the detectors click. The Wigner function
of the double click element $\Pi_{11}(\eta)$ of the POVM
[see Eq.~(\ref{povm11})] is given by \cite{ips:tele,cond:cola}
\begin{align}
W_{\eta}(\zeta,\xi) \equiv W[\Pi_{11}(\eta)](\zeta,\xi)
= \frac{1}{\pi^2}\{
1-Q_{\eta}(\zeta)-Q_{\eta}(\xi)
+Q_{\eta}(\zeta) Q_{\eta}(\xi)
\}\,,
\end{align}
with
\begin{equation}
Q_{\eta}(z) = \frac{2}{2-\eta}\,
\exp\Bigg\{-\frac{2\eta}{2-\eta}\, |z|^2 \Bigg\}\,.
\end{equation}
Using Eq.~(\ref{ptr}) and the phase-space expression of trace
for each mode [see Eq.~(\ref{Wtrace})], the Wigner function of
the output state, conditioned to the double click event, reads
\begin{equation}\label{w:ips:informal}
W_{r,\phi,\eta}(\alpha,\beta) =
\frac{f(\alpha,\beta)}{p_{11}
(r,\phi,\eta)}\,,
\end{equation}
where $f(\alpha,\beta) \equiv f_{r,\phi,\eta}(\alpha,\beta)$ with
\begin{equation}\label{w:ips:informal:f}
f(\alpha,\beta) =
\pi^2\,\int_{\mathbb{C}^2}d^2\zeta\,d^2\xi\,
\frac{4}{\pi^2}\,W_{r,\phi}(\alpha,\beta)\,
\sum_{k=1}^4 \frac{C_k(\eta)}{\pi^2}\,
G_{r,\phi,\eta}^{(k)}(\alpha,\beta,\zeta,\xi)\,,
\end{equation}
and $p_{11}(r,\phi,\eta)$ is the double-click probability reported above,
which can be written as function of $f(\alpha,\beta)$ as
follows
\begin{equation}\label{w:ips:informal:p}
p_{11}(r,\phi,\eta) =
\pi^2\,\int_{\mathbb{C}^2}d^2\alpha\,d^2\beta\,
f(\alpha,\beta)\,.
\end{equation}
The quantities $G_{r,\phi,\eta}^{(k)}(\alpha,\beta,\zeta,\xi)$ 
in Eq.~(\ref{w:ips:informal:f}) are given by 
\begin{multline}
G_{r,\phi,\eta}^{(k)}(\alpha,\beta,\zeta,\xi) =
\exp\big\{ -x_k |\zeta|^2 + (v + 2 B \xi \sin^2\phi)\zeta
+(\cv + 2 B \cxi \sin^2\phi)\czeta \big\} \\
\times\exp\left\{ -y_k |\xi|^2 + w \xi + \cw \cxi \right\}\,,
\label{meas:int}
\end{multline}
where the expressions of $x_k\equiv x_k(r,\phi,\eta)$,
$y_k\equiv y_k(r,\phi,\eta)$, and $C_k(\eta)$ are reported in Table \ref{t:jj}.
\begin{table}[t!]
\begin{center}
\begin{tabular}{|cccc|}
\hline\hline
\mcc{$k$} & \mcc{$x_k(r,\phi,\eta)$} & \mcc{$y_k(r,\phi,\eta)$} &
\mcc{$C_k(\eta)$}\vspace{0cm}\\ \hline
\mcc{1} &  \mcc{$a$}    & \mcc{$a$} & \mcc{1} \\
\mcc{2}   &  \mcc{${\displaystyle a + \frac{2}{2-\eta}}$} & \mcc{$a$} &
\mcc{${\displaystyle -\frac{2}{2-\eta}}$} \vspace{-.3cm}\\
\mcc{3}   &  \mcc{$a$}  & \mcc{${\displaystyle a + \frac{2}{2-\eta}}$} & 
\mcc{${\displaystyle -\frac{2}{2-\eta}}$} \vspace{-.3cm}\\
\mcc{4}   &  \mcc{${\displaystyle a + \frac{2}{2-\eta}}$} & 
\mcc{${\displaystyle a + \frac{2}{2-\eta}}$} &
\mcc{${\displaystyle \left( \frac{2}{2-\eta} \right)^2}$} \\
\hline\hline
\end{tabular}
\end{center}
\caption{Expressions of $C_k$, $x_k$, and $y_k$ appearing in
Eq.~(\ref{meas:int}). \label{t:jj}}
\end{table}
\par
The mixing with the vacuum in a beam splitter with transmissivity $\tau$
followed by on/off detection with quantum efficiency $\eta$ is equivalent
to mixing with an effective transmissivity \cite{ips:tele}
\begin{equation}
\tau_{\rm eff} \equiv
\tau_{\rm eff}(\phi,\eta) = 1 - \eta (1-\tau)
\label{taueff}
\end{equation}
followed by an ideal ({\em i.e.} efficiency equal to 1) on/off
detection. Therefore, the state (\ref{w:ips:informal}) can be studied
for $\eta = 1$ and replacing $\tau=\cos^2\phi=1-\sin^2\phi$
with $\tau_{\rm eff}$.
Thanks to this substitution, after the integrations we have
\begin{multline}
f(\alpha,\beta) =
\frac{1}{\pi^2}\,
\sum_{k=1}^4 \frac{16\, C_k}{x_k y_k - 4 B^2 (1-\tau_{\rm eff})^2}\\
\times\exp\{
(f_k-b)|\alpha|^2 + (g_k-b)|\beta|^2
+(2 B \tau_{\rm eff} + h_k)(\alpha\beta + \calpha\cbeta)\}
\end{multline}
and
\begin{equation}
p_{11}(r,\tau_{\rm eff}) =
\sum_{k=1}^4 \frac{16\, [x_k y_k - 4 B^2 (1-\tau_{\rm eff})^2]^{-1}\, C_k}{
(b-f_k)(b-g_k)-(2 B \tau_{\rm eff} + h_k)^2}
\,,
\end{equation}
where we defined $C_{k}\equiv C_{k}(1)$ and
\begin{subequations}
\begin{align}
f_k &\equiv f_k(r,\tau_{\rm eff}) =
N_k
\, [x_k B^2 + 4 B^2 (1-A) (1-\tau_{\rm eff}) + y_k (1-A)^2]\,,\\
g_k &\equiv g_k(r,\tau_{\rm eff}) =
N_k
\, [x_k (1-A)^2 + 4 B^2 (1-A) (1-\tau_{\rm eff}) + y_k B^2]\,,\\
h_k &\equiv h_k(r,\tau_{\rm eff}) =
N_k
\, [(x_k + y_k) B (1-A)
+ 2 B (B^2 + (1-A)^2) (1-\tau_{\rm eff})]\,,\\
N_k &\equiv  N_k(r,\tau_{\rm eff}) =
{\displaystyle
\frac{4 \tau_{\rm eff}\, (1-\tau_{\rm eff})}{x_k y_k - 4 B^2
(1-\tau_{\rm eff})^2}\,.
}
\end{align}
\end{subequations}
In this way, the Wigner function of the IPS state can be rewritten as
\begin{eqnarray}\label{ips:wigner}
W_{\hbox{\rm\tiny IPS}}(\alpha,\beta) =
\frac{4}{\pi^2}\frac{1}{p_{11}(r,\tau_{\rm eff})}
\sum_{k=1}^4 {\cal C}_k\,W_{k}(\alpha,\beta)\,,
\end{eqnarray}
where we introduced
\begin{equation}
{\cal C}_k\equiv{\cal C}_k(r,\tau_{\rm eff})=
\frac{4\, C_k}
{x_k y_k - 4 B^2 (1-\tau_{\rm eff})^2}\,,
\end{equation}
and defined
\begin{equation}
W_{k}(\alpha,\beta) =
\exp\{ (f_k-b) |\alpha|^2 +(g_k-b) |\beta|^2
+ (2B\tau_{\rm eff} + h_k) (\alpha\beta + \calpha\cbeta)\}\,.
\end{equation}
Finally, the density matrix corresponding to $W_{\hbox{\rm\tiny
IPS}}(\alpha,\beta)$ reads as follows \cite{ips:tele}
\begin{multline}
\label{ips:fock}
{\varrho}_{\hbox{\rm\tiny IPS}} =
  \frac{1-\lambda^2}{p_{11}(r,\tau_{\rm eff})}
  \sum_{n,m=0}^{\infty} (\lambda\, \tau_{\rm eff})^{n+m} \\
  \times \sum_{h,k=0}^{{\rm Min}[n,m]}
  \left(\frac{1-\tau_{\rm eff}}{\tau_{\rm eff}}\right)^{h+k}
  \sqrt{ {n \choose h}{n \choose k}{m \choose h}{m \choose k} } 
  \times \ket{n-k}_a \ket{n-h}_b {_b}\bra{m-h} {_a}\bra{m-k}\,,
\end{multline}
with $\lambda = \tanh r$.
\par
The state given in Eq.~(\ref{ips:wigner}) is no longer a Gaussian state.
Its use in the enhancement of the nonlocality \cite{nha,sanchez,ips:nonloc} 
and in the improvement of CV teleportation \cite{ips:tele} will be
investigated in Chapter \ref{ch:nonloc} and \ref{ch:tele}, respectively.
\subsection{De-Gaussification of tripartite state: the TWBA
state}\label{ss:TWBA}
\index{degaussification!three-mode}
In this Section we consider the tripartite state $|T\rangle$ given in
\refeq{T} as a source of two-mode states. In particular, we analyze 
two-mode non-Gaussian state obtained by a conditional measurement performed 
on it. Due to the structure of the state $|T\rangle$, its vacuum component
can be subtracted by a conditional measurement on mode $a_3$, the same
observation being valid for mode $a_2$. Let us consider on/off detection
performed on mode $a_3$. The three-mode two-valued POVM is 
$\{\Pi_0^{(3)}(\eta),\Pi_1^{(3)}(\eta)\}$, with the
element associated to the ``no photons'' result given by
\begin{align}
\label{c3:NoPh} \Pi_0^{(3)}(\eta)\doteq \iid_1 \otimes \iid_2
\otimes \sum_{n=0}^{\infty} (1-\eta)^n |n\rangle_3{}_3\langle n| \:.
\end{align}
The probability of a ``click'' is
\begin{align}
  \label{c3:P1} P_1 \equiv P_1(N_3,\eta)
  = \hbox{Tr}_{123} \left[|T \rangle\langle T|\:
  \Pi_1^{(3)}(\eta)\right] = \frac{\eta N_3}{(1+\eta N_3)} \:,
\end{align}
while the conditional output state reads as follows
\begin{align}
  \label{c3:TWBA} \varrho_{\hbox{\tiny TWBA}} &= \frac{1}{P_1}
  \hbox{Tr}_3\left[| T \rangle\langle T |\:
  \Pi_1^{(3)}(\eta)\right] \nonumber \\ 
  \hspace{-50pt}
  &=\frac{1+\eta N_3}{(1+N_1+N_2)\eta N_3}\, 
  \sum_{p=1}^\infty
  \left(\frac{N_3}{1+N_1}\right)^p \frac{1-(1-\eta)^p}{p!} 
  (a^\dagger_1)^p\,
  |\Lambda\rangle\rangle_{12}{}_{21}\langle\langle \Lambda| a^p_1 
  \:,
\end{align}
where we denote by $|\Lambda\rangle\rangle_{12}$ the TWB state of the modes
$a_1$ and $a_2$ with parameter $\lambda=\sqrt{N_2/(1+N_1)}$ (see
Section \ref{ss:opa}).  We indicated this state with a subscript TWBA ({\em
i.e.} TWB {\em added}) since it corresponds to a mixture of TWBs with
additional photons in one of the modes. In order to evaluate its Wigner
function we use (\ref{ChiPi0}) for the characteristic function  
of $\Pi_1^{(3)}(\eta)$, hence the characteristic function of 
$\varrho_{\hbox{\tiny TWBA}}$ is given by
\begin{multline}
\label{c3:RhoTWBAChi}
\chi[\varrho_{\hbox{\tiny TWBA}}](\lambda_1,\lambda_2)=\frac{1}{P_1}
\Bigg\lbrace
\chi[| T \rangle\langle T |](\lambda_1,\lambda_2,0) 
-\frac1\eta\int_\cc \frac{d^2\mu}{\pi}
\chi[| T \rangle\langle T |](\lambda_1,\lambda_2,\mu)
\exp\left\{-\frac{2-\eta}{2\eta}\,|\mu|^2\right\}
\Bigg\rbrace\;.  
\end{multline}
After some algebra the Wigner function associated with state
$\varrho_{\hbox{\tiny TWBA}}$ can now be calculated. It reads as follows
\begin{multline}
W_{\hbox{\tiny TWBA}}(\bmY) = 
\frac{1+\eta N_3}{4\eta N_3} \:
\left(\frac{2}{\pi}\right)^2
\Bigg\lbrace\frac{1}{\sqrt{\det [\bmV_T']}}
\exp\left\{ - \bmY^{\sT} \left(\bmV_T'\right)^{-1} \bmY \right\} \\
- \frac{1}{\eta} \frac{2}{\sqrt{\det [\bmD]}}
\exp\left\{ - \bmY^{\sT} \left({\bmD}^{-1}\right)' \bmY\right\}
\Bigg\rbrace
\label{c3:WTWBA}\;,
\end{multline}
where $\bmY = (x_1,x_2,y_1,y_2)^{\sT}$, and
${\bmD}={\bmV_T}+{\rm Diag}
(0,0,\frac{2-\eta}{\eta},0,0,\frac{2-\eta}{\eta} )$,
$\bmV_T$ being defined in Eq.~(\ref{c3:matV}).
In order to simplify the notation we have indicated with ${\bmO}'$ 
the $4\times4$ matrix obtained from the $6\times6$ matrix ${\bmO}$ 
deleting the elements corresponding to the third mode (3-rd row/column 
and 6-th row/column), due to the trace over the 3-rd mode.
Nonlocality properties of the TWBA state will investigated in Chapter
\ref{ch:nonloc}.
\section{Homodyne detection}\label{s:hom}
\index{homodyne detection}
Homodyne detection schemes are devised to provide 
the measurement of a single-mode quadrature $x_\phi$ through the 
mixing of the signal under investigation with a highly excited 
classical field at the same frequency, referred to as the 
{\em local oscillator} (LO). Homodyne detection was proposed 
for the radiation field in Ref. \cite{yuenchan}, and 
subsequently demonstrated in Ref. \cite{abbas}. For the radiation 
field quadrature measurements can be achieved by balanced and unbalanced 
homodyne schemes, whereas realizations for atomic systems have
also been proposed \cite{wil90}.
\subsection{Balanced homodyne detection}
\index{homodyne detection!balanced}
\begin{figure}
\vspace{-.2cm}
\begin{center}
\includegraphics[width=0.35\textwidth]{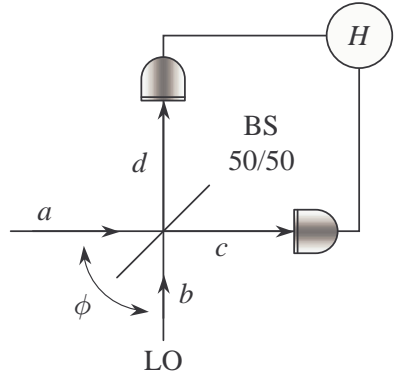}
\end{center}
\vspace{-.4cm}
\caption{Schematic diagram of the balanced homodyne
detector.\label{f:homodyne}}
\end{figure}
The schematic diagram of a balanced homodyne detector is 
reported in Fig.~\ref{f:homodyne}. The signal mode $a$ interferes 
with a second mode $b$ excited in a coherent semiclassical state 
({\em e.g.} a laser beam) in a balanced (50/50) beam splitter 
(BS). The mode $b$ is the LO mode of the detector. It operates 
at the same frequency of $a$, and is excited in a coherent 
state $|z\rangle$ with large amplitude $z$. 
The BS is tuned to have real coupling, hence no additional 
phase-shift is imposed on the reflected and transmitted beams.
Moreover, since in all experiments that use homodyne 
detectors the signal and the LO beams are generated by a common source, 
we assume that they have a fixed phase relation. In this case the LO 
phase provides a reference for the quadrature measurement, namely  
we identify the phase of the LO with the phase difference between 
the two modes.
As we will see, by tuning $\phi =\arg[z]$ we can measure 
the quadrature $x_\phi $ at different phases $\phi$.
After the BS the two modes are detected by two identical
photodetectors (usually linear photodiodes), and finally the
difference of photocurrents at zero frequency is electronically
processed and rescaled by $2|z|$. According to Eqs.~(\ref{EvolMode}) 
and (\ref{mixD}), denoting by $c$ and $d$ the output mode from the beam 
splitter, the resulting {\em homodyne photocurrent} $H$
is given by
\begin{eqnarray}
H={\frac{c^{\dag}c- d^{\dag}d }{2|z|}}=\frac{{a^\dag } b+b^{\dag}a}{2|z|}\;.
\end{eqnarray}
Notice that the spectrum of the operators ${a^\dag } b+b^{\dag}a$ is
discrete and coincides with the set ${\mathbb Z}$ of relative integers.
Therefore the spectrum of the homodyne photocurrent $H$ is discrete too, 
approaching the real axis in the limit of highly excited LO ($|z|\gg 1$).
We now exploit the assumption of a LO excited in a strong semiclassical
state, {\em i.e.} we neglect fluctuations of the LO and make the 
substitutions $b\rightarrow z$, $b^\dag \rightarrow z^*$. 
The moments of the homodyne photocurrent are then given by
\begin{align}
H = x_\phi\,, \qquad H^2 = x^2_\phi + \frac{a^\dag a}{4|z|^2}\,,  \qquad 
\cdots \mbox{} \qquad 
H^n = x^{2n-2}_\phi\: \left(x^2_\phi + \frac{a^\dag a}{4|z|^2}\right)
\label{momquad}\;,
\end{align}
which coincide with the quadrature moments for signals satisfying 
$\langle a^\dag a \rangle \ll 4|z|^2$. In this limit the distribution
of the outcomes $h$ of the homodyne photocurrent is equal to that 
of the corresponding field quadratures. The POVM $\{\Pi_h\}$ of 
the detector coincides with the spectral measure of the quadratures
\begin{eqnarray}
\Pi_h \xrightarrow{|z|\gg 1} \Pi (x) = |x\rangle_\phi {}_\phi\langle x|
\equiv \delta(x_\phi -x)\:, \label{pomH}
\end{eqnarray}
{\em i.e.} the projector on the eigenstate of the quadrature 
$x_\phi$ with eigenvalue $x$. In conclusion, the balanced 
homodyne detector achieves the ideal measurement of the quadrature 
$x_\phi $ in the strong LO regime. In this limit, which summarizes the two
conditions i) $|z|\gg 1$ to have a continuous spectrum and ii) 
$|z|^2\gg \langle a^\dag a \rangle$ to neglect extra terms in the 
photocurrent moments, the probability 
distribution of the output photocurrent $H$ approaches 
the probability distribution $p(x,\phi) = {}_\phi \langle x|\varrho 
|x \rangle_\phi$ of the quadrature $x_ \phi$ for  
of the signal mode $a$. 
The same result \cite{rev} can be obtained by evaluating the moment 
generating function $M_H(\mu)=
\hbox{Tr}\left[\varrho\otimes |z\rangle \langle z|\:
e^{i \mu H} \right]$. Using the disentangling formula for $\rmSU(2)$ 
(\ref{BCHJubs}) we have
\begin{eqnarray}
M_H(\mu)=\left\langle e^{
i\tan\left({\frac {\mu}{2|z|}}\right)b^{\dag}
a }\left[\cos\left({\frac{\mu}{2|z|}}\right)\right ]^{{a^\dag } a
-b^{\dag}b}e^{i\tan\left({\frac{\mu}{2|z|}}\right){a^\dag } b }\right\rangle _{ab}\;.
\end{eqnarray}
Since mode $b$ is in a coherent state $|z \rangle $ 
the partial trace over $b$ can be evaluated  as follows
\begin{equation}
M_H(\mu) =\left\langle e^{i\tan\left({\frac{\mu}{2|z|}}\right)
z^*a} \left[\cos\left({\frac{\mu}{2|z|}}\right)
\right]^{{a^\dag } a}e^{i\tan\left({\frac {\mu}{2|z|}}\right)z{a^\dag }}
\right\rangle _a 
\left\langle z\Bigg|\left[\cos\left({\frac {\mu}{2|z|}}
\right)\right]^{-b^{\dag}b}\Bigg|z\right\rangle \;.\label{xx1}
\end{equation}
Now, rewriting (\ref{xx1}) in normal order with respect to mode $a$ 
we have
\begin{eqnarray}
M_H(\mu)=
\left\langle e^{iz\sin\left({\frac {\mu}{2|z|}}\right)
{a^\dag } }
\exp\left\{-2\sin^2\left({\frac{\mu}{4|z|}}\right)({a^\dag }
a+|z|^2)\right\}
e^{iz^*\sin\left({\frac {\mu}{2|z|}}\right) a}
\right\rangle _a\!\!.\label{anbra}
\end{eqnarray}
In the strong LO limit (\ref{anbra}) becomes
\begin{eqnarray}
\lim_{z\to\infty}M_H(\mu)=
\left\langle e^{i{\frac {\mu} {2}}e^{i\phi }{a^\dag }}\,
\pexp{\frac{-
\mu^2} {8}}\,e^{i{\frac {\mu} {2}}e^{-i\phi}a }
\right\rangle _a= \left\langle \pexp{i\mu x_\phi}
\right\rangle _a\;.\label{wwww}
\end{eqnarray}
The generating function in  (\ref{wwww}) then corresponds to the POVM
\begin{eqnarray}
\Pi (x)= \int_{\rr} {\frac {d \mu }{2\pi}} \,
\exp\{i\mu  (x_\phi -x)\}=
\delta(x_\phi -x)\equiv |x\rangle _\phi {}_\phi \langle x|\;,\label{pmx}
\end{eqnarray}
which confirm the conclusions drawn in Eq.~(\ref{pomH}).
\par 
In order to take into account non-unit quantum
efficiency at detectors we employ the model introduced in 
the previous Sections, {\em i.e.} each inefficient detector 
is viewed as an ideal detector preceded by a beam splitter of
transmissivity $\eta$ 
with the second port left in the vacuum. The homodyne photocurrent 
is again formed as the difference photocurrent, now rescaled 
by $2|z|\eta$. We have 
\begin{eqnarray}
H_\eta \simeq  \frac{1}{2|z|}\left\{\left[ a+\sqrt{\frac{1-\eta
    }{2\eta }}(u +  v )\right]b^\dag + h.c. \right\}
\;,\label{IDeta}
\end{eqnarray}
where only terms containing the strong LO mode $b$ are retained, 
and $u$ and $v$ denote the additional vacuum modes introduced to 
describe loss of photons.
The POVM is obtained by replacing 
\begin{eqnarray}
x_\phi \rightarrow x_\phi +\sqrt{\frac{1-\eta }{2\eta
}}(u_\phi +v_\phi)\;
\end{eqnarray}
in Eqs.~(\ref{pmx}), with $w_\phi =\frac12 (w^\dag e^{i\phi} 
+we^{-i\phi}) $, $w=u,v$, and tracing the vacuum modes
$u$ and $v$. One then obtains
\begin{align}
\Pi _{\eta}(x) &=
\frac{1}{\sqrt{2\pi\delta^2_{\eta}}}\:
\pexp{-\frac{(x_\phi- x )^2}{2
\delta_{\eta}^2}} \nonumber \\ &= 
\frac{1}{\sqrt{2\pi\delta^2_{\eta}}}
\int_\cc dy\, 
\pexp{-\frac{(x-  y)^2}{2\delta_{\eta}^2}}
|y\rangle _\phi {}_\phi \langle y|
\;,\label{OM}
\end{align}
where 
\begin{eqnarray}
\delta_{\eta}^2=\frac{1-\eta}{4\eta}\;.\label{Deltaeta}
\end{eqnarray}
Thus the POVMs, and in turn the probability distribution of the output 
photocurrent, are just the Gaussian convolutions of the ideal ones.
\par
The Wigner functions of the homodyne POVM is given by
\begin{eqnarray}
W[\Pi_\eta(x)](\alpha) = 
\frac{1}{\sqrt{2\pi\delta^2_{\eta}}}
\exp\left\{ -\frac{[x-\frac12(\alpha e^{-i\phi} + 
\alpha^* e^{i\phi})]^2}{2\delta_{\eta}^2}\right\}
\:,
\end{eqnarray}
which leads to $W[\Pi(x)](\alpha)\xrightarrow{\eta\rightarrow 1}
\delta(x-\frac12(\alpha e^{-i\phi} + 
\alpha^* e^{i\phi}))$ in the limit of an ideal homodyne detector.
\subsection{Unbalanced homodyne detection}
\index{homodyne detection!unbalanced}
\begin{floatingfigure}[r]{0.4\textwidth}
\vspace{-1cm}
\begin{center}
\includegraphics[width=0.3\textwidth]{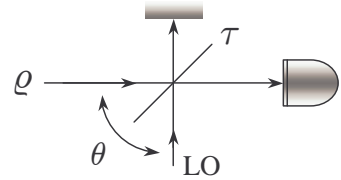}
\end{center}
\vspace{-.7cm}
\caption{Schematic diagram of unbalanced homodyne detector.}
\label{f:unbh}
\end{floatingfigure} \noindent
The scheme of Fig.~\ref{f:unbh} is known as unbalanced homodyne detector 
and represents an alternative method to measure the statistics of a 
field quadrature. The signal under investigation is mixed with the LO 
at a beam splitter with transmissivity $\tau=\cos^2\phi$. The reflected 
beam is then absorbed, whereas the transmitted beam is revealed 
through a linear photocounter. If $a$ is the signal mode and $b$ the LO mode
the transmitted mode $c$ can be written as $c=a\cos\phi+b\sin\phi$. 
The detected 
photocurrent is given by
\begin{align}
n_c  &\equiv c^\dag c = a^\dag a \cos^2\phi + b^\dag b \sin^2\phi 
+(a^\dag b + b^\dag a) \sin\phi\cos\phi
\label{unb1}\:
\end{align}
and the unbalanced homodyne photocurrent is obtained by a simple rescaling 
$$I_H = \frac{n_c}{2\sin^2\phi |z|}\:,$$ where $z$ is again the LO amplitude.
Upon tracing over the local oscillator, in the limit of 
$|z|\gg \langle a^\dag a\rangle$, we have for the first two moments
\begin{align}
\langle I_H \rangle &= \frac12 + \frac{1}{|z|\tan\phi} \langle x_\theta \rangle
+ O (|z|^{-2})\,,  \\
\langle I_H^2 \rangle &= \frac14 + \frac {1}{|z|\tan\phi} \langle x_\theta \rangle
+ \frac{1}{|z|^2\tan^2\phi} \langle x^2_\theta\rangle
+ O (|z|^{-2}) \label{unb2}\;,
\end{align}
and therefore $\langle \Delta I^2_H \rangle = (|z|^2 \tan^2\phi)^{-1} 
\langle\Delta x^2_\theta \rangle$, where $\theta$ is the shift between 
signal and LO. 
The procedure can be 
generalized to higher moments, thus concluding that through 
unbalanced homodyne one can recover the statistics of the field
quadratures. In order to minimize the effect of LO, the regime 
$\phi\ll 1$ with $|z|\phi$ finite should be adopted.
\subsection{Quantum homodyne tomography}\label{ss:QHT}
\index{homodyne detection!quantum tomography}
\index{quantum homodyne tomography}
The measurement of the field quadrature $x_\phi$ for {\em all} values 
of the phase $\phi$ provides the complete knowledge of the state 
under investigation, {\em i.e.} the expectation values of any quantity
of interest (including quantities not directly observable). This 
kind of measurement is usually referred to as {\em quantum homodyne 
tomography} \cite{rev,LNP} for reasons that will be explained at the end 
of this Section.
\par
In order to see how the knowledge of
$p(x,\phi)={}_\phi\bra{x}\varrho\ket{x}_\phi$ allows the
reconstruction of any expectation value let us rewrite the Glauber 
formula (\ref{GlauberF}) changing to polar variables $\lambda = (-i/2)k
e^{i\phi}$
\begin{equation}
  O = \int^{\pi}_0\frac{d\phi}{\pi}\int_\rr
\frac{d k\, |k|}{4}\,\hbox{Tr} [  O\;  e^{ik x_{\phi}}]\, 
e^{-ik x_{\phi}}\; , \label{op}
\end{equation}
which shows explicitly the dependence on the quadratures $ x_\phi $.
Taking the ensemble average of both members and evaluating the trace
over the set of eigenvectors of $ x_{\phi}$, one obtains
\begin{eqnarray}
\langle   O  \rangle = 
\int^{\pi}_0\frac{d\phi }{\pi}  \int_\rr dx\;
p(x,\phi) 
\;  {\cal R}[O] (x,\phi)\,, \label{qht1}
\end{eqnarray}
The function ${\cal R}[O] (x,\phi)$ is known as {\em kernel} or {\em pattern} 
function for the operator $ O$, its trace form is given by 
${\cal R}[O](x,\phi) = \hbox{Tr} [  O K(x_{\phi}-x)]$ where $K(x)$ 
writes as 
\begin{eqnarray}
K(x) \equiv \int_\rr \frac{dk}4|k|e^{ikx}=
\frac 12\Re\mbox{e}\int_0^{+\infty} dk\; k\, e^{ikx}
\;.\label{kernelaa}
\end{eqnarray} 
Therefore, upon calculating the corresponding pattern function, any
expectation value can be evaluated as an average over homodyne data.
Remarkably, tomographic reconstruction is possible also taking into 
account nonunit quantum efficiency of homodyne detectors, {\em i.e.} 
upon replacing  $p(x,\phi)$ with $p_\eta(x,\phi)$. Indeed, we have 
\begin{eqnarray}
\langle   O  \rangle 
=\int_0^\pi\frac{d\phi}\pi\int_\rr
dx\; p_\eta(x,\phi)\;  {\cal R}_\eta[O] (x,\phi) \label{qht2}\;,
\;\label{formtomhometa}
\end{eqnarray}
where the pattern function is now 
${\cal R}_\eta[  O](x,\phi) =
\hbox{Tr} [  O\:K_\eta(x_{\phi}-x)]$, with 
\begin{eqnarray}
K_\eta(x)=\frac 12\Re\hbox{e}\int_{0}^{+\infty}
dk\;k\,\pexp{\frac{1-\eta}{8\eta}k^2+ikx}
\;\label{kerneleta1}.
\end{eqnarray}
Notice that the anti-Gaussian in  (\ref{kerneleta1}) causes a slower 
convergence of the integral (\ref{qht2}) and thus, in order to achieve 
good reconstructions with non-ideal detectors, one has to collect a 
larger number of homodyne data.
As an example, the kernel functions for the normally ordered 
products of mode operators are given by \cite{rich,tokyo}
\begin{eqnarray}
{\cal R}_{\eta}[a^{\dag}{}^n a^m](x,\phi)=e^{i(m-n)\phi}
\frac{H_{n+m}(\sqrt{2\eta }x)}{\sqrt{(2\eta)^{n+m}}{{n+m}\choose n}}\;,\label{Rnm}
\end{eqnarray}
where $H_n(x)$ is the $n$-th Hermite polynomials, whereas the reconstruction
of the elements of the density matrix in the number representation 
$\varrho_{nm}=\hbox{Tr}[\varrho\: |n\rangle\langle m|]$ corresponds to 
averaging the kernel
\begin{eqnarray}
{\cal R}_\eta[|n\rangle \langle  n+d|](x,\phi)=
e^{id(\phi+\frac{\pi}{2})}\sqrt{\frac{n!}{(n+d)!}}\int_\rr
dk\,|k| e^{\frac{1-2\eta}{2\eta}k^2-i2kx} k^d L_n^d(k^2)\:,
\end{eqnarray}
where $L_n^d(x)$ denotes the generalized Laguerre polynomials. Notice
that the estimator is bounded only for $\eta >1/2$, and below
the method would give unbounded statistical errors.
\par
The name quantum tomography comes from the first proposal of using homodyne
data for state reconstruction. For a single mode a relevant property of the 
Wigner function $W[\varrho](\alpha)$ is expressed by the following formula 
\begin{eqnarray}
p(x,\phi) \equiv {}_\phi\langle x | \varrho |x\rangle_\phi 
= \int_\rr \frac{dy}\pi
W[\varrho]\left((x+iy)e^{i\phi}\right)
\:,\label{margW}
\end{eqnarray}
which says that the marginal probability obtained from the 
Wigner function integrating over a generic direction in the complex 
plane coincides with the distribution of a field quadrature. 
In conventional medical tomography,
one collects data in the form of marginal distributions of the mass
function $m(x,y)$. In the complex plane the marginal $r(x,\phi)$ is a
projection of the complex function $m(\alpha)\equiv m(x,y)$ on the 
direction indicated by the angle $\phi\in[0,\pi]$, namely
\begin{eqnarray}
r(x,\phi)= \int_\rr \frac{dy}\pi\,
m\left((x+iy)e^{i\phi}\right).
\;\label{defmarginale}
\end{eqnarray}
The collection of marginals for different $\phi$ is called ``Radon
transform''. The tomographic reconstruction essentially consists in the
inversion of the Radon transform (\ref{defmarginale}), in order to
recover the mass function $m(x,y)$ from the marginals
$r(x,\phi)$. 
Thus, by applying the same procedure used in medical imaging 
Vogel and Risken \cite{vogel} proposed a method to recover the
Wigner function {\it via} an inverse Radon transform from the
quadrature probability distributions $p(x,\phi)$, namely
\begin{eqnarray} 
W(x,y)= \int_0^\pi \frac{d\phi}\pi \int_\rr dx'
\; p(x',\phi)  \int_\rr \frac{dk}4\;|k|
\;e^{ik(x'-x\cos\phi-y\sin\phi)}
\;\label{inverseradon}.
\end{eqnarray}
In this way the Wigner function, and in turn any quantity of interest, 
would have been reconstructed by the {\em tomography} of the Wigner 
obtained through homodyne detection. However, this first method is 
unreliable for the reconstruction
of unknown quantum states, since there is an intrinsic unavoidable
systematic error. In fact the integral over $k$ in
 (\ref{inverseradon}) is unbounded. In order to use the inverse
Radon transform, one would need the analytical form of the marginal
distribution of the quadrature $p(x,\phi)$. This can be obtained by
collecting the experimental data into histograms and splining these
histograms. This is not an unbiased procedure since the degree of
splining, the width of the histogram bins and the number of different
phases on which the experimental data should be collected are
arbitrary parameters and introduce systematic errors whose effects
cannot be easily controlled.  For example, the effect of using high
degrees of splining is the wash--out of the quantum features of the
state, and, {\it vice-versa}, the effect of low degrees of splining is
to create negative bias for the probabilities in the reconstruction
(see Refs.~\cite{rev,LNP} for details).  On the other hand, the procedure 
outlined above allows the reconstruction of  the mean values of
arbitrary operators directly from the data, abolishing all the
sources of systematic errors. Only statistical errors are present, and
they can be reduced arbitrarily by collecting more experimental data.
\section{Two-mode entangled measurements}
\index{entangled measurements}
\label{s:two}
In this Section we describe in some details three different  
schemes achieving the joint measurement of the real and the 
imaginary part of the complex normal operators $Z_\pm=a \pm b^\dag$, $a$ 
and $b$ being two modes of the field. The POVMs (actually spectral measures
since $Z_\pm$ are normal) of this class of 
detectors are entangled, {\em i.e.} consist of projectors over 
a set of maximally entangled states, and thus represent the 
generalization to CV systems of the so-called {\em Bell measurement}.
Detection of $Z$ has been realized in different contexts, {\em e.g.} the 
double-homodyne scheme has been employed in the experimental demonstration 
of CV quantum teleportation \cite{furu}.
\par
In the next three Sections we address double (eight-port) homodyne,
heterodyne, and six-port homodyne respectively, whereas in
Section \ref{s:2pom} the common two-mode POVM is evaluated. In
Section \ref{s:2pom} we also derive the single mode POVMs corresponding to
situations in which the quantum state of one of the mode is known and used
as a probe for the other one. 
\subsection{Double-homodyne detector}\label{s:eig}
\begin{floatingfigure}[r]{.42\textwidth}
\vspace{-.3cm}
\begin{center}
\includegraphics[width=.42\textwidth]{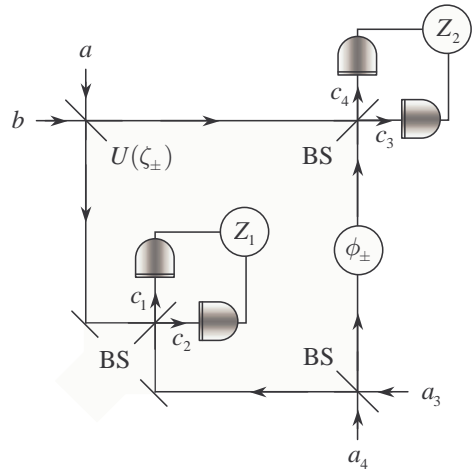}
\end{center}
\vspace{-.8cm}
\caption{Schematic diagram of an eight-port homodyne detector.}
\label{f:eig}
\end{floatingfigure} 
Double homodyne, also called eight-port homodyne, detector is known 
for a long time for the joint determination of phase and amplitude 
of the field in microwave domain, and it was subsequently 
introduced in the optical domain \cite{wa3}. 
\par
\index{homodyne detection!eight-port}
\index{two-mode measurements!eight-port homodyne detection}
A schematic diagram of the experimental setup is reported in
Fig.~\ref{f:eig}.
There are four input modes, which are denoted by $a$, $b$, $a_3$, and $a_4$,
whereas the output modes, {\em i.e.} the modes that are detected, are denoted 
by $c_k$. There are four identical photodetectors whose 
quantum efficiency is given by $\eta$. The {\em noise} modes used to take 
into account inefficiency are denoted by $u_k$. The mixing 
among the modes is obtained through four balanced beam splitters: three of 
them (denoted by BS in Fig.~\ref{f:eig}) have real coupling $\zeta=\pi/4$, 
{\em i.e.} they do not impose any additional phase, whereas the fourth 
has evolution operator [see Eq.~(\ref{ubs})] given by 
\begin{subequations}
\begin{align}
&U(\zeta_{\pm})=\pexp{\zeta_{\pm}a^\dag b - \zeta_{\pm}^{*}a b^\dag}\\
&\zeta_{\pm}=\frac{\pi}{4}\,
\pexp{i\left(\frac{\pi}{2}-\phpm\right)}\,,
\end{align}
\end{subequations}
where $\phpm = \pm\pi/2$ is the phase-shift imposed by a shifter 
(a quarter-wave plate) inserted in one arm.  
We consider $a$ and $b$ 
as signal modes. The mode $a_4$ is
unexcited, whereas $a_3$ is placed in a highly excited coherent state
$|z \rangle$ provided by an intense laser beam, and represents the local
oscillator of the device. The detected photocurrents are $ I_k=c^{\dag}_k
c_k$, which form the eight-port homodyne observables
\begin{eqnarray}
 {Z}_{1} = \frac{ I_2 -  I_1}{2\eta |z|}\,, \qquad 
 {Z}_{2} = \frac{ I_3 -  I_4}{2\eta |z|}
\label{8phot}\:.
\end{eqnarray}
The latter are derived by rescaling the difference photocurrent, each
of them obtained in an homodyne scheme. 
In Eq.~(\ref{8phot}) $\eta$ denotes the quantum efficiency of the
photodetectors whereas $|z|$ is the intensity of the local oscillator.
In order to obtain $ {Z}_1 $  and $ {Z}_2$
in terms of the input modes we first note that the input-output mode
transformation is necessarily linear, as only passive components are
involved in the detection scheme. Thus, we can write
\begin{eqnarray}
c_k =  \sum_{l=1}^{4} M_{kl} a_l\,,
\qquad
{\bmM} = \frac{1}{2}
\left(
\begin{array}{cccc}
1 &  e^{i\thpm} &  -1 &   1 \\
1 &  e^{i\thpm} & 1 &  -1 \\
-e^{-i\thpm} & 1 &  e^{i\phpm} &  e^{i\phpm} \\
e^{-i\thpm} & -1 & e^{i\phpm} & e^{i\phpm}
\end{array}
\right)
\label{linear}\:,
\end{eqnarray}
where $a_1=a$, $a_2=b$, $\thpm=\frac{\pi}{2}-\phpm$, and the transformation
matrix $\bmM$ can be computed starting from the corresponding
transformations for the beam splitters and the phase shifter.
Eq.~(\ref{linear}) together with the equivalent scheme for the inefficient
detection leads to the following expression for the output modes, namely
\begin{eqnarray}
c_k =  \sqrt{\eta} \,\sum_{l=1}^{4} M_{kl} a_l + \sqrt{1-\eta}\: u_k
\label{output}\:.
\end{eqnarray}
Upon inserting Eqs.~(\ref{output}) in Eq.~(\ref{8phot}), and by 
considering the limit of highly excited local oscillator, we obtain the
two photocurrents in terms of the input modes. If we set the phase shifter
at $\phi_{+}$ and tune the fourth beam splitter accordingly we have
\begin{subequations}
\label{zphot+}
\begin{align}
 {Z}_{1\eta+}&=
 q_{a} +  q_{b}
+ \sqrt{\frac{1-\eta}{\eta}} \left[ q_{u_1} - q_{u_2}\right] 
+ O(|z|^{-1})\,,\\
{Z}_{2\eta+}&=
p_{a} - p_b  + \sqrt{\frac{1-\eta}{\eta}} \left[ p_{u_4}-  p_{u_3}
\right] + O(|z|^{-1})\:,
\end{align}
\end{subequations}
while if we choose $\phi_{-}$ we obtain
\begin{subequations}
\label{zphot-}
\begin{align}
 {Z}_{1\eta-}&=
 q_{a} -  q_{b}
+ \sqrt{\frac{1-\eta}{\eta}} \left[ q_{u_1} + q_{u_2}\right] 
+ O(|z|^{-1})\,,\\
{Z}_{2\eta-}&=
p_{a} + p_b  + \sqrt{\frac{1-\eta}{\eta}} \left[ p_{u_4} + p_{u_3}
\right] + O(|z|^{-1})\:,
\end{align}
\end{subequations}
where $q_k$ and $p_k$ in Eqs.~(\ref{zphot-}) denotes quadratures 
of the different modes for specific phases as following (we assume
$\kappa_1=1$)
\begin{eqnarray}
q \equiv x_0 = \frac12 (a^\dag + a )\,, \qquad
p \equiv x_{\pi/2}= \frac{1}{2i}(a-a^\dag) 
\label{specquad}\;.
\end{eqnarray}
Using Eq.~(\ref{specquad}) we may write the complex photocurrent 
${Z}={Z}_1 +i{Z}_2$ as follows
\begin{eqnarray}
{Z_-}=a - b ^{\dag}\, \quad \mbox{or}\quad
{Z_+}=a + b ^{\dag}
\label{8zed}\:,
\end{eqnarray}
whereas, for non unit quantum efficiency, it becomes a Gaussian convolution
of Eq.~(\ref{8zed}), as we will discuss in detail in Section \ref{s:2pom}.
\par
It is worth noticing here that the mode transformation defined by
Eq.~(\ref{linear}) is distinctive for a canonical
$4\times 4$-port linear coupler as defined in Refs.~\cite{igg}. It has
been rigorously shown \cite{zei} that a $N\times N$-port linear coupler
can always be realized in terms of a number of beam splitters and
phase-shifters. However, this implementation is, in general,  not
unique. The interest of eight-port homodyne scheme 
lies in the fact it provides the minimal scheme for realizing a  
$4\times 4$-multiport.
\subsection{Heterodyne detector}\label{s:het}
\index{heterodyne detection}
\index{two-mode measurements!heterodyne detection}
Heterodyne detection scheme is known for a long time in radiophysics 
and it has been subsequently introduced in the domain of optics 
\cite{ysh}. The term ``heterodyne'' comes from the fact 
that the involved modes are excited on different frequencies.
\par\noindent
\begin{figure}[h]\setlength{\unitlength}{0.8cm}
\vspace{-.2cm}
\begin{center}
\includegraphics[width=.7\textwidth]{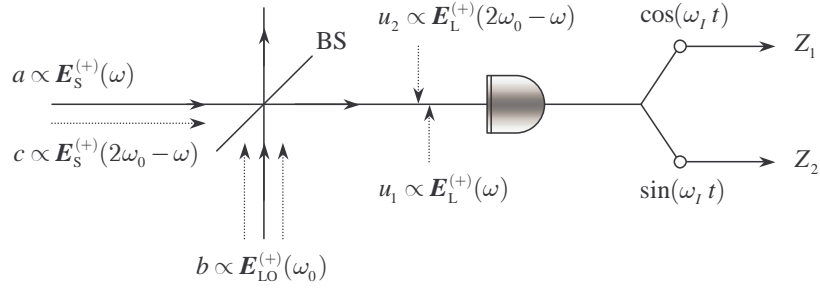}
\end{center}\vspace{-.6cm}
\caption{Schematic diagram of a heterodyne detection.
Relevant modes are pointed out.}\label{f:het}
\end{figure}
\par\noindent
In Fig.~\ref{f:het} we show a schematic diagram of the detector.
We denote by ${\E}_{\rm S}$ the signal field, whereas ${\E}_{\rm LO}$
describes the local oscillator. The field ${\E}_{\rm L}$ accounts for the
losses due to inefficient photodetection.
The input signal is excited in a single mode (say $a$) at the frequency
$\omega$, as well as the local oscillator which is excited at a mode
at the frequency $\omega_0$. This local oscillator mode is placed in
a strong coherent state $|z \rangle$ by means of an intense laser beam.
The beam splitter has a transmissivity given by $\tau$, whereas the
photodetectors shows quantum efficiency $\eta$.
The heterodyne output photocurrents are given by the real ${Z}_1$
and the imaginary ${Z}_2$ part of the complex photocurrent
${Z}$. The latter is obtained after the rescaling of the output
photocurrent $ I$, which is  measured at the intermediate frequency
$\omega_I = \omega -\omega_0$.
By Fourier transform of Eq.~(\ref{intensity}) we have
\begin{eqnarray}
 I (\omega_I )= \int_\rr d\omega ' \;
 {\E}^{(-)}_{\rm O} (\omega ' + \omega_I  ) \:{\E}^{(+)}_{\rm O} ( \omega ')
\;,\label{hetero1}
\end{eqnarray}
${\E}^{(\pm)}_{\rm O}$ being the positive and the negative part of the output
field. In terms of the input fields Eq.~(\ref{hetero1}) can be written as
\begin{multline}
 I (\omega_I ) =  \int_\rr \!\!\!d\omega '
\Big[\sqrt{\eta\tau} {\E}^{(-)}_{\rm S} (\omega ' + \omega_I  ) 
+ \sqrt{\eta (1-\tau)}{\E}^{(-)}_{\rm LO} (\omega ' + \omega_I  ) +
\sqrt{1-\eta}{\E}^{(-)}_{\rm L} (\omega ' + \omega_I  ) \Big] \\
\times \Big[\sqrt{\eta\tau} {\E}^{(+)}_{\rm S} (\omega ' ) +
\sqrt{\eta (1-\tau)}{\E}^{(+)}_{\rm LO} (\omega ' ) +
\sqrt{1-\eta}{\E}^{(+)}_{\rm L} (\omega ' ) \Big] 
\;.\label{hetero2}
\end{multline}
Heterodyne photocurrent is obtained by the following rescaling
\begin{eqnarray}
{Z} =\lim_{\substack{\tau \rightarrow 1\\ |z| \rightarrow \infty}}
\frac{ I (\omega_I )}{ |z| \eta \sqrt{\tau(1-\tau)}} \qquad \;
\mbox{(with $|z| \sqrt{1-\tau}$ constant)}
\;. \label{hetero3}
\end{eqnarray}
In practice, this definition corresponds to have a very intense local
oscillator, which is allowed only for a little mixing with the signal mode 
\cite{displa}. In this limit only terms containing the local oscillator
field $\E^{(\pm )}_{\rm LO} (\omega_0 )$ at the frequency $\omega_0$ can
survive in Eq.~(\ref{hetero2}), so that we have
\begin{eqnarray}
{Z}_{\eta-} =  {Z}_{1\eta-} + i {Z}_{2\eta-}
\;,\label{hetero4}
\end{eqnarray}
where
\begin{subequations}
\label{hetero5}
\begin{align}
 {Z}_{1\eta+}&=
 q_{a} +  q_{c}
+ \sqrt{\frac{1-\eta}{\eta}} \left[ q_{u_1} - q_{u_2}\right] 
+ O(|z|^{-1})\,,\\
{Z}_{2\eta+}&=
p_{a} - p_c  + \sqrt{\frac{1-\eta}{\eta}} \left[ p_{u_1}-  p_{u_2}
\right] + O(|z|^{-1})\:.
\end{align}
\end{subequations}
In writing Eq.~(\ref{hetero5}) we have substituted
\begin{eqnarray}
c \leftarrow {\E}^{(+)}_{\rm S} (2\omega_0-\omega )\,, \qquad 
u_1 \leftarrow {\E}^{(+)}_{\rm L} (\omega )\,, \qquad
u_2 \leftarrow {\E}^{(+)}_{\rm L} (2\omega_0-\omega )\,, \label{hetero6}
\end{eqnarray}
for the relevant modes involved.
Since $u_1$ and $u_2$ are not excited, they play the role of noise modes
accounting for the quantum efficiency of the photodetector. The expression 
(\ref{hetero5}) for the heterodyne photocurrents is thus equivalent to that 
of Eq.~(\ref{zphot+}) for the eight-port homodyne scheme. The full 
equivalence of the two detection  schemes has been thus proved.
Also for heterodyne detection, a simple rearrangements of phase-shifts
provides the measurement of the complex operators $Z_+$ instead of 
$Z_-$.
\subsection{Six-port homodyne detector}\label{s:tri}
\begin{figure}[h]
\vspace{-.2cm}
\begin{center}
\includegraphics[width=.8\textwidth]{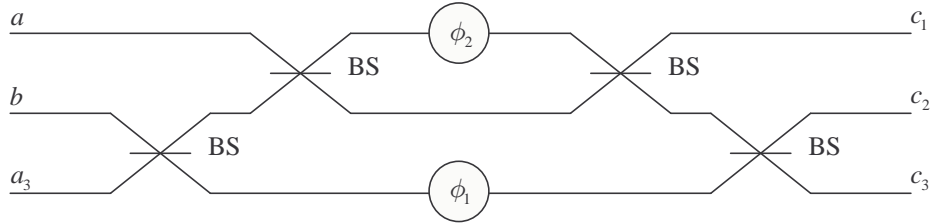}
\end{center}
\vspace{-.6cm}
\caption{Realization of a triple coupler in terms of 50/50 beam splitters
(BS) and phase shifters ``$\phi$''. In order to obtain a symmetric coupler
the following values has to be chosen: $\phi_1=\arccos (1/3)$ and
$\phi_2 = \phi_1 / 2$.}\label{f:TTT}
\end{figure}

\index{two-mode measurements!six-port homodyne detection}
\index{homodyne detection!six-port}
A linear, symmetric three-port optical coupler is a straightforward
generalization of the customary lossless symmetric beam splitter.
The three input modes $a_k$, $k=1,2,3$, are combined to form 3 output
modes $c_k$, $k=1,2,3$. In analogy to lossless beam splitters, which
are described by unitary 2$\times$2 matrices~\cite{cam},
any lossless triple coupler is characterized by a unitary
$3\times 3$ matrix~\cite{zap}. For the symmetric case we have the form
\begin{eqnarray}
{\bmT} = \frac{1}{\sqrt{3}}\left(\begin{array}{ccc}
1 & 1 & 1 \\ 1 & \xi &
\xi^* \\1 & \xi^*
& \xi\end{array}\right)\label{T3M}\;,
\end{eqnarray}
where $\xi=\pexp{i\frac{2}{3}\pi}$ and each matrix element $T_{hk}$
represents the transmission amplitude from the $k$-th input port to the
$h$-th output port, namely $c_h = \sum_{k=1}^3 T_{hk} a_k$. Such devices
have already been implemented in single-mode optical fiber technology and
commercial triple coupler are available \cite{she}.  Any triple coupler can
be also implemented by discrete optical components using symmetric beam
splitters and phase shifters only \cite{zap}. As it has already mentioned
in Section \ref{s:eig}, this is due to the fact that that any unitary
$m$-dimensional matrix can be factorized into a sequence of 2-dimensional
transformations plus phase-shifts \cite {zei}.  Moreover, this
decomposition is not, in general, unique. In Fig.~\ref{f:TTT} we sketch
a possible implementation of a triple coupler where the input
modes are $a_1=a$, $a_2=b$, and $a_3$.
Experimental realizations of triple couplers  has been reported for both cases,
the passive elements case and the optical fiber one \cite{zap,3zz}.
\par
\begin{figure}[t]
\vspace{-.2cm}
\begin{center}
\includegraphics[width=.6\textwidth]{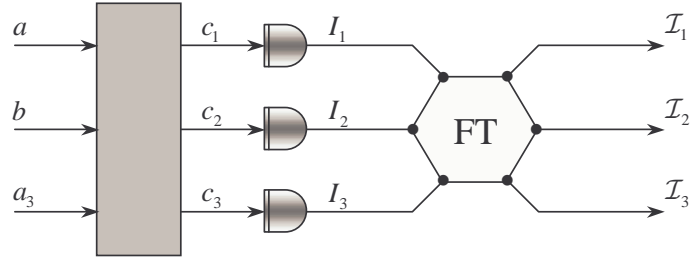}
\end{center} \vspace{-.5cm}
\caption{Outline of triple coupler homodyne detectors: the hexagonal box
symbolizes the electronically performed Fourier transform (FT).}\label{f:THD}
\end{figure}

Let us now consider the measurement scheme of Fig.~\ref{f:THD} \cite{chi}.
The three input modes are mixed by a triple coupler and the resulting output
modes are subsequently detected by three identical photodetectors. The
measured photocurrents are proportional to $ I_n$, $n=1,2,3$, given by
\begin{eqnarray}
 I_n = c^{\dag}_n c_n = \frac{1}{3} \sum_{k,l=1}^3
\exp\left\{i\theta_n (l-k)\right\} a_k^{\dag} a_l\;, \qquad
\theta_n=\frac{2\pi}{3}(n-1)
\label{pht}\;,
\end{eqnarray}
where $a_1 = a$ and $a_2 = b$.
After photodetection a Fourier transform (FT) on the photocurrents is 
performed
\begin{eqnarray}
 {\cal I}_s \equiv {\rm FT}( I_1, I_2, I_3) 
= \frac{1}{\sqrt{3}} \sum_{n=1}^3  I_n
\exp\left\{-i\theta_n (s-1)\right\}\;  \qquad (s=1,2,3) \, .
\label{FT1}
\end{eqnarray}
\par
This procedure is a straightforward generalization of the customary
two-mode balanced homodyning technique. In that case, in fact, the sum
and the difference of the two output photocurrents are considered, which
actually represent the Fourier transform in a two-dimensional space.
By means of the identity
\begin{eqnarray}
\delta_3 (s-1)=\frac{1}{3}\sum_{n=1}^3\exp\left\{i\frac{2\pi}{3}n(s-1)\right\}
\label{FT2}\;,
\end{eqnarray}
for the periodic (modulus 3) Kronecker delta  $\delta_3 $, we  
obtain our final expressions for the Fourier transformed photocurrents:
\begin{equation}
\begin{array}{c}
{\cal I}_1 = {\displaystyle \frac{1}{\sqrt{3}}} \left(
a^{\dag} a + b^{\dag} b + a^{\dag}_3 a_3\right)\,,\\ [2ex]
{\cal I}_2 = {\displaystyle \frac{1}{\sqrt{3}}} \left(
a^{\dag} b + b^{\dag} a_3 + a^{\dag}_3 a\right)\,, \qquad 
{\cal I}_3 = {\displaystyle \frac{1}{\sqrt{3}}} 
\left( a^{\dag} a_3 +b^{\dag} a + a^{\dag}_3 b
\right)\,.
\end{array}
\label{FT3}
\end{equation}
$ {\cal I}_1 $ gives no relevant information as it is insensitive to the phase
of the signal field, whereas $ {\cal I}_2 $ and $ {\cal I}_3$
are hermitian conjugates of each other and contain the relevant information
in their real and imaginary part. 
In the following let us assume $a$ and $b$ as the signal modes and $a_3$
fed by a highly excited coherent state $| z \rangle$ representing 
the local oscillator. For large $|z|$ the output photocurrents are intense
enough to be easily detected. They can be combined to give the reduced
photocurrents
\begin{subequations}
\label{THP}
\begin{align}
 {Z}_{1+} &= \sqrt{3}\,\frac{ {\cal I}_2 +
 {\cal I}_3 }{2 |z|} =  q_a + q_b
+ O(|z|^{-1})\,, \\
{Z}_{2+} &= \sqrt{3}\,
\frac{ {\cal I}_3 -  {\cal I}_3 }{2 i|z|}= p_a - p_b 
+ O(|z|^{-1}) \;,
\end{align}
\end{subequations}
which we refer to as the {\em triple homodyne photocurrents}.
Again, the complex photocurrent $ {Z_+}=
 {Z}_{1+}+ i  {Z}_{2+}$ has the form
$ {Z_-}= a + b^{\dag}$,  $a$ and $b$ being two modes 
of the field. 
\par
When accounting for the non unit quantum efficiency $\eta$ of the
photodetectors the output modes are written as
$ c_h = \sqrt{\eta}\,\sum_{k=1}^3 T_{hk}\, a_k
+ \sqrt{1-\eta}\: u_h$, $h=1,2,3$,
so that the reduced photocurrents are now given by
\begin{subequations}
\label{6zphot}
\begin{align}
 {Z}_{1\eta+}&= \sqrt{3}\,\frac{ {\cal I}_2 +
 {\cal I}_3 }{2 \eta|z|} =  q_a + q_b
+ \sqrt{\frac{1-\eta}{\eta}} \left[ q_{u_1} + q_{u_2}\right] 
+ O(|z|^{-1})\,, \\  
 {Z}_{2\eta+}&=
\sqrt{3}\, \frac{ {\cal I}_3 -  {\cal I}_2 }{2 i\eta |z|}=
p_{a} - p_b
+ \sqrt{\frac{1-\eta}{\eta}} \left[p_{u_2} - p_{u_1}
\right] + O(|z|^{-1})\:.
\end{align}
\end{subequations}
\par
When, as it is the case, the modes $u_k$ are placed in the vacuum, the 
six-port photocurrents in Eq.~(\ref{6zphot}) leads to the same statistics of 
the eight-port photocurrents in Eq.~(\ref{zphot+}). Indeed, they describe 
different devices leading to the same amount of information on the signal 
modes. The measurements of $Z_-$ can also be achieved by six-port
homodyne by a suitable choice of the phase-shifts among the modes.
\subsection{Output statistics from a two-photocurrent device}\label{s:2pom}
Although the two pairs of single-mode quadratures 
$[q_k,p_k]= i/2$, where $k=a,b$ are two modes of the 
field, do not commute, the sum and difference quadratures 
do, and therefore can be measured in a single experiment.
Indeed, the three detection schemes analyzed in this Section provide the 
joint measurement of the operators $q_a+q_b$ and $p_a-p_b$, or 
$q_a-q_b$ and $p_a+p_b$. In turn, the two cases corresponds to 
the measurement of the real and the imaginary 
part of the complex photocurrents $Z_\pm=a\pm b^\dag$ respectively. 
In both cases we have that $[Z_\pm,Z_\pm^\dag]=0=[\Re\hbox{e}[Z_\pm],
\Im\hbox{m}[Z_\pm]]$,
{\em i.e.} $Z_\pm$ are normal operators, and therefore the 
spectral theorem holds
$$
Z_\pm=\int_{\mathbb C} d^2z \: z \: 
\kket{z}_\pm {}_\pm\bbra{z}\:,
$$ where $\kket{z}_\pm$ with $z\in 
{\mathbb C}$ are orthogonal eigenstates  of $Z_\pm$, respectively. 
\par
Let us first consider $Z_-=a-b^\dag$. Using the matrix notation introduced
in Section \ref{s:MatNot} we have that 
\begin{eqnarray}
\kket{z}_-\equiv \frac{1}{\sqrt{\pi}}\:\kket{D(z)}= 
\frac{1}{\sqrt{\pi}}\:D(z) \otimes \iid\, \kket{\ii}  
=\frac{1}{\sqrt{\pi}}\: \iid \otimes D (-z^*) \kket{\ii}
\label{Zeigen}\;,
\end{eqnarray}
where $D(z)$ is the displacement operator and $\kket{\ii}=\sum_n
|n\rangle\otimes |n\rangle$. In fact
\begin{align}
Z_- \kket{z}_- & =\frac{1}{\sqrt{\pi}}
\:D(z)D^\dag(z)(a-b^\dag)D(z)\kket{\ii} \nonumber \\ 
& =\frac{1}{\sqrt{\pi}}\:D(z)(a+z-b^\dag)\kket{\ii} =
\frac{1}{\sqrt{\pi}}\:D(z)\:z\kket{\ii}=z\kket{z}_-\:,
\end{align}
where we have used the fact that 
$a\otimes \iid\, \kket{\ii}=\iid\otimes b^\dag \kket{\ii}$.
Orthogonality of $\kket{z}_-$'s follows from that of displacement
operators 
\begin{eqnarray}
{}_-\bbra{w}z\rangle\rangle_- = \frac1{\pi}\hbox{Tr}\left[D^\dag(w)D(z)\right]
=\delta^{(2)}(z-w)
\label{ortZ}\;.
\end{eqnarray}
Notice that the eigenstates of the complex photocurrent $Z_+=a+b^\dag$ may be
analogously written as 
\begin{eqnarray}
\kket{z}_+ = 
\frac{1}{\sqrt{\pi}}\:D(z) \otimes \iid\, \kket{\JJ} 
\label{Zprime}\;,
\end{eqnarray}
where $[\JJ]_{pq}=(-)^p\: \delta_{pq}$, {\em i.e.} $\kket{\JJ}=\sum_p (-)^p\:
|p\rangle\otimes |p\rangle$.
If $R$ is the density matrix describing the quantum state of modes $a$ 
and $b$ the statistics of the measurement is described by the probability
density 
$$ K_\pm(z)=\hbox{Tr}_{ab} \left[R\: E_\pm(z)\right]\,,$$
with $E_\pm(z)=\kket{z}_\pm{}_\pm\bbra{z}$ denoting the overall 
POVM of the detector.
\par
Let us now consider the effects of nonunit quantum efficiency. 
The measured photocurrents are given in Eqs.~(\ref{zphot+}), 
(\ref{zphot-}), (\ref{hetero5}) or (\ref{6zphot}); using 
(\ref{pomMGF2}) the POVM
$\Pi_\eta(z)$ is obtained upon tracing over the vacuum modes 
used to simulate losses: for either $\Pi_{\eta\pm}(z)$ we have
\begin{eqnarray}
\Pi_{\eta} (z) &=& \int_\cc \frac{d^2\gamma}{\pi^2} {}_{u_1u_2}\langle\langle
00|\pexp{\gamma (Z_\eta^\dag -z^*) - \gamma^* (Z_\eta -z)}|00\rangle\rangle_{u_1u_2}
\nonumber \\ &=& 
\int_\cc \frac{d^2\gamma}{\pi^2}\:
\pexp{\gamma (Z_+ -z^*) - \gamma^* (Z_-
-z)}\:\pexp{-\frac{1-\eta}{\eta}|\gamma|^2} 
\nonumber \\
&=& \frac{\eta}{\pi(1-\eta)}\pexp{-\frac{\eta}{1-\eta}|Z-z|^2}
\nonumber \\ &=&
\int_\cc \frac{d^2 w}{\pi \Delta^2_\eta}
\pexp{-\frac{|z-w|^2}{\Delta^2_\eta}}\: E(z)
\label{etaPovm2m}\;,
\end{eqnarray}
where 
\begin{eqnarray}
\Delta^2_\eta = \frac{1-\eta}{\eta}\:.
\end{eqnarray}
The characteristic function of the POVM, for unit quantum  efficiency, 
is given by
\begin{align}
\chi[E_-(z)](\lambda_1,\lambda_2) & =  
{}_-\bbra{z}D(\lambda_1)\otimes D(\lambda_2)\kket{z}_-
\nonumber \\ & =\frac{1}{\pi}
e^{\lambda_1 z^* - \lambda_1^* z}\: 
\bbra{\ii}D(\lambda_1)\otimes D(\lambda_2)\kket{\ii}
\nonumber \\ & = \frac{1}{\pi}
e^{\lambda_1 z^* - \lambda_1^* z} \:
\langle\langle\ii\kket{D(\lambda_1)D^{\sT}(\lambda_2)}
\nonumber \\ & = \frac{1}{\pi}
e^{\lambda_1 z^* - \lambda_1^* z}\: \hbox{Tr}[D(\lambda_1)D^{\sT}(\lambda_2)]
\nonumber \\ & = 
  e^{\lambda_1 z^* - \lambda_1^* z}\: \delta^{(2)} (\lambda_1 - \lambda_2^*)
\label{chiEz-}\:.
\end{align}
Analogously, 
\begin{align}
\chi[E_+(z)](\lambda_1,\lambda_2) & =  
{}_+\bbra{z}D(\lambda_1)\otimes D(\lambda_2)\kket{z}_+
\nonumber \\ & =\frac{1}{\pi}
e^{\lambda_1 z^* - \lambda_1^* z}\: 
\bbra{\jj}D(\lambda_1)\otimes D(\lambda_2)\kket{\jj}
\nonumber \\ & = \frac{1}{\pi}
e^{\lambda_1 z^* - \lambda_1^* z} \:
\langle\langle\jj\kket{D(\lambda_1)\:\jj\: D^{\sT}(\lambda_2)}
\nonumber \\ & = \frac{1}{\pi}
e^{\lambda_1 z^* - \lambda_1^* z}\: \hbox{Tr}[\gr{\Pi}\:
D(\lambda_1)\:\gr{\Pi}\:D^{\sT}(\lambda_2)]
\nonumber \\ & = 
  e^{\lambda_1 z^* - \lambda_1^* z}\: \delta^{(2)} (\lambda_1 + \lambda_2^*)
\label{chiEz+}\:,
\end{align}
where $\gr{\Pi}=\otimes_k (-)^{a^\dag_k a_k}\equiv (-)^{\sum_k a^\dag_k
a_k}$ is the multimode parity operator.
Using (\ref{chiEz-}) and (\ref{chiEz+}) we have
\begin{eqnarray}
W[E_\pm(z)](\gr{X}) = \frac{1}{\pi^2} \delta(x_1\pm x_2-x_z)
\delta(y_1\mp y_2-y_z)\:,
\end{eqnarray}
where $z=x_z+iy_z$ and we used the Cartesian form of the Wigner function 
for the sake of simplicity.
\par
\begin{figure}[h]
\begin{center}
\vspace{-0cm}
\includegraphics[width=0.6\textwidth]{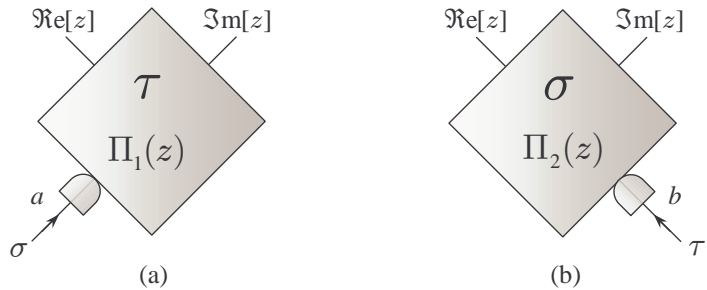}
\end{center}
\vspace{-0.7cm}
\caption{
(a): measurement of the two-mode POVM $E(z)$ viewed as a single-mode
measurement of the $\tau$-dependent POVM (\ref{2pom:Pi1})
on mode $a$; (b): the same for the POVM (\ref{2pom:Pi2}) 
on mode $b$.
\label{f:Pi:1:2}}
\end{figure}

For nonunit quantum efficiency $W[\Pi_{\eta\pm}](\gr{X})$
is given, according to Eq.~(\ref{etaPovm2m}), by a Gaussian convolution 
of $W[E_\pm(z)](\gr{X})$, {\em i.e.} 
\begin{eqnarray}
W[\Pi_{\eta\pm}](\gr{X}) = \frac{1}{2\pi\Delta^2_\eta}
\pexp{-\frac{(x_1\pm x_2-x_z)^2}{2\Delta_\eta^2}-
\frac{(y_1\mp y_2-y_z)^2}{2\Delta_\eta^2}} \:.
\end{eqnarray}
Let us now consider a situation in which $R$ is
factorized, namely $R=\sigma\otimes\tau$.
$\sigma$ is the state under investigation and $\tau$ a known reference state
usually referred to as the {\em probe} of the detector
(see Fig.~\ref{f:Pi:1:2}). 
The statistics of the outcomes, for unit quantum efficiency, may be described 
as follows
\begin{align} 
K(z)&=\hbox{Tr}_{ab} \left[\sigma\otimes\tau\:
\Pi(z)\right] = \frac1{\pi}
\hbox{Tr}_{a} \Big[\sigma\,
\hbox{Tr}_b\big[\iid\otimes\tau\: \kket{D(z)}\bbra{D(z)}\big]
\Big]
\nonumber \\
&= \frac1{\pi} \hbox{Tr}_{a} \Big[\sigma\,
\hbox{Tr}_b\big[
\kket{D(z)\tau^{\sT}}\bbra{D(z)}\big]
\Big]=
\hbox{Tr}_{a} \left[\sigma \: \Pi_1(z)
\right]\,, \label{povm}\; 
\end{align}
with 
\begin{equation}
\Pi_1(z) = \frac{1}{\pi} D(z)\, \tau^{\sT} D^\dag (z)\:,
\label{2pom:Pi1}
\end{equation}
which is the single mode POVM of the detector viewed as a measurement 
of the first mode probed by the second mode \cite{bpl}.
If $\tau=\ket{0}\bra{0}$ is the vacuum, then the POVM $\Pi_1(z)$ is the set 
of (nonorthogonal) projectors $| z \rangle\langle z |$ over coherent states,
and setup measures the $Q$-function
$Q(z)=\pi^{-1}\bra{z}\sigma\ket{z}$ of the state $\sigma$.
Notice that, as required for a POVM, $\Pi(z)$ is selfadjoint and normalized.
The first property follows from the fact that $\tau$ itself is selfadjoint.
In fact, $\tau^\dag=\tau$ implies that $\tau^{\sT}=\tau^*$ and therefore
$\Pi_1^\dag(z)=\pi^{-1}D(z)\,\tau^* D^\dag(z)=
\pi^{-1}D(z)\,\tau^{\sT} D^\dag (z)=\Pi_1(z)$.
Normalization follows from completeness of the set of displacement 
operators, and in particular from Eq.~(\ref{s2}).
The role of signal and probe may be exchanged, and the statistics can be
written as follows
\begin{align}
K(z)&= \frac1{\pi} \hbox{Tr}_{b} \Big[\tau\,
\hbox{Tr}_a\big[\sigma \otimes {\mathbb I}\: \kket{D(z)}\bbra{D(z)}\big]
\Big] \nonumber \\
&= \frac1{\pi} \hbox{Tr}_{b} \Big[\tau\, \hbox{Tr}_a\big[
\kket{\sigma D(z)}\bbra{D(z)}\big]\Big] = 
\hbox{Tr}_{b} \left[\tau \: \Pi_2 (z)\right]\,,
\end{align}
where the POVM acting on the mode $b$ is given by
\begin{equation}
\Pi_2 (z) = \frac{1}{\pi} D(-z^*)\, \sigma^{\sT} D^\dag (-z^*) = 
\frac{1}{\pi} D^{\sT}(z)\, \sigma^{\sT} D^* (z)\:.
\label{2pom:Pi2}
\end{equation}
\par\noindent
The action of $\Pi_1(z)$ and $\Pi_2(z)$ is depicted in
Fig.~\ref{f:Pi:1:2} (a) and (b) respectively.
The Wigner functions of the POVMs $\Pi_k(z)$, $k=1,2$ are
given by
\begin{align}
W[\Pi_1(z)](\alpha) &= \frac{1}{\pi} W[\tau^{\sT}](\alpha-z) = \frac1\pi
W[\tau](\alpha^*-z^*)\,, \nonumber \\
W[\Pi_2(z)](\alpha) &= \frac{1}{\pi} W[\sigma^{\sT}](\alpha+z^*) = \frac1\pi
W[\tau](\alpha^*+z)\,,
\end{align}
where we have used (\ref{Wdispla}) and the fact that transposition corresponds
to mirror reflection in the phase space [see Eq. (\ref{c4:TransPhSpace})]. 
For nonunit quantum efficiency the POVMs becomes Gaussian convolutions of 
the ideal POVM (with variance equal to $\Delta_\eta^2$). The Wigner functions 
modify accordingly.
   
\chapter{Nonlocality in continuous variable systems}\label{ch:nonloc}
In their famous paper of 1935 \cite{EPR35}, Einstein, Podolsky and
Rosen (EPR) introduced in quantum physics two strictly related
concepts: entanglement\footnote{\footnotesize The word ``entanglement'' was
introduced in this contest by Schr\"odinger in his reply \cite{Sch35} to EPR
paper.} and nonlocality, which afterward generated a longstanding debate on
the completeness of quantum mechanics. These two concepts have become more
and more important in the subsequent decades, as the recent progresses in
quantum information science definitely demonstrated.
\index{nonlocality!EPR}
\par
In Chapter \ref{SepGS}
quantum entanglement for continuous variable (CV) systems has been extensively
analyzed. We also pointed out that the concept of entanglement coincides
with nonlocality only for the simple case of bipartite pure states. As soon
as we deal with bipartite mixed states, entanglement can be found which do
not show properties of nonlocality (while the converse instead is always
true) \cite{wer89}.
\par
This Chapter will be devoted to the issue of
nonlocality for CV systems.
First of all, we recall what the {\em concept of nonlocality} means.
Usually two different notions are subsumed in it: (non-)locality and
(non-)realism. A theory is said to be local if no action at distance,
between two subsystems $A$ and $B$, is contemplated in it. Hence,
quoting Einstein \cite{Ein49}:
\begin{quote}
``The real factual situation of the system $B$ is
independent of what is done with the system $A$, which is spatially
separated from the former.''
\end{quote}
A realistic theory is a
theory able to assign a definite counterpart to every element of
reality and again following \cite{EPR35}:
\begin{quote}
``If, without in any way
disturbing a system, we can predict with certainty (i.e., with
probability equal one) the value of a physical quantity, then there
exist an element of physical reality corresponding to this physical
quantity.''
\end{quote}
A theory which is not a local realistic one is simply
incomplete, according to the spirit of EPR paper.
Let us now apply
this notions to a composite system of two distant particles, described
by the so-called EPR wave function, {\em i.e.} the TWB wave 
function (\ref{twbdef}) in the limit of infinite
squeezing\footnote{\footnotesize Using the expression (\ref{c3:TWBsigma})
for the TWB Wigner function it is immediate to see that the original EPR
state introduced in Ref.~\cite{EPR35} is recovered in the infinite squeezing
limit.}. By measuring, say, the position of one particle the position of
the other one can be predicted with certainty, as follows from the
correlations between the two. If there is no action at distance, this
prediction can be made without in any way disturbing the second particle.
Hence an element of physical reality must be assigned to its position. On
the other hand, the same argument apply to the measurement of momenta.
However, quantum theory precludes the simultaneous assignment of position
and momentum without uncertainty. So EPR conclude that quantum theory is
not complete.
\par
The debate about whether or not quantum mechanics is a local realistic
theory remained in the realm of philosophy, rather of physics, for
many years. The situation drastically changed when Bell proved that
EPR point of view leads to algebraic predictions (the celebrated
Bell's inequalities) that are contradicted by quantum mechanics
\cite{Bel64}. Bell formulated his inequalities in a dichotomized
fashion, suitable for a discrete variable setting rather than for the
original continuous variable one. In particular, Bell followed the
simple and elegant formulation given by Bohm to the EPR gedanken
experiment using spin-$\frac12$ particles. More recently, however, the 
increasing importance of CV systems leads many authors to
explore the nonlocality issue in its original setting, where
dichotomic observables to test Bell's inequalities are not uniquely
determined. The attempts to translate Bell's inequalities to
CV clarified the fact that crucial in a
nonlocality test is the existence of a set of dichotomized bounded
observables used to perform the test itself, from which the so called
{\em Bell operator} is derived. The more debated question has dealt with
the nonlocality of the normalized version of the original EPR state,
{\em i.e.} the TWB state of radiation. Nonlocality of the TWB was not
clear for along time. Using the Wigner function approach, Bell argued that the
original EPR state, and as a consequence the TWB too, does not exhibit
nonlocality because its Wigner function is positive, and therefore
represents a local hidden variable description \cite{Bel87}.
More recently, Banaszek and Wodkiewicz \cite{BW98} showed instead how to
reveal nonlocality of the EPR state through the measurement of
displaced parity operator. Furthermore a subsequent work of Chen {\em et
al.} \cite{CPH+02} showed that TWB's violation of Bell's inequalities may
achieve the maximum value admitted by quantum mechanics upon a
suitable choice of the measured observables. Indeed, the amount of
violation crucially depends on the kind of Bell operator adopted in
the analysis, ranging from no violation to maximal violation for the
same (entangled) quantum state.
\section{Nonlocality tests for continuous variables}\label{s:rev} 
\index{nonlocality!CV systems}
In this Section we recall the inequalities imposed by local realism to the
situations of our interest.  Let us start by focusing our attention on a
bipartite system. Let $ m(\alpha)$ denotes a one-parameter family of
single-system observable quantities with dichotomic spectrum. In the 
following $m(\alpha_1)=\pm 1$ and $m(\alpha_1')=\pm 1$ will denote the 
outcomes of the measurements on the first subsystem and,
similarly, $m(\alpha_2)=\pm 1$ and $m(\alpha_2')=\pm 1$ for the second
subsystem. The essential feature of this measurements is that they are
local, dichotomic and bounded.  The Bell's combination 
\begin{equation}
B_2 \equiv  m(\alpha_1)\otimes   m(\alpha_2)+ 
m(\alpha_1)\otimes   m(\alpha_2') 
+  m(\alpha_1')\otimes  
m(\alpha_2)-  m(\alpha_1')\otimes   m(\alpha_2')\,,
\label{c6:F2}
\end{equation} 
under the assumption of local realism, leads to the well known
Bell-CHSH inequality \cite{CHS+69}:
\begin{align}
{\cal B}_2 \equiv \left|\langle B_2 \rangle\right|= 
|E(\alpha_1,\alpha_2)+E(\alpha_1,\alpha_2')+
E(\alpha_1',\alpha_2)-E(\alpha_1',\alpha_2')| \le 2
\label{c6:BI2}\;,
\end{align} 
where $E(\alpha_1,\alpha_2)=\langle   m(\alpha_1)\otimes 
m(\alpha_2)\rangle$ is the correlation function among 
the measurement results. If we describe the system quantum mechanically,
then we have that
$$
E(\alpha_1,\alpha_2) \equiv {\rm Tr} [R\,   m(\alpha_1)\otimes  
m(\alpha_2)]\,,
$$ 
$R$ being the density matrix of the system under investigation. 
\par
Bipartite entangled pure states violate (\ref{c6:BI2}) for a suitable
choice of the observables $ m(\alpha)$ {\em and} of the values of the
parameters. Bipartite entangled mixed states may or may not violate 
(\ref{c6:BI2}).
Systems which involves only two parties are the simplest setting where
to study violation of local realism in quantum mechanics. A more
complex scenario arises if multipartite systems are considered.
Studying the peculiar quantum features of these systems is worthwhile
in view of their relevance in the development of quantum communication
technology, {\em e.g.}~to manipulate and distribute information in a
quantum communication network \cite{LB00,ATY+03}. Although the study of
multipartite nonlocality has originated without the use of
inequalities \cite{GHS+90}, an approach to derive Bell inequalities
(so called Bell-Klyshko inequalities) has been developed
\cite{Mer90,Kly93} also for these systems and applied to characterize
their entanglement properties \cite{GBP98}. Being originally
developed in the framework of discrete variables, these multiparty
Bell inequalities have found application also in the characterization
of continuous variable systems \cite{LB01,CZ02}. Bell-Klyshko
inequalities \cite{Mer90,Kly93,GBP98} provides a generalization of
inequality (\ref{c6:BI2}) and are based on the
following recursively defined Bell's combination (operator)
\index{Bell inequalities}
\begin{align}
B_n \equiv
\frac12\big[ m(\alpha_n)+ m(\alpha_n')\big]\otimes B_{n-1}+\frac12
\big[ m(\alpha_n)- m(\alpha_n')\big]\otimes B'_{n-1}
\label{c6:FN}\;,
\end{align}
where $B'_n$ denotes the same expression as $B_n$ but with all the
$\alpha_n$ and $\alpha_n'$ exchanged, and $m(\alpha_n)=\pm 1$,
$m(\alpha_n')=\pm 1$ denote the outcomes of the measurements on the $n$-th
party of the system.  Bell-Klyshko inequalities then read:
\begin{align}
\cB_n\equiv|\langle B_n \rangle|\le 2\,.
\label{c6:BIN}
\end{align}
In the case of a three-partite system, local realism assumption
imposes the following inequality from combination (\ref{c6:FN}):
\begin{align}
  {\cal B}_3 \equiv |E(\alpha_1,\alpha_2,\alpha_3')+E(\alpha_1,\alpha_2',
\alpha_3)+E(\alpha_1',\alpha_2,\alpha_3)-E(\alpha_1',\alpha_2',\alpha_3')| \le 2
\label{c6:BI3}\;,
\end{align}
where again $E(\alpha_1,\alpha_2,\alpha_3)$ is the correlation function 
between the measurement results.
\par
Quantum mechanical 
systems can violate inequalities (\ref{c6:BI2}) and(\ref{c6:BI3}) by a maximal
amount given by, respectively, ${\cal B}_2 \le 2\sqrt{2}$ and 
${\cal B}_3 \le 4$. In general $\cB_n^2\le 2^{n+1}$ holds (see, {\em e.g.}, 
Ref.~\cite{GBP98}).
\par
We now briefly review three different strategies to reveal quantum
nonlocality in the framework of continuous variables systems. These
{\em nonlocality tests} are the basis for the analysis the will be 
performed in the remaining of the Chapter. In order to introduce 
the argument, recall
that in the case of a discrete bipartite system, for example a
spin-$\frac12$ two particle system, the local dichotomic bounded
observable usually taken into account is the spin of the particle in a
fixed direction, say ${\bf d}$. Hence the correlation between two
measurements performed over the two particles is   $E({\bf d}_1,{\bf
d}_2)=\langle {\bf d}_1\cdot \boldsymbol\sigma\otimes {\bf d}_2
\cdot\boldsymbol\sigma\rangle$, where the operator $\boldsymbol
\sigma=(\bmsigma_x,\bmsigma_y,\bmsigma_z)$ is decomposed on the Pauli
matrices base and ${\bf d}_1,{\bf d}_2$ are two unit vectors.  The
Bell operator is then given by the expression:
\begin{equation}
B_{2\bmsigma} = 
{\bf d}_1\cdot\bmsigma \otimes {\bf d}_2\cdot \boldsymbol\sigma+ 
{\bf d'}_1\cdot\boldsymbol\sigma\otimes {\bf d}_2\cdot \boldsymbol\sigma+ 
{\bf d}_1\cdot\boldsymbol\sigma\otimes {\bf d'}_2\cdot\boldsymbol\sigma 
- {\bf d'}_1\cdot\boldsymbol\sigma\otimes {\bf d'}_2\cdot\boldsymbol\sigma
\label{c6:BO2Spin}\;.
\end{equation} 
\par
\index{nonlocality!$n$-partite systems}
Consider now a $n$-partite continuous variable system. Following the
original argument by EPR it is quite natural attempting to reveal the
nonlocality of this system trying to infer quadratures of one
subsystem from those of the others. From now on, we will refer to this
procedure as a {\em homodyne nonlocality test} ($H$), as quadrature
measurements of radiation field are performed through homodyne
detection. Here we identify the quadrature $x^k_\vartheta$, relative to
mode $k$, according to the definition (\ref{defquad}). As they are
local but neither bounded nor dichotomic, quadrature observables are
not immediately suitable to perform a nonlocality test based on Bell's
inequalities. The procedure to make them bounded and dichotomic is
quite arbitrary and consist in the assignment of two domains $D_+$ and
$D_-$ to each observable \cite{gil1}. When the result of a
quadrature measurement falls in the domain $D_\pm$ the value $\pm1$ is
associated to it. Usually the choice $D_\pm=\mathbb{R}^\pm$ is
considered, though a choice suitable to the system under investigation
may be preferable. Considering a bipartite system we can introduce the
following quantities
\begin{subequations}
\label{c6:PH}
\begin{align}
&P_{++}(x_\vartheta^1,x_\varphi^2)=\int_{D_+}dx_\vartheta^1
\int_{D_+}dx_\varphi^2 P(x_\vartheta^1,x_\varphi^2) \\ 
&P_{+-}(x_\vartheta^1,x_\varphi^2)=\int_{D_+}dx_\vartheta^1
\int_{D_-}dx_\varphi^2 P(x_\vartheta^1,x_\varphi^2) \\  
&P_{-+}(x_\vartheta^1,x_\varphi^2)=\int_{D_-}dx_\vartheta^1
\int_{D_+}dx_\varphi^2 P(x_\vartheta^1,x_\varphi^2) \\ 
&P_{--}(x_\vartheta^1,x_\varphi^2)=\int_{D_-}dx_\vartheta^1
\int_{D_-}dx_\varphi^2 P(x_\vartheta^1,x_\varphi^2) \;
\end{align}
\end{subequations}
where $P(x_\vartheta^1,x_\varphi^2)$ is the joint probability distribution of
the quadratures $x_\vartheta^1$ and $x_\varphi^2$. We can now identify the
homodyne correlation function $E_\sH(\vartheta,\varphi)$ as
\begin{align}
E_\sH(\vartheta,\varphi)=P_{++}(x_\vartheta^1,x_\varphi^2)+
P_{--}(x_\vartheta^1,x_\varphi^2)-P_{+-}(x_\vartheta^1,x_\varphi^2)
-P_{-+}(x_\vartheta^1,x_\varphi^2)
\label{c6:EH}\;,
\end{align} 
which can be straightforwardly used to construct the Bell combination
$\cB_{2\sH}$ of Eq.~(\ref{c6:BI2}) and to perform the
nonlocality test. The main problem of pursuing such a nonlocality test
is that it is not suitable in case of systems described by a positive
Wigner function, as the TWB state of radiation. Indeed, a positive
Wigner function can be interpreted as a hidden phase-space probability
distribution, preventing violation of Bell-CHSH inequality unless the
measured observables have an unbounded Wigner representation, which is
not the case of the dichotomized quadrature measurement described
above.  In fact $P(x_\vartheta^1,x_\varphi^2)$ can be
determined as a marginal distribution from the Wigner function. 
From Eqs.~(\ref{c6:PH}) and (\ref{c6:EH}) one has
\begin{align}
  E_\sH(\vartheta,\varphi)=\int_{\rr^4} dx_\vartheta^1\, dx_\varphi^2\, dx_{\vartheta
+\frac\pi2}^1\, dx_{\varphi+\frac\pi2}^2\,
{\rm sgn}\big[x_\vartheta^1\,x_\varphi^2\big] 
W(x_\vartheta^1,x_{\vartheta+\frac\pi2}^1,\, x_\varphi^2,x_{\varphi+\frac\pi2}^2)
\label{c6:EHWPos}\;,
\end{align} 
where the integration is performed over the whole phase-space and
without loss of generality we have considered $D_\pm={\mathbb R}^\pm$.
Eq.~(\ref{c6:EHWPos}) itself is indeed a local hidden variable description
of the correlation function, hence obeying inequality (\ref{c6:BI2}).
\par 
\index{nonlocality!Wigner function}
In order to overcome this obstacle different strategies have been
considered by many authors, based essentially on parity measurements.
Banaszek and Wodkiewicz \cite{BW98} have demonstrated the nonlocality of the
TWB considering as local observable on subsystem $k$ the parity
operator on the state displaced by $\alpha_k$ (hence we will refer to
this procedure as a {\em displaced parity} ($DP$) {\em test}), which is
dichotomic and bounded:
\begin{align}
\label{c6:DispParity}
\Pi(\boldsymbol\alpha ) =  \bigotimes_{k=1}^n D_k(\alpha_k)
(-1)^{n_k} D_k^{\dagger}(\alpha_k) .
\end{align}
In the above formula, $\boldsymbol\alpha=(\alpha_1,\ldots,\alpha_n)$, while
$n_k=a_k^{\dagger}a_k$ and $D_k(\alpha_k)$ denote the number operator and
the phase space displacement operator for the subsystem $k$, respectively.
Hence the correlation function reads:
\begin{align}
\label{c6:EDP}
E_{\sDP}(\boldsymbol\alpha) = \langle \Pi(\boldsymbol\alpha) \rangle , 
\end{align}
from which Bell's combinations ${\cal B}_{2\sDP}$ in Eq.~(\ref{c6:BI2})
and ${\cal B}_{3\sDP}$ in Eq.~(\ref{c6:BI3}) can be easily reconstructed
in the cases $n=2,3$. The reason why this procedure would be able to
reveal nonlocality also in case of quantum states characterized by a
positive Wigner function is clear using the following relation
[see \refeq{WdefTr}]:
\begin{align}
\label{c6:WandP}
W(\boldsymbol\alpha) = \left(\frac{2}{\pi}\right)^n 
\langle\Pi(\boldsymbol\alpha)\rangle\:.
\end{align}
Indeed, the analog of Eq.~(\ref{c6:EHWPos}) is:
\begin{align}
E_{\sDP}(\boldsymbol\alpha) = \int_{\cc^n} d^{2n}\boldsymbol\lambda  
\left(\frac{2}{\pi}\right)^n W(\boldsymbol\alpha)\,
\delta^{(2n)} (\boldsymbol\alpha-\boldsymbol\lambda) 
\label{c6:EDPWPos}\; .
\end{align} 
Since the Dirac-$\delta$ distribution is unbounded, then Ineqs.~(\ref{c6:BI2})
and (\ref{c6:BI3}) are no more necessarily valid for ${\cal B}_{2\sDP}$
and ${\cal B}_{3\sDP}$. 
\par
Another strategy, developed by Chen {\em et al.}~\cite{CPH+02}, shares a
similar behavior as the one described above, allowing to reveal
nonlocality for quantum states with positive Wigner function. This
type of nonlocality test will be referred to as {\em Pseudospin} ($PS$)
{\em nonlocality test}. It can be seen as a generalization to CV
systems of the strategy introduced by Gisin and Peres for the
case of discrete variable systems \cite{GP92}, hence, for the case of
a pure bipartite system, it is equivalent to an entanglement test
\cite{JSK+03}.  Let us consider the following set of operators, known
as {\em pseudospins} in view of their commutation relations,   ${\bf
s}_k=(s^k_x,s^k_y,s^k_z)$ acting on the $k$-th subsystem
\begin{subequations}
\label{c6:PS}
\begin{align}
&s^k_z = \sum_{n=0}^\infty\big(|2n+1\rangle_k{}_k\langle 2n+1|
-|2n\rangle_k{}_k\langle 2n|\big)\,, \\
&s^k_x\pm i s^k_y = 2s^k_\pm\,, \\
&{\bf d}_k {\bf s}_k = \cos\vartheta_k\,s^k_z +
\sin\vartheta_k\,(e^{i \varphi_k}s^k_-+e^{-i \varphi_k}s^k_+)\,,
\end{align}
\end{subequations}
where $s^k_-=\sum_{n} |2n\rangle_k{}_k\langle 2n+1|=(s^k_+)^\dagger$
and ${\bf d}_k$ is a unit vector associated to the angles $\vartheta_k$
and $\varphi_k$.  In analogy to the spin-$\frac12$ system mentioned
above and defining the vector ${\bf d}=({\bf d}_1,\ldots,{\bf d}_n)$ the
correlation function is simply given by:
\begin{align}
 E_{\sPS}({\bf d})=\langle \otimes_{k=1}^n {\bf d}_k {\bf s}_k \rangle    
\label{c6:EPS}\; ,
\end{align}  
from which the Bell combinations ${\cal B}_{2\sPS}$ and   ${\cal
B}_{3\sPS}$ are evaluated. Also different representations of the
spin-$\frac12$ algebra have been discussed in the recent literature
\cite{MFF02,PEW04,GKM+04}. In particular in Ref.~\cite{GKM+04} it has 
been pointed out that different representations lead to different
expectation values of the Bell operators. Hence, the violation of Bell
inequality for CV systems turns out to depend,
besides to orientational parameters, also to configurational ones.  
In the following Sections we will also consider the pseudospin operators
${\bf \Pi}_k=(\Pi_x^k,\Pi_y^k,\Pi_z^k)$ taken into account in 
Ref.~\cite{GKM+04}, which have the following Wigner representation 
(for $\kappa_1= 2^{-1/2}$):
\begin{equation}
  \label{c6:WPi}
  \begin{array}{c}
  W[\Pi_x^k](\alpha_k)={\rm sgn}\big[\real{\alpha_k}\big]\,,\qquad
  W[\Pi_y^k](\alpha_k)=-\delta^{(2)}\big(\real{\alpha_k}\big)~{\cal P}
  {\displaystyle \frac{1}{\immag{\alpha_k}}}\,,\\[2ex]
  W[\Pi_z^k](\alpha_k)=-\pi\delta^{(2)}(\alpha_k)\,,
  \end{array}
\end{equation}
where ${\cal P}$ stands for
the ``principal value''. The correlation function obtained using
operators ${\bf \Pi}_k$ will be indicated as ${\cal E}_{\sPS}({\bf d})=
\langle\otimes_{k=1}^n{\bf d}_k{\bf \Pi}_k\rangle$.
\section{Two-mode nonlocality}\label{s:2mNL}
In this Section we will analyze the nonlocality properties of two-mode
states. First we concentrate on the TWB state of radiation, then we
will consider non-Gaussian states and apply to them all the strategies
introduced in the preceding section.
\subsection{Twin-beam state}\label{ss:twbNL}
\index{twin-beam!nonlocality}
As already mentioned, the more debated question concerning nonlocality in
continuous variable systems involved the TWB state, due to its importance
both from an applicative point of view and from a fundamental perspective,
as it is a normalized version of the original EPR state. Since it is not
suitable for homodyne test, the TWB nonlocality will be investigated
exploiting the $DP$ and $PS$ tests.
\subsubsection{Displaced parity test}\label{sss:twbDP}
\index{displaced parity}
Let us follow Ref.~\cite{BW98}. Using \refeq{c3:TWBsigma}
for the Wigner function of the TWB and \refeq{c6:WandP}, it 
is immediate to evaluate the correlation function $E_{\sDP}(\alpha_1,\alpha_2)$ of \refeq{c6:EDP}. In Ref.~\cite{BW98} the following parameterization 
has been considered to construct the Bell combination $\cB_{2\sDP}$
\begin{align}
\alpha_1=\alpha_2=0\,, \qquad \alpha_1'=-\alpha_2'=\sqrt{\cJ} \;.
\label{c6:ParamBWDP}
\end{align}
It follows that
\begin{align}
\label{c6:B_BW}
\cB_{2\sDP}=1+2\exp\{-2\cJ\cosh (2r)\}-\exp\{-4\cJ e^{2r}\}\;.
\end{align}
\begin{figure}[h!]
\vspace{1.6cm}
\setlength{\unitlength}{0.5cm} 
\centerline{%
\begin{picture}(-13,-13) 
\put(0,0){\makebox(0,0)[c]{\includegraphics[width=55mm]{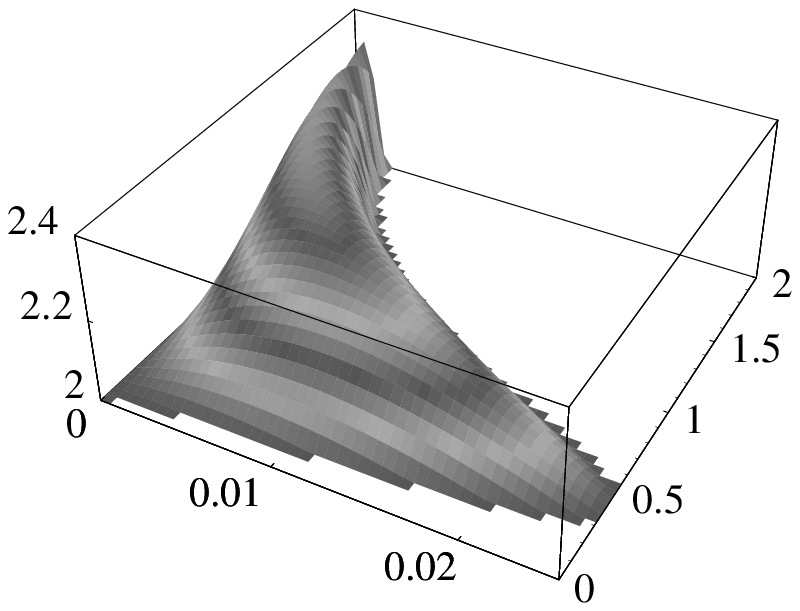}}}
\put(-2,-3.5){${\cal J}$}
\put(5,-2){$r$}
\put(-5,1.9){${\cal B}_{2\sDP}$}
\put(5.5,2.7){(b)}
\end{picture}
\begin{picture}(0,0) 
\put(0,0){\makebox(0,0)[c]{\includegraphics[width=55mm]{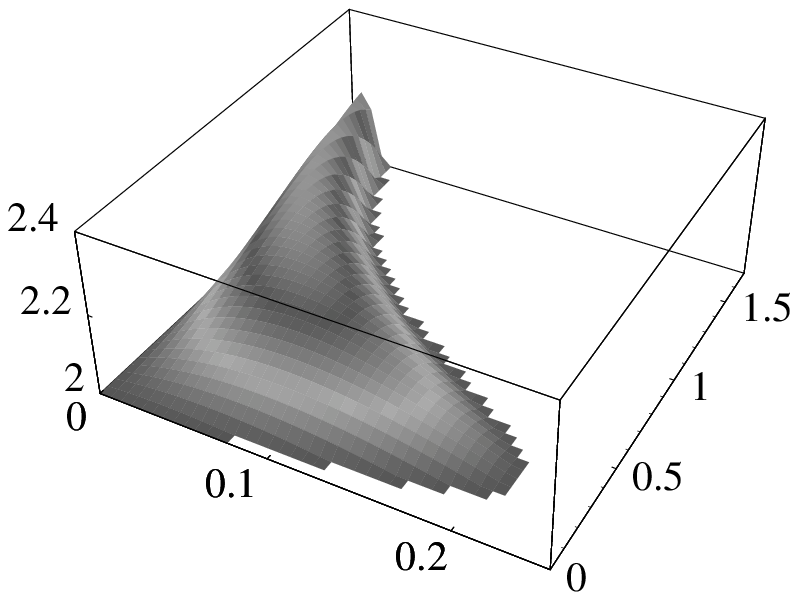}}}
\put(-2,-3.5){${\cal J}$}
\put(4.7,-1.7){$r$}
\put(-5,1.9){${\cal B}_{2\sDP}$}
\put(5.5,2.7){(a)}
\end{picture}}
\vspace{2cm}
\caption{ (a) Plot of the combination $\cB_{2\sDP}$ defined in 
Eq.~(\protect\ref{c6:B_BW}) and (b) according to the parametrization 
given by \refeq{c6:ParamGenDP}. Only values exceeding the bound
imposed by local theories are shown.}
\label{c6:B_2DPtwb}
\end{figure}
\par\noindent
As depicted in Fig.~\ref{c6:B_2DPtwb}(a), $\cB_{2\sDP}$ in (\ref{c6:B_BW})
violates the upper bound imposed by local theories. For increasing
$r$, the violation of the Bell's inequality is observed for smaller
${\cal J}$. Therefore an asymptotic analysis for
large $r$ and ${\cal J} \ll 1$ may be performed. Then a straightforward calculation
shows that the maximum value of ${\cal B}_2$ (for this particular
selection of coherent displacements) is obtained for
\begin{equation}
{\cal J} e^{2r} = \frac{1}{3} \ln 2,
\end{equation}
corresponding to ${\cal B}_{2\sDP} = 1 + 3 \cdot 2^{-4/3} \approx 2.19$.
Thus, in
the limit $r\rightarrow \infty$, when the original EPR state is
recovered, a significant violation of Bell's inequality takes place.
Notice that in order to observe the nonlocality of the EPR state, very
small displacements have to be applied, decreasing as ${\cal J}
\propto e^{-2r}$.
As pointed out in Ref.~\cite{BW98} the results above have been obtained
without any serious attempt to find the maximum violation. For this
purpose one should consider a general quadruplet of displacements. An
analysis to obtain the maximum violation of Bell inequalities within
this formalism is performed in \cite{WJK02}. Choosing
\begin{align}
 \alpha_1=-\alpha_2=i\sqrt{{\cal J}} \alpha_1'=-\alpha_2'=-3i \sqrt{{\cal J}}
\label{c6:ParamGenDP}
\end{align}
an asymptotic violation of $\cB_2\simeq2.32$ can be
obtained (see Fig.~\ref{c6:B_2DPtwb}(b)).
This shows that the EPR state does not maximally violate Bell's
inequalities in a DP test. The reason for this has been addressed in
Ref.~\cite{JSK+03}, and it is attributed to the fact that the displaced
parity operator does not completely flip the parity of the
entangled quantities characterizing the TWB ({\em i.e.}, the number
states). 
\subsubsection{Pseudospin test}\label{sss:twbPS}
\index{pseudospin}
Let us now focus on the ``pseudospin nonlocality test''. Considering a
TWB state, it is known that the correlation function (\ref{c6:EPS})
has the following expression (setting to zero the azimuthal angles)
\cite{CPH+02}:
\begin{align}
\label{c6:ETWBPS}
E_{\sPS}(\vartheta_1,\vartheta_2)=\cos\vartheta_1\cos\vartheta_2+f_{\hbox{\tiny TWB}}
\sin\vartheta_1\sin\vartheta_2 \:,
\end{align}
where $f_{\hbox{\tiny TWB}}=\tanh (2r)$.
Choosing $\vartheta_1=0$, $\vartheta_1'=\pi/2$ and $\vartheta_2=-\vartheta_2'$, we have
\begin{align}
\cB_2=2(\cos\vartheta_2+f_{\hbox{\tiny TWB}}\sin\vartheta_2)\;,
\end{align}
and, for this specific setting, the maximum of $\cB_2$ is
\begin{align}
\label{c6:B2PSTWB}
  \cB_2=2\sqrt{1+f_{\hbox{\tiny TWB}}}\,.
\end{align} 
It turns out that the TWB state violates the Bell's inequality
(\ref{c6:BI2}) for every $r\neq0$. The violation increases
monotonically to the maximum value of $2\sqrt2$ as the function
$f_{\hbox{\tiny TWB}}\to 1$, {\em i.e.}, as the squeezing parameter $r$
increases. This indicates that the EPR state maximally violate
Bell's inequality. Furthermore, $f_{\hbox{\tiny TWB}}$ may be regarded as a
quantitative measure of quantum nonlocality.
\par
In Ref.~\cite{GKM+04} different representations of $\rmSU(2)$ algebra
have been considered to exploit nonlocality of the TWB. In particular
using the operators given in \refeq{c6:WPi} one can show that the
correlation ${\cal E}_{\sPS}$ is still given by \refeq{c6:ETWBPS}, where now
the function $f_{\hbox{\tiny TWB}}$ is substituted by
\begin{align}
f'_{\hbox{\tiny TWB}}=\frac2\pi\arctan[\sinh(2r)]\;.
\end{align}
Therefore, the behavior of the Bell combination $\cB_2$ is the same as
above, but for any squeezing parameter $r$ it gives a lower violation
of local realism if compared to \refeq{c6:B2PSTWB}. In general, it is
possible to demonstrate that the configurational parameterization
given by \refeq{c6:PS} leads to maximal violation for all values of
$r$. Finally, we mention that besides the representations of $\rmSU(2)$
given by \refeq{c6:PS} and \refeq{c6:WPi}, different ones may be found
for which the Bell combination $\cB_2$ is not even a monotonic function of
$r$, {\em i.e.} is not a monotonic function of the entanglement.
\subsection{Non-Gaussian states}\label{ss:2mNG}
\index{nonlocality!non-Gaussian states}
\index{degaussification!nonlocality}
As already pointed out in Section \ref{s:NGS}, it is expected
that non-Gaussian states are characterize by a larger 
nonlocality. Let us now exploit this possibility using the non-Gaussian states
introduced in Section \ref{s:NGS}. Since the Wigner function of IPS and TWBA
states is non-positive, all the nonlocality tests introduced will be considered.
\subsubsection{Displaced parity test}\label{sss:2mNGDP}
In addressing nonlocality of IPS state, we will consider both the
parameterizations (\ref{c6:ParamBWDP}) and (\ref{c6:ParamGenDP}).
We denote $B(\cJ)\equiv\mathcal{B}_{2\sDP}$ for parameter \refeq{c6:ParamBWDP}, 
and $C(\cJ)\equiv\mathcal{B}_{2\sDP}$ for parameter \refeq{c6:ParamGenDP}.
As for a TWB, the violation of the Bell's inequality is observed
for small $r$ \cite{BW98}. For the rest of the section, we will refer to
$B(\cJ)$ as
$B^{\hbox{\tiny (TWB)}}(\cJ)$ when it is evaluated for a TWB
(\ref{twb:wig}), and as $B^{\hbox{\tiny (IPS)}}(\cJ)$
when we consider the IPS state (\ref{ips:wigner}). 
\begin{figure}[h!]
\begin{center}
\includegraphics[width=0.8\textwidth]{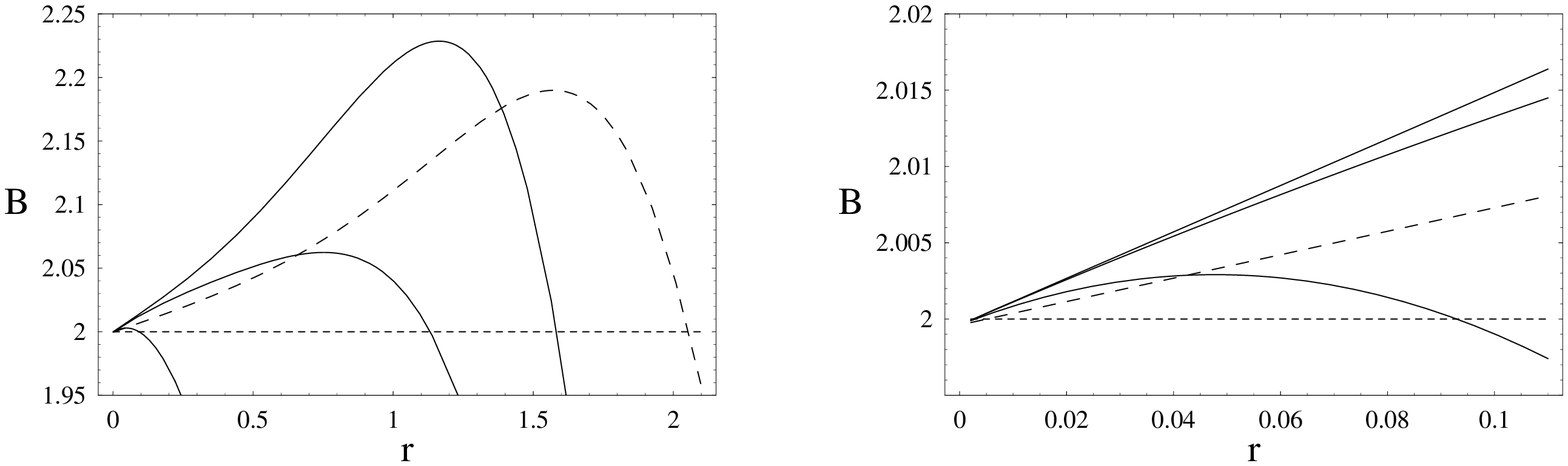}
\end{center}\vspace{-.5cm}
\caption{\label{f:ips:vs:twb:j} Plot of $B(\cJ)$ for
$\cJ = 10^{-2}$. The dashed line is
$B^{\hbox{\tiny (TWB)}}(\cJ)$, while the solid lines are
$B^{\hbox{\tiny (IPS)}}(\cJ)$ for different values of
$\tau_{\rm eff}$ (see the text): from top to bottom $\tau_{\rm eff} =
0.999, 0.99$ and $0.9$. When $\tau_{\rm eff} = 0.999$, the maximum of
$B^{\hbox{\tiny (IPS)}}(\cJ)$ is $2.23$. The plot on the right
is a magnification of the region $0\le r \le 0.11$ of the upper one.
Notice that for small $r$ there is always a region where
$B^{\hbox{\tiny (TWB)}}(\cJ) < B^{\hbox{\tiny (IPS)}}(\cJ)$.}
\end{figure}\par
\begin{figure}[h]
\begin{center}
\includegraphics[width=0.8\textwidth]{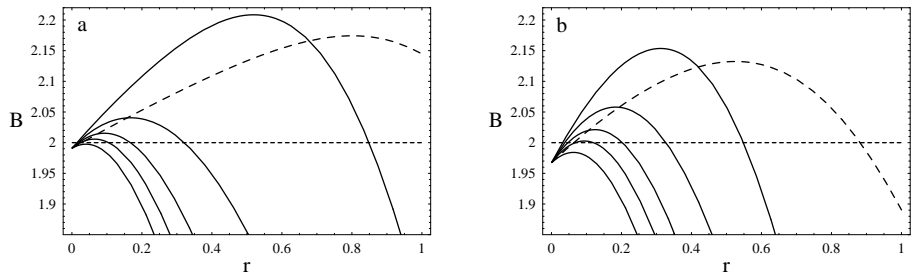}
\end{center}\vspace{-.5cm}
\caption{\label{f:ips:vs:twb} Plots of $B(\cJ)$ as a function of the 
  squeezing parameter $r$ for two different values of $J$: (a)
  $\cJ=5\cdot10^{-2}$ and (b) $\cJ=10^{-1}$.  In all the plots the dashed
  line is $B^{\hbox{\tiny (TWB)}}(\cJ)$, while the solid lines are
  $B^{\hbox{\tiny (IPS)}}(\cJ)$ for different values of
  $\tau_{\rm eff}$ (see the text): from top to bottom $\tau_{\rm eff}
  = 0.999, 0.9, 0.8, 0.7$ and $0.5$.  Notice that there is always a
  region for small $r$ where $B^{\hbox{\tiny (TWB)}}(\cJ) <
  B^{\hbox{\tiny (IPS)}}(\cJ)$.  When $\tau_{\rm eff} =
  0.999$ the maximum of $B^{\hbox{\tiny (IPS)}}(\cJ)$ is
  always greater than the one of $B^{\hbox{\tiny (TWB)}}(\cJ)$.}
\end{figure}

\noindent
We plot
$B^{\hbox{\tiny (TWB)}}(\cJ)$ and $B^{\hbox{\tiny
(IPS)}}(\cJ)$ in the Figs.~\ref{f:ips:vs:twb:j} and
\ref{f:ips:vs:twb} for different values of the effective
transmissivity $\tau_{\rm eff}$ and of the parameter $J$: for not
too big values of the squeezing parameter $r$, one has that
$2<B^{\hbox{\tiny (TWB)}}(\cJ)<
B^{\hbox{\tiny (IPS)}}(\cJ)$. Moreover, when
$\tau_{\rm eff}$ approaches unit, {\em i.e.} when at most one
photon is subtracted from each mode, the maximum of
$B{\hbox{\tiny (IPS)}}$ is always greater than the
one obtained using a TWB. A numerical analysis shows that in the
limit $\tau_{\rm eff} \to 1$ the maximum is $2.27$, that is
greater than the value $2.19$ obtained for a TWB \cite{BW98}. The
limit $\tau_{\rm eff} \to 1$ corresponds to the case of one single
photon subtracted from each mode \cite{opatr,coch}. Notice that
increasing $\cJ$ reduces the interval of the values of $r$ for which
one has the violation. For large $r$ the best result is thus
obtained with the TWB since, as the energy grows, more photons are
subtracted from the initial state \cite{ips:tele}. Since the
relevant parameter for violation of Bell inequalities is
$\tau_{\rm eff}$, we have, from Eq.~(\ref{taueff}), that the IPS
state is nonlocal also for low quantum efficiency of the IPS
detector.
\par\noindent
\par
\begin{figure}[h]
\begin{center}
\includegraphics[scale=.4]{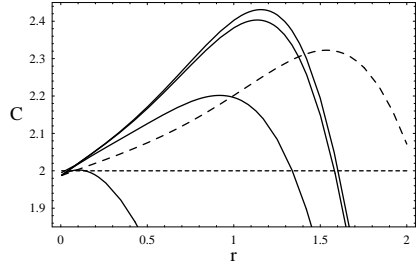}
\end{center}\vspace{-.5cm}
\caption{\label{f:ips:ale} Plots of $C(\cJ)$ as
a function of the squeezing parameter $r$ for $\cJ = 1.6\cdot 10^{-3}$.  In all
the plots the dashed line is $C^{\hbox{\tiny (TWB)}}(\cJ)$, while the
solid lines are $C^{\hbox{\tiny (IPS)}}(\cJ)$ for different
values of $\tau_{\rm eff}$ (see the text): from top to bottom $\tau_{\rm
eff} = 0.9999, 0.999, 0.99$ and $0.9$.  When $\tau_{\rm eff} = 0.9999$ the
maximum of $C^{\hbox{\tiny (IPS)}}(\cJ)$ is $2.43$.}
\end{figure}

The same conclusions holds when we consider the parametrization
of Eq.~(\ref{c6:ParamGenDP}). In Fig.~\ref{f:ips:ale} we plot
$C^{\hbox{\tiny (TWB)}}(\cJ)$ and $C^{\hbox{\tiny
(IPS)}}(\cJ)$, {\em i.e.} $C(\cJ)$ evaluated for the TWB and the IPS
state, respectively. The behavior is similar to that of $B(\cJ)$,
the maximum violation being now $C^{\hbox{\tiny
(IPS)}}(\cJ) = 2.43$ for $\tau_{\rm eff} = 0.9999$ and $\cJ = 1.6\cdot
10^{-3}$.
Finally, notice that the maximum violation using IPS states is
achieved (for both parameterizations) when $\tau_{\rm eff}$
approaches unit and for values of $r$ smaller than for TWB.
\par\noindent
\par
\begin{figure}[h]
\vspace{18mm}
\setlength{\unitlength}{0.5cm} 
\centerline{%
\begin{picture}(0,0) 
\put(0,0){\makebox(0,0)[c]{\includegraphics[width=6cm]{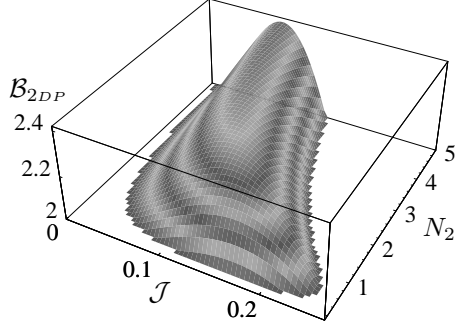}}}
\put(-2.3,-3.7){${\cal J}$}
\put(5,-2){$N_2$}
\put(-6,1.7){${\cal B}_{2\sDP}$}
\end{picture}}
\vspace{18mm}
\caption{ Bell combination obtained choosing optimized displacement 
parameters for TWBA state (\protect\ref{c3:TWBA}) (see text for details). Only values 
violating inequality (\protect\ref{c6:BI2}) are shown.}
\label{f:B2DPTWBA}
\end{figure}

Concerning the TWBA (\ref{c3:TWBA}), let us consider the case of large
$N_2$ and small $N_3$, say $N_3=10^{-2} (N_2)^{-1}$.  As in the analysis
of the entanglement properties of the tripartite state $|T\rangle$, the
phase coefficients $\phi_2$ and $\phi_3$ play no role in the
characterization of nonlocality. Using the parametrization
$\alpha_1=\frac12\alpha_2=\frac13\alpha_1'=i\sqrt{{\cal J}}$ and
$\alpha_2'=0$, an enhancement of the violation of Bell's inequality can be
observed with respect to the TWB. Indeed the asymptotic violation turns out
to be of ${\cal B}_{2\sDP} = 2.41$. It can be found, for large $N_2$, when
${\cal J} N_2 = 0.042$ (see Fig.~\ref{f:B2DPTWBA})\footnote{\footnotesize
The same analysis holds if we reverse the role of the two modes, provided
that the conditional measurement to obtain the TWBA is performed on mode
$a_2$ of the original tripartite state, rather then on mode $a_3$.}. 
\par\noindent
\par
Although the IPS and the TWBA states allow for an enhancement of
nonlocality with respect to the usual TWB state, they never reach the
maximum violation admitted by quantum mechanics. As already pointed
out the reason for this can be attributed to the fact that the
displaced parity operator does not completely flip the parity of the
entangled quantities characterizing the three states above ({\em
  i.e.} the number states). However, the maximum violation
of the Bell's inequality in the contest of a DP test could be
achieved if the following state $|\hbox{ECS}\rangle$ (entangled
coherent state) \cite{San92} could be produced experimentally
\begin{align}
  |{\rm ECS}\rangle={\cal N}(|\gamma\rangle|-\gamma\rangle
  -|-\gamma\rangle|\gamma\rangle) \;, \label{c3:ECS}
\end{align}
where ${\cal N}$ is a normalization factor and $|\gamma\rangle$ is a
coherent state with $\gamma\neq0$.
Its Wigner function read as follows:
\begin{multline}
W_{\hbox{\tiny ECS}}(\alpha,\beta) = 4{\cal N}^2\Big\{
\exp\{-2|\alpha-\gamma|^2-2|\beta+\gamma|^2\}
+\exp\{-2|\alpha+\gamma|^2-2|\beta-\gamma|^2\}\\
-\exp\{-2(\alpha-\gamma)(\alpha^*+\gamma)
-2(\beta+\gamma)(\beta^*-\gamma)-4\gamma^2\}\\
-\exp\{-2(\alpha^*-\gamma)(\alpha+\gamma)
-2(\beta^*+\gamma)(\beta-\gamma)-4\gamma^2\}\Big\}\;,
\label{c3:WECS}
\end{multline}
where $\gamma$ is assumed to be real for simplicity. Notice that the 
ECS state may be represented in the
$2\times2$-Hilbert space as
\begin{align}
|{\rm ECS}\rangle=\frac{1}{\sqrt{2}}(|e\rangle|d\rangle
 -|d\rangle|e\rangle),
\label{c3:ECS2}
\end{align}
where $|e\rangle={\cal N_+}(|\gamma\rangle+|-\gamma\rangle)$ and
$|d\rangle={\cal N_-}(|\gamma\rangle-|-\gamma\rangle)$ are even and
odd states with normalization factors ${\cal N_+}$ and ${\cal N_-}$.
Note that these states form an orthogonal basis, regardless of the
value of $\gamma$, which span the two-dimensional Hilbert space.  For
this state the maximum violation can be achieved due to the fact the
displaced parity operators act like an ideal rotation on the even and
odd microscopic states $|e\rangle$ and $|d\rangle$ in which the ECS
state may be decomposed [see \refeq{c3:ECS2}]. As a consequence the
parity of $|e\rangle$ and $|d\rangle$, which are the orthogonal
entangled elements in the entangled coherent state, can be perfectly
flipped by the displacement operator (for $\gamma\rightarrow\infty$),
allowing for the maximum violation of Bell's inequality \cite{JSK+03}.
\subsubsection{Pseudospin test}\label{sss:2mNGPS}
\begin{figure}
\vspace{3.2cm}
\setlength{\unitlength}{1mm}
\centerline{%
\begin{picture}(50,0)
\put(0,0){\includegraphics[width=5cm]{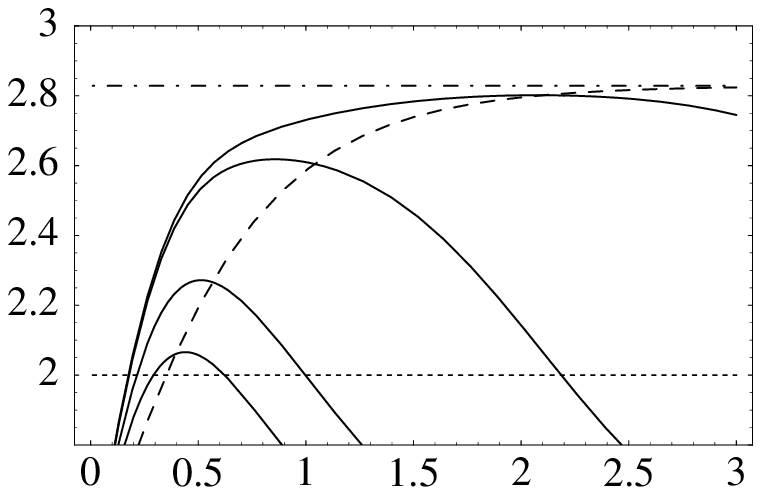}}
\put(27,-3){$r$}
\put(-4,15){${\cal B}_{\sPS}$}
\end{picture}
}
\vspace{0cm}
\caption{Plots of ${\cal B}_{PS}$:
the dashed line refers to the TWB, whereas the solid lines refer to the IPS
with, from top to bottom, $\tau_{\rm eff} = 0.9999, 0.99, 0.9$, and $0.8$.}
\label{f:PS:id}
\end{figure}
Now we investigate the nonlocality of the IPS state by means of the
pseudospin test considering the pseudospin operators given in
Eqs.~(\ref{c6:WPi}). If we set to zero the azimuthal angle, the correlation
function (\ref{c6:EPS}) reads
\begin{equation}
E_{\sPS}^{\hbox{\tiny (IPS)}}(\vartheta_1, \vartheta_2) =
\frac{1}{p_{11}(r,\tau_{\rm eff})} \sum_{k=1}^4 {\cal C}_k\,
\big[ \cos\vartheta_1 \cos\vartheta_2
+ f_{\hbox{\tiny IPS}}\, \sin\vartheta_a \sin\vartheta_b \big]\,,\label{c6:IPS:PS}
\end{equation}
where we defined
$$
f_{\hbox{\tiny IPS}} = \frac{8}{\pi{\cal A}_k}\,
\arctan\left( \frac{2 B \tau_{\rm eff} + h_k}{\sqrt{{\cal A}_k}} \right)\,,
$$
with
$${\cal A}_k=(b-f_k)(b-g_k)-(2 B \tau_{\rm eff} + h_k)^2\,,$$
and all the quantities appearing in Eq.~(\ref{c6:IPS:PS}) have been
defined in Section \ref{s:degauss}.
\par
In Fig.~\ref{f:PS:id} we plot ${\cal B}_{\sPS}$ for the TWB and IPS;
we set $\vartheta_{1}=0$, $\vartheta'_{1}=\pi/2$, and
$\vartheta_{2}=-\vartheta'_{2}=\pi/4$. As usual the IPS leads to better results
for small values of $r$. Whereas ${\cal B}_{\sPS}^{\hbox{\tiny (TWB)}} \to
2\sqrt{2}$ as $r\to \infty$, ${\cal B}_{\sPS}^{\hbox{\tiny (IPS)}}$ has a
maximum and, then, falls below the threshold $2$ as $r$ increases. It is
interesting to note that there is a region of small values of $r$ for which
${\cal B}_{\sPS}^{\hbox{\tiny (TWB)}}\le 2 < {\cal B}_{\sPS}^{\hbox{\tiny
(IPS)}}$, {\em i.e.} the IPS process can increases the nonlocal properties
of a TWB which does not violates the Bell's inequality for the pseudospin
test, in such a way that the resulting state violates it.  Note that the
maximal of violation for the IPS occur for a range of values $r$
experimentally achievable.
\par
\begin{figure}
\vspace{1.7cm}
\setlength{\unitlength}{0.5cm} 
\centerline{%
\begin{picture}(0,0)
\put(0,0){\makebox(0,0)[c]{\includegraphics[width=5cm]{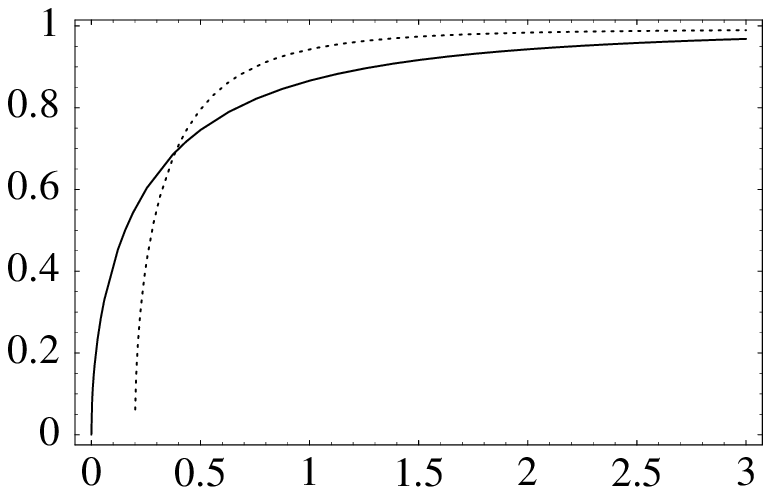}}}
\put(-5.5,0){$f$}
\put(0,-3.8){$N$}
\end{picture}}
\vspace{20mm}
\caption{Comparison between $f_{\hbox{\tiny
TWB}}$ (solid line) and $f_{\hbox{\tiny TWBA}}$ (dotted line)
as functions of the total number of photons $N$ (the summation has
been numerically performed for $\eta =0.8$ and $N_3=0.1$).}
\label{f6:B2PS}
\end{figure}

Concerning the TWBA state, a straightforward calculation shows
that an expression identical in form to Eq.~(\ref{c6:ETWBPS}), where the
following function $f_{\hbox{\tiny TWBA}}$ can be identified:
\begin{multline}
 f_{\hbox{\tiny TWBA}}=2\,{\sqrt{\frac{N_2}
        {1 + N_1}}}\,\frac{
    \left( 1 + N_3\,\eta  \right)}{N_3\,
    \left( 1 + N_1 \right) \,\eta } \\
    \times\sum_{k,p = 0}^{\infty }
         \frac{\left( 2\,k + p \right) !}
           {\left( 2\,k \right) !\,p!}\,
          {\sqrt{\frac{2\,k + p + 1}{2\,k + 1}}}\, 
	  \left( 1 - {\left( 1 - \eta  \right) }^p \right)
          {\left( \frac{N_3}
              {1 + N_1} \right) }^p\,
          {\left( \frac{N_2}
              {1 + N_1} \right) }^ {2\,k} \:.
\end{multline}
In order to compare the violations in the case of the TWB and the
TWBA, let us fix as in the previous subsection a small value for
$N_3$.
A plot of the functions $f_{\hbox{\tiny TWB}}$ and
$f_{\hbox{\tiny TWBA}}$ versus the total number of photons of the TWB, 
and the total number of photons of the initial
three-partite state, is given in Fig.~(\ref{f6:B2PS}).
It can be seen that the TWBA reaches large violations for smaller energies with
respect to the TWB.
As for the TWBA state, also in the ECS case an expression 
identical in form to Eq.~(\ref{c6:ETWBPS}) may be found, where now the
following function $f_{\hbox{\tiny ECS}}$ can be identified:
\begin{align}
f_{\hbox{\tiny ECS}}=\cosh (\gamma^2)\,\sinh (\gamma^2)\,
\left(\sum_{n=0}^\infty\frac{\gamma^{4n+1}}{\sqrt{(2n)!(2n+1)!}}\right)^{-2}
\;.\nonumber
\end{align}
A remarkable feature of this case is that it allows for a maximum
violation not only when $\gamma\rightarrow\infty$, but also when
$\gamma\rightarrow 0$.
\subsubsection{Homodyne test}\label{sss:2mNGH}
\index{homodyne detection!nonlocality}
The negativity of the Wigner function that may occur for non-Gaussian
states suggests to perform a nonlocality test based upon a homodyne
detection scheme. Indeed, homodyne nonlocality test has attracted much
attention in the recent years \cite{gil1,Mun99,nha,sanchez,ips:nonloc}, in
view of the high quantum efficiency achievable with homodyne detection,
which offers the possibility of a loop-hole free test of local realistic
theories \cite{gil1}. As we seen, the positivity of a Wigner function
prevents the violation of homodyne Bell inequality (\ref{c6:BI2}). On the
other hand, its negativity is not sufficient, in general, to ensure a
violation. Quantum states with negative Wigner function that doesn't
violate local realism with homodyne test are given for example in
Refs.~\cite{Mun99,MMS+03}. As shown in Ref.~\cite{gil1} also the ECS state
does not allow for any violation unless it is subjected to an additional
squeezing. The same situation occur if the TWBA is considered \cite{FP04}.
\par
Concerning the IPS state if one dichotomizes the
measured quadratures as described in Section (\ref{s:rev})
the Bell parameter reads $\cB_{2\sH}= E(\vartheta_1,\varphi_1) 
+ E(\vartheta_1,\varphi_2) + E(\vartheta_2,\varphi_1) - E
(\vartheta_2,\varphi_2)$  where $\vartheta_h$ and $\varphi_h$ 
are the phases of the two homodyne measurements at the modes $a$ 
and $b$, respectively. Eq.~(\ref{c6:EH}) can be rewritten as 
\begin{equation}
E(\vartheta_h,\varphi_k) =
\int_{\mathbb{R}^2} d x_{\vartheta_h}\,d x_{\varphi_k}\,
{\rm sign}[x_{\vartheta_h}\, x_{\varphi_k}]\,
P(x_{\vartheta_h}, x_{\varphi_k})\,,
\end{equation}
$P(x_{\vartheta_h}, x_{\varphi_k})$ being the joint
probability of obtaining the two outcomes
$x_{\vartheta_h}$ and $x_{\varphi_k}$. 
\begin{figure}[tb]
\vspace{1.8cm}
\setlength{\unitlength}{.46cm}
\centerline{
\begin{picture}(0,0)
\put(-6.2,0){\makebox(0,0)[c]{\includegraphics[width=55mm]{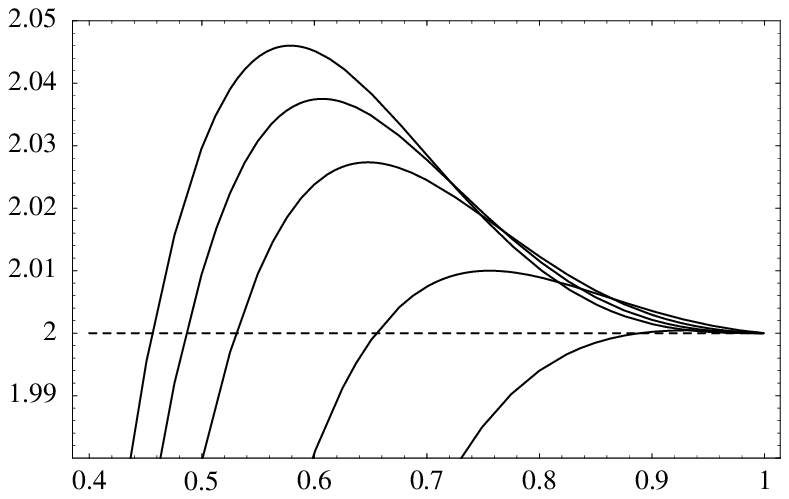}}}
\put(-10.6,2.7){\small a}
\put(-13,1){\small ${\cal B}_{2\sH}$}
\put(-6.7,-4.5){\small $\tanh r$}
\end{picture}
\begin{picture}(0,0)
\put(7,0){\makebox(0,0)[c]{\includegraphics[width=55mm]{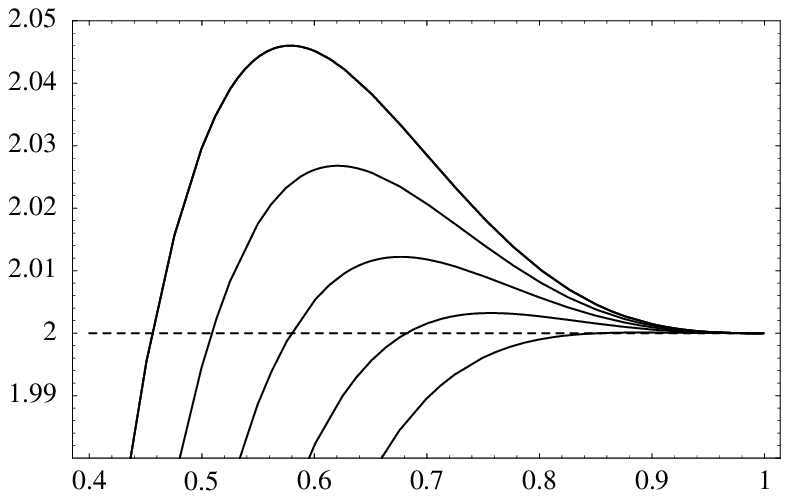}}}
\put(2.6,2.7){\small b}
\put(0.2,1){\small ${\cal B}_{2\sH}$}
\put(6.5,-4.5){\small $\tanh r$}
\end{picture}
}
\vspace{2cm}
\caption{\label{f:homo:id:eta} Plots of ${\cal B}_2$
as a function of $\tanh r$: (a) for different values of $\tau_{\rm eff}$ and
for ideal homodyne detection ({\em i.e.}~with quantum efficiency $\eta_{\rm
H}=1$): from top to bottom $\tau_{\rm eff} = 0.99, 0.95, 0.90, 0.80$ and
$0.70$; (b) with $\tau_{\rm eff} = 0.99$ and for different values of the
homodyne detection efficiency $\eta_{\hbox{\tiny H}}$: from top to bottom
$\eta_{\hbox{\tiny H}} = 1, 0.95, 0.90, 0.85$ and $0.80$. The maximum of
the violation decreases and shifts toward higher values of $r$ as
$\eta_{\hbox{\tiny H}}$ decreases. For smaller values of $\tau_{\rm eff}$
the violation is further reduced.}
\end{figure}
\par
In Fig.~\ref{f:homo:id:eta}~(a) we plot $\cB_{2\sH}$ for $\vartheta_1 = 0$,
$\vartheta_2 = \pi/2$, $\varphi_1 = -\pi/4$ and $\varphi_2 = \pi/4$: as
pointed out in Ref.~\cite{sanchez}, the Bell's inequality is violated
for a suitable choice of the squeezing parameter $r$. Moreover, when
$\tau_{\rm eff}$ decreases the maximum of violation shifts toward higher
values of $r$.
As one expects, nonunit quantum efficiency 
$\eta_{\hbox{\tiny H}}$ of the homodyne detection further reduces 
the violation [see Fig.~\ref{f:homo:id:eta}~(b)]. Notice that, 
when $\eta_{\hbox{\tiny H}}<1$, violation occurs for higher values 
of $r$, although its maximum is actually
reduced: in order to have a significant violation one needs a homodyne
efficiency greater than $80\%$ (when $\tau_{\rm eff}=0.99$).
On the other hand, the high efficiencies of this kind of detectors
allow a loophole-free test of hidden variable theories
\cite{gil1,gil2}, though the violations obtained are quite small.
This is due to the intrinsic information loss of the binning
process, which is used to convert the continuous homodyne data in
dichotomic results \cite{Mun99}. Better results, even if the
violation is always small, can be achieved using a {\em circle}
coherent state \cite{gil1,gil2} or a superposition of photon
number states \cite{Mun99}, while maximal violation, {\em i.e.} 
$\cB_{2\sH} = 2\sqrt{2}$, is obtained by means of a different binning process,
called root binning, and choosing a particular family of quantum
states \cite{aub,weng}.
\section{Three-mode nonlocality}\label{s:3mNL}
\index{nonlocality!three-mode}
As we have seen in Section \ref{s:rev} nonlocality of multipartite 
systems may be analyzed by means of Bell-Klyshko inequalities. In 
particular, we have explicitly considered the constraints (\ref{c6:BI3})
that every  tripartite system must respect in order to be described 
by a local realistic theory. The aim of this section is to analyze 
the violation of Ineq.~(\ref{c6:BI3}) in tripartite systems. Both 
the states introduced in Section \ref{s:3mGS} will be considered, 
as well as the parity-entangled GHZ state introduced in Ref.~\cite{CZ02}.
\subsection{Displaced parity test} \label{ss:3mDP}
\begin{figure}[h]
\vspace{2.5cm}
\setlength{\unitlength}{0.4cm} 
\centerline{%
\begin{picture}(-13,-13) 
\put(0,0){\makebox(0,0)[c]{\includegraphics[width=5cm]{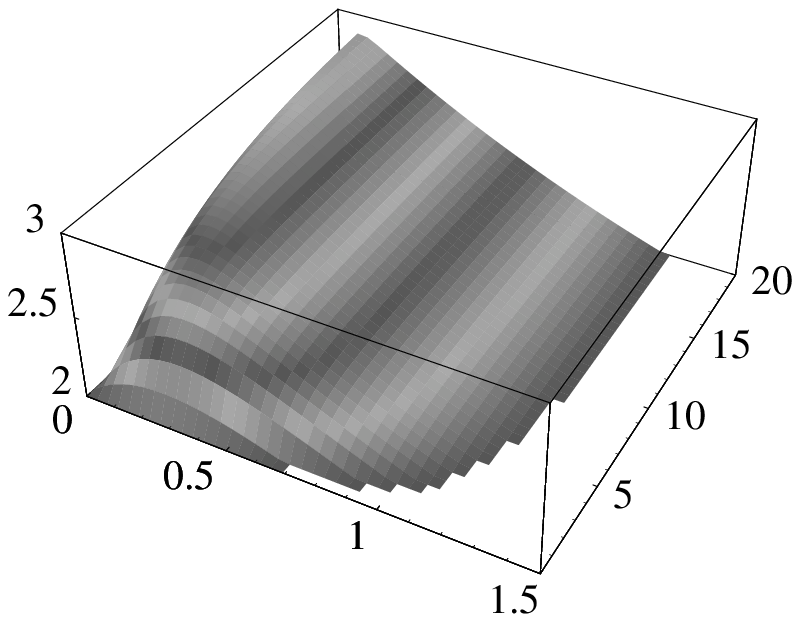}}}
\put(-2.3,-4){${\cal J}$}
\put(4.5,-2.9){$N$}
\put(-6.3,2.2){${\cal B}_{3\sDP}$}
\put(6,4){(b)}
\end{picture}
\begin{picture}(0,0) 
\put(0,0){\makebox(-3,0)[c]{\includegraphics[width=5cm]{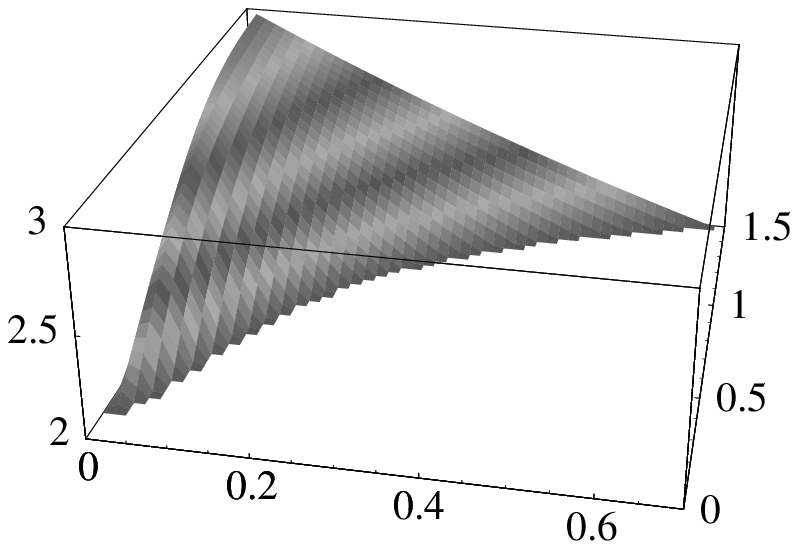}}}
\put(-3,-4.3){${\cal J}$}
\put(4,-2.8){$r$}
\put(-8.2,1.6){${\cal B}_{3\sDP}$}
\put(4,4){(a)}
\end{picture}
}
\vspace{1.8cm}
\caption{(a) Plot of the Bell combination (\ref{c6:B3DPVLBGen}) and (b) Bell
combination obtained choosing optimized displacement parameters for state
$|T\rangle$ (see text for details). Only values violating Bell Inequality
(\ref{c6:BI3}) are shown.} \label{f6:B3_VLandT}
\end{figure}

Let us start our study of nonlocality for tripartite systems  
using the displaced parity test. Considering the correlation function
$E_{\sDP}(\boldsymbol \alpha )$ given by Eq. (\ref{c6:EDP}), the state
$\bmV_{3}$ ({\em i.e.}, the state whose covariance matrix
$\bmV_{3}$ is given by \refeq{c3:matC}) was found in \cite{LB01}
to give a maximal violation of ${\cal B}_{3\sDP} \simeq 2.32$ in the limit
of large squeezing and small displacement. The study in \cite{LB01}
however was performed for a particular choice of displacement
parameters: $\alpha_1=\alpha_2=\alpha_3=0$ and
$\alpha_1'=\alpha_2'=\alpha_3'=i \sqrt{{\cal J}}$. One can identify a
number of parameterizations that allow a significantly higher
violation of Bell's inequality \cite{FP04}. Consider
the one given by $\alpha_1=\alpha_2=\alpha_3=i\sqrt{{\cal J}}$ and
$\alpha_1'=\alpha_2'=\alpha_3'=-2i \sqrt{{\cal J}}$ from which follows
that
\begin{align}
{\cal B}_{3\sDP}=3\exp \left\{-12e^{-2 r} {\cal J}\right\} - 
\exp\left\{24 e^{2 r} {\cal J}\right\}
\label{c6:B3DPVLBGen}\;.
\end{align}
The asymptotic value ${\cal B}_{3\sDP}=3$ is
found for large $r$ and ${\cal J} \ne 0$ [see
Fig.~\ref{f6:B3_VLandT} (a)]. The importance of a suitable choice of the
displacement parameters is apparent if this asymptotic value is
compared to the violations obtained in Ref.~\cite{LB01}. In that work 
in fact generalizations to more than
three modes of state $\bmV_{3}$ were also considered, giving an
increasing violation of Bell inequality as the number of modes
increases, but never founding a violation greater than $2.8$.
\par
Let us now consider the tripartite state $|T\rangle$, the correlation 
function is now given by Eq.~(\ref{c6:EDP}) with the covariance matrix 
$\bmV_{T}$ in \refeq{c3:matV}. The
symmetry of the state suggests a maximum violation of Bell inequality
for $N_2=N_3=\frac{1}{4}N $ [recall Eq.~(\ref{c3:Ndit})], while the
fact that the separability of the state doesn't depend on the phases
$\phi_2$ and $\phi_3$ suggests that they are not crucial
for the nonlocality test. If we consider the same parametrization leading
to Eq.~(\ref{c6:B3DPVLBGen}) and fix $\phi_2=\phi_3=\pi$, we
find:
\begin{align}
{\cal B}_{3\sDP}=\frac{-1 + e^{6\,{\cal J}\,\left( 1 + N + 
         2\,{\sqrt{2}}\,{\sqrt{N\,\left( 2 + N \right) }} \right)
         } + 2\,e^
      {\frac{3}{2}\,{\cal J}\,\left( 4 + 7\,N + 
            6\,{\sqrt{2}}\,{\sqrt{N\,\left( 2 + N \right) }}
            \right) }}{e^
    {4\,{\cal J}\,\left( 3 + 3\,N + 
        2\,{\sqrt{2}}\,{\sqrt{N\,\left( 2 + N \right) }} \right) 
      }}\;,
\label{c6:B3DPTVLB}
\end{align}
from which follows an asymptotic violation of Bell's inequalities of
${\cal B}_{3\sDP} = 2.89$, for large $N$ and small ${\cal J}$. A
slightly better result is found if a parametrization, more suitable
and numerically optimized for the state $|T\rangle$, is considered:
    $\alpha_1=\frac23\sqrt{{\cal
J}},\alpha_2=\alpha_3=\alpha_1'=0,\alpha_2'=- \sqrt{{\cal
J}},\alpha_3'=\sqrt{{\cal J}},\phi_2=0$ and $\phi_2=\pi$. The Bell
combination ${\cal B}_{3\sDP}$ for this choice of parameters is depicted
in Fig.~\ref{f6:B3_VLandT}(b). We note that in this case a larger
choice of angles allows the violation of Bell inequality if compared
with Fig.~\ref{f6:B3_VLandT} (a). The asymptotic violation of Bell's
inequality is now ${\cal B}_{3\sDP} = 2.99$.  Comparing the results
obtained for the two states $\bmV_{3}$ and $|T\rangle$ it is
possible to show that, even if the two states have quite the same
asymptotic violation, state $\bmV_{3}$ reaches it for lower
energies \cite{FP04}.
\subsection{Pseudospin test} \label{ss:PS3}
Consider now the pseudospin nonlocality test.
Let us calculate the expectation value of the correlation function
(\ref{c6:EPS}) for the state $|T\rangle$ (for simplicity we consider
$\phi_2=\phi_3=0$). The only non vanishing contributes are given by:
\begin{subequations}
\label{c6:c1c2c3}
\begin{align}
c_1 &= \langle s_z^1 \otimes s_x^2 \otimes s_x^3 \rangle=
      \langle s_z^1 \otimes s_y^2 \otimes s_y^3 \rangle\nonumber \\ 
    &=-\frac{\sqrt{{\cal N}_2 {\cal N}_3}}{2(1+N_1)}
      \sum_{s,t}{\cal N}_2^{2s}\,
      {\cal N}_3^{2t}\,\frac{(2s+2t+1)!}{(2s)!(2t)!\sqrt{(2s+1)(2t+1)}} 
      \;, \\ 
c_2 &= \langle s_x^1 \otimes s_z^2 \otimes s_x^3 \rangle =
     -\langle s_y^1 \otimes s_z^2 \otimes s_y^3 \rangle \nonumber \\ 
    &= \frac{\sqrt{{\cal N}_3}}{2(1+N_1)}\sum_{s,t}{\cal N}_2^{2s}
      \,{\cal N}_3^{2t}\,\frac{(2s+2t)!}{(2s)!(2t)!}
      \sqrt{\frac{2s+2t+1}{2t+1}}
      \;, \\
c_3 &= \langle s_x^1 \otimes s_x^2 \otimes s_z^3 \rangle =
     -\langle s_y^1 \otimes s_x^2 \otimes s_z^3 \rangle\nonumber \\ 
    &= \frac{\sqrt{{\cal N}_2}}{2(1+N_1)}\sum_{s,t} {\cal N}_2^{2s}
      \,{\cal N}_3^{2t}\,\frac{(2s+2t)!}{(2s)!(2t)!}
      \sqrt{\frac{2s+2t+1}{2s+1}}
      \;,
\end{align}
\end{subequations}
with ${\cal N}_k = N_k /(1 + N_1)$,
and by $\langle s_z^1 \otimes s_z^2 \otimes s_z^3 \rangle=1 $. The
correlation function then, according to
Eqs.~(\ref{c6:PS}) and (\ref{c6:EPS}), reads as follows:
\begin{multline}
\label{c6:EPS3ms}
E_{\sPS}({\bf d})=  \cos\vartheta_1\cos\vartheta_2\cos\vartheta_3
+c_1\cos\vartheta_1\sin\vartheta_2\sin\vartheta_3
\cos(\varphi_2-\varphi_3)\\
+c_2\cos\vartheta_2\sin\vartheta_1\sin\vartheta_3
\cos(\varphi_1+\varphi_3) 
+c_3\cos\vartheta_3\sin\vartheta_1\sin\vartheta_2
\cos(\varphi_1-\varphi_2)\,.
\end{multline}
Hence, without loss of generality, we can fix for example $\varphi_1=0$ and
$\varphi_2=\varphi_3=\pi$ and look for the maximum violation of Bell
inequality (\ref{c6:BI3}) constructed from Eq.~(\ref{c6:EPS3ms}). Notice
that if the coefficients $c_k$, $k=1,2,3$, were equal to $1$
then the maximum violation
admitted, ${\cal B}_{3\sPS}=4$, should be reached. Considering
Eqs.~(\ref{c6:c1c2c3}) two limiting cases can be studied. First, for
large $N_2$ and small $N_3$ (or {\em viceversa}) a numerical evaluation
of the coefficients
$c_k$ shows that $c_3 \rightarrow 1$, while the other two vanish.
Hence, considering $\vartheta_3=0$, the correlation function (\ref{c6:EPS3ms}) 
reduces to that of a TWB subjected to a pseudospin
nonlocality test [see Eq.~(\ref{c6:ETWBPS})], allowing an
asymptotic violation of ${\cal B}_{3\sPS}=2\sqrt2$. 
This result should be expected, since in this case the state 
(\ref{T}) reduces to a TWB for modes $a_1$ and $a_2$, 
while mode $a_3$ remains in the vacuum state and factors out.
Consider now the case in
which $N_2=N_3=\frac14 N$. A numerical evaluation shows that the coefficients
$c_4 \rightarrow \frac12$ for large $N$, hence also in this case the maximum
violation cannot be attained. The asymptotic violation turns out to be
${\cal B}_{3\sPS} = 2.63$.
\par
As already mentioned in Section \ref{s:rev} other representations for
the pseudospin operators can be considered. Using Eqs.~(\ref{c6:WPi}) and
the Wigner function associated to state $|T\rangle$ it is possible to 
calculate the correlation function
${\cal E}_{\sPS}({\bf d})$. Setting again the azimuthal angles $\varphi_k=0$,
the latter shows the same structure as $E_{\sPS}({\bf d})$ where now the
coefficient $c_k$ are replaced by
\begin{align}
  \label{c6:c1c2c3primes}
  c_1'=\frac{2\arctan\left({\displaystyle
  \frac{N}{2\sqrt{1+N}}}\right)}{\pi(1+N)}\,, \qquad 
  c_2'=c_3'=\frac{2\arctan \sqrt{N}}{\pi(1+\frac12 N)} \;.
\end{align}
An appropriate choice of angles leads to a violation of Bell's
inequality given by ${\cal B}_{3\sPS} = 2.22$, which is now reached
for $N \simeq 1$, value for which the coefficients $c_k'$ are
approximately near their maxima. As already pointed out, we see that
different representations of the pseudospin operators give rise to different
expectation values for the Bell operator.
\par
Applying now the same procedure to state $\bmV_{3}$ we find the same
structure for the correlation function $E_{\sPS}'$, where the
coefficients are now given by
\begin{align}
  \label{c6:c1c2c3primesVL}
  c_1'=c_2'=c_3'=\frac{-6\,\arctan \left[{\displaystyle
  \frac{4\,\cosh r\,\sinh r }
      {\,{\sqrt{3(2 + e^{4\,r})}}}}\right]}{\pi \,
    {\sqrt{5 + 4\,\cosh (4\,r)}}}
   \;.
\end{align}
After an optimization of the angles $\vartheta_k$ we obtain a maximal
violation of ${\cal B}_{3\sPS} = 2.09$, 
for $r \simeq 0.42$ ($N\simeq0.56 $) that maximizes the coefficients $c_k$.
\par
Finally, one may consider the nonlocality issue
in the general case of an $n$-party system. We recall that in the
case of discrete variable systems Mermin \cite{Mer90} showed that the
multipartite GHZ state, defined as 
\begin{align}
\ket{\hbox{GHZ}}_n=\frac{1}{\sqrt2} (\ket{+}_1\ldots\ket{+}_n-\ket{-}_1\ldots\ket{-}_n)
\label{c6:GHZ} \;,
\end{align} 
where $\ket{+}_k$ is the eigenvector with eigenvalue $+1$ of the Pauli
matrix $\bmsigma_z$ relative to the $k$-qubit, admits a violation of
local realism that exponentially grows with the number of party. The
first attempt to compare this behavior with continuous variables case
was performed in Ref.~\cite{LB01}. There, the violation of local
realism by the states $\bmV_{n}$, a straightforward
generalization to $n$ modes of state $\bmV_{3}$, has been
analyzed.  Considering the Bell combination $\cB_{n}$ given in
\refeq{c6:BIN} in a DP setting, it was found that the degree of
nonlocality of states $\bmV_{n}$ indeed grows with increasing
number of parties. This growth, however continuously decrease for
large $n$, as opposite to the qubit case.  Nevertheless, as already
pointed out in Ref.~\cite{LB01}, this analysis was performed for a
particular choice of displacement parameters $\boldsymbol \alpha$,
${\boldsymbol \alpha} '$ which unfortunately seems not to be optimal.
In fact, the maximum violation attained with that choice never reach 
the asymptotic value of $\cB_{3\sPS} = 3$
obtained with \refeq{c6:B3DPVLBGen} ({\em e.g.}, $\cB_{85} = 2.8$ in
\cite{LB01}). Another approach has been pursued in Ref.~\cite{CZ02},
where eigenstates of the pseudospin operator $s_z$ (parity-entangled
states) has been considered, in direct analogy to the $n$-party GHZ
states (\ref{c6:GHZ}). Due to this analogy it is straightforward to show
that these states lead to an exponential increase of the violation of
local realism. For example in the tripartite case a maximum violation
$\cB_3=4$ can be found. Hence a behavior identical to that of the
discrete variable systems is recovered. However, recall that to our
knowledge there is no proposal concerning a possible experimental
realization of the parity-entangled (non-Gaussian) states.

\chapter[Teleportation and telecloning]{Teleportation and
telecloning}\label{ch:tele} 
In this Chapter we deal with the transfer and the distribution of 
quantum information, {\em i.e.} of the information contained in a 
quantum state. At first we address {\em teleportation}, 
{\em i.e.} the transmission of an {\em unknown} quantum state 
from a sender to receiver that are spatially separated. 
Teleportation is achieved by means of a classical and a distributed 
quantum communication channel, realized by a suitably chosen 
nonlocal entangled state. Indeed, quantum teleportation has no 
classical analog: the use of entanglement permits to transmit 
an unknown signal without classically broadcasting the whole 
information about  its quantum state. 
On the other hand, we have that quantum information cannot be 
perfectly copied, even in principle. The {\em no-cloning} 
theorem follows from the linearity of quantum mechanics 
\cite{Die82,WZ82}, and forbids the existence of any device 
producing perfect copies of generic unknown quantum states.  
Only approximate clones can be realized, that can be subsequently 
distributed in a quantum network by means of teleportation. 
Alternatively, the entire process can be realized nonlocally 
exploiting multipartite entanglement which is shared among 
all the involved parties. The latter process is known as 
{\em telecloning}, and will analyzed in the second part
of this Chapter.
\section[Continuous variable quantum teleportation (CVQT)]{Continuous variable
quantum teleportation}\label{s:CVQT}
\sectionmark{Continuous variable quantum teleportation}
\index{teleportation}
In this Section we address continuous variable quantum 
teleportation (CVQT), where the goal is teleporting an 
unknown state $\sigma_1$ of a given mode 1, from Alice, 
the sender, to Bob, the receiver, {\em i.e.} reconstructing the 
quantum state onto another mode, on which Alice has no access. 
In the following we refer to the
CVQT protocol sketched in Fig.~\ref{f:telescheme}. Alice and Bob share an
entangled two-mode state of radiation described by the density matrix
$\varrho_{23}$, where the subscripts refer to modes 2 and 3, respectively:
mode 2 is own by Alice, the other by Bob. In order to implement the
teleportation, Alice performs a joint measurement, {\em i.e.} the measurement 
of the normal operator $Z$ on modes 1 and 2, getting as outcome a complex 
number $z$ (see Section \ref{s:eig}); then, she sends her result to Bob 
via a classical communication 
channel. Once received this classical information, Bob applies a displacement 
$D(z)$ to his mode 3 and obtains a quantum state
$\varrho_{\rm out}$ which, in the ideal case, is identical to the initial state
$\sigma_1$ \cite{braun1}.
\par
\begin{floatingfigure}{0.45\textwidth}
\vspace{-0.2cm}
\begin{center}
\includegraphics[width=0.43\textwidth]{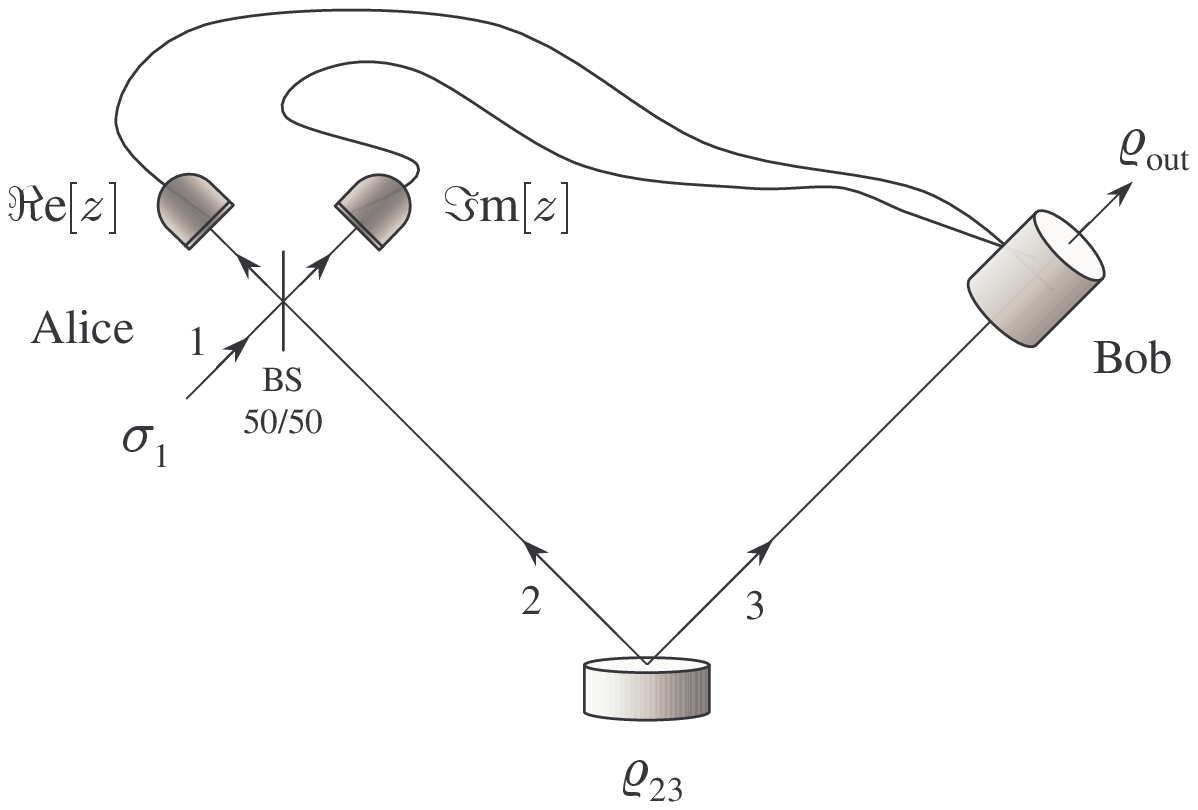}
\end{center}
\vspace{-0.7cm}
\caption{Continuous variable quantum teleportation: scheme 
of the optical realization.
\label{f:telescheme}}
\vspace{-0.2cm}
\end{floatingfigure}
\par
The original proposal for teleportation concerned states in a bidimensional
Hilbert space \cite{bennett}.  The  corresponding experiments have been
performed in the optical domain, using polarization qubit
\cite{boschi,tele:zei}, and for the state of a trapped ion \cite{recent}.
Also CVQT can be realized by optical means \cite{braun1}, and successful
teleportation of coherent states has been realized \cite{furu}.  In optical
CVQT, entanglement is provided by the twin-beam state (TWB) of radiation
$\ket{\Lambda}\rangle_{23} = \sqrt{1-\lambda^2}\sum_{n=0}^{\infty}\lambda^n
\ket{n}_2 \ket{n}_3$, $2$ and $3$ being two modes of the field and
$\lambda$ the TWB parameter.  We assume $\lambda$ as real $|\lambda|<1$.
TWBs $\ket{\Lambda}\rangle_{23}$ are produced by optical amplifiers (see
Section \ref{ss:opa}) and, being a pure state, their entanglement can be
quantified by the excess von Neumann entropy. 
We refer to Section \ref{c4:2pPure} for details and
just remind that the degree of entanglement is a monotonically increasing 
function of $\lambda$ (or equivalently of the average number of photons). As
we will see, the larger is the entanglement the higher (closer to unit) is 
the fidelity of teleportation based.
There are different ways to describe CVQT
\cite{braun1,tele:enk,noise:take}, but, in general, two of them are the
most common: the first makes use of photon number-state basis, the
other is in terms of Wigner functions. This Section addresses the
description of CVQT protocol following these two approaches and, in
particular, we derive the completely positive (CP) map $\mathcal{L}$
describing the teleportation process also in the presence of noise.
\subsection{Photon number-state basis representation}\label{s:tele:ph:numb}
\index{teleportation!photon number representation}
\index{twin-beam!teleportation}
Here we describe the teleportation protocol in the photon number basis. 
The state Alice wishes to teleport to Bob is described by the density matrix
\begin{eqnarray}
\sigma_{1} = \sum_{p,q}\: \sigma_{pq}\:\ket{p}_1 {}_1\bra{q}\,,
\label{sigma}\;
\end{eqnarray}
while the pure two-mode state they share is (in general)
\begin{eqnarray}
\varrho_{23} = \dket{\bmC}_{23} {}_{32}\dbra{\bmC}\,,\quad
\dket{\bmC}_{23} = \sum_{h,k}\: c_{hk}\:\ket{h}_2 \ket{k}_3
\label{shared:density}\;,
\end{eqnarray}
where we used the matrix notation introduced in Section \ref{s:MatNot}.
\par
The first step of the protocol consists in Alice's joint measurement on
modes 1 and 2, which corresponds to the measure of the complex photocurrent
$Z=a_1+a_2^{\dag}$ (see Section \ref{s:eig}). The whole measurement process
is described by the POVM
\begin{eqnarray}
\Pi_{12}(z) = \frac{1}{\pi}\:D_1(z) \dket{\ii}_{12}
 {}_{21}\dbra{\ii} D_{1}^{\dag}(z)\,,\label{het:POVM}\;
\end{eqnarray}
where $\dket{\ii}_{12} \equiv \sum_{v}\:\ket{v}_{1}\ket{v}_{2}$, and
$D_1(z)$ is a displacement operator on mode 1 (see Section \ref{s:2pom}).
The conditional state of mode 3 is then
\begin{align}
\varrho_3(z) &= \frac{1}{p(z)}\:{\rm Tr}_{12} \left[
\sigma_1\otimes\varrho_{23} \: \Pi_{12}(z) \otimes
\mathbb{I}_3 \right]\\
\mbox{} &= \frac{1}{\pi p(z)}\: {\rm Tr}_{12} \Bigg[ \Bigg( \sum_{p,q}\:
\sigma_{pq} \ket{p}_1 {}_{1}\bra{q} \Bigg) \otimes \Bigg(
\sum_{h,k}\:\sum_{n,m}\: c_{hk}\:c^*_{nm} \ket{h}_{2}\ket{k}_{3}
{}_{3}\bra{m} {}_{2}\bra{n} \Bigg) \nonumber\\
&\mbox{} \hspace{2cm} \times D_{1}(z) \Bigg( \sum_{v,w}\:
\ket{v}_{1} \ket{v}_{2} {}_{2}\bra{w} {}_{1}\bra{w} \Bigg)
D_{1}^{\dag}(z) \Bigg]\\
&=  \frac{1}{\pi p(z)}\: {\rm Tr}_{12} \Bigg[\sum_{p,q}\: \sum_{h,k}\:
\sum_{n,m}\: \sum_{v,w}\: \sigma_{pq}\:c_{hk}\:c^*_{nm}\:
\underbrace{{}_{2}\braket{n}{v}_{2}}_{\delta_{n,v}}\:
{}_{1}\bra{q}D_1(z)\ket{v}_{1} \nonumber\\
&\mbox{} \hspace{2cm} \times 
\ket{k}_{3}{}_{3}\bra{m} \otimes \ket{h}_{2}{}_{2}\bra{w} \otimes
\ket{p}_{1}{}_{1}\bra{w}D_{1}^{\dag}(z)\Bigg]\\
&= \frac{1}{\pi p(z)}\: \sum_{p,q}\: \sum_{h,k}\: \sum_{n,m}\:
\sigma_{pq}\:c_{hk}\:c^*_{nm}\: {}_{1}\bra{q}D_1(z)\ket{n}_{1}\:
{}_{1}\bra{h}D_1^{\dag}(z)\ket{p}_{1} \, 
\ket{k}_{3}{}_{3}\bra{m} \nonumber\\
&= \frac{1}{\pi p(z)}\: \bmC^{\sT}D^{\dag}(z)\:\sigma\:D(z)\:\bmC^{*}
\label{trace:2}\,,
\end{align}
where $(\cdots)^{\sT}$ denotes transposition, the
subscripts have been suppressed, and $p(z)$ is the double-homodyne 
probability density, given by
\begin{equation}
p(z) = {\rm Tr}_{123} \left[
\sigma_1\otimes\varrho_{23} \: \Pi_{12}(z) \otimes \mathbb{I}_3\right]\,.
\label{DH:prob}
\end{equation}
After the measurement, Alice sends her result to Bob through a classical
channel and, then, he applies a displacement $D(z)$ to his mode, in formula:
\begin{eqnarray}
\varrho_3(z) \rightarrow \varrho'_{3}(z) \equiv D(z)\:\varrho_{3}(z)\:D^{\dag}(z)\,.
\label{bubba}
\end{eqnarray}
\par
If the entangled channel is provided by the TWB, then 
$\bmC = (1-\lambda^2)^{1/2}\:\lambda^{a^{\dag}a}$ and
Eq.~(\ref{bubba}) rewrites as 
\begin{eqnarray}\label{implicit:output}
\varrho'_{3}(z) = \frac{(1-\lambda^2)}{\pi p(z)}\:  D(z)\: \lambda^{a^{\dag}a}
\:D^{\dag}(z)\: \sigma\:D(z)\:\lambda^{a^{\dag}a} \:D^{\dag}(z)\,.
\end{eqnarray}
Now, using the operatorial identity (\ref{other}),
Eq.~(\ref{implicit:output}) can be reduced to
\begin{multline}
\varrho'_{3}(z) = \frac{(1-\lambda^2)}{\pi p(z)}\:
\int_{\mathbb{C}^2}
\frac{d^2 w\: d^2 v}{\pi^2 (1-\lambda)^2}
\:\exp\left\{-\frac12 \frac{1+\lambda}{1-\lambda}\:
\big(|w|^2+|v|^2\big)
\right\}\\
\times D(z)\:D(w)\:D^{\dag}(z)\:\sigma\:D(z)\:D^{\dag}(v)\:D^{\dag}(z)\,,
\label{integral:output}\;
\end{multline}
which, thanks to (\ref{compD}), becomes
\begin{multline}
\varrho'_{3}(z) = \frac{(1+\lambda)}{\pi p(z)}\: 
\int_{\mathbb{C}^2}
\frac{d^2 w\: d^2 v}{\pi^2 (1-\lambda)}
\:\exp\left\{-\frac12 \frac{1+\lambda}{1-\lambda}\big( |w|^2+|v|^2 \big)
\right\}\\
\times \:\exp\left\{ (w-v)^{*} z - (w-v) z^{*}\right\}
\: D(w)\:\sigma\:D^{\dag}(v)\,.
\label{integral:output:2}
\end{multline}
\par
The final output of CVQT is obtained integrating $\varrho'_{3}$ over all the
possible outcomes $z$ of the double-homodyne detection, {\em i.e.}
\begin{eqnarray}
\varrho_{\rm out} = \int_{\mathbb{C}} d^2 z\: p(z)\: \varrho'_3 (z)\,,
\label{integral:output:3}
\end{eqnarray}
and, remembering the definition (\ref{defDeltaCmpl}) of the delta function
$\delta^{(2)}(\xi)$, we obtain
\begin{equation}
\varrho_{\rm out} = \int_{\mathbb{C}}
\frac{d^2 w}{4 \pi \sigma_{-}^2}\:
\exp\left\{-\frac{|w|^2}{4 \sigma_{-}^2}\right\}\: D(w)\:\sigma\:D^{\dag}(w)\,,
\label{fock:CVQT:output}\;
\end{equation}
with
$$\sigma_{-}^2 \equiv \frac14 \frac{1-\lambda}{1+\lambda} = \frac14\,
e^{-2r}\,,$$
$r = \tanh^{-1} \lambda$ being the squeezing parameter of the TWB. 
Eq.~(\ref{fock:CVQT:output})
corresponds to an overall Gaussian noise with parameter $\sigma_{-}^2$ (see
Section \ref{gauss:noise}): in
this way the CVQT protocol can be seen as a thermalizing quantum channel
\cite{banjpa}. Notice that $\varrho_{\rm out}$ approaches the input state
$\sigma$ only for $\lambda \to 1$ (or $r \to \infty$), {\em i.e.} for a
TWB with infinite energy.
\subsection{The completely positive map of CVQT}
\index{teleportation!CP map}
CVQT corresponds to a Gaussian completely positive (CP) map and,
as we will see in Section \ref{s:tele:noise}, this result holds also in the
presence of noise. If $\varrho_{\rm in}$ is
the state at Alice's side, the state at Bob's side will be 
\begin{equation}
\varrho_{\rm out} =
\mathcal{L}\varrho_{\rm in} \equiv \int_{\mathbb{C}}
d^2 w\:
\frac{\exp\left\{-\bmw^{\sT}\,\bmSigma^{-1}\,\bmw \right\}}
{\pi \sqrt{{\rm Det}[\bmSigma]}}
\: D(w)\:\varrho_{\rm in}\:D^{\dag}(w)
\label{tele:map2}\;,
\end{equation}
$\bmw$ denoting the vector $(\real{w},\immag{w})^{\sT}$, and $\bmSigma$ is a
$2 \times 2$ matrix describing a Gaussian noise, as we have addressed in
Section \ref{gauss:noise}. Notice that
Eq.~(\ref{tele:map2}) provides already the Kraus diagonal form of the
teleportation map. In Section \ref{s:tele:noise} we will explicitly derive
the map (\ref{tele:map2}) for teleportation in the presence of noise. In
the case of Eq.~(\ref{fock:CVQT:output}) one has
\begin{equation}
\bmSigma =
\bmSigma_0 \equiv 4
\left(
\begin{array}{cc}
\sigma_{-}^2 & 0 \\
0 & \sigma_{-}^2
\end{array}
\right)\,.
\label{Sigma:map:0}
\end{equation}
\subsection{CVQT as conditional measurement on the TWB}
As we have seen in Section \ref{s:2pom}, when the modes in the measurement
of $Z$ are initially excited in a factorized state, then we can write the 
POVM as a single-mode POVM depending on the state of the other mode. This 
is the case of CVQT, which can be seen as the measurement of the POVM [see
Eq.~(\ref{2pom:Pi2})]
\begin{equation}
\Pi_{2}(z) = \frac{1}{\pi}\,  D_2^{\sT}(z)\:\sigma^{\sT}\:D^{*}_2(z)\,,
\label{POVM:2:het}
\end{equation}
acting on the mode 2: in this way CVQT is reduced to a conditional
measurement on the TWB followed by a displacement.
Let $\sigma$ be again the state we wish to teleport and let the TWB
be the entangled state shared between Alice and Bob.
The conditioned state of mode 2 is then (we put 
$\varrho_{\hbox{\tiny TWB}}\equiv\varrho_{23}$)
\begin{align}
\varrho_3(z) &= \frac{1}{p(z)}\:{\rm Tr}_{2} \left[ \varrho_{\hbox{\tiny
TWB}}\:\Pi_2(z) \otimes \iid \right]\nonumber\\
          &= \frac{(1-\lambda^2)}{\pi p(z)}\:{\rm Tr}_{2} \left[
\sum_{h,k}\:\lambda^{h+k}\:\ket{h}_2 \ket{h}_3 {}_{3}\bra{k}{}_2\bra{k}
D_2^{\sT}(z)\:\sigma^{\sT}\:D^{*}_{2}(z)\right]\nonumber\\
          &= \frac{(1-\lambda^2)}{\pi p(z)}\:\sum_{h,k}\:\lambda^{h+k}\:
{}_{2}\bra{k}D_2^{\sT}(z)\:\sigma^{\sT}\:D^{*}_2(z)\ket{h}_{2}\:
\ket{h}_{3} {}_{3}\bra{k}\,,\label{conditioned}
\end{align}
with the double-homodyne density probability distribution $p(z)$ given by
\begin{equation}
p(z) = {\rm Tr}_{23} \left[ \varrho_{\hbox{\tiny TWB}}\:
\Pi_2(z) \otimes \iid \right]\,.
\end{equation}
Since $ [D^{\sT}(z)]^{*} = D^{\dag}(z)$, in Eq.~(\ref{conditioned})
we can write
\begin{align}
{}_{2}\bra{k}D_2^{\sT}(z)\:\sigma^{\sT}\:D^{*}_2(z)\ket{h}_{2} &=
{}_{2}\bra{k}\big[ D_2^{\dag}(z)\:\sigma\:D_2(z) \big]^{\sT}\ket{h}_{2}
\nonumber\\
&= {}_{2}\bra{h}D_2^{\dag}(z)\:\sigma\:D_2(z)\ket{k}_{2}\,,
\label{transposition}\;
\end{align}
and, suppressing all the subscripts, we have
\begin{equation}
\varrho_3(z) = \frac{(1-\lambda^2)}{\pi p(z)}\:\sum_{h,k}\:\lambda^{h+k}\:
\bra{h} D^{\dag}(z)\:\sigma\:D(z)\ket{k}\: \ket{h} \bra{k}\,.
\label{conditioned:T}
\end{equation}
In order to have a full quantum teleportation, we must displace the state
(\ref{conditioned:T}) applying $D(z)$, obtaining
\begin{align}
\varrho'_{3}(z) &=  D(z)\:\varrho_3(z)\:D^{\dag}(z)\nonumber\\
&= \frac{(1-\lambda^2)}{\pi p(z)}\:\sum_{h,k}\:\lambda^{h+k}\:
\bra{h} D^{\dag}(z)\:\sigma\:D(z)\ket{k}\: D(z)\ket{h} \bra{k}
D^{\dag}(z)\label{conditioned:D:1}\\
&= \frac{(1-\lambda^2)}{\pi p(z)}\:\sum_{h,k}\:\lambda^{h+k}\:
\bra{\psi_h(z)}\sigma\ket{\psi_k(z)}\:
\ket{\psi_h(z)}\bra{\psi_k(z)}\,, \label{conditioned:D}
\end{align}
where we defined the new s.o.n.c. $\left\{ \ket{\psi_h(z)} \right\}$, with 
$\ket{\psi_h(z)}\equiv D(z)\ket{h}$. Notice that Eq.~(\ref{conditioned:D:1})
can be written in operational form as
\begin{align}
\varrho'_{3}(z) &= \frac{(1-\lambda^2)}{\pi
p(z)}\:\sum_{h,k}\:\lambda^{h+k}\: D(z)\ket{h} \bra{h}
D^{\dag}(z)\:\sigma\:D(z)\ket{k}\: \bra{k}
D^{\dag}(z)\nonumber\\
&= \frac{(1-\lambda^2)}{\pi
p(z)}\:\sum_{h,k}\:\: D(z)\ket{h}\lambda^{a^{\dag} a} \bra{h}
D^{\dag}(z)\:\sigma\:D(z)\ket{k}\: \bra{k}\lambda^{a^{\dag} a}
D^{\dag}(z)\nonumber\\
&= \frac{(1-\lambda^2)}{\pi p(z)}\:  D(z)\: \lambda^{a^{\dag}a}
\:D^{\dag}(z)\: \sigma\:D(z)\:\lambda^{a^{\dag}a} \:D^{\dag}(z)\,,
\end{align}
which is the same as in Eq.~(\ref{implicit:output}). Finally, $\varrho_{\rm
out}$ is obtained by means of Eq.~(\ref{integral:output:3}).
\subsection{Wigner functions representation}\label{s:tele:wig}
\index{teleportation!Wigner representation}
This Section addresses CVQT described in terms of Wigner functions.
We first derive the teleported state Wigner function when the
shared state has the general form given in Eq.~(\ref{shared:density}),
then we specialize the results to the case of a TWB.
\par
Let $W[\sigma](\alpha_1)$ and $W[\varrho_{23}](\alpha_2,\alpha_3)$ be the
Wigner functions associated to the states (\ref{sigma}) and
(\ref{shared:density}), respectively, where $\alpha_h = \kappa_2(x_h+iy_h)$
(see Chapter \ref{ch:basics}).  Since the Wigner function corresponding 
to the POVM of ideal double- homodyne detection on mode 1 and 2 is
\begin{equation}
W[\Pi_{12}(z)](\alpha_1,\alpha_2) = \frac{1}{\pi^2}\:
\delta^{(2)}\big( (\alpha_1-\alpha_2^*)-z \big)\,,
\label{wig:het}\;
\end{equation}
with $z=\kappa_2(x+iy)$, using Eq.~(\ref{Wtrace}) the double-homodyne
density probability distribution (\ref{DH:prob}) reads
\begin{equation}
p(z) = \pi^3\,
\int_{\mathbb{C}^3}d^2\alpha_1\,d^2\alpha_2\,d^2\alpha_3\,
W[\sigma](\alpha_1)\, W[\varrho_{23}](\alpha_2,\alpha_3)
\, W[\Pi_{12}(z)](\alpha_1,\alpha_2)\:
W[\mathbb{I}_3](\alpha_3)\,,
\label{wigner:het:prob}\;
\end{equation}
while the conditioned state of mode 3 is
\begin{equation}
W[\varrho_3(z)](\alpha_3) = \frac{\pi^2}{p(z)}\: 
\int_{\mathbb{C}^2}d^2\alpha_1\,d^2\alpha_2 \:
W[\sigma](\alpha_1)\: W[\varrho_{23}](\alpha_2,\alpha_3)
\, W[\Pi_{12}(z)](\alpha_1,\alpha_2)\,
W[\mathbb{I}_3](\alpha_3)\,,
\label{wig:rho:cond}
\end{equation}
where $W[\mathbb{I}_3](\alpha_3) = \pi^{-1}$. Thanks to Eq.~(\ref{wig:het})
and after the integration over $\alpha_1$, Eq.~(\ref{wig:rho:cond}) becomes
\begin{align}
W[\varrho_3(z)](\alpha_3) &= \frac{1}{\pi p(z)}\: 
\int_{\mathbb{C}}d^2\alpha_2\, \: W[\sigma](\alpha_2^*+z)\:
W[\varrho_{23}](\alpha_2,\alpha_3)\nonumber\\
&= \frac{1}{\pi p(z)}\:
\int_{\mathbb{C}}d^2\alpha_2 \: W[\sigma](\alpha_2)\:
W[\varrho_{23}](\alpha_2^*-z^*,\alpha_3)\,.
\label{wig:rho:cond:2}
\end{align}
Now we perform the displacement $D(z)$, on mode 3, obtaining
\begin{equation}
W[\varrho'_3(z)](\alpha_3) = \frac{1}{\pi p(z)}\: 
\int_{\mathbb{C}}d^2\alpha_2 \: W[\sigma](\alpha_2)\:
W[\varrho_{23}](\alpha_2^*-z^*,\alpha_3-z)\,.
\label{wig:rho:cond:3}
\end{equation}
with $\varrho'_{3}(z) \equiv D(z)\:\varrho_{3}(z)\:D^{\dag}(z)$, and where
we used the property (\ref{Wdispla}). The output state of CVQT is obtained
integrating Eq.~(\ref{wig:rho:cond:3}) over all the possible outcomes of
the double homodyne detection, namely
\begin{equation} W[\varrho_{\rm out}](\alpha_3)=\int_{\mathbb{C}}
d^2 z\: p(z)\:W[\varrho'_3(z)](\alpha_3)\,.
\end{equation}
\par
If the shared state is the TWB, the Wigner function reads as
follows
\begin{equation}\label{c12:twb:wig}
W[{\varrho_{23}}](\alpha_2,\alpha_3)
= \frac{\exp \left\{ -\frac12\, \bmalpha_{23}^{\sT}\,\bmsigma_{\bmalpha}^{-1}
\,\bmalpha_{23}\right\}}{(2\,\pi)^2\sqrt{{\rm Det}[\bmsigma_{\bmalpha}]}}
\end{equation}
with $\bmalpha_{hk} \equiv
\big(\real{\alpha_h},\immag{\alpha_h},\real{\alpha_k},
\immag{\alpha_k}\big)^{\sT}$ and
\begin{equation}\label{c12:twb:CM}
\bmsigma_{\bmalpha} = \frac{1}{2}
\left(
\begin{array}{cc}
(\sigma_{+}^2 + \sigma_{-}^2)\, {\mathbbm 1}_2&
(\sigma_{+}^2 - \sigma_{-}^2 )\,\bmsigma_3 \\ [1ex]
(\sigma_{+}^2 - \sigma_{-}^2)\,\bmsigma_3  &
(\sigma_{+}^2 + \sigma_{-}^2)\,{\mathbbm 1}_2
\end{array}
\right)\,,
\end{equation}
${\mathbbm 1}_2$ being the $2\times 2$ identity matrix, $\sigma_{\pm}^2 =
\frac14 e^{\pm 2r}$, $\bmsigma_3 = {\rm Diag}(1,-1)$ is a Pauli matrix, and
$r = \tanh^{-1} \lambda$ is the squeezing parameter of the TWB.
Substituting Eq.~(\ref{c12:twb:wig}) into
Eq.~(\ref{wig:rho:cond:3}) and integrating over $z$ one has
\begin{align}
W[\varrho_{\rm out}](\alpha_3) &=
\int_{\mathbb{C}}
\frac{d^2\alpha_2}{4 \pi
\sigma_{-}^2}\:
\exp\left\{ - \frac{|\alpha_2-\alpha_3|^2}{4 
\sigma_{-}^2}\right\}\: W[\sigma](\alpha_2) \\
&= \int_{\mathbb{C}} d^2 w\:
\frac{\exp\left\{ -
\bmw^{\sT}\,\bmSigma^{-1}\,\bmw\right\}}
{\pi\sqrt{{\rm Det}[\bmSigma]}}\:
W[D(w)\:\sigma\:D^{\dag}(w)](\alpha_3)\,, \label{wig:output}
\end{align}
with $w \equiv \alpha_2$, $\bmw = (\real{w},\immag{w})^{\sT}$, and $\bmSigma
\equiv \bmSigma_0$ is given in Eq.~(\ref{Sigma:map:0}). Finally, using
Eq.~(\ref{WigF}), we obtain the same density matrix as in
Eq.~(\ref{tele:map2}).
\subsection{Teleportation fidelity}\label{s:fid:tele}
\index{teleportation!fidelity}
Teleportation has occurred when the output signal $\varrho_{\rm out}$ is in the
same quantum state of the unknown input $\sigma$. Therefore, we need 
to define a quantity which gauges the similarity between $\sigma$ and
$\varrho_{\rm out}$. This task is achieved using the so called ``fidelity'' or
``average fidelity'' between the input and output state. When the input
signal is a pure state $\sigma = \ket{\psi}\bra{\psi}$ the fidelity 
is defined in the following way \cite{fuchs:fid}\footnote{When $\sigma$ is not a pure 
state, a good measure for the fidelity is given by
$\overline{F} = {\rm Tr}\left[ \sqrt{\sqrt{\sigma}\,
\varrho_{\rm out} \, \sqrt{\sigma}} \right]$.}
\begin{subequations}
\label{grp:fidelity}
\begin{align}
\overline{F} &\equiv {\rm Tr} [ \sigma\: \varrho_{\rm out} ] = 
\langle \psi | \varrho_{\rm out} | \psi\rangle\:,
\label{fock:fid} \\
\overline{F} &\equiv \pi
\int_{\mathbb{C}}d^2\alpha\:
W[\sigma](\alpha)\: W[\varrho_{\rm out}](\alpha)\,,
\label{wigner:fid}\;
\end{align}
\end{subequations}
in terms of density matrix and Wigner function representation,
respectively. $\overline{F}$ has the property that it
equals 1 if and only if $\sigma$ is a pure state and $\varrho_{\rm out} =
\sigma$; on the other hand, it equals 0 if and only if the input and output
states can be distinguished with certainty by some quantum measurement. In
particular the average fidelity evaluates the extent at which all possible
measurement statistics produceable by the output state match the
corresponding statistics produceable by the input state. In order to
explain this last consideration, let us consider the generic POVM
$\{\Pi_{\alpha}\}$, describing a certain observable, with measurement
outcomes $\alpha$. If the observable were performed on the input system,
it would give a probability density for the outcomes $\alpha$ given by
\begin{equation}
P_{\rm in}(\alpha) = {\rm Tr} [ \sigma\: \Pi_{\alpha} ]\,;
\label{p:fid:in}\;
\end{equation}
if the same observable were performed on the output system, it would give,
instead, the probability density
\begin{equation}
P_{\rm out}(\alpha) = {\rm Tr} [ \varrho_{\rm out}\: \Pi_{\alpha} ]\,.
\label{p:fid:out}\;
\end{equation}
Now, a natural way to gauge the similarity of these two probability
densities is by their overlap $\mathcal{Q}$, defined as follows
\begin{eqnarray}
\mathcal{Q} = \int_{\mathbb{C}}
d^2\alpha\: \sqrt{P_{\rm in}(\alpha)\:
P_{\rm out}(\alpha)}
\label{overlap}\;.
\end{eqnarray}
It turns out that regardless of which observable is being considered
$\mathcal{Q} \geq \overline{F}$ and, moreover, one can show
\cite{fuchs1} that there exists an observable that gives precise
equality in this expression.
\par
When the shared state is the TWB, substituting
Eq.~(\ref{fock:CVQT:output}) into the Eq.~(\ref{fock:fid}), one
straightforward obtains
\begin{eqnarray}
\overline{F}_{\hbox{\tiny TWB}}(\lambda) \equiv
\overline{F}(\lambda) = \frac{1}{1+4\:\sigma_{-}^2} = \frac{1+\lambda}{2}
\label{ave:fid:TWB}\;.
\end{eqnarray}
\par
The maximum average fidelity achievable by means of some classical ({\em
local}) procedure to teleport a state is known as {\em classical
limit}.  This procedure should be characterized by a local measurement on
the state to be teleported, a classical communication of the result, say
$\mathcal{R}$, and, finally, a preparation stage at the receiver, according
to a rule that associates a certain output state to $\mathcal{R}$ such that
fidelity is maximized.
\par
Let us suppose that Alice wishes to transmit to Bob an unknown coherent
state without the resource of entanglement, {\em i.e.} by no means of a
shared entangled state. In such a case, we are interested in evaluating the
maximum average fidelity achievable. First of all we assume that the coherent 
state is drawn from the set $\cal S$ constituted by the coherent states
$\ket{\beta}$,where the complex parameter $\beta$ is distributed according
to the Gaussian distribution
\begin{equation}
p(\beta)=\frac{\Omega}{\pi}\,e^{-\Omega|\beta|^2}\;,
\end{equation}
$\Omega$ being a real, positive parameter. Ultimately, of course,
we would like to consider the case where Alice
and Bob have no information about the drown coherent state:
this is simply described by taking the limit $\Omega\rightarrow 0$.
\par
Alice's measurement for estimating the unknown parameter $\beta$ when it is
distributed according to a Gaussian distribution \cite{Yuen73} is the POVM
$\{\Pi_ z\}$ constructed from the coherent state projectors according to
$\Pi_ z=\pi^{-1}| z\rangle\langle z|$: this kind of measurement is
equivalent to optical heterodyning described in Section \ref{s:2pom},
where we send the vacuum in the other detector input port
[see Eq.~(\ref{2pom:Pi1}) with $\tau=\ket{0}\bra{0}$].
As in the case of the teleportation protocol, Alice's measurement outcome
$z$ is classically sent to Bob, that generates a new quantum state
according to the rule $ z \rightarrow |f_ z\rangle$.  Let us make no {\it a
priori\/} restrictions on the states $|f_ z\rangle$.
\par
Now, we find the maximum average fidelity $\overline{F}_{\rm
max}(\Omega)$ Bob can achieve for a given $\Omega$. For a given strategy
$ z \rightarrow |f_ z\rangle$, the achievable average fidelity is
\cite{fuchs:fid}
\begin{align}
\overline{F}(\Omega) &=
\int_{\mathbb{C}}d^2\beta\, p(\beta)
\int_{\mathbb{C}}d^2 z\, p( z|\beta)\, |\langle
f_ z|\beta\rangle|^2 
=\frac{\Omega}{\pi^2}\int_{\mathbb{C}^2}d^2 z\, d^2\beta\,
e^{-\Omega|\beta|^2}\,
e^{-| z-\beta|^2}|\langle f_ z|\beta\rangle|^2 \nonumber\\
&= \frac{\Omega}{\pi^2} \int_{\mathbb{C}} d^2 z\,
e^{-| z|^2}\langle f_ z| {\cal O}_{ z}(\Omega) |f_ z\rangle
\;, \label{ShutUp--GoAway}
\end{align}
where $p( z|\beta) = {\rm Tr}[ |\beta\rangle \langle\beta|\,
\Pi_{ z} ]$ is the heterodyne probability density distribution and we 
defined the positive semi-definite Hermitian operator
\begin{equation}
{\cal O}_ z(\Omega) \equiv \int_{\mathbb{C}} d^2\beta\,
\exp\Big\{-(1+\Omega)|\beta|^2+ 2\Re{\rm e}
[ z^*\beta]\Big\}\, |\beta\rangle\langle\beta|\;,
\end{equation}
that depends only on the real parameter $\Omega$ and the complex parameter
$ z$. It follows that
\begin{equation}
\langle f_ z|{\cal O}_ z(\Omega)| f_ z\rangle\le
\max\big[{\cal O}_ z(\Omega)\big]\;,
\label{Macaroni}
\end{equation}
where $\max[X]$ denotes the largest eigenvalue of the operator
$X$.
\par
Now, for each $z$, Bob adjusts the state $| f_ z\rangle$ to be the
eigenvector of ${\cal O}_ z(\Omega)$ with the largest eigenvalue. Then
equality is achieved in Eq.~(\ref{Macaroni}) and it is just a question
of being able to perform the integral in Eq.~(\ref{ShutUp--GoAway}).
The first step in carrying this out is to find the eigenvector and the
eigenvalue achieving equality in Eq.~(\ref{Macaroni}). This is most
easily evaluated by unitarily transforming ${\cal O}_ z(\Omega)$ into a
new operator diagonal in the number basis, picking off the largest
eigenvalue and transforming back to get the optimal $| f_ z\rangle$ (we
remember that eigenvalues are invariant under unitary transformations).
\par
In order to find the largest eigenvalue of ${\cal O}_ z(\Omega)$, we consider 
the positive operator
\begin{equation}
P = \int_{\mathbb{C}}d^2\beta\,
e^{-(1+\Omega)|\beta|^2} |\beta\rangle\langle\beta|\,.
\label{fid:class:auxilium}
\end{equation}
Since the number basis expansion of $P$ has the following diagonal form
\begin{align}
P
&= \int_{\mathbb{C}}d^2\beta\, e^{-(1+\Omega)|\beta|^2}\,
\sum_{n,m} \frac{e^{-|\beta|^2}}{\sqrt{n!\, m!}}\,
\beta^{n}\, (\beta^*)^{m}\, \ket{n}\bra{m}
\nonumber\\
&= \sum_{n,m} \frac{1}{\sqrt{n!\, m!}}
\underbrace{\int_{0}^{2\pi} d\phi\, e^{i \phi (n-m)}}_{\displaystyle 2 \pi
\delta_{n,m}}
\int_{0}^{\infty} d\rho \,
\rho^{n+m+1}\, e^{-(2 + \Omega)\rho^2}\, \ket{n}\bra{m}
\nonumber\\
&= \sum_n \frac{2 \pi}{n!} \underbrace{\int_{0}^{\infty} d\rho \,
\rho^{2n+1}\, e^{-(2 + \Omega)\rho^2}}_{\displaystyle \mbox{$\frac12$} n!\,
(2 + \Omega)^{-(n+1)}}
\, \ket{n}\bra{n}
= \pi\sum_{n=0}^\infty (2+\Omega)^{-(n+1)}
|n\rangle\langle n|\;,
\end{align}
its eigenvalues are $\{ \pi\, (2 + \Omega)^{-(n+1)} \}$ and, then, 
$\mu(P)=\pi/(2+\Omega)$, {\em i.e.} the vacuum state's eigenvalue.
\par
Now consider the displaced operator
\begin{equation}
Q_ z(\Omega) = D\!\left(\frac{ z}{1+\Omega}\right) P
\,D^\dag\!\left(\frac{ z}{1+\Omega}\right)\;,
\end{equation}
where $D(\nu)$ is the standard displacement operator.
Using Eq.~(\ref{fid:class:auxilium}), one finds
\begin{align}
Q_ z (\Omega) &=
\int_{\mathbb{C}}d^2\beta\,
\pexp{-(1+\Omega)|\beta|^2}\left|\beta + \frac{ z}{1+\Omega}\right\rangle
\left\langle\beta + \frac{ z}{1+\Omega}\right|
\nonumber\\
&= \int_{\mathbb{C}}d^2\xi\,
\exp\left\{-(1+\Omega)\left|\xi-\frac{ z}{1+\Omega}
\right|^2\,\right\}|\xi\rangle\langle\xi|
\nonumber\\
&= \exp\left\{-\frac{| z|^2}{1+\Omega}\right\}
\int_{\mathbb{C}}d^2\xi\,
\exp\Big\{-(1+\Omega)|\xi|^2+ 2\Re{\rm e}[ z^*\xi]\Big\}
|\xi\rangle\langle\xi|\,
\nonumber\\
&=
\exp\left\{-\frac{| z|^2}{1+\Omega}\right\}\, {\cal O}_ z(\Omega)\;,
\end{align}
and, substituting this into Eq.~(\ref{ShutUp--GoAway}), we have
\begin{align}
\overline{F}(\Omega) &=
\frac{\Omega}{\pi^2}\int_{\mathbb{C}}d^2 z\,
\exp\left\{-\left(1-\frac{1}{1+\Omega} \right)| z|^2\right\}\langle f_ z|
Q_ z (\Omega)|f_ z\rangle
\nonumber\\
&\le
\frac{1}{\pi}\frac{\Omega}{2+\Omega}\int_{\mathbb{C}}d^2 z\,
\exp\left\{-\frac{\Omega}{1+\Omega}| z|^2\right\}
= \frac{1+\Omega}{2+\Omega}\;.
\end{align}
Equality is obviously achieved in the previous expression by taking
\begin{equation}
|f_ z\rangle=
D\!\left(\frac{ z}{1+\Omega}\right)|0\rangle=
\left|\frac{ z}{1+\Omega}\right\rangle\,;
\end{equation}
therefore the maximum average fidelity is given by
\begin{equation}
\overline{F}_{\rm max}(\Omega)=\frac{1+\Omega}{2+\Omega}\;.
\end{equation}
For $\Omega \rightarrow \infty$ we have $\overline{F}_{\rm
max}(\Omega)\rightarrow 1$ since this situation corresponds 
to the teleportation of a single {\em known} coherent state, 
a task that can be achieved classically by transmitting
the value of the amplitude. On the other hand, in the limit 
$\Omega\rightarrow 0$, {\em i.e.} when the coherent
state to be sent is drawn from a uniform distribution, we have
$\overline{F}_{\rm max}(\Omega)\to 1/2$.
\par
It should be noted that nothing in this argument depended upon the
mean of the Gaussian distribution being $\beta=0$: Bob would need to
minimally modify his strategy to take into account Gaussians with a
non-vacuum state mean, but the optimal fidelity would remain the
same.
\subsection{Effect of noise}\label{s:tele:noise}
\index{teleportation!effect of noise}
\index{noisy channels!teleportation}
In this Section we study CVQT assisted by a TWB propagating through a
squeezed-thermal environment. Taking into account the results obtained
in Section \ref{prop:two:gauss}, the teleported state is now obtained from
Eq.~(\ref{wig:output}) with
\begin{equation}
\bmSigma \equiv \bmSigma(\Gamma,N_{\rm th},N_{\rm s}) = 4
\left(
\begin{array}{cc}
\Sigma_{3}^2 & 0 \\
0 & \Sigma_{2}^2
\end{array}
\right)\,,
\end{equation}
$\Sigma_2^2$, $\Sigma_3^2$ being given in Eqs.~(\ref{sigma:1234}).
\par
Finally, non-unit quantum efficiency $\eta$ in the joint measurement
modifies the POVM, which becomes a Gaussian convolution of the ideal one,
as pointed out in Chapter~\ref{ch:detection}. In this case the output state
is given by Eq.~(\ref{wig:output}) where
\begin{equation}
\bmSigma \equiv \bmSigma(\Gamma,N_{\rm th},N_{\rm s},\eta)=
\left(
\begin{array}{cc}
4\,\Sigma_{3}^2 + D_\eta^2 & 0 \\
0 & 4\,\Sigma_{2}^2 + D_\eta^2
\end{array}
\right)\,,
\end{equation}
with $D_\eta^2 = (1-\eta)/\eta$ \cite{cond:cola}.
\subsection{Optimized teleportation in the presence of noise}
\index{teleportation!optimization}
In order to use CVQT as a resource for quantum information processing,
we look for a class of squeezed states which achieves an average
teleportation fidelity greater than the one obtained teleporting coherent
states in the same conditions. If the input squeezed
state to be teleported is $\sigma = \ket{\alpha,\xi} \bra{\alpha,\xi}$,
$\ket{\alpha,\xi}=D(\alpha)S(\xi) |0\rangle$, then the teleported
state given by Eq.~(\ref{wig:output}) has 
average teleportation fidelity [see Eq.~(\ref{wigner:fid})]
\begin{equation}
\overline{F}_{\xi}(\lambda)
= \Big( \sqrt{(e^{2\xi} + 4\: \Sigma_{2}^2)
(e^{-2\xi} + 4\: \Sigma_{3}^2)} \Big)^{-1}\,,
\label{squeezed:fid:gen}
\end{equation}
which attains its maximum
\begin{eqnarray}\label{squeezed:fid:max}
\overline{F}(\lambda) =(1+4\:\Sigma_2\:\Sigma_3)^{-1}\,,
\end{eqnarray}
when
\begin{eqnarray}\label{zeta:max}
\xi = \xi_{\rm max} \equiv
\frac12 \ln \left( \frac{\Sigma_2}{\Sigma_3} \right)\,.
\end{eqnarray}
\par
For non squeezed environment $N_{\rm s} \rightarrow 0$ we 
have $\Sigma_2 = \Sigma_3$, and thus then $\xi_{\rm max}\rightarrow 0$, 
{\em i.e.} the input state that maximizes the average fidelity 
(\ref{squeezed:fid:gen}) reduces to a coherent state. In other words, 
in a non squeezed environment the teleportation of coherent states 
is more effective than that of squeezed states. Moreover, Eq.~(\ref{squeezed:fid:max})
shows that meanwhile the TWB becomes
separable, {\em i.e.} $\Sigma_2^2\:\Sigma_3^2 \ge 16^{-1}$
[see Eqs.~(\ref{20})], one has $\overline{F}\leq
0.5$. Finally, the asymptotic
value of $\overline{F}$ for $\Gamma t \rightarrow
\infty$ is
\begin{eqnarray}
\overline{F}^{(\infty)} = [2\: (1 + N_{\rm th})]^{-1}\,,
\end{eqnarray}
which does not depend on the number of squeezed photons and is
equal to $0.5$ only if $N_{\rm th} = 0$. This last result is
equivalent to say that in the presence of a zero-temperature
environment, no matter if it is squeezed or not, the TWB is
non-separable at every time.
\begin{figure}[tb]
\begin{center}
\includegraphics[width=0.8\textwidth]{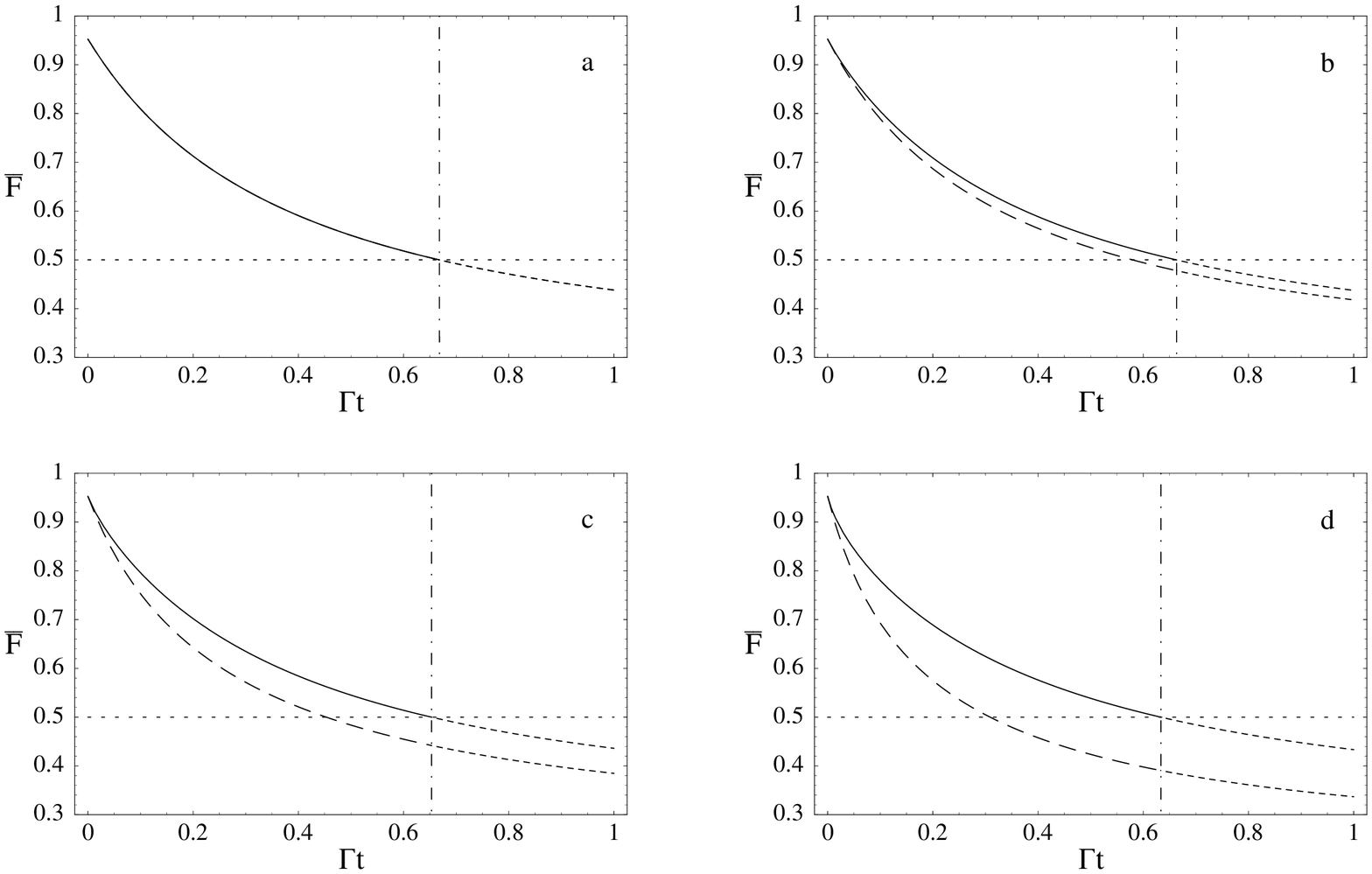}
\caption{Plots of the average teleportation fidelity. The solid and the
dashed lines represent squeezed and coherent state fidelity, respectively,
for different values of the number of squeezed photons $N_{\rm s}$: (a)
$N_{\rm s} = 0$, (b) $0.1$, (c) $0.3$, (d) $0.7$. In all the plots we put
the TWB parameter $\lambda = 1.5$ and number of thermal photons $N_{\rm th}
= 0.5$.  The dot-dashed vertical line indicates the threshold $\Gamma
t_{\rm s}$ for the separability of the shared state: when $\Gamma t >
\Gamma t_{\rm s}$ the state is no more entangled. Notice that, in the case
of squeezed state teleportation, the threshold for the separability
corresponds to $\overline{F} = 0.5$.}\label{f:fidelity:noise}
\end{center}
\end{figure}
In Fig.~\ref{f:fidelity:noise} we plot $\overline{F}_{\rm tele}$ as a function
of $\Gamma t$ for different values of $\lambda$, $N_{\rm th}$ and $N_{\rm
s}$.  As $N_{\rm s}$ increases, the nonclassicality of the thermal bath
starts to affect the teleportation fidelity and we observe that the
best results are obtained when the state to be teleported is the squeezed
state that maximizes (\ref{squeezed:fid:gen}). Furthermore the difference
between the two fidelities increases as $N_{\rm s}$ increases.  Notice that
there is an interval of values for $\Gamma t$ such that the coherent state
teleportation fidelity is less than the classical limit $0.5$, although
the shared state is still entangled.
\subsection{Teleportation improvement}
\index{degaussification!teleportation improvement}
TWBs are produced either by degenerate (with additional beam splitters) or
nondegenerate optical parametric amplifiers. The TWB parameter $\lambda=\tanh
r$ depends on the physical parameters as $r \propto \chi^{(2)}L$, $\chi^{(2)}$
being the nonlinear susceptibility of the crystal used as amplifying medium
and $L$ the effective interaction length. For a given amplifier, the TWB
parameter and thus the amount of entanglement are fixed. Therefore, since
nonlinearities are small, and the crystal length cannot be increase at will,
it is of interest to devise suitable quantum operations to increase
entanglement and in turn to improve teleportation fidelity.
\par
In Section \ref{ss:twbNL} we have seen that the nonlocal correlations of 
TWB are enhanced for small energies by means of the IPS process described 
in Section \ref{s:degauss}: motivated by this
result, we will use the IPS state (\ref{ips:wigner}) or, equivalently,
(\ref{ips:fock}) as shared entangled state between Alice and Bob. In this
case, Eqs.~(\ref{grp:fidelity}) lead to the following expression for the
average teleportation fidelity of coherent states\cite{ips:tele}
\begin{eqnarray}\label{ave:summed}
  \overline{F}(\lambda,\tau_{\rm eff}) = \frac12
  \frac{(1+\lambda)(1+\lambda \tau_{\rm eff})(1-\lambda^2\tau_{\rm
  eff})[2-2\lambda \tau_{\rm eff} + \lambda^2
  \tau_{\rm eff}]}{(1+\lambda^2
  \tau_{\rm eff})[1+(1-\tau_{\rm eff})\lambda]\{ 2 - [2+(1-\tau_{\rm eff})
  \lambda]\lambda \tau_{\rm eff} \}}\:,
\end{eqnarray}
where $\tau_{\rm eff} = 1 - \eta (1-\tau)$ (see Section \ref{s:degauss}).
\begin{figure}[tb]
\begin{center}
\includegraphics[width=0.45\textwidth]{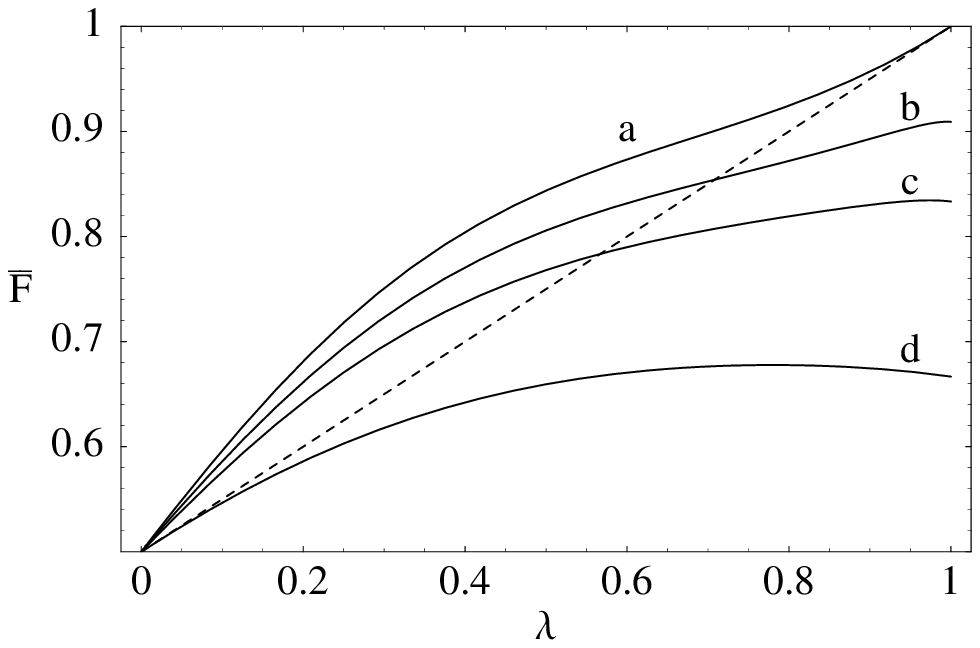}
\includegraphics[width=0.45\textwidth]{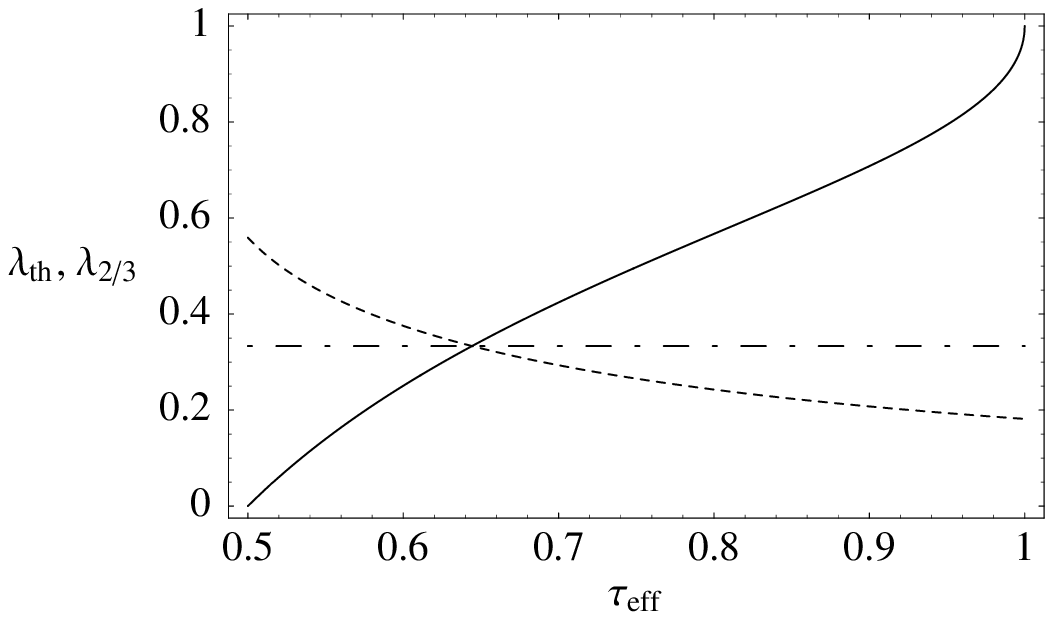}
\end{center}\vspace{-.5cm}
\caption{ On the left: IPS average fidelity $\overline{F}(\lambda,\tau_{\rm
eff})$ as a function of the TWB parameter $\lambda$ for different values of
$\tau_{\rm eff} = 1 - \eta(1-\tau)$: (a) $\tau_{\rm eff}=1$, (b) $0.9$, (c)
$0.8$, and (d) $0.5$; the dashed line is the average fidelity
$\overline{F}_{\tinyTWB}(\lambda)$ for teleportation with TWB. On the
right: Threshold value $\lambda_{\rm th}(\tau_{\rm eff})$ on the TWB
parameter $x$ (solid line): when $\lambda < \lambda_{\rm th}$ we have
$\overline{F}(\lambda,\tau_{\rm eff}) > \overline{F}_{\tinyTWB}(\lambda)$
and teleportation is improved. The dot-dashed line is $\lambda=1/3$, which
corresponds to $\overline{F}_{\tinyTWB}=2/3$: when fidelity is greater than
2/3 Bob is sure that his teleported state is the best existing copy of the
initial state \cite{gross}.  The dashed line represents the values
$\lambda_{2/3}(\tau_{\rm eff})$ giving an average fidelity
$\overline{F}(\lambda,\tau_{\rm eff})=2/3$.  When
$\lambda_{2/3}<\lambda<\lambda_{\rm th}$ both the teleportation is improved
and the fidelity is greater than $2/3$.}
\label{f:2figs}
\end{figure}
In Fig.~\ref{f:2figs} we plot the average fidelity for
different values of $\tau_{\rm eff}$: the IPS state improves the
average fidelity of quantum teleportation when the energy of the
incoming TWB is below a certain threshold, which depends on
$\tau_{\rm eff}$ and, in turn, on $\tau$ and $\eta$.
When $\tau_{\rm eff}$ approaches unit
(when $\eta \rightarrow 1$ and $\tau \rightarrow 1$),
Eq.~(\ref{ave:summed}) reduces to the result obtained by Milburn
\emph{et al.} in Ref.~\cite{coch} and the IPS average fidelity (line
labeled with ``a'' in Fig.~\ref{f:fidelity:noise}) is always greater
than the fidelity $\overline{F}_{\hbox{\tiny TWB}}(\lambda)$ obtained
with the TWB state [see Eq.~(\ref{ave:fid:TWB})].
However, a threshold value, $\lambda_{\rm th}(\tau_{\rm eff})$, for the
TWB parameter $\lambda$ appears when $\tau_{\rm eff} < 1$: only if
$\lambda$ is below this threshold the teleportation is actually improved
[$\overline{F}(\lambda,\tau_{\rm eff}) > \overline{F}_{\tinyTWB}(\lambda)$], as
shown in Fig.~\ref{f:2figs}. Notice that, for $\tau_{\rm eff}
< 0.5$, $\overline{F}(\lambda,\tau_{\rm eff})$ is always below
$\overline{F}_{\tinyTWB}(\lambda)$.
\par
Ralph \emph{et al.} demonstrated that entanglement is needed to achieve a
fidelity greater than $1/2$ \cite{ralph} and, using both the TWB
and the IPS state (\ref{ips:fock}), this limit is always reached
(see Fig.~\ref{f:2figs}). Nevertheless, we remember that in teleportation
protocol the state to be teleported is destroyed during the measurement
process performed by Alice, so that the only remaining \emph{copy} is that
obtained by Bob. When the initial state carries reserved information, it is
important that the only existing copy  will be the Bob's one. On the other
hand, using the usual teleportation scheme, Bob cannot avoid the presence
of an eavesdropper, which can clone the state, obviously introducing some
error \cite{cerf}, but he is able to to verify if his state was duplicated.
This is possible by the analysis of the average teleportation fidelity:
when fidelity is greater than $2/3$, Bob is sure that his state was not
cloned \cite{cerf,gross}. The dashed line in Fig.~\ref{f:2figs} (right) 
shows the values $\lambda_{2/3}(\tau_{\rm eff})$ which give an average fidelity
(\ref{ave:summed}) equal to $2/3$: notice that when $\lambda_{2/3} <
\lambda < \lambda_{\rm th}$ both the teleportation is improved and the the
fidelity is greater than $2/3$. Moreover, while the condition
$\overline{F}_{\tinyTWB}(\lambda) > 2/3$ is satisfied only if $\lambda >
1/3$, for the IPS state there exists a $\tau_{\rm eff}$-dependent interval
of $\lambda$ values ($\lambda_{\rm 2/3} < \lambda < 1/3$) for which
teleportation can be considered \emph{secure}
[$\overline{F}(\lambda,\tau_{\rm eff}) > 2/3$].
\section{Quantum cloning}\label{c7:Clo}
\index{cloning}
A fundamental difference between classical and quantum information is
that the latter cannot be perfectly copied, even in principle. This
means that there exist no physical process that can produce perfect
copies of generic unknown quantum states. This so called {\em
  no-cloning} theorem emerges as an immediate consequence of the
linearity of quantum dynamics \cite{Die82,WZ82}. Remarkably, if
cloning was permitted, the Heisenberg uncertainty principle would
be violated by measuring conjugate observables on many copies of a
single quantum state. Nevertheless, even if perfect cloning is not
possible, one may attempt to attain imperfect copies of generic
unknown quantum states. With an abuse of language, this is what is
generally referred to as {\em cloning} process. With an $n$ to $m$
cloning process it is thus meant that $m$ imperfect copies are
produced from $n$ identical original states ($m>n$).  In this
section we address the cloning issue for Gaussian states, first
recalling the bounds that no-cloning theorem imposes in this case,
then investigating some local and nonlocal (telecloning) cloning
protocols.
\subsection{Optimal universal cloning} \label{c7:OptClo}
The first concept to introduce is the {\em universality} of a cloning
machine \cite{BH96}. By universality we mean that the quality of the clones
should be independent on the original states. As we have already seen
in Section \ref{s:fid:tele}, we may use the fidelity as a measure of
the similarity between two states, hence in a universal cloning machine
every clone has the same fidelity with respect to
the original state, independently of the original state itself.
Furthermore, when the clones are equal one each other then we deal
with {\em symmetric cloning}, while if we admit differences in the
copies we have {\em asymmetric cloning}. Consider
for the moment the first scenario. Universal cloning machines have
been extensively studied for the case of discrete variables, for which
it has been shown that the optimal $n$ to $m$ universal cloning
machine of $d$-dimensional systems yields the fidelity \cite{Wer98}:
\be
\label{c7:QditFid} F=\frac{n(d-1)+m(n+1)}{m(n+d)}\,.
\ee 
\par
In the limit of infinite dimensional systems one can show that the
optimal universal cloner reduces to a classical probability
distributor, attaining $F=n/m$ \cite{BBH01}, consistent with the
$d\rightarrow \infty$ limit of \refeq{c7:QditFid}. This means that a
universal continuous variable cloner behaves like a simple classical
device that distributes the $n$ original input states into $n$ output
states, chosen by chance between the possible $m$ outputs and
disregarding the remaining states. Nevertheless, a more interesting
situation occur if we restrict the input states to the class of
Gaussian states. Consider for the moment $n$ identical arbitrary
coherent states. The imperfection of the $m$ copies may be regarded as
an excess noise variance $\sigma_{n,m}^2$ in the quadratures, due to
the $n$ to $m$ cloning process. Then, a procedure similar to what was
done for qubits \cite{BEM98} allows one to estimate a lower bound
$\overline{\sigma}_{n,m}^2$ on the noise variance \cite{CI00}. In
fact, make use of the property that cascading two cloning processes
results in a single cloning process whose excess noise variance is
simply the sum of the variances of the two cloning.  Then, the
variance $\overline{\sigma}_{n,l}^2$ of an optimal $n$ to $l$ cloning
must satisfy $\overline{\sigma}_{n,l}^2\le
\sigma_{n,m}^2+\sigma_{m,l}^2$ ($n \le m \le l$). In particular, if
the $m$ to $l$ cloner is itself optimal and $l \rightarrow \infty$, we have
\be
\overline{\sigma}_{n,\infty}^2\le
\sigma_{n,m}^2+\overline{\sigma}_{m,\infty}^2\,.
\ee
\par
As a matter of
fact, a cloner that allows to build infinitely many copies corresponds
to an optimal measurement of the original states, hence, with the aid
of quantum estimation theory, one may identify
$\overline{\sigma}_{n,\infty}^2=1/n$ (we put $\kappa_1=2^{-1/2}$).  As
a consequence, the lower bound we were looking for is given by
\be
\label{c7:VarBound}
\overline{\sigma}_{n,m}^2=\frac{m-n}{m\,n}\,.
\ee
This result implies that the {\em optimal cloning fidelity} for coherent
states is bounded by \cite{CI00}
 \be
\label{c7:OptFid}
F_{n,m}=\frac{m\,n}{m\,n+m-n}\,.
\ee
Notice that this result does not depend on the amplitude of the input
coherent states. If general squeezed states are considered, optimality
can then be achieved only if the excess noise variance is squeezed by
the same amount as the initial state, thus making the cloner
state-dependent.
\par
\begin{figure}
\begin{center}
\vspace{-3mm}
\includegraphics[width=0.5\textwidth]{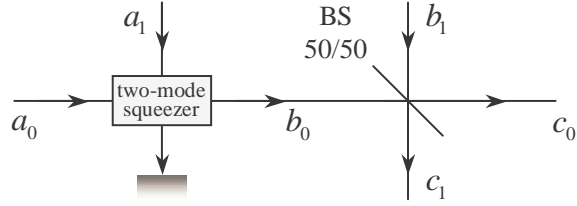}
\end{center}
\vspace{-0.7cm}
\caption{Scheme of local $1$ to $2$ cloning.}
\label{f7:LocalClo}
\end{figure}

Remarkably, optimal cloners achieving the fidelity given in
\refeq{c7:OptFid} may be implemented in an optical framework with the
aid of only a phase-insensitive two-mode squeezer and a sequence of
beam splitters \cite{BCI+01,Fiu01}. As an example, consider the $1$ to
$2$ cloner depicted in Fig.~\ref{f7:LocalClo}. Mode $a_0$, excited in the state
to be cloned, is sent to a two-mode squeezer with an ancillary mode
$a_1$. Using the notation introduced in Section \ref{ss:opa}, we have
that the output mode $b_0$ is given by $b_0=\mu a_0+\nu a_1$. Then a
linear mixing of modes $b_0$ and $b_1$ in a phase insensitive balanced
beam splitter give rise to
\be
c_0=\frac{1}{\sqrt2}(\mu a_0 + \nu a_1^\dagger + b_1)\,, \qquad
c_1=\frac{1}{\sqrt2}(\mu a_0 + \nu a_1^\dagger - b_1).  
\ee
Now, considering vacuum inputs for modes $a_1$ and $b_1$, and
squeezing parameters $\mu=\sqrt2$, $\nu=1$, it follows that $\langle
c_0 \rangle=\langle c_1 \rangle=\langle a_0 \rangle$. As a
consequence, the scheme considered allows to copy the amplitude of the
original mode $a_0$. Thus, if the latter is excited in a coherent
state two clones are produced at the output modes $c_0$ and $c_1$.
The optimality of the cloner follows from the fact that the two-mode
squeezing chosen yields an excess noise variance $\sigma^2=1/2$
\cite{Cav82}.  Finally, the generalization of this method allows to
realize an optimal $n$ to $m$ {\em local} cloning machine.
\subsection{Telecloning} \label{c7:TeleClo}
\index{telecloning}
\index{coherent states!telecloning}
The cloning process described above, even if local, may be applied in
order to distribute quantum information among many distant parties, in
what is called a {\em quantum information network}.  Suppose that one
wants to distribute the information stored into $n$ states to $m$
receivers. This may be achieved by two steps.  One may first produce
locally $m$ copies of the original states by means of the cloning
protocol presented above. Then, the teleportation of each copy,
following the scheme described in Section \ref{s:CVQT}, allows
to attain the transfer of information \cite{MJP+99}. This strategy has
the obvious advantage to use only bipartite entangled sources.
However, even in the absence of losses, it does not leave the
receivers with $m$ optimum clones of the original states, due to the
non-unitary fidelity of the teleportation protocol in case of finite
energy.  This problem may be circumvented by pursuing a one-step
strategy consisted of a {\em nonlocal} cloning. By this we mean that
the cloning process is supported by a multipartite ($m+n$) entangled
state which is distributed among all the parties involved. This so
called {\em telecloning} process is thus nonlocal in the sense that it
proceed along the lines of a natural generalization of the teleportation
protocol to the many-recipient case \cite{MJP+99}. To clarify this
second scenario, let us describe now in details a $1$ to $2$
telecloning process based on the tripartite state $|T\rangle$
introduced in \refeq{T} \cite{FPB+04}.
\par
\begin{figure}[tb]
\begin{center}
\includegraphics[width=0.6\textwidth]{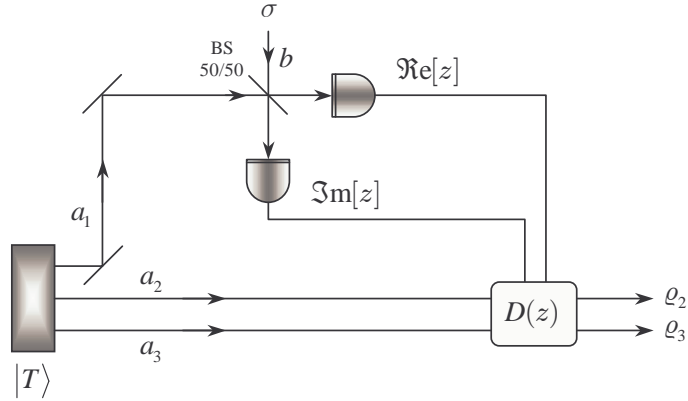} 
\vspace{-.4cm}
\caption{Schematic diagram of the $1\rightarrow 2$ telecloning scheme 
described in the text.}
\label{f7:telfig}
\end{center}
\end{figure}
A schematic diagram of the telecloning process is depicted in
Fig.~\ref{f7:telfig}.  After the preparation of the state $|T\rangle$, a
joint measurement is made on the mode $a_1$ and the mode $b$ to be
telecloned, which corresponds to the measure of the complex photocurrent $Z
= b + a_1^\dag$, as in the case of the
teleportation protocol.
The whole measurement is described by the POVM
(\ref{2pom:Pi1}), acting on the mode $a_1$, namely $\Pi (z) =
\pi^{-1} D(z)\: \sigma^{\sT} D^\dag (z)$, where $\sigma$ is the
the state to be teleported and cloned.
The probability distribution of the outcomes is given by
\begin{align}
P (z) &= \hbox{Tr}_{123} \left[|T\rangle\langle T|\:
\Pi(z) \otimes
\iid_2 \otimes \iid_3\right] \nonumber \\
&= \frac{1}{\pi(1+N_1)} \sum_{pq} \frac{N_2^p
N_3^q}{(1+N_1)^{p+q}} \frac{(p+q)!}{p!\: q!} \: \langle
p+q|D(z)\, \sigma^{\sT} D^\dag (z) |p+q\rangle \label{Palfa}\;.
\end{align}
The conditional state of the mode $a_2$ and $a_3$ after the
outcome $z$ is given by
\begin{align}
\varrho_z &= \frac{1}{P(z)}\: \hbox{Tr}_{1}
\left[|T\rangle\langle T|\:
\Pi(z) \otimes \iid_2 \otimes \iid_3\right] \nonumber \\
&= \frac{1}{P(z)}\frac{1}{\pi(1+N_1)} \sum_{p,q}\sum_{k,l}
\sqrt{\frac{N_2^{p+k}\, N_3^{q+l}}{(1+N_1)^{p+q+k+l}}}\,
\sqrt{
\frac{(p+q)!}{p!\,q!}
\frac{(k+l)!}{k!\,l!}}\,
\nonumber\\
&\hspace{4.5cm}\times
\langle k+l|D(z)\, \sigma^{\sT} D^\dag (z) |p+q\rangle
\: |p,q\rangle\langle k,l| \label{Rhoalfa}\;.
\end{align}
After the measurement, the conditional state should be transformed by
a further unitary operation, depending on the outcome of the
measurement. In our case, this is a two-mode product displacement
$U_z = D_2^{\sT}(z)\otimes D_3^{\sT}(z)$. This is a
local transformation which generalizes to two modes the procedure
already used in the original CVQT
protocol described in Section \ref{s:CVQT}. The overall
state of the two modes is obtained by averaging over the possible
outcomes
$$
\varrho_{23}=\int_{\mathbb C} d^2 z\: P (z) \:
\tau_z\:.$$ where $\tau_z=U_z\: \varrho_z\:
U_z^\dag$.
\par
If $b$ is excited in a coherent state $\sigma=|\alpha\rangle\langle
\alpha|$, then the probability of the outcomes is given by
\begin{equation}
P_\alpha(z) = \frac{1}{\pi(1+N_1)} \: \exp
\left\{-\frac{|z+\alpha^*|^2}{1+N_1}\right\}
\label{PalfaCoh}\;.
\end{equation}
Moreover, since the POVM is pure also the conditional state is
pure. Is this way we have that $\varrho_z=
|\psi_z\rangle\rangle\langle\langle \psi_z|$ is the product of two states,
namely
\begin{equation}
|\psi_z\rangle\rangle =
| (\alpha + z^*)\,\varepsilon_2 \rangle \otimes
|  (\alpha + z^*)\,\varepsilon_3\rangle \label{PsiBeta}\;,
\end{equation}
where
\begin{equation}
\varepsilon_h = \sqrt{\frac{N_h}{1+N_1}}\qquad (h=2,3) \label{kappas}\;.
\end{equation}
Correspondingly, we have $\tau_z=U_z\:|\psi_z\rangle\rangle\langle\langle
\psi_z| \: U_z^\dag $ with
\begin{equation}
U_z\:|\psi_z\rangle\rangle=
|\alpha\varepsilon_2+z^*\,(\varepsilon_2-1)\rangle \otimes
|\alpha\varepsilon_3+z^*\,(\varepsilon_3-1)\rangle \:.
\label{out33}
\end{equation}
The partial traces
$\varrho_2=\hbox{Tr}_3[\varrho_{23}]$ and
$\varrho_3=\hbox{Tr}_2[\varrho_{23}]$ read as follows
\begin{eqnarray}
\varrho_h = \int_{\mathbb C} d^2z\: P_\alpha (z) \:
|\alpha\varepsilon_h+z^*\,(\varepsilon_h-1)\rangle\langle
\alpha\varepsilon_h+z^*\,(\varepsilon_h-1) | \label{clones}\;.
\end{eqnarray}
From the teleported states in (\ref{clones}) we see that, depending on
the values of the coupling constants of the Hamiltonian (\ref{intH})
the two clones can either be equal one to each other or be different.
In other words, a remarkable feature of this scheme is that it is
suitable to realize both symmetric, when $N_2=N_3=N$, and
asymmetric cloning, $N_2\neq N_3$. This arise as a consequence of the
possible asymmetry of the state that supports the teleportation.
\par
Let us first consider the symmetric cloning. According to Eq.~(\ref{c3:Ndit})
the condition $N_2=N_3=N$ holds when
\begin{equation}
\cos(\Omega t) = \frac{|\gamma_1|^2}{2|\gamma_2|^2-|\gamma_1|^2}
\;, \qquad
N =
\frac{4|\gamma_1|^2|\gamma_2|^2}{(2|\gamma_2|^2-|\gamma_1|^2)^2}
\label{Nequal}\;.
\end{equation}
Since
$|\langle z^{'}| z^{''}\rangle|^2=\exp\{-| z^{'}- z^{''}|^2\}$,
the fidelity of the clones is given by (we put $\varepsilon_2=\varepsilon_3
=\varepsilon$)
\begin{align}
F &= \int_{\mathbb C}
\frac{d^2z}{\pi (2N+1)}\:
\exp\left\{-\frac{|\alpha+z^*|^2}{2N+1}\right\}\:
\exp\left\{-|\alpha+z^*|^2 (\varepsilon-1)^2\right\} \nonumber \\
&= \left(2 + 3N - 2\sqrt{N(2N+1)}\right)^{-1} \label{fid}\;.
\end{align}
As we expect from a proper cloning machine, the fidelity is
independent of the amplitude of the initial signal and for $0<N<4$ it
is larger than the classical limit $F=1/2$. Notice that the
transformation $U_z$ performed after the conditional measurement,
is the only one assuring that the output fidelity is independent of
the amplitude of the initial state. Exploiting \refeq{fid} we can see
that the fidelity reaches its maximum $F=2/3$ for $N=1/2$ which means,
according to Eq.~(\ref{Nequal}), that the physical system allows an
optimal cloning when its coupling constants are chosen in such a way that
$|\gamma_1/\gamma_2|=(6-\sqrt{32})^{1/2}\simeq 0.586$ . The total mean
photon number required to reach the optimal telecloning is thus
$N_1+N_2+N_3=2$, hence, as we claimed above, it can be achieved
without the need of infinite energy. The scheme presented is analog to
that of Ref.~\cite{LB01} in the absence of an amplification process
for the signal. There, the telecloning is supported by a state similar
to the one given in \refeq{c3:matC}, where at the input ports of the
tritter only two squeezed states are involved, the third mode being a
vacuum state. Both the protocols described here and in
Ref.~\cite{LB01} achieve the optimality relying on minimal energetic
resources, {\em i.e.} the total mean photon number is $2$ in both
cases. Notice that a generalization to realize a $1\rightarrow m$ telecloning 
machine can be realized upon the implementation of $\rmSU(p,1)$ Hamiltonian 
introduced in Section \ref{ss:Hpq}. In fact, having at disposal a 
$1+m$ multipartite entangled state of the form (\ref{Cp1}), it is
straightforward to show that a measurement of $Z$ on the mode to be 
telecloned and the sum-mode of (\ref{Cp1}), followed by a local 
multimode displacement operation provides optimal clones in the remaining
$m$ modes.
\par
Let us now consider the asymmetric case. For $N_2 \neq N_3$ the
fidelities of the two clones (\ref{clones}) are given by
\begin{equation}
F_h = \left(2 + N_h +2N_k - 2\sqrt{N_h(N_1+1)}\right)^{-1}
\label{FidAsym}\;,
\end{equation}
where $h,k=2,3$ ($h\neq k$). A question arises whether it is possible to
tune the coupling constants so as to obtain a fidelity larger than the
bound $F=2/3$ for one of the clones, say $\varrho_2$, while accepting a
decreased fidelity for the other clone.  In particular if we impose
$F_3=1/2$, {\em i.e.} the minimum value to assure the genuine quantum
nature of the telecloning protocol, we can maximize $F_2$ by varying the
value of the coupling constants $\gamma_1$ and $\gamma_2$. The maximum
value turns out to be $F_{2,{\rm max}}=4/5$ and it corresponds to the choice
$N_3=1/4$ and $N_2=1$. More generally one can fix $F_3$, then the maximum
value of $F_2$ is obtained choosing $N_2=(1/F_3-1)$ and $N_3= (4/F_3-4)^{-1}$.
The relation between the fidelities is then
\begin{equation}
F_2 = 4 \frac{(1-F_3)}{(4-3F_3)}\:, \label{AsymFid}
\end{equation}
which shows that $F_2$ is a decreasing function of $F_3$ and that
$2/3 <F_2< 4/5$ when $1/2 <F_3< 2/3$ . The sum of the two
fidelities $F_2+F_3=1+ 3F_2 F_3/4$ is maximized in the symmetric
case in which optimal fidelity $F_2=F_3=2/3$ can be reached. The
role of $\varrho_2$ and $\varrho_3$ can be exchanged, and the
above considerations still hold.

\chapter[State engineering]{State engineering}\label{ch:StEng}
\index{state engineering}
In this Chapter we analyze the use of conditional
measurements on entangled twin-beam state (TWB) of radiation 
to engineer quantum
states, {\em i.e.} to produce, manipulate, and transmit
nonclassical light. In particular, we will focus our attention on
realistic measurement schemes, feasible with current technology,
and will take into account imperfections of the apparata such
as quantum efficiency and finite resolution.
\par
The reason to choose TWB as {\em entangled resource} for
conditional measurements is twofold. On one hand, TWBs are the
natural generalization to continuous variable (CV) systems of Bell
states, {\em i.e.} maximally entangled states for qubit systems.
On the other hand TWBs are CV entangled states that can be reliably 
produced with current technology, either by parametric downconversion 
of the vacuum in a nondegenerate parametric amplifier \cite{kum0}, or 
by mixing two squeezed vacua from a couple of degenerate parametric 
amplifiers in a balanced beam splitter \cite{furu,joint}. 
\par
The first kind of measurement we analyze is {on/off}
photodetection, which provides the generation of
conditional {\em nonclassical mixtures}, which are not destroyed by
decoherence induced by noise and permits a robust test of the quantum
nature of light.
\par
The second apparatus is homodyne detection, which represents a tunable
source of squeezed light, with high conditional probability and
robustness to experimental imperfections, such non-unit quantum
efficiency and finite resolution.
\par
The third kind of measurement is that of the normal operator 
$Z=b+c^\dag$, $b$ and $c$ being two modes of the field, as described 
in Section \ref{s:two}. In our case one of the two modes is a beam 
of the TWB, whereas the second one, usually referred to as the probe 
of the measurement, is excited in a given reference state. This
approach allows to describe CV quantum teleportation as a
conditional measurement, and to easily evaluate the degrading
effects of finite amount of entanglement, decoherence due to
losses, and imperfect detection \cite{cond:cola}.
\section{Conditional quantum state engineering}
\label{ch:eng::main}
\begin{figure}[t]
\begin{center}
\includegraphics[width=.7\textwidth]{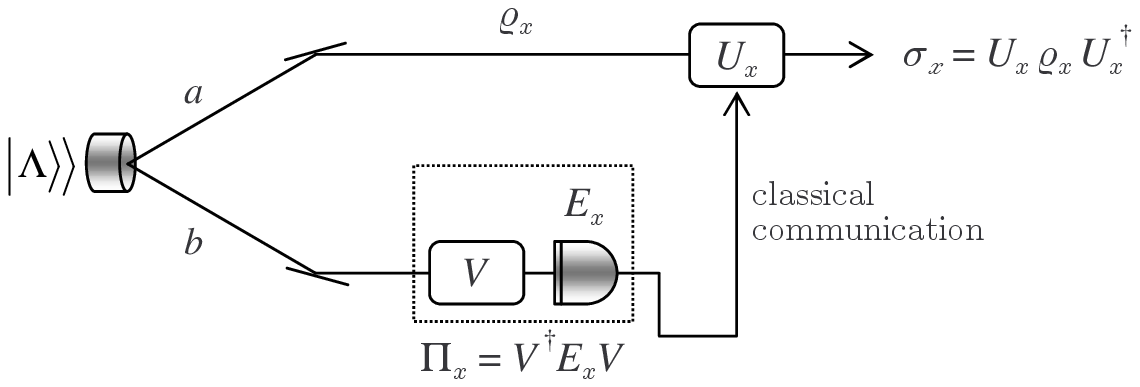}
\end{center}\vspace{-.5cm}
\caption{Scheme for quantum state engineering assisted by 
entanglement.}\label{f:eng:scheme}
\end{figure}
The general measurement scheme we are going to consider is schematically
depicted in Fig.~\ref{f:eng:scheme}.
The entangled state subjected to the conditional measurement is the TWB
$|\Lambda \rangle\rangle$, $\Lambda = \sqrt{1-\lambda^2} \lambda^{a^\dag
a}$, with $\lambda = \tanh r$ assumed as real. 
\index{conditional measurements}
A measurement, performed on one of the two modes, {\em reduces} the
other one accordingly to the projection postulate. Each possible
outcome $x$ of such a measurement occurs with probability $P_x$,
and corresponds to a conditional state $\sigma_x$ on the other subsystem
(Fig.~\ref{f:eng:scheme}). Upon denoting by $\Pi_x$ the POVM of the 
measurement\footnote{\footnotesize In this Chapter, in order to simplify 
notation, we denote the dependence of the element of the POVM 
$\left\{\Pi_x\right\}_{x \in {\cal X}}$ on the outcome $x$ as a 
subscript rather than on parenthesis as in Section 
\ref{s:pom}.} we have
\begin{align}
P_x &=  \hbox{Tr}_{ab} \Big[
|\Lambda\rangle\rangle\langle\langle \Lambda|\: \iid \otimes
\Pi_x\Big] 
(1-\lambda^2) \sum_{q=0}^{\infty} \: \lambda^{2q}\: \langle q|
\Pi_x | q\rangle = (1-\lambda^2)\,\hbox{Tr}_b\big[ \lambda^{2 b^\dag
b} \: \Pi_x^{\sT} \big]\,, \label{pygen}
\end{align}
and
\begin{align}
\varrho_x &= \frac{1}{P_x}\, \hbox{Tr}_b \Big[
|\Lambda\rangle\rangle\langle\langle \Lambda|\: \iid \otimes
\Pi_x\Big] 
\frac{1-\lambda^2}{P_x} \sum_{p,q} \: \lambda^{p+q} \:
\langle p | \Pi_x^{\sT} | q\rangle \: |p\rangle\langle q| =
\frac{\displaystyle
\lambda^{a^\dag a} \:\Pi_x^{\sT}\: \lambda^{a^\dag a}}{\displaystyle 
\hbox{Tr}_b \big[
\lambda^{2 b^\dag b} \: \Pi_x^{\sT} \big]}\,.\label{rhygen}
\end{align}
Notice that in the second line of Eq.~(\ref{rhygen})
$\Pi_x$ should be meant as an operator acting on
the Hilbert space ${\cal H}_a$ of the mode $a$.
Our scheme is general enough to include the
possibility of performing any unitary operation on the beam
subjected to the measurement. In fact, if $E_x$ is the original
POVM and $V$ the unitary, the overall measurement process is
described by $\Pi_x=V^\dag E_x V$, which is again a POVM. In the following we
always consider $V=\iid$, {\em i.e.} no transformation before the measurement.
A further generalization consists in sending the result of the
measurement (by classical communication) to the reduced state
location and then performing a conditional unitary operation $U_x$
on the conditional state, eventually leading to the state
$\sigma_x = U_x \varrho_x U_x^\dag$. This degree of freedom will
be used in Section \ref{ss:heter}, where we re-analyze CV quantum
teleportation as a conditional measurement.
\subsection{On/off photodetection}
\label{ss:eng:onoff}
\index{photodetection!on/off}
By looking at the expression of TWB in the Fock
basis, 
$\dket{\Lambda} = \sqrt{1-\lambda^2}\sum_{q} \lambda^q
\ket{q}\ket{q}$,
or at Eq.~(\ref{rhygen}) 
it is apparent that ideal photocounting on one of the two
beams, described by the POVM $\Pi_k=|k\rangle\langle k|$, is a
conditional source of Fock number state $|k\rangle$, which would
be produced with a conditional probability $P_k=(1-\lambda^2)
\lambda^k$. However, realistic photocounting can be very challenging 
experimentally, therefore we 
consider the situation in which one of the two beams, say mode
$b$, is revealed by an avalanche {on/off} photodetector (see Section
\ref{ss:onoff}). The action of
an {on/off} detector is described by the two-value POVM $\{\Pi_0(\eta),
\Pi_1(\eta)\}$ given in Eq.~(\ref{onoffPOM}). The outcome ``$1$''
({\em i.e.} registering a ``click'' corresponding to one or more incoming
photons) occurs with probability
\begin{align}
P_1 = \langle\langle\Lambda | \iid \otimes  \Pi_1(\eta)
| \Lambda\rangle\rangle = \frac{\eta\, \lambda^2}{1-\lambda^2 (1-\eta)}
 = \frac{\eta\, \ntwb}{2+\eta\, \ntwb}\label{probs}\;,
\end{align}
with $\ntwb = 2 \lambda^2 / (1-\lambda^2)$,
and correspondingly, the conditional output state for
the mode $a$ is given by \cite{robust}
\begin{eqnarray}
\varrho_1 = \frac{1-\lambda^2}{P_1}
\sum_{k=1}^\infty \lambda^{2k} \left[1-(1-\eta)^k\right]\: |k
\rangle\langle k|\label{fock}\;.
\end{eqnarray}
The density matrix in Eq.~(\ref{fock}) describes a mixture: a {\em
pseudo}-thermal state where the vacuum component has been removed
by the conditional measurement. Such a state is highly
nonclassical, as also discussed in Ref.~\cite{mandel}. Notice that
the nonclassicality is present only when the state exiting the
amplifier is entangled.  In the limit of low TWB energy the
conditional state $\varrho_1$ approaches the number state
$|1\rangle\langle 1|$ with one photon.
\par
The Wigner function $W[\varrho_1](\alpha)$ of $\varrho_1$ 
exhibits negative values for any value of $\lambda$ and $\eta$.
In particular, in the origin of the phase space we have
\begin{eqnarray}
W[\varrho_1](0) =-\frac{2}{\pi}\:\frac{1}{1 + \ntwb}\:\frac{2 +
\eta\,\ntwb}{2(1+\ntwb) - \eta\,\ntwb}
\label{wig0}\;.  \end{eqnarray}
One can see that also the generalized Wigner function for $s$-ordering
$$
W_s[\varrho_1](\alpha) = -\frac2{\pi s}\int_{\cc} d^2 \gamma\:
W[\varrho_1](\gamma) \,\exp\left\{\:-\frac2s\:
|\alpha -\gamma|^2\right\}\:,
$$
shows negative values for $s \in (-1,0)$. In particular
one has
\begin{eqnarray}
W_s[\varrho_1](0) = -
\frac{2(1+s)(2 + \eta\,\ntwb)}{\pi(1+\ntwb-s)\left[2(1+\ntwb-s)-
\eta\, \ntwb (1+s)\right]}
\label{ws0}\;.
\end{eqnarray}
A good measure of nonclassicality is given by the lowest index
$s^\star$ for which $W_s$ is a well-behaved probability, {\em
i.e.} regular and positive definite \cite{Lee:PRA:91}. Eq.~(\ref{ws0})
says that for $\varrho_1$ we have $s^\star=-1$, that is
$\varrho_1$ describes a state as nonclassical as a Fock number
state.
\par
Since the Fano factor of $\varrho_1$ is given by
\begin{eqnarray}
F\equiv \frac{\left\langle[b^\dag b - \langle b^\dag b\rangle]^2\right\rangle}
{\langle b^\dag b\rangle}
=\frac{\left( 2 + \ntwb \right)}{2} \left[ 1 + {\frac{2}{2 +
\eta\,\ntwb}} - {\frac{4\,\left( 2 + \ntwb \right) }{4 +
\ntwb\,\left( 4 + \eta\,\ntwb
\right) }} \right] \label{fano}\;,
\end{eqnarray}
we have that the beam $b$ is always subPossonian 
for (at least) $\ntwb < 2$.  The verification of
nonclassicality can be performed, for any value of the gain, by
checking the negativity of the Wigner function through quantum
homodyne tomography \cite{robust}, and in the low gain regime,
also by verifying the subPoissonian character by measuring the
Fano factor via direct noise detection \cite{kum,garbo}.
\par
Note that besides quantum efficiency, {\em i.e.} lost photons, the
performance of a realistic photodetector may be degraded by the
presence of dark-counts, {\em i.e.} by ``clicks'' that do not correspond to
any incoming photon. In order to take into account both these
effects we should describe the detector by the POVM (\ref{gendark})
rather than (\ref{onoffPOM}). 
However, at optical frequencies the number of dark counts is
small and we are not going here to take into account this effect,
which have been analyzed in details in Ref.~\cite{robust}.
\subsection{Homodyne detection}
\index{homodyne detection}
\label{ss:homod}
In this Section we consider the kind of conditional state that can be
obtained by homodyne detection on one of the two beams of the TWB . We
will show that they are squeezed states. We first consider ideal homodyne
detection described by the POVM $\Pi_x=|x\rangle\langle x|$ where 
$|x\rangle$ denotes the eigenstate (\ref{eigenquad})
of the quadratures $x=\frac12(a+a^\dag)$ (throughout the Section we use
$\kappa_1=\kappa_2=1$) and where, without loss of generality, we have
chosen a zero reference phase (see Section \ref{s:hom} for details about
homodyne detection).  Then, in the second
part of the Section we will consider two kinds of imperfections: non-unit
quantum efficiency and finite resolution. As we will see, the main effect
of the conditional measurement, {\em i.e} the generation of squeezing,
holds also for these realistic scenarios.
\par
The probability of obtaining the outcome $x$ from a homodyne detection on
the mode $b$ is obtained from Eq.~(\ref{pygen}). We have
\begin{eqnarray}
P_{x} = (1-\lambda)^2 \sum_{q=0}^\infty \lambda^{2q}\:
\left|\langle x | q \rangle \right|^2
=\frac{1}{\sqrt{2\pi\sigma_{\lambda}^2}}\
\exp\left\{-\frac{x^2}{2\sigma_{\lambda}^2}\right\}
\label{pyeta1}\;,
\end{eqnarray}
where
\begin{eqnarray}
\sigma_{\lambda}^{2}= \frac14 \frac{1+\lambda^{2}}{1-\lambda^{2}}
= \frac14 (1+\ntwb)\:.
\end{eqnarray}
$P_x$ is Gaussian with variance that increases as $\lambda$ is approaching
unit. In the (unphysical) limit $\lambda \rightarrow 1$, {\em i.e.}
infinite gain of the amplifier, the distribution for $x$ is uniform over the
real axis. The conditional output state is given by Eq.~(\ref{rhygen}),
and, since $\Pi_x$ is a pure POVM, it is a pure state
$\varrho_x=|\psi_x\rangle\langle \psi_x|$ where
\begin{eqnarray}
|\psi_x\rangle = \sqrt{\frac{1-\lambda^2}{P_x}}\,
\lambda^{a^\dag a} \: |x\rangle = \sum_{k=0}^{\infty} \psi_k \,| k \rangle
\label{rheta1}\;.
\end{eqnarray}
The coefficients of $|\psi_x\rangle$ in the Fock basis are given by
\begin{eqnarray}
\psi_k = (1-\lambda^4)^{1/4}
\left(\frac{\lambda^2}{2}\right)^{k/2}
\frac{H_k (\sqrt{2}x)}{\sqrt{k!}}\,
\pexp{-\frac{2\lambda^2 x^2}{1+\lambda^2}}\,,
\end{eqnarray}
which means that $|\psi_x\rangle$ is a squeezed state of the form
\begin{eqnarray}
|\psi_x\rangle  =  D(\alpha_x) S(\zeta) |0\rangle \label{sqy}
\:,\end{eqnarray}
where
\begin{subequations}
\begin{align}
\alpha_x  &=   \frac{2x\lambda}{1+\lambda^2}
= \frac{x\sqrt{\ntwb(\ntwb+2)}}{1+\ntwb} \\
\zeta &= \tanh^{-1} (\lambda^2) = \tanh^{-1}
\left(\frac{\ntwb}{\ntwb+2}\right)\:,
\end{align}
\end{subequations}
and the quadrature fluctuations are given by
\begin{eqnarray}
\Delta x_a^2 = \frac14 \frac1{1+\ntwb}\,, \qquad
\Delta y_a^2 = \frac14 (1+\ntwb) \label{fluct}\:.
\end{eqnarray}
Notice that (i)
the amount of squeezing is independent on the outcome of the
measurement, which only influences the coherent amplitude; (ii)
according to Eq.~(\ref{pyeta1}) the most probable conditional
state is a  squeezed vacuum. The average number of photon of the
conditional state is given by
\begin{eqnarray}
N_x=\langle \psi_x| a^\dag a | \psi_x\rangle =
x^2\,\frac{\ntwb(2+\ntwb)}{(1+\ntwb)^2}
+\frac14\, \frac{\ntwb^2}{1+\ntwb}\:.
\end{eqnarray}
The conservation of energy may be explicitly checked by averaging over
the possible outcomes, namely
\begin{eqnarray}
\int_{\rr} dx \: P_x \: N_x = \frac14 \frac{\ntwb^2}{1+\ntwb} +
\sigma_\lambda^2\,
\frac{\ntwb(2+\ntwb)}{(1+\ntwb)^2} = \frac{1}{2}\,\ntwb \label{checkE}\:,
\end{eqnarray}
which correctly reproduces the number of photon pertaining each part
of the TWB.
\par
We now take into account the effects of non-unit quantum efficiency $\eta$
at the homodyne detector on the conditional state. We anticipate that
$\varrho_{x\eta}$ will be no longer pure states, and in particular they
will not be squeezed states of the form (\ref{sqy}). Nevertheless, the
conditional output states still exhibit squeezing, {\em i.e.} quadrature
fluctuations below the coherent level, for any value of the outcome $x$,
and for $\eta > 1/2$.
The POVM of a homodyne detector with quantum efficiency $\eta$ is given 
in Eq.~(\ref{OM}). Since the nonideal POVM is a Gaussian convolution 
of the ideal POVM, the main effect is that $\Pi_{x\eta}$ is no longer a 
pure orthogonal POVM. The probability $P_{x\eta}$ of obtaining the 
outcome $x$ is still a Gaussian with variance
\begin{equation}
\Delta^{2}_{\lambda\eta} = \sigma_{\lambda}^{2} + \delta_{\eta}^{2}\:,
\label{totdl}
\end{equation}
where $\delta_\eta^2$ is given in Eq. (\ref{Deltaeta}).
The conditional output state is again given by Eq.~(\ref{rhygen}).
After some algebra we get the matrix element in the Fock basis
\begin{multline}
\langle n | \varrho_{x\eta} | m \rangle = \frac{\left(1-\lambda^2\right)
\lambda^{n+m}}{\sqrt{2^{n+m}n!m!}}
\sqrt{\frac{\eta\,[2-\eta(1-\lambda^2)]}{1-\lambda^2}}
\exp \left\{-\frac{4 \eta^2 \lambda^2 x^2}
{1-\lambda^2  (1-2\eta)}\right\}\\  
\times\sum_{k=0}^{{\rm min}[m,n]}\!\!\!
2^{k}k {m \choose k} {n \choose
k}\,\sqrt{\eta ^{m+n-2k}}\, H_{m+n-2k}\left(\sqrt{2\eta}\:x \right)\:,
\end{multline}
where $H_n(x)$ is the $n$-th Hermite polynomials.
The quadrature fluctuations are now given by
\begin{equation}
\Delta x_a^2 = \frac14 \frac{1+\ntwb(1-\eta)}{1+\eta\, \ntwb}\,, \qquad
\Delta y_a^2 = \frac14 (1+\ntwb) \label{fleta}\:.
\end{equation}
As a matter of fact, $\Delta y_a^2$ is independent on $\eta$, whereas
$\Delta x_a^2$ increases for decreasing $\eta$. Therefore, the conditional
output $\varrho_{x\eta}$ is no longer a minimum uncertainty state. However,
for $\eta$ large enough we still observe squeezing in the direction
individuated by the measured quadrature. We have that the conditional
state is a general Gaussian state of the form (\ref{c3:rho:1m})
with an average number of thermal photons given by
\begin{eqnarray}
N_{\rm th} =
\frac12 \left\{\sqrt{\frac{(1+\ntwb)[1+\ntwb(1-\eta)]}{1+\eta\,\ntwb}} -
1\right\}
\label{ntheta}\;,
\end{eqnarray}
and with amplitude and squeezing parameters 
\begin{align}
\alpha_{x\eta} =
\frac{\eta\sqrt{\ntwb(\ntwb+2)}}{1+\eta\, \ntwb}\: x\,, \qquad 
\xi_{\eta} = \frac14 \ln \left[
\frac{(1+\ntwb)(1+\eta\, \ntwb)}{1+\ntwb(1-\eta)}
\right]
\label{rxeta}\;.
\end{align}
From Eqs.~(\ref{fleta}) and (\ref{rxeta}) we notice that $\varrho_{x\eta}$
shows squeezing if $\eta > 1/2$, independently on the actual value $x$
of the homodyne outcome.
\par
The outcome of homodyne detection is, in principle, continuously
distributed over the real axis. However, in practice, one has
always to discretize data, mostly because of finite experimental
resolution. The POVM describing homodyne detection with binned
data is given by
\begin{eqnarray}
\Pi_{x\eta}(\delta) = \frac1\delta \int_{x-\delta/2}^{x+\delta/2} dt \:
\Pi_{t\eta} \label{binnedPOVM}\;,
\end{eqnarray}
where $\Pi_{t\eta}$ is given in Eq.~(\ref{OM}), and $\delta$ is
the width of the bins. The probability distribution is now given by
\begin{align}
P_{x\eta}(\delta) &= \frac{1}{2\delta}
\left[ \hbox{Erf}\left(\frac{x+\mbfrac\delta}
{\sqrt{2\Delta^2_{\lambda\eta}}}\right)
-\hbox{Erf}\left(\frac{x-\mbfrac\delta}{\sqrt{2\Delta^2_{\lambda\eta}}}\right)
\right] \\
&=\frac{1}{\sqrt{2\pi\Delta^2_{\lambda\eta}}}\,\exp\left\{
-\frac{x^2}{2\Delta^2_{\lambda\eta}}\right\}\,
\left(1-\frac{x^2-\Delta^2_{\lambda\eta}}{24\Delta^2_{\lambda\eta}}
\:\delta^2 \right) + O(\delta^3)\label{probrefin}\;
\end{align}
where $\Delta_{\lambda\eta}^2$ is given in Eq.~(\ref{totdl}) and
$$
\hbox{Erf}(x)=\frac{2}{\sqrt{\pi}}\int_0^x dt\, e^{-t^2}
$$
denotes the error function. The conditional
state is modified accordingly. Concerning the quadrature
fluctuations of the conditional state we have, up to second order
in $\delta$,
\begin{eqnarray}
\Delta x_a^2(\delta) = \Delta x_a^2 +
 \frac{\delta^2}{12} \frac{\eta^2 \ntwb (2+\ntwb)}{(1+\eta\, \ntwb)^2}\,x^2
\label{flbin}\;,
\end{eqnarray}
which is below the coherent level for $\eta > 1/2$ and for
\begin{eqnarray}
|x| < x_\delta \equiv
\frac1\delta \sqrt{\frac{3(1+\eta\,\ntwb)(2\eta-1)}{\eta^2(\ntwb+2)}}
\label{xdel}\;.
\end{eqnarray}
Therefore, the effect of finite resolution is that the conditional
output is squeezed only for the subset $|x|< x_\delta$ of the possible
outcomes which, however, represents the range where the probability is
higher \cite{cond:cola}.
\subsection{Joint measurement of two-mode quadratures}\label{ss:heter}
\index{teleportation!conditional measurement}
In this Section we assume that mode $b$ is
subjected to the measurement of the the real and the imaginary
part of the complex operator $Z=b+c^\dag$, where $c$ is an
additional mode excited in a reference state $S$. 
As we have seen in Section \ref{s:two} this kind of 
measurement is described by the POVM 
\begin{equation}\label{POVM:tele}
{\Pi}_\alpha = \frac{1}{\pi}
{D}(\alpha) \, {S}^{\sT} \, {D}^{\dag}(\alpha)\:.
\end{equation}
The present scheme is equivalent to that
of CV teleportation, which, as pointed out in Section \ref{s:tele:ph:numb},
can be viewed as a conditional measurement, with the state to be teleported
playing the role of the reference state $S$ of the apparatus. In order to complete the
analogy we assume that the result of the measurement is
classically transmitted to the receiver's location, and that a
displacement operation $D^\dag(\alpha)$ is performed on the conditional
state $\varrho_\alpha$.  Eqs.~(\ref{pygen}) and (\ref{rhygen}) are
rewritten as follows
\begin{align}
p_\alpha &= (1-\lambda^2) \hbox{Tr}^{2} \left[
\lambda^{2 a^\dag a} \: \Pi_\alpha^{\sT}
\right] \\
\varrho_\alpha &= \frac{\displaystyle \lambda^{a^\dag a}\:\Pi_\alpha^{\sT}\:\lambda^{a^\dag
a}}{\displaystyle \hbox{Tr}^{2}\left[\lambda^{2 a^\dag a}\:\Pi_\alpha^{\sT}\right]} \\
\sigma_\alpha &= D^{\dag}(\alpha)\,\varrho_\alpha\, D (\alpha)
= \frac{D^\dag(\alpha)\lambda^{a^\dag a}\: D(\alpha)\, S\, D^\dag (\alpha )\:\lambda^{a^\dag
a} D(\alpha)}{\displaystyle\hbox{Tr}^{2}\left[\lambda^{2 a^\dag a}\:\Pi_\alpha^{\sT}\right]}
\label{condalpha}\;,
\end{align}
while the teleported state is the average over all the possible
outcomes, {\em i.e.}
\begin{equation}
\sigma = \int_{\cc} d^2\alpha \, p_\alpha \, \sigma_\alpha
 \label{teleported,} =  \int_{\cc} d^2\alpha\,
D^\dag (\alpha) \,
\langle\langle \Lambda | \iid \otimes \Pi_\alpha
|\Lambda \rangle\rangle\, D (\alpha)\;.
\end{equation}
After performing the partial trace, and some algebra, one has
\begin{eqnarray}
\sigma = \int_{\cc} \frac{d^2\alpha}{\pi \sigma_{-}^2}\,
\pexp{-\frac{|\alpha|^2}{\sigma_{-}^2}}\, D(\alpha)S D^{\dag} (\alpha)
\label{tel}\;,
\end{eqnarray}
where $\sigma_{-}^2=1+\ntwb-\sqrt{\ntwb(\ntwb+2)}$, {\em i.e.}
the result of Section \ref{s:CVQT}.
\newpage
$ $

\renewcommand{\chaptermark}[1]{\markboth{#1}{}}
\renewcommand{\sectionmark}[1]{\markright{\thesection\ \hspace{.2cm} #1}}
\addcontentsline{toc}{chapter}{{}Bibliography}
\chaptermark{Bibliography}

%
%
%
%
%
%
%
%
%
%

\newpage
$ $
\newpage
\addcontentsline{toc}{chapter}{{}Index}{}
\chaptermark{Index}
\input{Napoli04.ind}
\end{document}